\pdfoutput=1

\RequirePackage[l2tabu, orthodox]{nag}

\documentclass[11pt, a4paper, oneside]{Thesis} 
\graphicspath{{Figures/}} 

\hypersetup{urlcolor=blue, colorlinks=true, breaklinks=true} 

\usepackage[usenames,dvipsnames,svgnames,table]{xcolor}

\usepackage{afterpage} 

\usepackage{lscape} 

\usepackage{pdflscape} 

\usepackage{wrapfig} 

\usepackage[T1]{fontenc}

\usepackage[utf8]{inputenc}

\usepackage{csquotes}

\usepackage[english]{babel}

\usepackage[version=4]{mhchem}

\usepackage{tikz}
\usetikzlibrary{matrix}

\usepackage{siunitx}

\usepackage{microtype}

\usepackage{rotating}

\usepackage{multirow}

\usepackage{bookmark}

\usepackage{physics}

\usepackage[makeroom]{cancel}

\usepackage{notoccite}

\usepackage[numbers,super,comma,sort&compress]{natbib}				
\newcommand{\Ham}{\mathcal{H}}
\newcommand{\Fcal}{\mathcal{F}}

\newcommand{\diff}{\,d}

\newcommand{\vect}[1]{\boldsymbol{\mathbf{#1}}}
\newcommand{\rvec}{\vect{r}}
\newcommand{\kvec}{\vect{k}}
\newcommand{\pvec}{\vect{p}}
\newcommand{\f}{\hat}

\title{\ttitle} 

\begin{document}

\frontmatter 

\setstretch{1.3} 

\fancyhead{} 
\rhead{\thepage} 
\lhead{} 

\pagestyle{fancy} 

\newcommand{\HRule}{\rule{\linewidth}{0.5mm}} 

\hypersetup{pdftitle={\ttitle}}
\hypersetup{pdfsubject=\subjectname}
\hypersetup{pdfauthor=\authornames}
\hypersetup{pdfkeywords=\keywordnames}


\begin{titlepage}
\begin{center}
\begin{figure}[h!]
\centering
\includegraphics[width=0.5\textwidth]{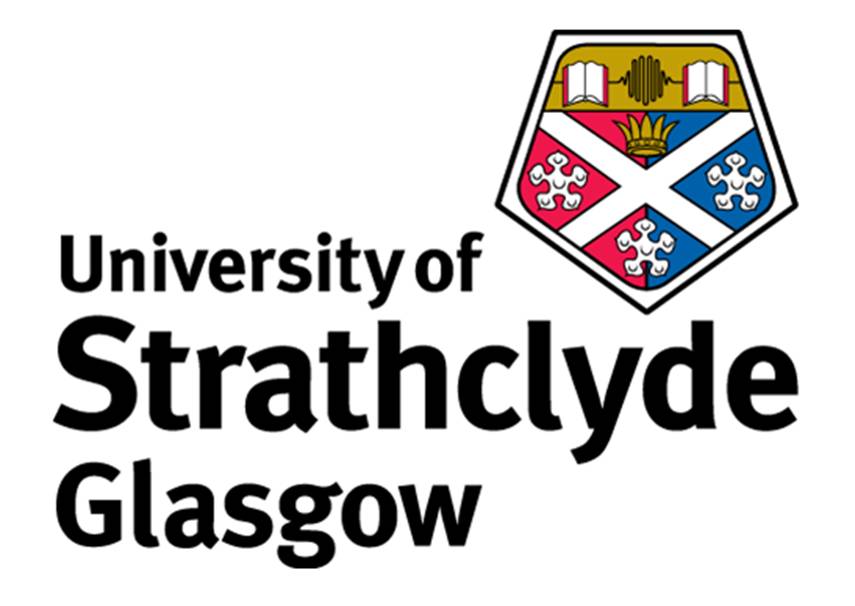}
\end{figure}
 
\textsc{\Large PhD Thesis}\\[0.5cm] 

\HRule \\[0.4cm] 
{\huge \bfseries \ttitle}\\[0.4cm] %

\HRule \\[1.5cm] 
 
\begin{minipage}{0.4\textwidth}
\begin{flushleft} \large
\emph{Author:}\\
{\authornames} 
\end{flushleft}
\end{minipage}
\begin{minipage}{0.4\textwidth}
\begin{flushright} \large
\emph{Supervisor:} \\
{\supname} 
\end{flushright}
\end{minipage}\\[3cm]
 
\large \textit{A thesis submitted to the University of Strathclyde\\ in partial fulfilment of the requirements\\ for the degree of\\ \degreename}\\[0.3cm] 
\textit{}\\[0.4cm]
\deptname\\[2cm] 
 
{\large \today}\\[4cm] 

\vfill
\end{center}

\end{titlepage}


\Declaration{

\addtocontents{toc}{\vspace{1em}} 

This thesis is the result of the author’s original research. It has been composed by the author and has not been previously submitted for examination which has led to award of a degree.

The copyright of this thesis belongs to the author under the terms of the United Kingdom Copyright Acts as qualified by University of Strathclyde Regulation 3.50. Due acknowledgement must always be made of the use of any material contained in, or derived from, this thesis.
\vspace{10cm}

Signed:

Date: \today
}

\clearpage 


\addtotoc{Abstract} 

\abstract{\addtocontents{toc}{\vspace{1em}} 
The thesis focuses on the prediction of solvation thermodynamics using integral equation theories. 
Our main goal is to improve the approach using a rational correction.
We achieve it by extending recently introduced pressure correction, and rationalizing it in the context of solvation entropy.
The improved model (to which we refer as advanced pressure correction) is rather universal.
It can accurately predict solvation free energies in water at both ambient and non-ambient temperatures, is capable of addressing ionic solutes and salt solutions, and can be extended to non-aqueous systems.
The developed approach can be used to model processes in biological systems, as well as to extend related theoretical models further.\\
}

\clearpage 


\setstretch{1.3} 

\acknowledgements{\addtocontents{toc}{\vspace{1em}} 
	First of all, I would like to thank my supervisors, Prof. Maxim V. Fedorov and Dr. David S. Palmer, who made this Ph.D. an amazing experience. They offered me guidance and support, helping me not only to start but also to finish this project. I would like to acknowledge two visiting students Petteri A. Vainikka, and Samuel W. Coles, who turned out to be wonderful collaborators. Various Strathclyde postgraduate students: Ivor Kresic, Samiul M. Ansari, Benjamin R. Smith, Sean O'Connor, Rosemary Orr, helped me through stimulating discussions. My special thank you goes to Dr. Vladislav Ivanistsev and his group at the University of Tartu, who provided a wonderful and supportive environment during my time there.
	
	I would also like to thank my mother and grandmother, without their love and support I would not have come this far. Most of all, I would like to thank my loving, encouraging, and patient Anna, whose support during this Ph.D. is so appreciated. Thank you.
}
\clearpage 


\pagestyle{fancy} 

\lhead{\emph{Contents}} 
\tableofcontents 

\lhead{\emph{List of Figures}} 
\listoffigures 

\clearpage 
\setstretch{1.5}  








\mainmatter 

\pagestyle{fancy} 



\chapter{Introduction} 

\label{Chapter1} 

\lhead{Chapter 1. \emph{Introduction}} 



What is the amount of reversible work needed to bring a molecule from gas phase to solvent?
It turns out that this question is not just a matter of scientific curiosity.
Accurate predictions of phenomena such as solubility, partition coefficients, substrate binding, acid dissociation constants, all in one way or another depend on how accurately we can measure or predict this quantity \cite{Thompson2003tki,Palmer2008ujv,Palmer2012wtb,Casasnovas2014vba,Garrido2009vzo,Ratkova2015teb,Alongi2010upu,Gilson2007vgc,Takeuchi2012trn,Ratkova2011vje,Zhang2015wrv}.
The aim of this thesis is to develop a theory-based computational method for computing this amount of work.

Formally, the problem is to predict the free energy change, $\Delta G_A$, occurring upon transfer of molecule from gas phase to liquid $A$ \cite{Ben-naim2006vpu}.
This free energy tells us the probability $P$ to find a molecule in a specific phase compared to gas: $P_{A}/P_{gas} = \exp\left[-\Delta G_A/(RT)\right]$, where $T$ is the temperature and $R$ is the gas constant.

\begin{figure}
	\centering
	\includegraphics[width=0.7\linewidth]{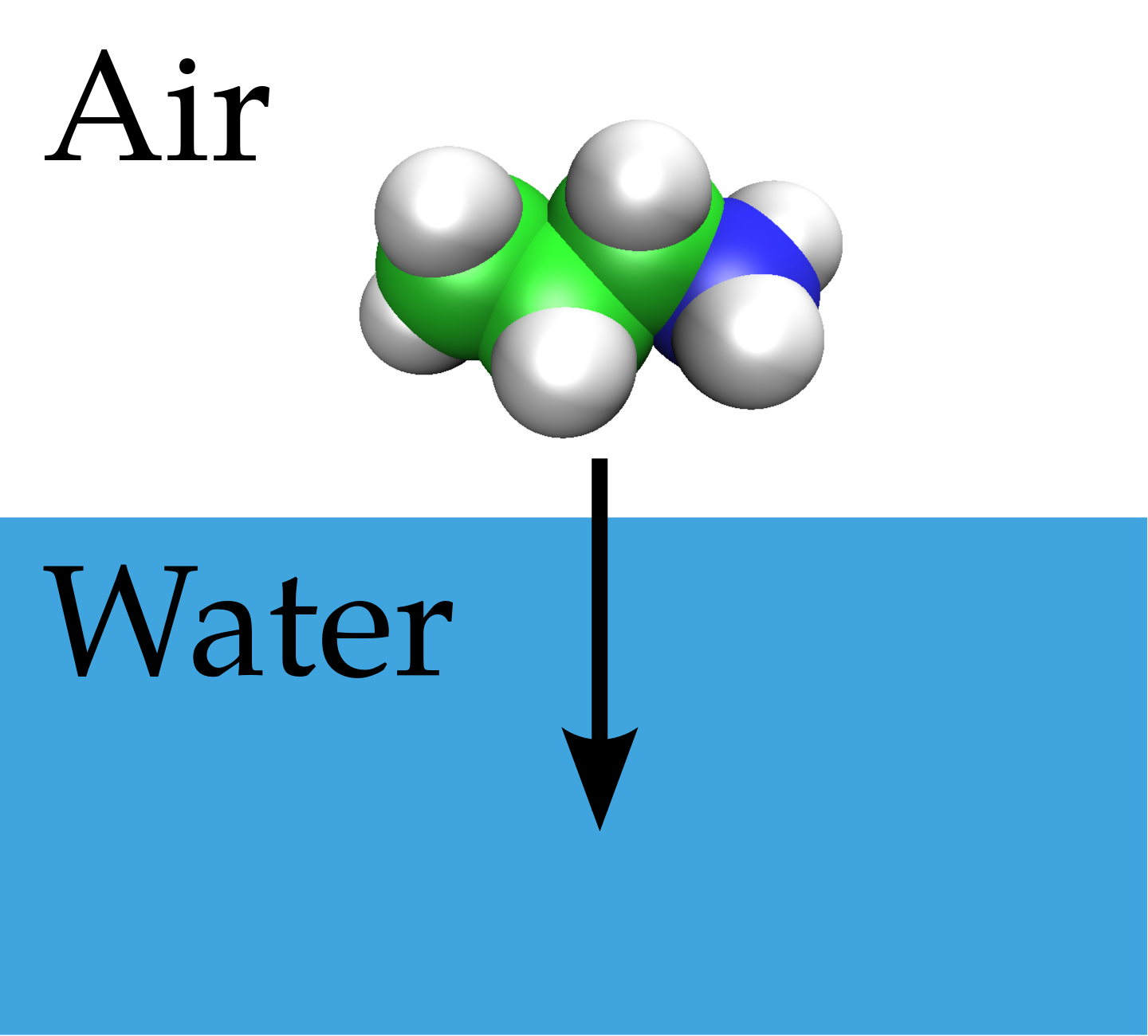}
	\caption{The solvation free energy is equal to the reversible work required to bring molecule from (ideal) gas phase to solvent. Alternatively, it can be computed from its equilibrium concentration ratio in two phases.}
	\label{fig:solvation_free_energy}
\end{figure}

There are multiple factors which make prediction of solvation free energies difficult \cite{Mobley2009wdk,Mobley2014tsl}.
First of all, a good approximation of intermolecular potentials between molecule (solute) and liquid (solvent) is required \cite{Shivakumar2012vcr}.
Second, one has to take into account all possible conformations in which solute can exist in the solution and the gas phase \cite{Klimovich2010tyx}.
Finally, and most importantly, one also has to consider all possible configurations of solvent molecules around the solute; after all, the affinity of a molecule to the phase is determined not only by the solute-solvent interactions, but also by the solvent-solvent ones \cite{Klimovich2015wbq,Pohorille2010uob,Shirts2013wmv}.
The nature of all involved forces is quantum-mechanical; this puts \textit{ab initio} prediction of $\Delta G_A$ into the category of (practically) unsolvable quantum many-body problems.

To make progress, we need to make some approximations.
To an extent, each approximation we make is a trade-off between the speed of the model and its domain of applicability.
A classical molecular dynamics simulation, which uses only a few approximations, is a very general tool that can be applied to systems where quantum effects are not relevant to the motion of particles \cite{Cramer2004vxu}.
At the same time, methods based on statistical learning, such as quantitative structure-property relationships (QSPR), are less generally applicable and are usually limited to systems and compounds that are sufficiently similar to the training data \cite{Stenzel2013wpe,Van_noort2010uti}.
As one would expect, molecular dynamics uses a large number of computational resources, while QSPR calculations are practically effortless.

A family of methods based on classical density functional theory and related integral equation theories offer an attractive balance between speed and generality \cite{Ratkova2015teb,Jeanmairet2015tgx,Liu2013vac,Evans2016tct}.
The idea behind these methods is to ignore unimportant degrees of freedom in a solvent and view it as a local density field $\rho(\rvec)$.
In the absence of an external potential, the solvent will be homogeneous with the value of local density in each point being equal to the bulk number density $\rho(\rvec)=\rho$.
However, bringing a solute molecule in a solvent introduces an external field $\phi(\rvec)$, which breaks the symmetry.
As a result, solvent re-distributes itself around a solute, giving rise to a new density distribution $\rho(\rvec; \phi)$ that is uniquely determined by the external potential.
Moreover, a new density field will be such that the total free energy of the system will be minimised.
The last two statements are the key results of density functional theory, known as the Hohenberg–Kohn–Mermin theorems \cite{Hansen2000utg}.


Similarly to the electronic version of the density functional theory, the classical functional that relates the system's density field $\rho(\rvec)$ to its free energy is unknown.
Therefore, one has to use approximations, tailored to a given problem.
Moreover, often, such approximations lead to non-trivial results and are best understood through the applications of theory to specific problems \cite{Henderson1992vag}.

\begin{figure}
	\centering
	\includegraphics[width=0.7\linewidth]{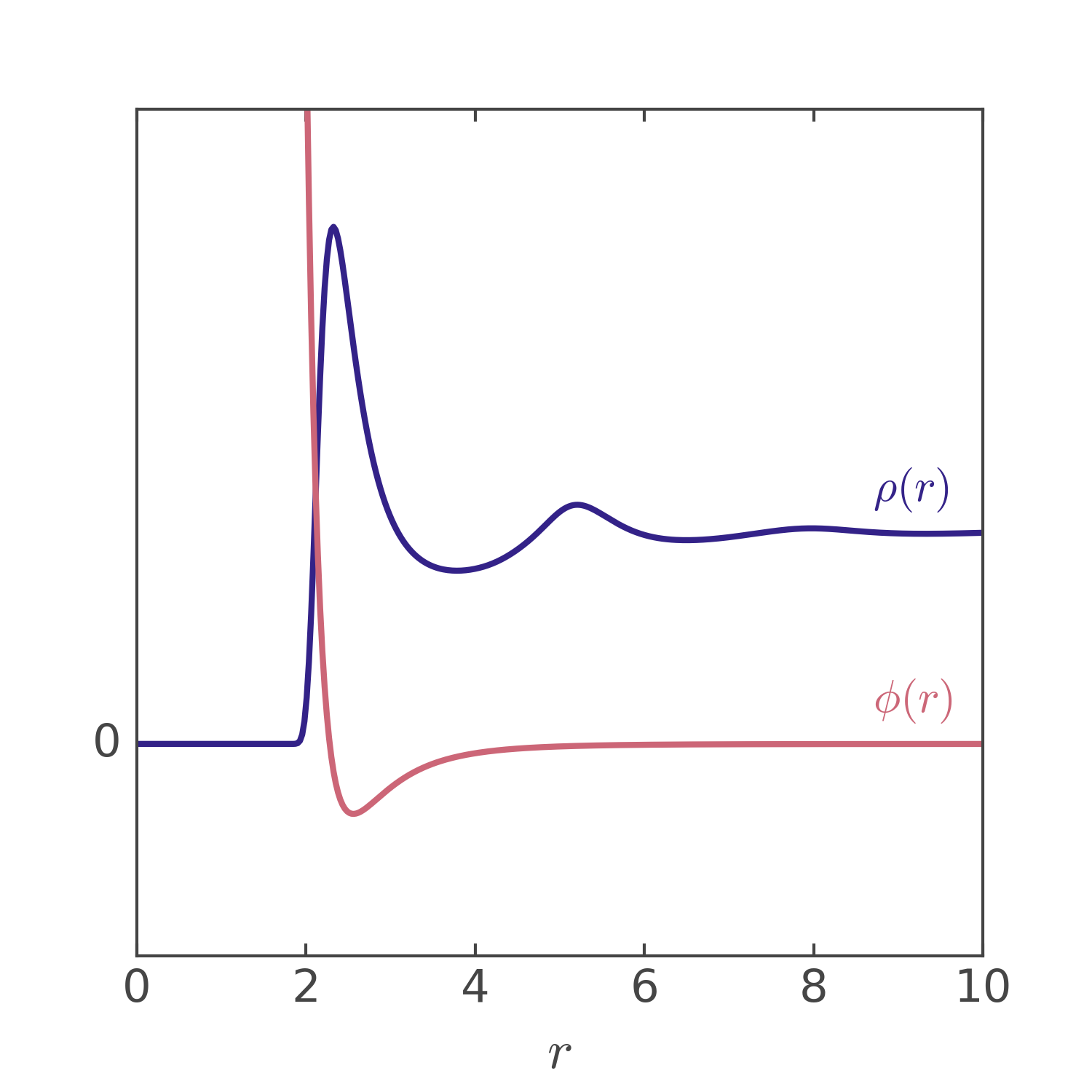}
	\caption{A particle with a spherically symmetric potential $\phi(r)$ is placed at the origin. The surrounding solvent rearranges, forming multiple solvation shells. The dependence of its average density $\rho(r)$ can be obtained using classical density functional theories. The units are arbitrary.}
	\label{fig:potentials}
\end{figure}

In this thesis, we focus on one of the most popular approaches, called three-dimensional reference interaction site model or 3D-RISM \cite{Beglov1997vrd,Kovalenko1998wmo,Hirata2003tpg,Ratkova2015teb}.
It is commonly categorised as an integral equation theory we will see that it is naturally derived and analysed from the viewpoint of classical density functional theory.
The approach owes its popularity to its simplicity, robustness, speed, as well as the fact that it can be applied to biologically relevant solutions such as water with dissolved electrolytes.
In 3D-RISM, the response of the solvent to external perturbation is essentially linear.
This makes the approach accurate for small perturbations, but it ultimately fails to describe larger ones, such as the creation of a cavity in a bulk solvent.

However, recently, some studies have shown that with a few empirical corrections, 3D-RISM can also be used to predict solvation free energies of small molecules with a good accuracy \cite{Palmer2010upn,Palmer2010wgf,Ratkova2010tvr,Palmer2011wlg,Truchon2014vty,Palmer2015vaf}.
Later, Sergiievskyi and co-workers demonstrated that these corrections are related to the overestimated bulk pressure of the liquid, found in 3D-RISM \cite{Sergiievskyi2014vax,Sergiievskyi2015tab}.
They came up with a theory-based pressure correction to 3D-RISM (PC), which, while removing a need to introduce empirical adjustments to the model, was not as accurate as empirical alternatives.
Building upon the work by Sergiievskyi et al., we introduced what is now called an advanced pressure correction (PC+) \cite{Misin2015wdu,Misin2016tqn,Misin2016wlh,Misin2016vee}.
While this correction was initially suggested based on empirical observation, it is now clear that it has a physical basis, as we discuss in the second half of the thesis.

\textbf{The main goal of the thesis is to investigate the accuracy and the scope of 3D-RISM advanced pressure correction, PC+.}
We show that while PC+ suffers from a range of problems related to its approximations, ultimately, for molecular solutes, it can predict solvation free energies with a good accuracy in water as well as in a range of nonpolar solvents.
Since the model does not have adjustable parameters, it can be applied to both pure liquids as well as mixtures with comparable accuracy.
Additionally, it can be useful for prediction of the first derivatives of free energy such as entropy or enthalpy, although, the accuracy is lower.
Through the thesis, we compare model performance with other approaches and discuss their advantages and downsides.

\section{Structure}

The thesis is split into two halves, with the first mostly dedicated to the review of the background theory and previous results, and the second concerned with the new findings, analyses, and discussion.
Specifically, we dedicate the second chapter to solvation thermodynamics, third to the exact results in simple liquids theory, and fourth to approximations necessary for computing liquids structure and free energy.
The second part of the thesis starts with the fifth chapter in which we introduce pressure corrections and present a number of analytical results that justify it and establish its limits.
The sixth chapter is mostly concerned with the accuracy of the model when applied to water, while the seventh chapter discusses non-aqueous solvents.
We wrap everything up with the conclusion that summarizes the main findings and suggests directions for future research.

The thesis is supplemented by two appendices dedicated to the methodology of the calculations performed throughout the thesis as well as additional figures and tables that did not fit into the main part of the work.


\part{Theoretical background}

\chapter{Solvation thermodynamics} 

\label{Chapter2}

\lhead{Chapter 2. \emph{Solvation thermodynamics}} 

The first half of the chapter discusses a number of useful results from statistical mechanics with derivation based on the Ref. \citenum{Ben-naim2006vpu}.
The second half of the chapter is largely based on the recent research literature on solvation thermodynamics and contains a couple of original results.

\section{Classical partition function}

In this thesis, we will be mainly concerned with solvation in common solvents near room temperatures. 
Typically, the behaviour of atomic nuclei in such systems can be described sufficiently accurately using classical statistical mechanics \cite{Hansen2000utg,Ben-naim2006vpu,Diraison1999ukc}. 
Electrons in these systems do behave quantum mechanically, but in many cases, their interactions can be reasonably well approximated by empirical potentials.

Equilibrium properties of the system can be conveniently described using partition functions.
For simplicity, we assume that particle interactions do not depend on their orientation \footnote{The orientation dependant potentials do not change much conceptually, but make notation more cumbersome.}.
Then the partition function of $N$ indistinguishable classical particles in the canonical ensemble takes the following form:
\begin{equation}
\label{eq:canonical_partition_function}
    Q_N = \frac{q^N}{h^{3N} N!} \iint \exp \left[-\beta \Ham\left(\pvec_1\cdots \pvec_N, \rvec_1 \cdots \rvec_N\right) \right] \diff \rvec_1 \cdots \diff \rvec_N  \diff \pvec_1 \cdots  \diff \pvec_N,
\end{equation}
where $h$ is Planck's constant, $\Ham$ is the Hamiltonian of the system, $\beta=1/kT$, with $k$ being the Boltzmann constant and $T$ is temperature. $\rvec$ is the position vector, in 3 dimensions given by $\rvec = \left[x, y, z\right]$ and $\pvec$ is the momentum vector, given by $\pvec = \left[p_x, p_y, p_z\right]$.
The integrals are taken over all possible positions and momenta for each particle.
To simplify notation, whenever the limits on the integral sign are omitted, it is implied that integration is performed over the whole range of possible values.

Another quantity appearing in the equation, $q$ is the single-particle partition function, containing degrees of freedom which we assume to be entirely independent of particle's position and interaction with other particles.
For molecules, it can be typically factored into a product of the partition functions for each degree of freedom $q = q_{rot} q_{vib} q_{el}$, with subscripts $rot$, $vib$, and $el$ representing rotational, vibrational, and electronic degrees of freedom.
The exact form of these functions depends on the molecule in question \cite{Mcquarrie1997vac}.

For our systems of interest, kinetic and potential energy are independent so that Hamiltonian can be split into two parts
\begin{equation}
\label{eq:hamiltonian}
    \Ham\left(\pvec_1\cdots \pvec_N, \rvec_1 \cdots \rvec_N\right) = \sum\limits_{i=1}^N \frac{\pvec^2_i}{2 m_i} + U(\rvec_1 \cdots \rvec_N),
\end{equation}
where $m_i$ is the mass of the $i$-th particle and $U$ is the potential energy of the whole system.
Substituting the above expression into \ref{eq:canonical_partition_function} and integrating over momenta we get
\begin{equation}
\label{eq:partiton_function-Z}
    Q_N = \frac{q^N}{N! \Lambda^{3N}} Z_N.
\end{equation}
Here we introduced the so-called thermal de Broglie wavelength, which is defined by
\begin{equation}
    \Lambda = \sqrt{\frac{h^2}{2\pi m k T}}.
\end{equation}
Another new quantity in the equation, $Z$, is called the configurational partition function and is defined as
\begin{equation}
    Z_N = \int \exp \left[-\beta  U(\rvec_1 \cdots \rvec_N)\right] \diff \rvec_1 \cdots \diff \rvec_N.
\end{equation}

For the majority of systems, the configurational partition function cannot be evaluated analytically.
One notable exception is an ideal gas for which $U=0$ everywhere.
The ideal gas partition function is then
\begin{equation}
    \label{eq:Q_ig_deBrogile}
    Q^{ig}_N = \frac{q^N}{N! \Lambda^{3N}} \int \diff \rvec_1 \cdots \diff \rvec_N = \frac{q^N V^N} {N! \Lambda^{3N}},
\end{equation}
where $V$ is volume.

An average value of some observable quantity $A(\pvec_1\cdots\pvec_N,\rvec_1\cdots\rvec_N)$ can be conveniently expressed via a partition function.
Note that the probability density $\mathrm{Pr}$ to find system in a state $\rvec_1 \cdots \rvec_N, \pvec_1 \cdots \pvec_N$ is
\begin{equation}
\mathrm{Pr} =  \frac{1}{Q_N}\exp \left(-\beta \Ham \right)\diff \rvec_1 \cdots \diff \rvec_N  \diff \pvec_1 \cdots  \diff \pvec_N,
\end{equation} 
where we omitted the dependence of $\Ham$ on the phase space position for clarity.
It follows that an average of some quantity $A$ is
\begin{equation}
\label{eq:ensemble_average}
\left\langle A \right\rangle = \frac{1}{Q_N}\iint A\left(\rvec_1 \cdots \rvec_N, \pvec_1 \cdots \pvec_N\right) \exp \left(-\beta \Ham \right) \diff \rvec_1 \cdots \diff \rvec_N  \diff \pvec_1 \cdots  \diff \pvec_N,
\end{equation}
where $\left\langle \cdots \right\rangle$ here denote ensemble average.

In addition to the canonical ensemble (which has constant $N$, $V$, and $T$), one can also define the isobaric-isothermal ensemble that has constant pressure $P$, number of particles $N$, and temperature $T$.
For such systems the partition function becomes
\begin{equation}
\Delta_N = \frac{1}{V_0} \int\limits_{0}^{\infty} \exp \left( -\beta PV \right) Q_N \diff V,
\end{equation}
where $V_0$ is a unit volume that is used to make $\Delta$ dimensionless.
In the grand canonical ensemble, the constant quantities are chemical potential $\mu$, temperature $T$, and volume $V$.
The number of particles, $N$, is allowed to vary.
The corresponding partition function is
\begin{equation}
\Xi = \sum\limits_{N=0}^{\infty} Q_N \exp \left(\beta N \mu \right).
\label{eq:grand_partition}
\end{equation}


\section{Free energies}
\label{sec:connection_to_thermodynamics}

The second law of thermodynamics states that the equilibrium state of an isolated thermodynamic system maximizes its total entropy \cite{Callen1985vqb}.
However, in practice, we rarely deal with isolated systems.
For systems in contact with some external reservoirs this law can be reformulated in the following way: at equilibrium, the thermodynamic system minimizes its corresponding thermodynamic potential, which depends on the systems constraints.
For systems subject to $NVT$ conditions (canonical ensemble) the appropriate potential is Helmholtz free energy $F$, for $NPT$ condition (isobaric-isothermal ensemble) it is Gibbs free energy $G$, and for $\mu VT$ conditions (grand canonical ensemble) it is the grand potential $\Omega$.

Importantly, the thermodynamic potentials (free energies) are linked to partition functions in the following way
\begin{equation}
\begin{split}
F &= -k T \ln Q_N,\\
G &= -k T \ln \Delta,\\
\Omega &= -k T \ln \Xi.
\end{split}
\label{eq:HelmholtzFE-Q}
\end{equation}
These relationships provide a link between thermodynamics and statistical mechanics.

Similarly to partition functions, free energies can be decomposed into contributions from kinetic energy (ideal gas) and from potential energy (usually called excess free energy).
These parts can be expressed in terms of partition functions using equations \ref{eq:HelmholtzFE-Q}, \ref{eq:Q_ig_deBrogile}, and \ref{eq:partiton_function-Z}
\begin{equation}
    F = -kT \ln Q^{ig}_N - kT \ln \frac{Z}{V^N} = F^{ig} + F^{ex},
\end{equation}
where superscript $ig$ indicates ideal part and $ex$ excess part of the free energy.
$F^{ig}$ can be readily evaluated using \ref{eq:Q_ig_deBrogile} and Stirling's approximation
\begin{equation}
	\begin{split}
	F^{ig} &= -NkT\ln\frac{qV}{\Lambda^3} + kT\ln N!\\
	&\approx -NkT\ln\frac{qV}{\Lambda^3} +kT(N\ln N - N)\\
	&= NkT\left( \ln \Lambda^3 \rho q^{-1} - 1\right).\\
	\end{split}
\end{equation}

%

%


\section{Chemical potential}
\label{sec:chemical_pot}

The chemical potential of the $i$-th type of particle in a multicomponent mixture is defined as:
\begin{equation}
    \mu_i = \left( \frac{\partial F}{ \partial N_i} \right)_{T, V, N_{j \neq i}}.
\end{equation}
The above expression is perfectly valid for other thermodynamic potentials as long as appropriate constraints are chosen (for example, the derivative can be taken with respect to Gibbs free energy $G$ if pressure instead of volume is fixed).

The chemical potential, despite often being considered as a somewhat mysterious quantity \cite{Baierlein2001wyy}, actually, has a straightforward physical meaning.
Similarly to how thermodynamic free energies can be viewed as generalizations of the potential of classical systems (hence the name thermodynamic potential), the derivatives of the said potentials are similar to forces.
$\mu_i$ can be viewed as a force exerted by the system on the particle of type $i$, with negative and positive signs corresponding to particles being driven into and out of the system respectively.

Two, more formal definitions of chemical potentials within statistical mechanics, will be presented below.
Both derivations will be presented in the canonical ensemble, but almost identical results hold in systems with other constraints.
To simplify notation, we will be viewing solvation in a one-component system where the inserted molecule is identical to other particles (the same results hold true where the inserted molecule is distinct).
This subject is given much attention since the main goal of the thesis is an estimation of chemical potentials in various systems.

It can be shown \cite{Ben-naim2006vpu} that in the thermodynamic limit where $N \rightarrow \infty$ (or in other words, when the insertion of a new particle does not change the composition of the system):
\begin{equation}
\label{eq:mu_deltaF}
    \mu = F(N + 1) - F(N)
\end{equation}
with all other variables kept constant. Combining this equation with \ref{eq:HelmholtzFE-Q} we get
\begin{equation}
\label{eq:widow-deriv0}
    \mu = -kT \ln \frac{Q_{N+1}}{Q_N} = -kT \ln \left[\frac{q^{N+1} N! \Lambda^{3N}}{q^N N!(N+1) \Lambda^{3(N+1)}} \frac{Z_{N+1}}{Z_N}\right] .
\end{equation}
Then
\begin{equation}
\label{eq:widow-deriv1}
    \mu = -kT \ln \left[\frac{q}{(N+1) \Lambda^3} \frac{ \int \diff \rvec_1 \cdots \diff \rvec_N \diff \rvec_{N+1} \exp (-\beta U_{N+1} ) }{ \int \diff \rvec_1 \cdots \diff \rvec_N \exp (-\beta U_{N} ) }\right] .
\end{equation}

To move forward we need to split the potential energy of system with added particle $U_{N+1}$ into two parts:
\begin{equation}
    U_{N+1} = U_N (\rvec_1\cdots \rvec_N) + E^{uv}(\rvec_1\cdots \rvec_N,\rvec_{N+1})
\end{equation}
where $E^{uv}$ is the interaction (binding) energy of the particle $N+1$ to the rest of the system.
The superscript $uv$ denotes interactions between the sol$u$te, the particle being inserted, and sol$v$ent, the medium.
Substituting the above back into equation \ref{eq:widow-deriv1} we get:
\begin{equation}
    \mu = -kT \ln \left[\frac{q}{(N+1) \Lambda^3} \frac{ \int \diff \rvec_1 \cdots \diff \rvec_N \diff \rvec_{N+1}\exp(-\beta E^{uv}) \exp (-\beta U_{N} ) }{ \int \diff \rvec_1 \cdots \diff \rvec_N \exp (-\beta U_{N} )} \right],
\end{equation}
where dependencies of potential energies on positions of all particles were omitted for clarity.
This equation can be further simplified by noting that in a homogeneous liquid the potential energy depends only on the relative positions of particles.
By setting $\rvec^{\prime}_i = \rvec_i - \rvec_{N+1}$ for position vectors $1 \cdots N$ and integrating out $\rvec_{N+1}$ we get
\begin{equation}
    \mu = -kT \ln \left[\frac{qV}{(N+1) \Lambda^3} \left\langle \exp (- \beta E^{uv}) \right \rangle_0 \right],
\end{equation}
where $\left\langle \cdots \right\rangle_0 $ denote the averaging over positions of all particles (the chemical potential does not depend on the moment, so their values are unimportant).
In the thermodynamic limit $V/(N+1) \approx V/N$:
\begin{equation}
\label{eq:widom-mu}
    \mu = kT \ln \left( \rho \Lambda^3 q^{-1} \right) - kT \ln \left\langle \exp (- \beta E^{uv} ) \right\rangle_0 
\end{equation}
This important result was first discovered by Widom in 1963 \cite{Widom1963txu,Widom1982thj}.
In the above equation the chemical potential is clearly split into ideal and non-ideal parts, with
\begin{equation}
\label{eq:mu-ig}
    \mu^{ig} = kT \ln \left( \rho \Lambda^3  q^{-1} \right).
\end{equation}
and
\begin{equation}
\label{eq:mu-mu_ig-mu_ex}
    \mu^{ex} = - kT \ln \left\langle \exp (- \beta E^{uv} ) \right\rangle_0.
\end{equation}

There is an alternative and equally useful statistical mechanics expression for chemical potential that was first derived by Kirkwood in 1935 \cite{Ben-amotz2005txu,Kirkwood1935wpl}.
Imagine insertion of a single particle into the system as a continuous process during which particle-system interactions are slowly turned on.
To characterise such a process we introduce a modified Hamiltonian $\Ham(\lambda)$, with $\lambda=0$ representing an uncoupled state in which $N+1$-th particle does not interact with the rest of the system, and $\lambda=1$ being a final state in which all particle-system interactions are turned on.
The approach is quite general since any continuous function $\Ham(\lambda)$ satisfying the above requirements would suffice.

The derivative of free energy with respect to coupling parameter is given by:
\begin{equation}
\begin{split}
\frac{\diff F}{\diff \lambda} &= -kT \frac{\diff}{\diff \lambda} \ln  \iint \exp \left[-\beta \Ham\left(\lambda\right) \right] \diff \rvec_1 \cdots \diff \rvec_{N+1}  \diff \pvec_1 \cdots  \diff \pvec_{N+1}\\
&= -kT \frac{  \iint -\frac{\diff \Ham(\lambda)}{\diff \lambda}\exp \left[-\beta \Ham\left(\lambda\right) \right] \diff \rvec_1 \cdots \diff \rvec_{N+1}  \diff \pvec_1 \cdots  \diff \pvec_{N+1}}{  \iint \exp \left[-\beta \Ham\left(\lambda\right) \right] \diff \rvec_1 \cdots \diff \rvec_{N+1}  \diff \pvec_1 \cdots  \diff \pvec_{N+1}}.
\end{split}
\end{equation}
Expressing the above in terms of an ensemble average we get:
\begin{equation}
\frac{\diff F}{\diff \lambda} = \left\langle \frac{\diff \Ham(\lambda)}{\diff \lambda} \right\rangle_{\lambda},
\end{equation}
which after applying the fundamental theorem of calculus becomes
\begin{equation}
\Delta F = \int\limits_{0}^{1}  \left\langle \frac{\diff \Ham(\lambda)}{\diff \lambda} \right\rangle_{\lambda} \diff \lambda.
\label{eq:kirkwood_mu}
\end{equation}
In the case of a linear coupling of solute-solvent potential energy $\Ham(\lambda) = \Ham(0) + \lambda E^{uv}$, equation \ref{eq:kirkwood_mu} reduces to
\begin{equation}
\mu^{ex} = \int\limits_{0}^{1}\left\langle E^{uv} \right\rangle_{\lambda} \diff \lambda,
\label{eq:linear_coupling_mu}
\end{equation}
where we necessarily get $\mu^{ex}$, since the mass of the inserted particle, and thus, the total kinetic energy of the system was unchanged.
Practically, the linear coupling is not always convenient, but from the theory perspective, it gives exact results.

Assuming a linear dependence of interaction energy on the coupling strength (linear response) $\langle E^{uv} \rangle_{\lambda} = \langle E^{uv} \rangle_{\lambda=1} \lambda + \langle E^{uv} \rangle_{\lambda=0} (1 - \lambda)$, we can obtain a useful rough estimate of the free energy change
\begin{equation}
\Delta F \approx \int\limits_0^1 \diff \lambda \langle E^{uv} \rangle_{\lambda=1} \lambda = \frac{1}{2} \langle E^{uv} \rangle_{\lambda=1} + \frac{1}{2}  \langle E^{uv} \rangle_{\lambda=0},
\end{equation} 
which is just an average of initial and final particle system interaction energies.
For chemical potential, this equation is not very useful, since $\langle E^{uv} \rangle_{\lambda=0}$ is not well defined, but for some smaller perturbations, such as adding electrostatic charge to the inserted formula, this equation is quite accurate \cite{Ben-amotz2008tzp}.

Comparing the two obtained equations for the chemical potential, it may appear that Widom's formula (\ref{eq:widom-mu}) is computationally more convenient as the interactions are computed only in one state, as opposed to equation \ref{eq:kirkwood_mu} in which one has to compute $\diff\Ham(\lambda)/\diff\lambda$ in a number of systems with varying $\lambda$.
However, in the case of dense systems, Kirkwood's equation is advantageous.
For most liquids random insertions of a particle will cause overlap, resulting in a significant number of trials required for expression $\left\langle \exp (-\beta E^{uv}) \right\rangle_0$ to converge \cite{Leach2001vev}.
On the other hand, using Kirkwood's formula, one can start with a system in which particle is already fully coupled and then slowly decouple it, letting the surroundings relax.
Such an approach is guaranteed to yield a good estimate of $\mu$ independent of the density of the system.
For this reason, the majority of chemical potential calculations employ Kirkwood's formula.

Finally, it is useful to consider the relationship between the chemical potential and the ensemble from a thermodynamic perspective.
Intuitively, it seems reasonable that as long as macroscopic thermodynamic parameters such as pressure or density are identical, and the system is sufficiently large, the chemical potential would be independent of the types of system constraints.
However, if we write out the equations for insertion explicitly, we will notice that this independence is realised through the cancellation of ensemble specific contributions \cite{Qian1996wbl}.

Let's denote the chemical potential obtained by differentiating Helmholtz free energy at constant volume as $\mu_V$ and the one obtained by differentiating Gibbs free energy at constant pressure as $\mu_P$.
To see the relationship between these quantities we right out
\begin{equation}
    G(T,P,N) = F(T,V(T,P,N),N) + PV(T,P,N)
\end{equation}
and then using a chain rule \cite{Qian1996wbl}:
\begin{align}
    \mu_{P} &= \left( \frac{\partial G}{\partial N'} \right)_{T, P, N} \notag\\
    &= \left( \frac{\partial F}{\partial N'} \right)_{T, P,N} + P \left( \frac{\partial V}{\partial N'} \right)_{T, P,N} \notag\\
    &= \left( \frac{\partial F}{\partial N'} \right)_{T, V, N} + \left( \frac{\partial F}{\partial V} \right)_{T, N} \left( \frac{\partial V}{\partial N'} \right)_{T, P, N} + P \left( \frac{\partial V}{\partial N'} \right)_{T, P, N} \notag\\
    &=\left( \frac{\partial F}{\partial N'} \right)_{T, V, N} = \mu_{V},
\end{align}
where in the third equality we used the identity $P = -\left( \partial F / \partial V \right)_{T, N}$ and we used superscript $N'$ to separate introduced particle from the rest of the system.
The partial derivative appearing in the above expression $\left(\partial V/\partial N' \right)_{T, P, N}=\bar{V}$ is called the partial molar volume.
It indicates by how much system volume changes when we introduce a small number of new particles; we will discuss it in more detail in later section.
At constant pressure, the extra work required to increase system size by $P\bar{V}$ is compensated by a decrease in free energy due to the system expansion.

An analogous procedure can be performed in the case of the Grand potential.
We start by writing
\begin{equation}
\Omega(T,V,\mu) = F(T,V,N(T,V,\mu)) - \mu N(T,V,\mu).
\end{equation}
Assuming the solvated particle is distinct from the rest
\begin{align}
\mu_{\mu} &= \left( \frac{\partial \Omega}{\partial N'} \right)_{T, V, \mu} \notag\\
&= \left( \frac{\partial F}{\partial N'} \right)_{T, V, \mu} - \mu \left( \frac{\partial N}{\partial N'} \right)_{T, V, \mu} \notag\\
&= \left( \frac{\partial F}{\partial N'} \right)_{T, V, N} + \left( \frac{\partial F}{\partial N} \right)_{T, V} \left( \frac{\partial N}{\partial N'} \right)_{T, V,\mu} - \mu \left( \frac{\partial N}{\partial N'} \right)_{T, V,\mu} \notag\\
&=\left( \frac{\partial F}{\partial N'} \right)_{T, V, N} = \mu_{V},
\end{align}
where $\mu = \left( \partial F / \partial V \right)_{T, V, N}$ was utilised.
The quantity $\left(\partial N/\partial N'\right)_{T, V, \mu}$ is directly related to partial molar volume.
In appendix \ref{sec:pmv_gc} we show that $\left(\partial N/\partial N'\right)_{T, V, \mu} = -\rho \bar{V}$.
Then in a grand canonical ensemble, we obtain extra $\mu \rho \bar{V}$ energy from the particle bath, but it gets compensated due to a decrease of particle number inside the system.


\section{Solvation free energy}

To discuss solvation thermodynamics we first need to define a term solution. 
According to IUPAC "Gold Book" \cite{Mcnaught1997uze}: "Solution is a liquid or solid phase containing more than one substance".
A dominant component of the solution is usually called solvent, while minor components are referred to as solutes.
In most of our discussion, we will be dealing with infinitely dilute solutions in which concentrations of solutes, as well as their mutual interactions, tend to zero.

According to Ben-Naim, the solvation may be defined as the process of transferring a solute from a fixed position in an ideal gas phase into a fixed position in the solvent \cite{Ben-naim2006vpu,Ratkova2015teb}.
To express this definition analytically, we need to introduce the pseudo-chemical potential, which is a chemical potential associated with a stationary particle.
It is typically denoted as $\mu^{\ast}$ and is given by
\begin{equation}
    \label{eq:pseudo_chemical_pot}
    \mu^{\ast} = \mu - kT \ln \rho \Lambda^3.
\end{equation}

The solvation free energy is defined as
\begin{equation}
    \label{eq:ben_naims_deltaG}
    \Delta G^{\ast} = \mu^{\ast l} - \mu^{\ast ig} = \Delta \mu^{\ast},
\end{equation}
where superscript $l$ indicates solvent and $\ast$ refers to Ben-Naim's definition.
As we saw in the previous section, the value of chemical potential is ensemble-independent and thus $\Delta G^{\ast}=\Delta F^{\ast} = \Delta \Omega^{\ast}$.
However, in practice, the symbol $\Delta G$ is used to indicate that the system is connected to a constant pressure and temperature bath.

Expressing everything in terms of $\mu^{ex}$ (defined via the equation \ref{eq:widom-mu}) we can rewrite the previous equation to get:
\begin{equation}
    \Delta G^{\ast} = \mu^{ex} - kT\ln \frac{q^{l}}{q^{ig}}.
\end{equation}
Clearly, if the internal partition function of the molecule is unaffected by the phase transfer, $\Delta G^{\ast}$ is equal to the coupling work of the solvent or $\mu^{ex}$.

Ben-Naim's convention for solvation free energies is not the only one in use.
Another commonly used way of expressing solvation free energies is based on standard states.
The process of solvation is then described as a transfer of a compound at a standard gaseous state (a hypothetical state of pure substance at which it exhibits ideal gas behaviour and has standard pressure $P^o=\SI{1}{\bar}$) to the standard solution state (a hypothetical state of an ideal solution at standard pressure $P^o$ and molality $b^o = \SI{1}{\mol\per\kilogram}$).
Solvation free energies corresponding to this process are denoted as $\Delta G^o$.

While the use of standard states has many advantages, for the process of solvation they are not very convenient.
Within the standard-state approach, solute molecules change their density during the transfer, which leads to an artificial dependence of derivatives of solvation free energy on quantities such as thermal expansion or compressibility.
Additionally, one molal standard state is far from infinite dilution; defining such state as an ideal solution in which solute molecules do not interact with each other is not physically meaningful.

Due to the above reasons, in this thesis, we will be primarily using Ben-Naim's definition of solvation.
For simplicity, we will also drop the unnecessary $\ast$ and will simply denote corresponding solvation free energies as $\Delta G$, $\Delta F$, or $\Delta \Omega$, depending on the system.
To convert from one type of definition to another, we can use equation \ref{eq:ben_naims_deltaG}. Then
\begin{equation}
\begin{split}
\Delta G^o &= \Delta G^{\ast} -kT \ln \frac{\rho_l}{\rho_g}\\
&= \Delta G^{\ast} -kT \ln \frac{b^o_u M_v \rho_v kT}{P^o},
\end{split}
\end{equation}
where $M_v$ is the molar mass and $\rho_v$ is the density of the solvent.
For water at \SI{298.15}{K} and standard pressure, this corresponds to $\Delta G^{\ast} = \Delta G^o - \SI{1.9}{kcal\per\mol}$.


\section{Decomposing solvation free energy}
\label{sec:solv_thermodynamics}

A lot of insight can be obtained by examining various decompositions of solvation free energy into different components.
In this section, we will demonstrate the separation of excess chemical potential into energetic and entropic components, following the approach that is commonly used in thermodynamic and statistical mechanics treatments of the subject, and in section \ref{sec:numerical_exp} we will approach this task from the simulations perspective.

Ben-Amotz et al. have demonstrated that one can formally decompose $\mu^{ex}$ into two equivalent representations \cite{Ben-amotz2005ubb,Ben-amotz2008tzp}
\begin{equation}
\begin{split}
\mu^{ex} & = \left\langle E^{uv} \right\rangle_{\lambda=1} + \beta \int\limits_{0}^{1} \diff \lambda \lambda \Big[ \left\langle \left(E^{uv}\right)^{2} \right\rangle_{\lambda} - \left\langle E^{uv} \right\rangle_{\lambda}^{2} \Big]\\
& = \left\langle E^{uv} \right\rangle_{\lambda=1} + \frac{1}{\beta} \ln \left \langle \exp \left[\beta (E^{uv} - \langle E^{uv}\rangle) \right] \right \rangle_{\lambda=1}\,,
\end{split}
\end{equation}
where brackets $\left\langle\cdots\right\rangle_{\lambda}$ denote averaging in the ensemble of interest at a particular coupling strength $\lambda$.
The first term in the both equations is the strength of interactions between solvent and fully coupled solute, and it represents an enthalpic contribution to the solvation free energy.
Both second terms represent an entropic contribution to the solvation free energy and will be denoted as $-TS^{uv}$; they are both proportional to fluctuations of solute-solvent interaction energy.
Integrating the first of the above equation by parts we get
\begin{equation}
\mu^{ex} = \left\langle E^{uv} \right\rangle_{\lambda=1} + \frac{\beta}{2} \Big[ \left\langle \left(E^{uv}\right)^{2} \right\rangle_{\lambda=1} - \left\langle E^{uv} \right\rangle_{\lambda=1}^{2} \Big] + \cdots,
\end{equation}
where $\cdots$ represent higher order cumulants that disappear if the fluctuations of solute-solvent energy are Gaussian.

From the previous paragraph, we can see that $\mu^{ex} = E^{uv} - TS^{uv}$, which is a very convenient decomposition from the theoretical point of view.
The entropic contribution is always positive $-TS^{uv} \geq 0$ and energetic is negative for the absolute majority of solutes.
Note that within the linear response approximation, discussed in section \ref{sec:chemical_pot}, $-T S^{uv} =  -1/2 E^{uv} + 1/2 E^{uv}_0$.


We can also decompose the excess chemical potential of solvation $\mu^{ex}$ using more conventional definitions of solvation energy and entropy; however, these values have solvent-solvent contributions that will cancel each other out.
We start by using the following definition of chemical potential
\begin{equation}
    \mu^{ex} = \Delta U - T \Delta S,
\end{equation}
where $\Delta S$ is the excess solvation entropy given by
\begin{equation}
    \Delta S = - \left(\frac{\partial \mu^{ex}}{\partial T}\right)_{V}
\end{equation}
and $\Delta U$ is the change in system's excess internal energy, given by
\begin{equation}
\label{eq:gibbs-helmholtz}
    \Delta U = \left[\frac{\partial \left(\mu^{ex}/T\right)}{\partial \left( 1/T\right)}\right]_{V}.
\end{equation}
In the above equations, subscripts indicating that we are dealing with the excess quantities were dropped for clarity.


Let us first look at the change in excess internal energy $\Delta U$.
Similarly to equation \ref{eq:mu_deltaF}, the following result holds $\Delta U = U^{ex}(N_v,N_u=1) - U^{ex}(N_v,N_u=0)$.
Then we can formally write
\begin{equation}
\Delta U = \left\langle U^{vv} + E^{uv} \right\rangle_{\lambda=1} - \left\langle U^{vv} \right\rangle_{\lambda=0} = E^{uv}  + \Delta \left\langle U^{vv}, \right\rangle
\end{equation}
where $U^{vv}$ is the interaction energy of solvent atoms and $\Delta U^{vv}=\left\langle  U^{vv} \right\rangle_{\lambda=1} - \left\langle U^{vv} \right\rangle_{\lambda=0}$ is called the solvent reorganization energy.

The above equation contains a term accounting for the solute-solvent and the solvent-solvent interactions.
Both terms can be either positive or negative.
It also should be noted that the first term is relatively easy to compute, as for typical potentials $\left\langle  E^{uv} \right\rangle_{\lambda=1}$ is short ranged.
On the other hand, $\Delta \left\langle U^{vv} \right\rangle$ is the difference of two large interaction energies and is usually difficult to evaluate.

To obtain an expression for solvation entropy $\Delta S$, we note that
\begin{equation}
\label{eq:TS_kirkwood}
    T \Delta S = -T \left(\frac{\partial \mu^{ex}}{\partial T}\right)_{V} = \beta \left(\frac{\partial \mu^{ex}}{\partial \beta}\right)_{V}= \beta \int\limits_{0}^{1} \diff \lambda \frac{ \partial \left\langle E^{uv} \right\rangle_{\lambda}}{\partial \beta},
\end{equation}
where we used the Kirkwood-Buff expression for chemical potential (equation \ref{eq:linear_coupling_mu}) to obtain the final equality.
We can simplify the integrand in the above by writing out its definition
\begin{equation}
\begin{split}
\frac{ \partial \left\langle E^{uv} \right\rangle_{\lambda}}{\partial \beta} &= \frac{ \partial }{\partial \beta} \frac{\int \diff \rvec_1 \cdots \diff \rvec_N \diff \rvec_{N+1} E^{uv}  \exp \left[ -\beta U_{N+1} (\lambda) \right]}{Z(\lambda)} \\
&= \frac{-\left\langle U_{N+1}(\lambda) \right\rangle \, \left\langle E^{uv} \right\rangle \, Z^{2} (\lambda) + \left\langle E^{uv}\, U_{N+1}(\lambda) \right\rangle \, Z^{2} (\lambda)}{Z^{2}(\lambda)} \\
&= -\left\langle U_{N+1}(\lambda) \right\rangle \, \left\langle E^{uv} \right\rangle  + \left\langle E^{uv}\, U_{N+1}(\lambda) \right\rangle\\
&= -\left\langle U^{vv} \right\rangle_{\lambda} \, \left\langle E^{uv} \right\rangle_{\lambda} - \lambda \left\langle E^{uv} \right\rangle_{\lambda}^{2} + \left\langle E^{uv} \, U^{vv} \right\rangle_{\lambda} + \lambda \left\langle (E^{uv})^{2} \right\rangle_{\lambda}\\
&= -\left\langle U^{vv} \right\rangle_{\lambda} \, \left\langle E^{uv} \right\rangle_{\lambda} - \left\langle E^{uv} \, U^{vv} \right\rangle_{\lambda} - \lambda \left[\left\langle E^{uv} \right\rangle_{\lambda}^{2}  -  \left\langle (E^{uv})^{2} \right\rangle_{\lambda} \right],\\
\end{split}
\end{equation}
where we used $\langle U_{N+1}\rangle_{\lambda} = \langle U^{vv} \rangle_{\lambda} + \lambda E^{uv}$ to obtain the fourth equality.
Returning back to equation \ref{eq:TS_kirkwood} we get
\begin{equation}
    T \Delta S = -\beta \int\limits_{0}^{1} \diff \lambda \lambda \Big[\left\langle E^{uv} \right\rangle_{\lambda}^{2}  -  \left\langle (E^{uv})^{2} \right\rangle_{\lambda} \Big] - \beta \int\limits_{0}^{1} \diff \lambda \Big[  \left\langle U^{vv} \right\rangle_{\lambda} \, \left\langle E^{uv} \right\rangle_{\lambda} - \left\langle E^{uv} \, U^{vv} \right\rangle_{\lambda} \Big].
    \label{eq:suv}
\end{equation}
Ben-Amotz, and earlier Yu and Karplus, have shown that \cite{Ben-naim2006vpu,Yu1988utv}
\begin{equation}
    \Delta U^{vv} = -  \beta \int\limits_{0}^{1} \diff \lambda \Big[  \left\langle U^{vv} \right\rangle_{\lambda} \, \left\langle E^{uv} \right\rangle_{\lambda} - \left\langle E^{uv} \, U^{vv} \right\rangle_{\lambda} \Big].
\end{equation}
This equation combined with the result above gives us the final expression
\begin{equation}
\label{eq:canonical_solvation_entropy}
    T \Delta S = T \Delta S^{uv} + \Delta U^{vv}.
\end{equation}

Notice that both internal energy change $\Delta U$ and solvation entropy $\Delta S$ contain contributions due to solvent reorganization energy $U^{vv}$.
However, it cancels out when we add individual derivatives $\mu^{ex} = \Delta U-T\Delta S = E^{uv}-T\Delta S^{uv}$.

Following Ben-Amotz, we can extend these results to other ensembles
\begin{equation}
\begin{split}
\Delta G_s & = E^{uv}_P - T S^{uv}_P + P \bar{V},\\
\Delta \Omega_s & = E^{uv}_\mu -T S^{uv}_\mu + \mu \rho \bar{V}.
\end{split}
\end{equation}
The rationale behind $\mu \rho \bar{V}$ term was given at the end of section \ref{sec:chemical_pot}.
Note that since the averaging is done in different ensembles, generally $E^{uv}_P \neq E^{uv}_V \neq E^{uv}_\mu$, with the same holding true for entropy.
Similarly, derivatives of free energy are not necessarily equal in different ensembles.
Even though the chemical potential is ensemble-independent, its decompositions are not.


\section{Numerical experiments}
\label{sec:numerical_exp}

The dynamics of the majority of liquids at room temperature can be well approximated using Newton's equations of motion \cite{Hansen2000utg}.
Thus, in principle one could simulate the liquid by putting a sufficient number of molecules in a box, giving them initial velocities according to the Maxwell-Boltzmann distribution and then updating their positions and velocities using the force $\vect{F} = - \nabla U(\rvec_1\cdots\rvec_N)$.
This is the basic idea behind molecular dynamics (MD) simulations, which is an extremely powerful tool for studying liquids and their solutions.

The success of molecular dynamics simulation largely depends on the quality of the approximation of intermolecular potential $U$.
In all of the simulations performed in this thesis, we assumed that $U$ is pair decomposable:
\begin{equation}
U = \sum\limits_{i,j} u_{ij}(r_{ij}),
\end{equation}
where $\sum\limits_{i,j}$ is the sum over all pairs of interacting sites (particles), $r_{ij}$ is the distance between them, and $u_{ij}$ is the pair potential.

The form of $u_{ij}$ depends on whether sites $i$ and $j$ are part of the same or different molecules.
If sites $i$ and $j$ are both located on the same molecules, the interaction between them will depend on the types of bonds present in the molecule.
For $i$ and $j$ which are parts of different molecules, or are separated by a sufficiently large number of bonds, the pair potential is usually given by the sum of short-ranged and electrostatic potentials:
\begin{equation}
\begin{split}
u_{ij}(r) &= u^{LJ}_{ij}(r) + u^{el}_{ij}(r),\\
u^{LJ}_{ij}(r) &= 4 \epsilon_{ij} \left[\left(\frac{\sigma_{ij}}{r_{ij}}\right)^{12} - \left(\frac{\sigma_{ij}}{r_{ij}}\right)^6\right],\\
u^{el}_{ij}(r) &= \frac{q_i q_j}{4\pi \epsilon_0 r_{ij}},
\end{split}
\label{eq:lj_elec_pot}
\end{equation}
where $\epsilon_{ij}$ is the Lennard-Jones well depth, $\sigma_{ij}$ is the Lennard-Jones diameter, $q$ is the partial charge, and $\epsilon_0$ is the vacuum permittivity.
The 12-6 Lennard-Jones potential is the most commonly used model to approximate short-ranged repulsive forces that originate due to repulsion of electronic clouds as well as somewhat longer range (decaying as $r^{-6}$) attractive forces due to dispersion interactions.

The equation \ref{eq:lj_elec_pot} defines the most standard and commonly used form of the intramolecular potential \cite{Allen1987uqm}.
Potentials of this form are robust and fast.
However, they ignore a number of potentially significant effects such as polarization, charge transfer, multi-body interactions, etc.

The process of finding the interaction parameters describing each site is largely empirical.
The Lennard-Jones parameters are usually fit to reproduce macroscopic parameters such as density or viscosity.
In case of the water, there are a number of models with varying sophistication and number of sites.
For a typical organic solute, one can take parameters from various force fields.
More simple ones, such as general Amber force field (GAFF) \cite{Wang2004vlm}, will have a single set of Lennard-Jones parameters per element, while more advanced ones, such as the optimised potential for liquid simulations (OPLS), have a number of different force field constants depending on the elements bonding \cite{Jorgensen1983wtl,Banks2005vwm,Harder2016vvq}.

To estimate Lennard-Jones interactions between different types of atoms, one can use various combination rules.
For all simulations in the thesis, we will be using so-called Lorentz-Berthelot rules \cite{Allen1987uqm}
\begin{align}
\sigma_{ij} = \frac{\sigma_i + \sigma_j}{2} && \epsilon_{ij} = \sqrt{\epsilon_i \epsilon_j},
\end{align}
where $\sigma_i$ and $\epsilon_i$ are input taken from the force fields, and $\sigma_{ij}$ and $\epsilon_{ij}$ are inputs fore equations \ref{eq:lj_elec_pot}.

An accurate set of partial charges should ideally reproduce an electrostatic potential field produced by the real molecule.
Additionally, partial charges should be compatible with Lennard-Jones parameters.
The solvent models typically come with specifically adjusted charges.
For solutes, one commonly has to perform an electronic structure calculation to find the molecules electron density distribution and then fit partial charges to it.
In the case of GAFF, a typical procedure is to use Austin model one (AM1) method to estimate the distribution of valence electrons approximately and then correct it using semi-empirical bond charge corrections (BCC) scheme, abbreviated as AM1-BCC \cite{Jakalian2002uyj}.
This approach has been well established and is known to yield good estimates of non-bonding parameters.
More sophisticated schemes, such as charge model five (CM5), use full electronic structure calculations with large basis sets and hybrid electronic density functional theories.

Once parameters have been fit, the solvation free energy can be calculated by slowly coupling (or decoupling) the solute to the solvent.
The procedure is split into two stages.
At the first stage, one turns on the solute Lennard-Jones parameters, at the second -- partial charges.
Thus, the solvation free energy can be formally split into:
\begin{equation}
\Delta F = \Delta F^{LJ} + \Delta F^{el}.
\end{equation}

The above equation presents an alternative scheme for decomposing solvation free energies (besides thermodynamic decomposition into enthalpic-entropic parts), useful in computational chemistry.
Since electrostatic free energy can be calculated relatively accurately by modelling solvent as a dielectric continuum, one can combine it with some empirical way of approximating $\Delta F^{LJ}$ to obtain a computationally cheap method for deducing solvation free energies without any solvent modelling.
Simple models that use partial charges for the electrostatic part include the generalised Born and Poisson-Boltzmann models.
Alternatively, models such as SMD or SM-12 combine an empirical $\Delta F^{LJ}$ term with continuum charge distributions derived from quantum mechanics.


\section{Ionic solvation}
\label{sec:ionic_solvation}

In the preceding sections, we discussed solvation in the bulk phase of a solvent, ignoring any effects caused by the liquid-gas boundary.
For situations when solutes are neutral such an approach is entirely justified since the range of the interfacial effects is small.
However, in the case of the charged solutes (for which the sum of atomic charges is non-zero), the nature of the liquid interface actually does affect the solvation free energy, although, the forces inside the bulk are independent of it \cite{Hunenberger2011vfc}.

Any homogeneous bulk phase $P$ will have a characteristic constant potential $\phi_G^{P}$, called the Galvani potential.
It can be expressed as a sum of the Volta potential $\psi^P$ and interfacial potential $\chi^P$:
\begin{equation}
\phi_G^P = \psi^P + \chi^P.
\end{equation}
Typically, all three potentials would depend on both the other phases in contact with the bulk phase as well as the shape of the surface.
However, an isolated phase with a homogeneous surface polarization has $\psi_P=0$, so $\phi_P = \chi_P$.
In this case, the Galvani potential is uniquely determined by properties of the phase and is independent of its shape or size \cite{Hunenberger2011vfc}.
This is the situation we will be primarily concerned with, so the terms Galvani and surface potential are going to be used interchangeably.

\begin{figure}
\centering
\includegraphics[width=0.7\linewidth]{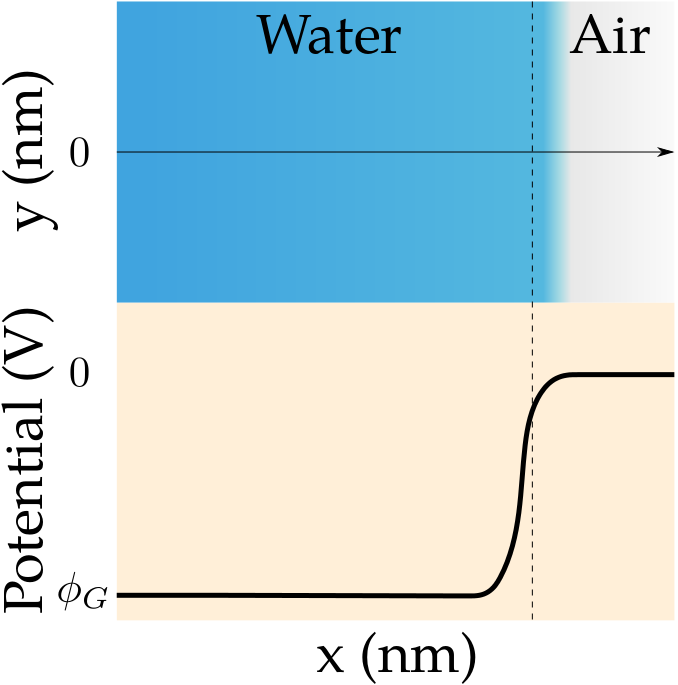}
\caption{A schematic drawing demonstrating the dependence of electrostatic potential on the distance from the water-air interface.}
\label{fig:galvani_surf}
\end{figure}

In the bulk phase, the Galvani potential is constant and thus does not affect the forces between the particles in any way.
However, it has a noticeable effect on the insertion of ions, shifting their intrinsic chemical potentials $\bar{\mu}$ by an amount proportional to the charge \footnote{The name intrinsic comes from the fact that this is a chemical potential arising purely due to the interactions within the system, without any contribution from the external field, in this case, $\phi_G^P$.}:
\begin{equation}
\mu = \bar{\mu} + q \phi_G^P
\end{equation}
where $q$ is the total charge of the ion.
In the electrochemical literature, the quantity $\mu$ is often called the "real" or electrochemical potential and intrinsic chemical potential $\bar{\mu}$ is called the "chemical potential" \cite{Cheng2012uqa}.
On the other hand, in the theory of liquids, the situation is reversed: the real chemical potential is called the "chemical potential" and is denoted as $\mu$, whereas intrinsic chemical potential is usually given a special symbol.
Throughout the thesis, we will be using the latter notation \cite{Hansen2000utg}.

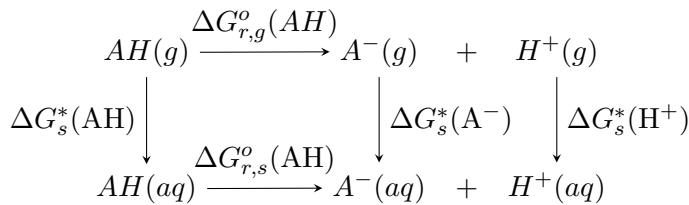
\begin{figure}[h]
    \centering
    \begin{tikzpicture}
    \matrix (m) [matrix of math nodes,row sep=3em,column sep=4em,minimum width=2em]
    {
        AH (g) & A^-(g)  &[-15mm] + &[-15mm]  H^+ (g) \\
        AH (aq)& A^-(aq) & + & H^+ (aq) \\
    };
    \path[-stealth] 
    (m-1-1) edge node [left] {$\Delta G^{\ast}_s (\ce{AH})$} (m-2-1)
    edge node [above] {$\Delta G^{o}_{r,g} (AH)$} (m-1-2)
    (m-2-1) edge node [above] {$\Delta G^{o}_{r,s} (\ce{AH})$} (m-2-2)
    (m-1-2) edge node [right] {$\Delta G^{\ast}_s (\ce{A-})$} (m-2-2)
    (m-1-4) edge node [right] {$\Delta G^{\ast}_s (\ce{H+})$} (m-2-4);
    \end{tikzpicture}
    \caption{Thermodynamic cycle relating solvation free energy of an ion \ce{A-} to the solvation free energies of acid AH, its dissociation free energies $\Delta G_{r}^o$ in both gas and solvent, and solvation free energy of the proton. Since free energy of the system is independent of the path, we have $\Delta G^{\ast}_s (\ce{AH}) + \Delta G^{o}_{r,s} (\ce{AH}) = \Delta G^{o}_{r,g} (AH) + \Delta G^{\ast}_s (\ce{A-}) + \Delta G^{\ast}_s (\ce{H+})$, from which $\Delta G^{\ast}_s (\ce{A-})$ can be deduced.}
    \label{fig:a-_solvation}
\end{figure}

The Galvani potential $\phi_G^P$, and thus $\bar{\mu}$, are experimentally inaccessible quantities.
Even though $\mu$ is in principle measurable using certain electrochemical techniques (for example by measuring absolute electrode potentials of the metal in solvent \cite{Hush1948vrm,Randles1956uyd,Hunenberger2011vfc}), a commonly involved approach to measuring chemical potentials (solvation free energies) of ions is done in a different way.
One usually measures a hydration free energy of proton and then obtains solvation free energies of other ions from appropriate thermodynamic cycles involving the dissolution of neutral compounds.
For example, the solvation free energy of an ion \ce{A-} can be obtained from the cycle shown on the figure \ref{fig:a-_solvation}.
In case an ion \ce{B+} cannot be easily protonated or does not have protons, its solvation free energy can be deduced from the solvation free energy of the ionic pair \ce{AB}, where \ce{A-} is some anion with a known free energy of transfer.

Note that the sum of solvation free energy of a pair of ions \ce{AB} is independent of the phase's Galvani potential:
\begin{equation}
\begin{split}
\mu (\ce{A^{+n}})  + \mu (\ce{B^{-n}}) &= \bar{\mu} (\ce{A^{+n}}) + n\phi_G^P  + \bar{\mu} (\ce{B^{-n}}) - n\phi_G^P\\
&= \bar{\mu} (\ce{A^{+n}})  + \bar{\mu} (\ce{B^{-n}}) = \Delta G (\ce{AB}),\\
\end{split}
\end{equation}
where $n$ is the charge on the ion.
Thus, the type (intrinsic or real) of solvation free energy of the ion is determined by the type of solvation free energy of a proton.
A number of recent articles suggest that the commonly used value for the solvation free energy of proton in water, \SI{265.9}{kcal\per\mol} by Tissandier et al. \cite{Tissandier1998tvt}, contains a contribution from surface potential \cite{Asthagiri2003uwh,Beck2013vnn,Kelly2006tom}.
It follows that hydration free energies of single ions, evaluated using Tissandier's value of proton's hydration free energy, are also "real".


\section{Applications}
\label{sec:solv_free_energy_app}


\begin{tikzpicture}

\node (dG) at (0:0) 
[align=center,font=\bfseries] {$\boldsymbol{\Delta G_s}$\\ {\scriptsize Solvation free energy}};

\node (Kh) at (-180:5) 
[align=center] {$K_H$\\ {\scriptsize Henry's Law Constant}};

\node (activity) at (-135:5)
[align=center] {$\gamma$\\ {\scriptsize Activity coefficient}};

\node (Keq) at (-90 :5)
[align=center] {$K^l_{eq}$ \\ {\scriptsize Liquid phase equilibrium}};

\node (logP) at (-45 :5)
[align=center] {$\log P$\\ {\scriptsize Water-Octanol partition}};

%

\node (Ks) at (0:5) 
[align=center] {$K_S$\\ {\scriptsize Salting out coefficient}};

\draw [->] (dG) -- (Kh) 
node[pos=.5, above,sloped,font=\tiny] {$K_H = \exp -\Delta G_s/RT$};
\draw [->] (dG) -- (activity)
node[pos=.5, above,sloped,font=\tiny] {$\gamma = \exp \Delta \Delta G_s/RT$};
\draw [->] (dG) -- (Keq)
node[pos=.5, above,sloped,font=\tiny] {$K^l_{eq} = K_{eq}^{ig} \exp \sum \upsilon \Delta G_{s,i}$};
\draw [->] (dG) -- (logP)
node[pos=.5, above,sloped,font=\tiny] {$\log P = -\Delta \Delta G_S/RT$};
\draw [->] (dG) -- (Ks)
node[pos=.5, above,sloped,font=\tiny] {$K_S = \Delta G_S + T\Delta S_S$};

\end{tikzpicture}

In this section, we will briefly overview the relation of solvation free energy (chemical potential) to other solution thermodynamic quantities that present interest to chemical engineers, as well as environmental and life scientists.

The solvation free energy is directly related to Henry's law.
Henry's law states that the amount of dissolved gas is proportional to its partial pressure above the solution \cite{Ben-naim2006vpu}.
The proportionality factor is called the Henry's law constant $H$.
This constant is used in various areas of environmental research since it describes the distribution of species between air and liquid cloud droplets, rivers, wastewaters, as well as other naturally occurring liquid reservoirs.
In the literature, a number of different definitions of Henry's law exist, with one of the most common involving dimensionless Henry constant $H^{cc} = c(solution)/c(gas)$, where $c$ stands for the molar concentration\cite{Sander2015uov}.
It can be related to Ben-Naim's solvation free energy through:
\begin{equation}
\label{eq:henrys_const}
\Delta G^{\ast} = - RT \ln H^{cc},
\end{equation}
where $R$ is the universal gas constant.
From the previous equation, it follows that solvation free energy and Henry's law constant are two different names for the same quantity.

Knowing solvation free energies at different concentrations allows one to compute solute activities, which are useful for understanding the properties of concentrated solutions.
Assuming finite number density of solute $\rho_u$, we can recast expression \ref{eq:widom-mu} into the following form \cite{Ben-naim2006vpu}
\begin{equation}
\mu = \mu^{o} + kT \ln \rho_u + kT \ln \gamma^{D,\rho},
\end{equation}
where  $\gamma^{D,\rho}$ is the activity coefficient and $\mu^{o}$ is the standard chemical potential.
Above quantities are measured on the number density $\rho$ scale, but can be converted to more commonly used molar or molal scales \cite{Ben-naim2006vpu}.
Relating this equation to our previous results one finds that
\begin{equation}
\mu^{\ast}_{\rho_u} - \mu^{\ast}_{0} = \Delta G^{\ast}_{\rho_u} - \Delta G^{\ast}_{0} = kT \ln \gamma^{D,\rho},
\end{equation}
where subscripts $\rho_u$ and $0$ correspond to finite and infinitely dilute concentrations of solute respectively.

Finally, solvation free energies can be used for predicting the equilibrium state of reactions and complex formations in different mediums.
Thus, they present a considerable interest for areas of chemistry, biology, and material science that are concerned with the formation of various compounds in solutions.
Consider the following reaction:
\begin{align}
\ce{A <--> B}.
\end{align}
The equilibrium constant $K$ and Gibbs free energy are related as $K^{ig} = \exp ( - \beta \Delta G^{ig}_r)$, where superscript $ig$ indicates that reaction takes place in ideal gas phase.
To obtain the equilibrium constant in the liquid phase we construct a thermodynamic cycle similar to figure \ref{fig:a-_solvation}. Then:
\begin{equation}
K^{l} = \exp [ - \beta ( \Delta G^{ig}_r - \Delta G^{\ast}_A + \Delta G^{\ast}_B)] = K^{ig} \exp[-\beta(\Delta G^{\ast}_B - \Delta G^{\ast}_A)].
\end{equation}
Knowledge of solvation free energies allows one to compute reaction equilibria in any medium from the free energy of the gas phase reaction, which can often be computed relatively accurately using quantum chemistry based methods or estimated on the basis of the bond strengths.

\chapter{Theory of simple liquids} 

\label{Chapter3} 

\lhead{Chapter 3. \emph{Simple liquids}}

In this chapter, we overview the main aspects of classical density functional theory. 
The focus is on the exact results that can be obtained for simple liquids.
Both the derivations and the structure of the chapter are based on Ref. \citenum{Hansen2000utg}.

\section{Particle densities and distributions}
\label{sec:particle_densities}

Condensed matter systems such as liquids and solids are difficult to study.
They consist of a large number of electrons and atomic nuclei interacting with each other in a complicated manner.
To describe such systems one usually has to ignore unimportant degrees of freedom.
For example, when studying molecules at room temperature vibrations of bonds and angles can be neglected.

A systematic approach towards the reduction of degrees of freedom is called coarse-graining.
The basic idea is to replace values of a rapidly varying observable with its average value over certain volume \cite{Gorban2006tse}.
As a result, we get a continuous and smoothly varying function that is called an order parameter field.

For the description of liquids, a natural order parameter is an ensemble average of the instantaneous density $\rho^I(\rvec)$:
\begin{equation}
\rho(\rvec)  = \left \langle \rho^I (\rvec) \right\rangle = \left \langle \sum\limits_{k=1}^{N} \delta (\rvec - \rvec_k) \right \rangle,
\end{equation}
where $\delta(\rvec)$ is Dirac's delta function and $N$ here is the total number of atoms in a system.
$\rho(\rvec)$ is usually called a single-particle density distribution, or local density, and it shows an average density of atoms in the volume element $\diff\rvec$.
The product $\rho(\rvec) \diff \rvec$ thus shows an average number of particles at that position.

A related quantity, called the pair distribution function, describes the correlation between single-particle densities:
\begin{equation}
\begin{split}
\rho^{(2)}(\rvec_1, \rvec_2) &=  \left\langle \sum\limits_{l=1}^{N}\sum_{\substack{m=1 \\ m\neq l}}^{N} \delta (\rvec_1 - \rvec_l) \delta (\rvec_2 - \rvec_m) \right\rangle\\
&= \left\langle \rho^I(\rvec_1) \rho^I(\rvec_2) \right\rangle -  \rho (\rvec_1) \delta(\rvec_1 - \rvec_2).\\
\end{split}
\label{eq:pair_distribution}
\end{equation}
Note that due to the $m \neq l$ condition, two particles found in positions $\rvec_1$ and $\rvec_2$ must be different.
Thus, $\rho^{(2)}(\rvec_1, \rvec_2)\diff\rvec_1\diff\rvec_2$ can be interpreted as an average number of pairs formed by particles in volume element $\diff\rvec_1$ with particles in a volume element $\diff\rvec_2$ with a condition that two particles in a pair must be different \cite{Ben-naim2006vpu}.

It is useful to consider integrals of particle distribution functions.
Using the property of the delta function
\begin{equation}
\label{eq:density_normalization}
\int_V \rho(\rvec) \diff \rvec = \left\langle N \right\rangle_V,
\end{equation}
where $\left\langle N \right\rangle_V$ is the total number of particles in the box $V$.
In homogeneous liquids, local density should be constant by the definition.
It follows that $\rho(\rvec) = N/V = \rho$. 
Similarly, we can take a double integral of pair distribution function to find
\begin{equation}
\iint_V \rho^{(2)}(\rvec_1,\rvec_2)\diff\rvec_1\diff\rvec_2 = \left\langle N^2 \right\rangle_V -\left\langle N \right\rangle_V.
\end{equation}
Then, the volume average of the pair distribution function is
\begin{equation}
\rho^{(2)}_{avg,V} = \frac{\left\langle N^2 \right\rangle_V -\left\langle N \right\rangle_V}{V^2} =\rho^2 \left(1 - \frac{1}{N}\right) \approx \rho^2.
\end{equation}
Since in the ideal gas there are no correlations, the value of pair distribution function in it must be constant throughout the box, or in other words: $\rho^{(2)}_{id}(\rvec_1,\rvec_2) = \rho^{(2)}_{avg}$.

\begin{figure}
	\centering
	\includegraphics[width=0.6\linewidth]{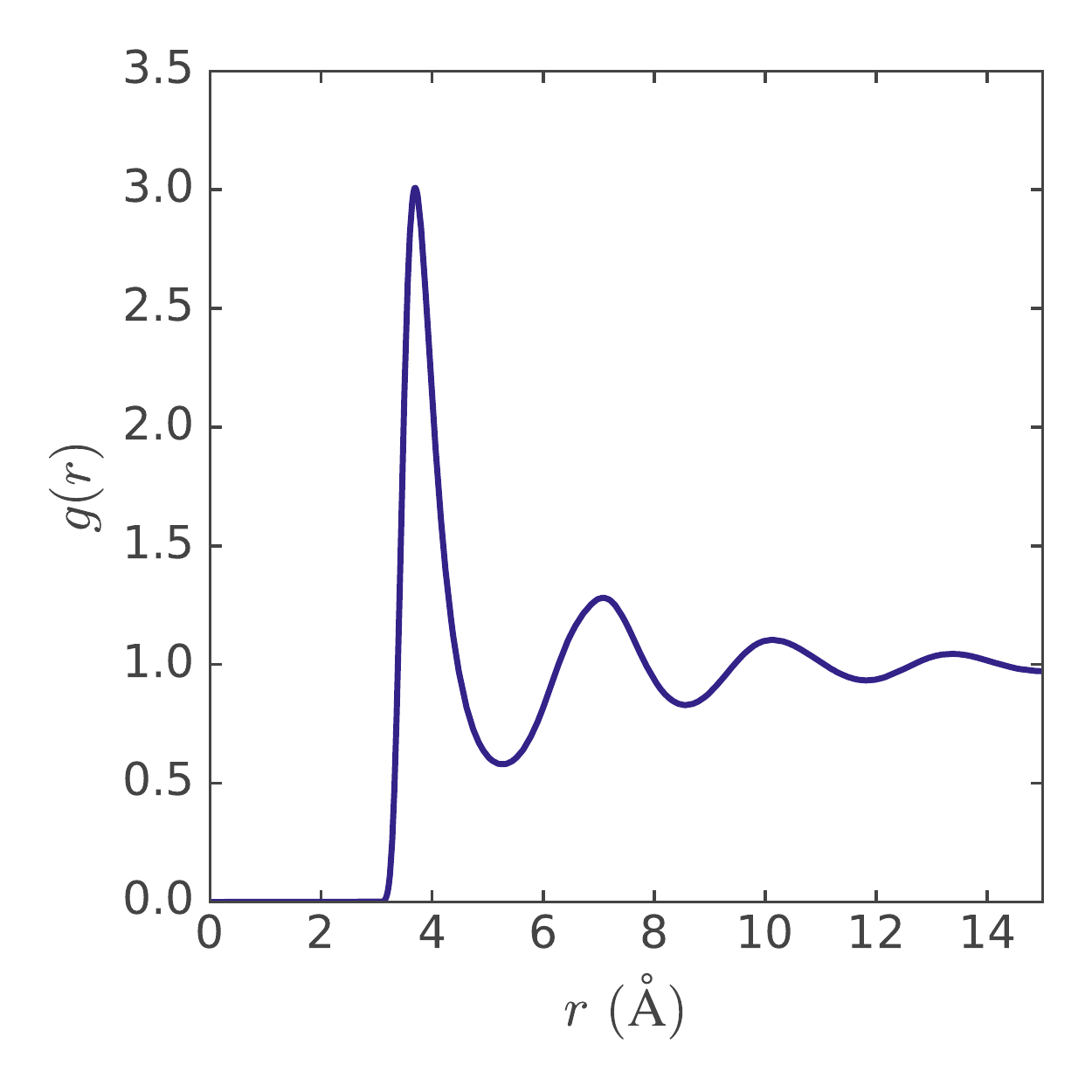}
	\caption{Radial distribution function of \ce{Ar} near its boiling point obtained via molecular dynamics simulations using parameters from Ref. \citenum{Toth2001vdf}.}
	\label{fig:argon_gr}
\end{figure}

In real liquids, as the distance between $\rvec_1$ and $\rvec_2$ increases, the value of the pair distribution function tends to the ideal limit: $\rho^{(2)}(\rvec_1,\rvec_2) \to \rho(\rvec_1) \rho(\rvec_2)$.
Hence, it is useful to define a pair correlation function
\begin{equation}
g(\rvec_1, \rvec_2) = \frac{\rho^{(2)}(\rvec_1, \rvec_2)}{\rho(\rvec_1) \rho(\rvec_2)},
\label{eq:pari_distr_func}
\end{equation}
which quantifies deviation of $\rho^{(2)}(\rvec_1,\rvec_2)$ from its large distance behaviour.
In isotropic liquids, the pair distribution function only depends on the distance between two particles $g(\rvec_1, \rvec_2) = g(|\rvec_1 - \rvec_2|) = g(r)$, with $g(r)\to 1$ as $r \to \infty$.
This spherically averaged pair distribution function is often referred to as the radial distribution function.
An example of this function for argon near boiling point is shown in figure \ref{fig:argon_gr}.

Finally, following Ben-Naim, we also define conditional local density:
\begin{equation}
\rho(\rvec_2/\rvec_1) = \frac{\rho^{(2)}(\rvec_1,\rvec_2)}{\rho(\rvec_1)},
\end{equation}
which shows the local density at $\rvec_2$, given a particle in $\rvec_1$.
From equation \ref{eq:pari_distr_func} it follows that $g(\rvec_1,\rvec_2) = \rho(\rvec_2/\rvec_1)/\rho(\rvec_2)$ for isotropic systems $g(r) = \rho(\rvec_2/\rvec_1)/\rho$.
These equations suggest an interpretation of the pair correlation function as a scaled local density of a system, in which a single particle is fixed at $\rvec_1$ and other particles are moving in its field.
Using this interpretation, we can view both inhomogeneous and homogeneous liquids under the same framework.

\section{Free energy functionals}
\label{sec:free_energy_functionals}

The basic ideas behind the classical density functional theory of liquids can be illustrated in the following manner.
Suppose we have (in general) an inhomogeneous system of interacting particles.
Such system can be split into small volume elements containing $n_j = \Delta V_j \rho_j$ number of particles.
As particles are free to move from one compartment to another, the free energy of each individual volume element is best described using the grand potential $\Omega_j = F_j - \mu n_j$, where $\mu$ is the chemical potential, constant in each compartment.
To satisfy grand canonical ensemble condition, the whole system must be connected to external heat and particle reservoirs.


Assume that there is a spatially varying external potential $\phi_j$ that interacts with particles.
Then, the Helmholtz free energy of each volume component is given by $F_j = \Fcal^{ig} + \Fcal^{ex} + n_j \phi_j$, where $\Fcal$ is an intrinsic Helmholtz free energy and superscripts $ig$ and $ex$ denote ideal and excess contributions.
Taking the derivative of the above expression with respect to the number of particles in a box, we get:
\begin{equation}
\left(\frac{\partial F_j}{\partial n_j}\right)_{V,T} = \mu = \bar{\mu}^{ig}_j + \bar{\mu}^{ex}_j + \phi_j,
\end{equation}
where $\bar{\mu}$ stands for the intrinsic chemical potential, familiar from the section \ref{sec:ionic_solvation}.
Notice that while the total chemical potential $\mu$ is constant throughout the system, the quantity $\bar{\mu}_j = \bar{\mu}^{ig} + \bar{\mu}^{ex}_j$ is spatially varying.
Also, since $\mu^{ig} = \bar{\mu}^{ig}$, we will ignore the superscript.


The simple arguments described in the previous two paragraphs can be made formal by shrinking the volume of each box element to an arbitrarily small value and substituting $\rho_j$ with single particle density, introduced in the previous section.
The total grand potential becomes a functional of density \cite{Hansen2000utg}
\begin{equation}
\label{eq:grand_functional}
\Omega [\rho] = \Fcal[\rho] + \int \rho (\rvec) \phi(\rvec) \diff \rvec - \mu \int \rho (\rvec) \diff \rvec,
\end{equation}
with $\Fcal$ being an intrinsic free energy functional related to Helmholtz free energy $F$ in the following way
\begin{equation}
\Fcal[\rho] = F[\rho] - \int \rho (\rvec) \phi(\rvec) \diff \rvec.
\end{equation}
The intrinsic Helmholtz free energy turns out to be a much more useful quantity than normal Helmholtz free energy for the description of these inhomogeneous systems.

Similarly to what we had before, $\Fcal$ can be split into two parts $\Fcal = \Fcal^{id} + \Fcal^{ex}$, where $\Fcal^{id}$ is an ideal part, given by
\begin{equation}
\Fcal^{id} = kT \int \rho(\rvec) \left\{\ln\left[\Lambda^3\rho(\rvec)\right] - 1\right\}\diff\rvec.
\end{equation}
This is the same equation as we found in section \ref{sec:connection_to_thermodynamics} and $\Fcal^{ex}$ is an excess contribution (relative to an ideal gas).
Notice that we dropped an internal partition function $q$ since we are going to deal with liquids of particles without any internal structure in this section.

While the explicit form of $\Fcal^{ex}[\rho]$ is usually unknown, it is possible to compute the change of $\Fcal^{ex}$ relative to some reference system. 
Similarly to the approach presented in section \ref{sec:chemical_pot}, one first splits the pair potential between densities into reference and perturbation parts:
\begin{equation}
u_{\lambda} (\rvec_1, \rvec_2) = u_0 (\rvec_1, \rvec_2) + \lambda u (\rvec_1, \rvec_2).
\end{equation}
By gradually increasing the perturbation part of the interaction between particles, we can find the change in the excess free energy:
\begin{equation}
\Fcal^{ex}[\rho] - \Fcal^{ex}[\rho_0] = \frac{1}{2} \int\limits_{0}^{1} \diff \lambda \iint \rho^{(2)}(\rvec_1,\rvec_2; \lambda) u(\rvec_1, \rvec_2) \diff \rvec_1 \rvec_2.
\end{equation}
The formula above forms the basis for various perturbation theories and simplifications.

Together with n-particle densities, free energy functionals form a useful set of tools to study liquid systems.
The particle densities describe the structure of the liquid, while free energy functionals incorporate energetic information.
A rigorous basis for these ideas, called density functional theory, is summarised in two results, called Hohenberg–Kohn–Mermin theorems \cite{Hansen2000utg,Hohenberg1964tzx,Mermin1965wln}.

The first theorem states that for a given $\mu$, $T$ and $V$,the  equilibrium density distribution $\rho_{eq}(\rvec)$ is uniquely determined by an external potential $\phi(\rvec)$ acting on the system.
Thus, the equilibrium particle distribution $\rho_{eq}(\rvec; \phi)$ is a unique functional of the external potential.
As a result, it follows that the intrinsic free energy functional $\Fcal[\rho]$ is a unique functional of the single particle density $\rho(\rvec)$.

The second theorem states that equilibrium density $\rho_{eq}$ minimizes the grand potential $\Omega$ for a given external field $\phi$:
\begin{equation}
\left(\frac{\delta \Omega}{\delta \rho(\rvec)}\right)_{\rho=\rho_{eq}, \phi} = 0
\end{equation}
and
\begin{equation}
\Omega [\rho; \phi] \ge \Omega,
\end{equation}
where equality only applies when $\rho(\rvec) = \rho_{eq}$.

These theorems and density functional theory, in general, can be applied to both quantum and classical systems.
Since its original formulation in the 1960s, the theory has been mostly applied in many-electron systems \cite{Labanowski1991tge,Giustino2014wwd}, although quite a lot of work has also been done in the field of classical liquids \cite{Henderson1992vag,Evans2016tct}.
The field is too broad to cover completely.
Thus, we will only cover the results associated with 3D-RISM and related theories, leaving more advanced approaches mostly for future work.

\section{Functional derivatives and correlations}
\label{sec:func_deriv_and_cor}

Particle densities and correlations can be naturally obtained as derivatives of free energy functionals.
This approach provides more insight into their relation with each and with various system properties.
In this section we define a number of useful functions that can be obtained by differentiating free energy functionals and relate them using Ornstein-Zernike equation.

We start by rewriting equation \ref{eq:grand_functional}:
\begin{equation}
\begin{split}
\Omega [\rho] &= \Fcal[\rho] + \int \rho (\rvec) \left[\phi(\rvec) - \mu \right] \diff \rvec\\
&= \Fcal[\rho] - \int \rho (\rvec) \bar\mu(\rvec) \diff \rvec.\\
\end{split}
\end{equation}
The intrinsic free energy functional does not explicitly depend on $\bar\mu$.
It follows then that:
\begin{equation}
\frac{\delta \Omega [\rho]}{\delta \bar{\mu}(\rvec)} = -\rho(\rvec),
\label{eq:first_omega_derivative}
\end{equation}
where $\delta$ indicates a functional derivative.
This equation can also be used as an alternative definition of a single particle density \cite{Hansen1996tmk}.

After lengthy algebraic manipulations, it is possible to show that the second derivative of the grand potential with respect to intrinsic chemical potential gives a correlation between density fluctuations \cite{Hansen2000utg}:
\begin{equation}
\begin{split}
- kT\frac{\delta^2 \Omega[\rho]}{\delta \bar\mu(\rvec_1) \delta \bar\mu(\rvec_2) } & = \chi(\rvec_1,\rvec_2)\\
&  = \left\langle \left[\rho^I(\rvec_1) - \rho(\rvec_1) \right] \left[ \rho^I(\rvec_2) - \rho(\rvec_2) \right]\right\rangle\\
& = \left\langle  \rho^I(\rvec_1) \rho^I(\rvec_2)\right\rangle - \rho(\rvec_1) \rho(\rvec_2)\\
& =  \rho^{(2)}(\rvec_1, \rvec_2) + \rho(\rvec_1)\delta(\rvec_2 - \rvec_1)- \rho(\rvec_1) \rho(\rvec_2).\\
\end{split}
\label{eq:density-density_correlation_f}
\end{equation}
The last equality in the above equation follows from equation \ref{eq:pair_distribution}.
The function $\chi(\rvec_1,\rvec_2)$ is called a density-density correlation function.
Note that using equation \ref{eq:first_omega_derivative} we can rewrite the above functional derivative in a number of different ways that will be useful in the later chapters
\begin{equation}
\chi(\rvec_1,\rvec_2) = - \frac{\delta^2 \beta \Omega[\rho]}{\delta \beta \bar\mu(\rvec_1) \delta \beta \bar\mu(\rvec_2) } = \frac{\delta \rho(\rvec_1)}{\delta\beta \bar{\mu}(\rvec_2)}.
\label{eq:chi_alternative}
\end{equation}
It is clear that at large separations where $r = | \rvec_2 - \rvec_1 |$ goes to $\infty$, the fluctuations of density become independent of one another and $\chi(\rvec_1,\rvec_2)$ should approach $0$.

Let us define a total correlation function $h(\rvec_1, \rvec_2)$, given by
\begin{equation}
h(\rvec_1, \rvec_2) = \frac{\rho^{(2)}(\rvec_1, \rvec_2) - \rho(\rvec_1) \rho(\rvec_2)}{\rho(\rvec_1) \rho(\rvec_2)} = g(\rvec_1, \rvec_2) - 1.
\label{eq:h_r_def}
\end{equation}
Then using equations \ref{eq:density-density_correlation_f} and \ref{eq:h_r_def}
\begin{equation}
\begin{split}
\chi(\rvec_1,\rvec_2) = \rho(\rvec_1) \rho(\rvec_2)h(\rvec_1, \rvec_2) + \rho(\rvec_1) \delta (\rvec_2 - \rvec_1)
\end{split}
\label{eq:density-density_cor}
\end{equation}
In the absence of any inter-particle correlations (ideal gas), $\rho^{(2)}(\rvec_1,\rvec_2) \approx \rho(\rvec_2)\rho(\rvec_1)$ for any two points, and thus $h(\rvec_1,\rvec_2) = 0$.
The total correlation function contains only non-ideal (excess) pair correlations.

An alternative family of correlation functions emerges if we take the functional derivatives of the intrinsic free energy with respect to density.
Writing the first derivative we get
\begin{equation}
\frac{\delta \Fcal}{\delta \rho (\rvec)} = \frac{\delta \Fcal^{id}}{\delta \rho (\rvec)} + \frac{\delta \Fcal^{ex}}{\delta \rho (\rvec)} = \mu^{ig}(\rvec) + \bar{\mu}^{ex} (\rvec).
\label{eq:intrinsic_derivatives}
\end{equation}
Ignoring the ideal part, we define the derivative of excess free energy as a direct correlation function:
\begin{equation}
c(\rvec) = -\beta \frac{\delta \Fcal^{ex}[\rho]}{\delta \rho (\rvec)} = -\beta \bar{\mu}^{ex} (\rvec).
\end{equation}

The direct correlation function incorporates the effects of many-body interactions within the system.
One can find its relationship with single particle density by considering derivative of grand potential $\frac{\delta \Omega}{\delta \rho(\rvec)} = kT \ln \left[\Lambda^3 \rho(\rvec)\right] - kTc(\rvec) + \phi(\rvec) = 0$. 
Rearranging the last result we get $\Lambda^3 \rho(\rvec) = \exp \left[\beta \phi (\rvec) + c(\rvec) \right]$, which reveals the direct correlation function as a type of generalised potential.

Analogously, we can define a two-particle direct correlation function:
\begin{equation}
c^{(2)}(\rvec_1,\rvec_2) = - \beta \frac{\delta^2 \Fcal^{ex}}{\delta \rho(\rvec_1) \delta\rho(\rvec_2)} = \frac{\delta c(\rvec_1)}{\delta \rho(\rvec_2)},
\end{equation}
which reveals the effect of density change at $\rvec_2$ on the direct correlation function at $\rvec_1$.
The higher order direct correlation functions can be obtained in a similar way.

\begin{figure}
\centering
\includegraphics[width=\linewidth]{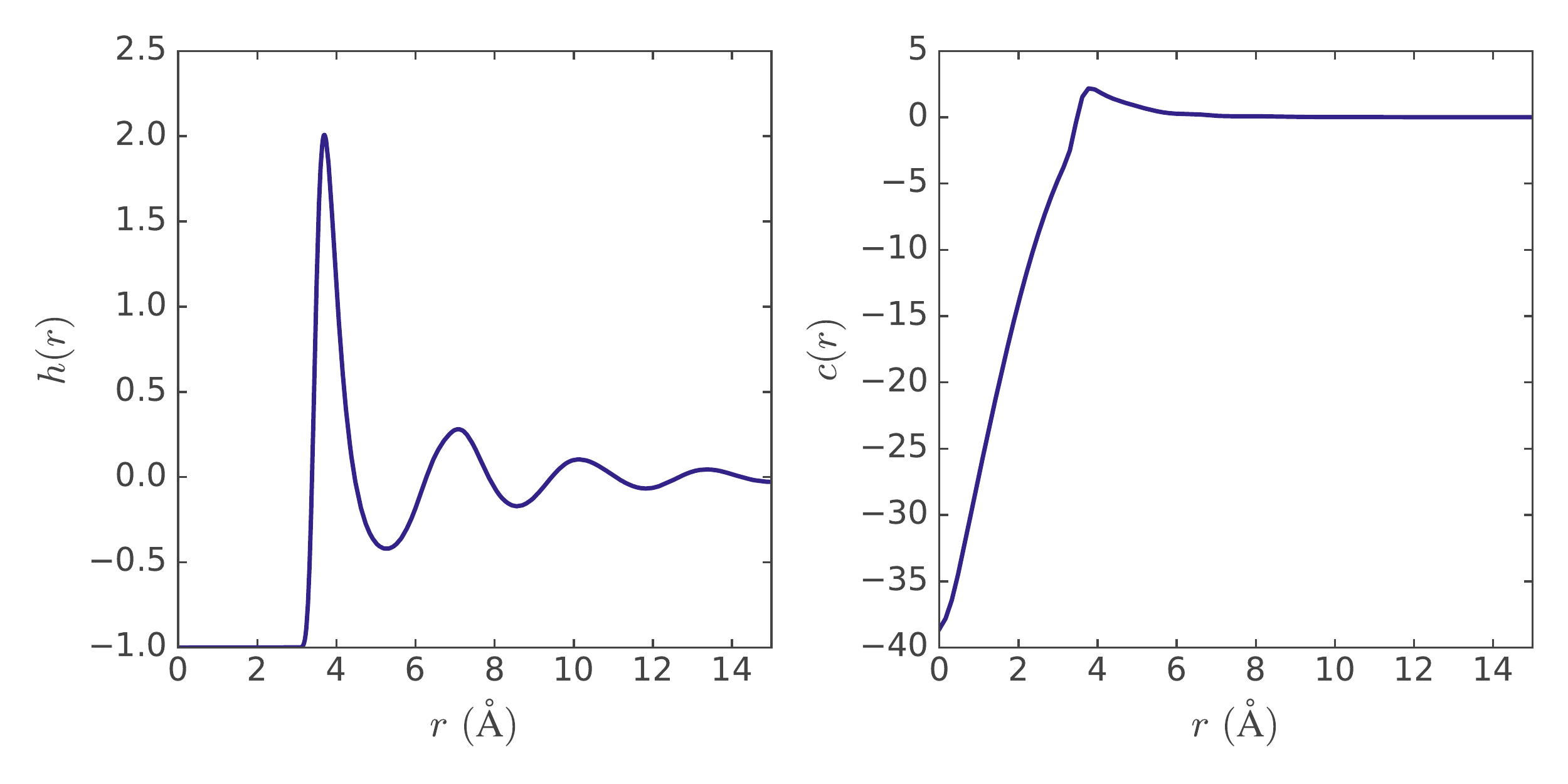}
\caption{Total correlation function (left) and direct correlation function (right) of liquid argon near its boiling point. The correlation functions were extracted from molecular dynamics simulations performed with parameters taken from Ref. \citenum{Toth2001vdf}.}
\label{fig:h_and_c_of_r}
\end{figure}

In isotropic liquids, the total and direct correlation functions depend only on the separation between particles.
Both of them also tend to $0$ as $r \to \infty$.
An example of a typical behaviour of these functions for simple liquids is shown in figure \ref{fig:h_and_c_of_r}.

Is it possible to relate the two types of correlation functions we defined above?
The answer is yes.
To obtain a meaningful relation, we start by writing down the identity that follows from the equation \ref{eq:chi_alternative} \cite{Hansen2000utg}:
\begin{equation}
\int \chi(\rvec_1,\rvec_3) \chi^{-1}(\rvec_3,\rvec_2)\diff \rvec' = 
\int \frac{\delta \rho(\rvec_1)}{\delta \beta \bar\mu(\rvec_3)} \frac{\delta \beta \bar\mu(\rvec_3)}{\delta \rho (\rvec_2)} \diff \rvec_3 = \delta(\rvec_2 - \rvec_1),
\label{eq:func_inverse_identity}
\end{equation}
where $\chi^{-1}(\rvec_1,\rvec_2)$ is the functional inverse of the density-density correlation function.
Using equation \ref{eq:intrinsic_derivatives} we can express it via direct correlation function:
\begin{equation}
\label{eq:intrinsic_chem_pot_derivative}
\begin{split}
\chi^{-1}(\rvec_1,\rvec_2) & = \frac{\delta\beta\bar\mu(\rvec_1)}{\delta \rho(\rvec_2)} = \frac{\delta \ln \Lambda^3 \rho(\rvec_1)}{\delta \rho(\rvec_2)} -  \frac{\delta c(\rvec_1)}{\delta \rho(\rvec_2)}\\
&=  \frac{1}{\Lambda^3 \rho(\rvec_1)}\frac{\delta \Lambda^3 \rho(\rvec_1)}{\delta \rho(\rvec_2)} -  \frac{\delta c(\rvec_1)}{\delta \rho(\rvec_2)}\\
& =\frac{1}{\rho(\rvec_1)}\delta(\rvec_1 - \rvec_2) - c(\rvec_1, \rvec_2).
\end{split}
\end{equation}
Plugging this result back into \ref{eq:func_inverse_identity} we find:
\begin{equation}
\label{eq:oz_functional_inverse}
\begin{split}
\int [&\rho(\rvec_1) \rho(\rvec_3)h(\rvec_1, \rvec_3) + \rho(\rvec_1) \delta (\rvec_3 - \rvec_1)] \left[\frac{\delta(\rvec_3 - \rvec_2)}{\rho(\rvec_3)} - c(\rvec_3, \rvec_2)\right] \diff \rvec_3\\
=& \int \Big[ \rho(\rvec_1) h(\rvec_1, \rvec_3) \delta(\rvec_3-\rvec_2) - \rho(\rvec_1) \rho(\rvec_3) h(\rvec_1,\rvec_3) c(\rvec_3, \rvec_2)\\
&+ \frac{\rho(\rvec_1)}{\rho(\rvec_3)} \delta (\rvec_3-\rvec_2)\delta(\rvec_3-\rvec_1) - \rho(\rvec_1)\delta(\rvec_3 - \rvec_1) c(\rvec_3,\rvec_2)\Big] \diff \rvec_3\\
=& \rho(\rvec_1)h(\rvec_1, \rvec_2) - \rho(\rvec_1) c(\rvec_1,\rvec_2) - \int \rho(\rvec_1)\rho(\rvec_3) h(\rvec_1,\rvec_3) c(\rvec_3, \rvec_2) \diff \rvec_3\\
&+ \delta(\rvec_2 - \rvec_1) =  \delta(\rvec_2 - \rvec_1).\\
\end{split}
\end{equation}
Cancelling delta functions and dividing everything by $\rho(\rvec_1)$ we get the famous Ornstein-Zernike relation:
\begin{equation}
h(\rvec_1, \rvec_2) - c(\rvec_1,\rvec_2) = \int \rho(\rvec_3) h(\rvec_1,\rvec_3) c(\rvec_3, \rvec_2) \diff \rvec_3
\end{equation}
that will be the topic of next section.

\section{Ornstein-Zernike and mixtures}
\label{sec:oz_and_mixtures}

The physical meaning of the Ornstein-Zernike equation can be understood by expressing the total correlation function in terms of direct correlation functions:
\begin{equation}
\begin{split}
h(\rvec_1, \rvec_2) &= c(\rvec_1, \rvec_2) + \int c(\rvec_1, \rvec_3) \rho(\rvec_3) c(\rvec_3, \rvec_2) \diff \rvec_3 \\
&+ \iint c(\rvec_1, \rvec_3) \rho(\rvec_3) c(\rvec_3,\rvec_4) \rho(\rvec_4) c(\rvec_4,\rvec_2) \diff \rvec_3 \diff \rvec_4 + \cdots
\end{split}
\end{equation}
The total correlation of densities (particles) at point $1$ and point $2$ is given by direct correlation of densities plus correlation through one intermediate point, two intermediate points and so on.

For homogeneous and isotropic fluids the equation can be simplified:
\begin{equation}
h(r) = c(r) + \rho \int c(|\rvec' - \rvec|) h(r') \diff \rvec'.
\label{eq:isotropic_OZ}
\end{equation}
We can see that as $\rho\to0$, $h(r)\approx c(r)$, indicating that at low densities the correlation between particles is purely due to "direct" interactions between them, while as densities become larger, indirect interactions start playing a greater role.

To simplify equation \ref{eq:isotropic_OZ} further, we need to use a Fourier transform, which we define in the next two paragraphs.
In this thesis, we will use the following convention of a Fourier transform $F$
\begin{equation}
F\left\{ f \right\} (\kvec) = \f f (\kvec) = \int f(\rvec) \exp(-i \kvec \cdot \rvec) \diff \rvec,
\end{equation}
where $\cdot$ stands for dot product.
The inverse Fourier transform is given by
\begin{equation}
F^{-1}\left\{ \f f \right\} (\rvec) = f(\rvec) = \frac{1}{(2\pi)^3} \int \f f(\kvec) \exp(i \kvec\cdot\rvec) \diff \kvec.
\end{equation}
Generally, $\f f(\kvec)$ is a complex valued function. However, the majority of transforms in this thesis are going to be performed on spherically symmetric functions for which $\f f(\kvec)$ is strictly real.
Moreover, in that case, the Fourier transform simplifies and we have \cite{Sergiievskyi2013upd}:
\begin{equation}
\f f(k) = \frac{4\pi}{k} \int\limits_{0}^{\infty} f(r) r \sin(kr) \diff r
\end{equation}
and
\begin{equation}
f(r) = \frac{1}{2\pi r} \int\limits_{0}^{\infty} \f f(k) k \sin(kr) \diff k.
\end{equation}
We will be mostly applying Fourier transforms on convolutions since they simplify considerably in k-space:
\begin{equation}
F\left\{ \int f(\rvec') g(\rvec - \rvec')\diff\rvec \right\} (\kvec) = \hat{f}(\kvec) \hat{g} (\kvec).
\label{eq:convol_theorem}
\end{equation}

\begin{figure}
	\centering
	\includegraphics[width=\linewidth]{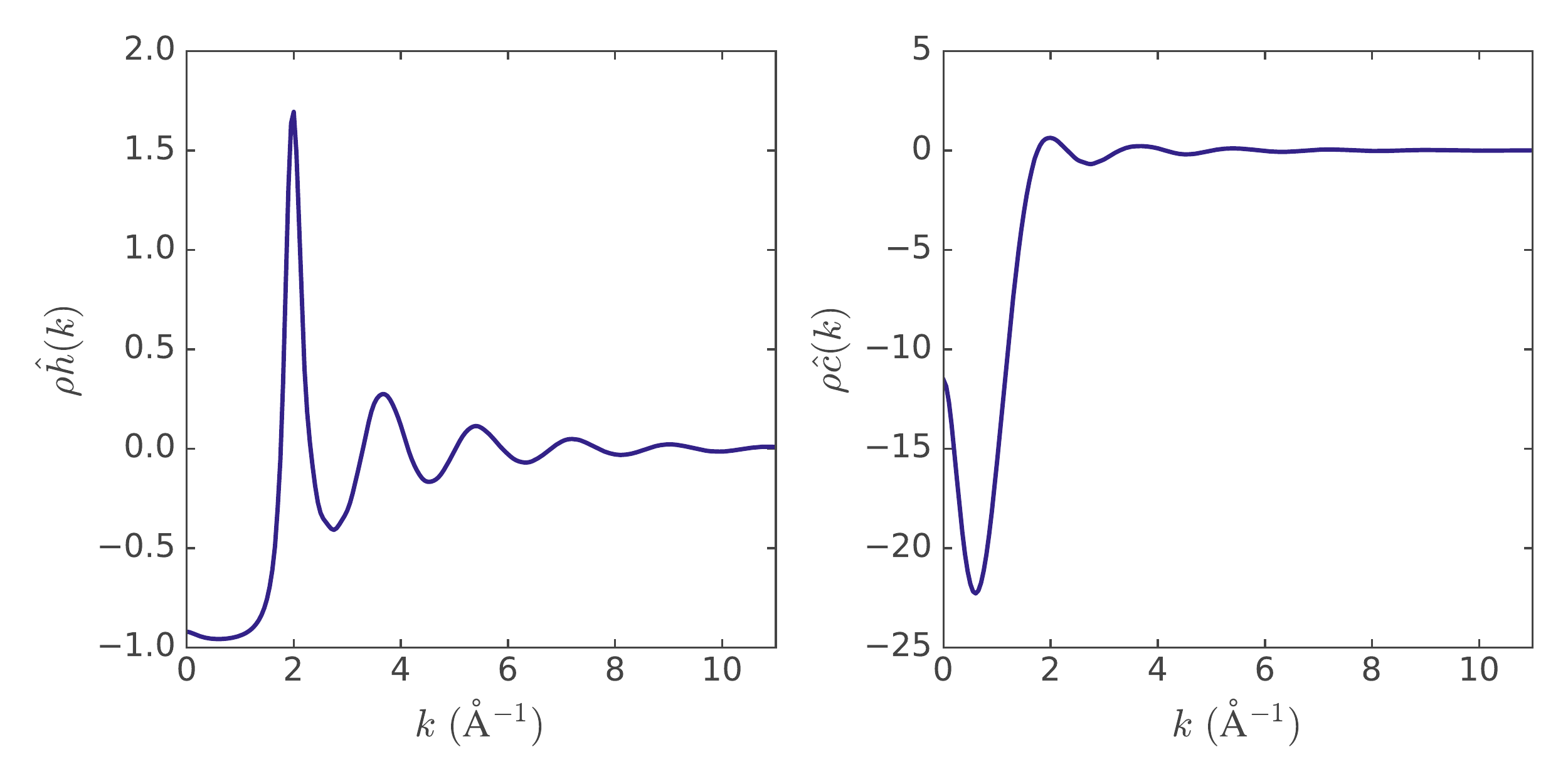}
	\caption{Fourier transforms of the total correlation function (left) and the direct correlation function (right) from figure \ref{fig:h_and_c_of_r}.}
	\label{fig:hk_and_ck_of_k}
\end{figure}

Coming back to equation \ref{eq:isotropic_OZ} and taking Fourier transform of both sides we can write down:
\begin{equation}
\hat{h}(k) = \hat{c}(k) + \rho \hat{c}(k) \hat{h}(k).
\label{eq:oz_k_space}
\end{equation}
We obtain a simple algebraic equation, from which we can obtain expressions for $h(k)$ or $c(k)$
\begin{align}
\hat{h}(k) = \frac{\hat c(k)}{1 - \rho \hat c(k)} && \hat{c}(k) = \frac{\hat h(k)}{1 + \rho \hat h(k)}.
\end{align}

The above results can be readily extended to mixtures.
Consider a system containing $n$ different types of particles labelled $i=1\ldots n$.
The average density of a particle of type $i$ is given by
\begin{equation}
\rho_i = \frac{N_i}{V} = x_i \rho,
\label{eq:mol_ratio}
\end{equation}
where $N_i$ is the total number of particles of type $i$ and $x_i$ is their mole ratio.

While the definition of single particle density $\rho_i(\rvec)$ in mixtures remains the same, the expression for two-particle density $\rho^{(2)}_{ij}(\rvec_1,\rvec_2)$ depends on whether $i$ and $j$ are the same species.
If $i=j$, equation \ref{eq:pair_distribution} still holds, but for $i\neq j$ we no longer have to worry about the correlation of particle with itself:
\begin{equation}
\rho^{(2)}_{ij}(\rvec_1,\rvec_2) = \left\langle \sum\limits_{l=1}^{N_i} \sum\limits_{m=1}^{N_j} \delta(\rvec_1 - \rvec_l) \delta(\rvec_2 - \rvec_m)\right\rangle = \left\langle \rho_i^I(\rvec_1) \rho_j^I(\rvec_2) \right\rangle.
\end{equation}
These two conditions can be summarised in a single equation using the Kronecker delta $\delta_{ij}$:
\begin{equation}
\rho^{(2)}_{ij} (\rvec_1, \rvec_2) = \langle \rho^I_i (\rvec_1) \rho^I_j(\rvec_2)\rangle - \delta_{ij} \rho(\rvec_1)\delta(\rvec_1 - \rvec_2).
\end{equation}
Similarly, for the density-density correlation function we have
\begin{equation}
\chi_{ij}(\rvec_1,\rvec_2) = \frac{\delta\rho_i(\rvec_1)}{\delta\bar{\mu}_j(\rvec_2)} = h_{ij}\rho_i(\rvec_1) \rho_j(\rvec_j) + \delta_{ij} \delta(\rvec_1 - \rvec_2) \rho_i(\rvec_1).
\end{equation}

The multicomponent Ornstein-Zernike equation is similar to its single component version:
\begin{equation}
h_{ij}(\rvec_1, \rvec_2) = c_{ij}(\rvec_1,\rvec_2) + \sum\limits_{m=1}^{n} \rho_m (\rvec_3) c_{im}(\rvec_1,\rvec_3)h_{mj}(\rvec_3,\rvec_2) \diff \rvec_3,
\label{eq:oz_simple_mixtures}
\end{equation}
where $n$ is the number of components in mixture. 
The main difference is that in the multicomponent case we have to account for interactions through other types of species.
For a homogeneous mixture we have:
\begin{equation}
h_{ij}(r) = c_{ij}(r) + \sum\limits_{m=1}^{n} \rho_k c_{im}(|\rvec_1 - \rvec_2|) h_{mj}(r_2) \diff \rvec_2
\label{eq:oz_simple_homo_mixtures}
\end{equation}
in real space, and
\begin{equation}
h_{ij}(k) = c_{ij}(k) + \sum\limits_{m=1}^{n} \rho_m c_{im}(k) h_{mj}(k)
\end{equation}
in Fourier space. Summation over indices suggests a convenient matrix form:
\begin{equation}
\vect{H}(k) = \vect{C}(k) + \vect{D} \vect{C}(k)\times \vect{H}(k),
\end{equation}
where $\vect{D}$ is the diagonal matrix of species densities:
\begin{equation}
\vect{D} = \begin{bmatrix}
\rho_{1}       & 0 & 0 & \dots & 0 \\
0       & \rho_{2} & 0 & \dots & 0 \\
\vdots & \vdots & \vdots & \ddots & \vdots \\
0      & 0 & 0 & \dots & \rho_n
\end{bmatrix},
\label{eq:density_matrix}
\end{equation}
and correlation functions are grouped into matrices in which $ij$-th element describes the correlation between particles of type $i$ and $j$: $\vect{C}(k) = \left[c_{ij}(k)\right]$, and $\vect{H}(k) = \left[ h_{ij}(k)\right]$.


\section{Linear response}
\label{sec:linear_response}

In section \ref{sec:solv_thermodynamics} we discussed the linear response in the context of solvation free energy.
The solute-solvent energy depended linearly on the coupling strength.
Here we take a more general and microscopic approach, additionally considering the spatial dependence of the response.

Consider a uniform liquid exposed to a weak external potential $\delta\phi(\rvec)$ (here we use $\delta$ to indicate that the field is small) that couples to local density in the usual way:
\begin{equation}
\Omega_{\delta\phi} = \Omega_0 + \int \delta\phi(\rvec) \rho(\rvec) \diff \rvec.
\end{equation}
For a weak perturbation we would expect that the density response can be described using the first order Taylor expansion:
\begin{equation}
\label{eq:density_linear_response}
\delta \rho(\rvec) = \int \left.\frac{\delta \rho(\rvec)}{\delta \beta \phi(\rvec')}\right|_{\phi=0} \beta \delta \phi(\rvec')\diff \rvec'\,,
\end{equation}
where $\delta \rho(\rvec_1) = \rho_{\delta \phi}(\rvec_1) - \rho_0$.
The density response is linear, but nonlocal.
Notice that since $\phi(\rvec) = \mu - \bar\mu(\rvec)$, we can use equation \ref{eq:chi_alternative} to obtain
\begin{equation}
\label{eq:density_susceptibility}
\begin{split}
-\frac{\delta\rho(\rvec_1)}{\delta\beta\phi(\rvec_2)}  & = \frac{\delta\rho(\rvec_1)}{\delta\beta\bar\mu(\rvec_2)} = \chi(\rvec_1,\rvec_2)\\
& = \rho(\rvec_1) \rho(\rvec_2)h(\rvec_1, \rvec_2) + \rho(\rvec_1) \delta (\rvec_2 - \rvec_1),
\end{split}
\end{equation}
where the derivatives are assumed to be taken in the unperturbed system with $\phi=0$.
The above equation shows a connection between a density-density correlation and liquid susceptibility to an external field.
This result is quite general and is valid for all classical systems \cite{Zwanzig2001ukg}.
Note that in statistical mechanics texts the static susceptibility is typically defined as $\chi = \beta \left\langle \delta A \delta A \right\rangle$, but in the reference interaction site model (RISM) literature the above definition is more widespread \cite{Chandler1986tqr,Beglov1997vrd,Ratkova2015teb}.

For simplicity, in the following, we focus on the response of an isotropic reference system.
In that case liquid susceptibility depends only on the separation between two points $\chi(\rvec_1,\rvec_2) = \chi(|\rvec_2 - \rvec_1|) = \chi(r)$. 
Applying the convolution theorem (equation \ref{eq:convol_theorem}) we get
\begin{equation}
\label{eq:density_linear_response_k}
\widehat{\delta \rho}(\kvec) = -\beta \f\chi(k)  \widehat{ \delta \phi}(\kvec),
\end{equation}
where $k=|\kvec|$. 
This is a remarkable result that allows us to calculate the perturbation of liquid density due to a field with a certain periodicity.
Since $\f \chi(k) = \rho \left[\f h(k)\rho + 1\right]$, we can see that response is quite sensitive to the wavenumber, and can be both amplified or weakened (see figure \ref{fig:hk_and_ck_of_k}).

Equation \ref{eq:density_linear_response_k} suggests that there is a dual relationship between the density perturbation and the potential.
We can rewrite the previous equation as
\begin{equation}
\beta \widehat{\delta\phi}(\kvec) = - \f \chi^{-1}(k)\widehat{\delta \rho}(\kvec)
\label{eq:inverse_susceptibility}
\end{equation}
in which the inverse of susceptibility $\chi^{-1}(k)$ determines a field created by a periodic density modulation $\Delta \rho(\kvec)$.
Similarly, since $\delta \phi = - \delta \bar{\mu}$ we have
\begin{equation}
\beta \widehat{\delta\bar \mu}(\kvec) = \f \chi^{-1}(k)\widehat{\delta \rho}(\kvec)
\label{eq:chem_pot_lr}
\end{equation}
that describes the effect of density modulation on the intrinsic chemical potential.

Using equation \ref{eq:chem_pot_lr} we can express the inverse of susceptibility as a functional derivative
\begin{equation}
\begin{split}
\chi^{-1}(|\rvec_2 - \rvec_1|) & = -\frac{\delta \beta \phi(\rvec_1)}{\delta \rho(\rvec_2)}  = \frac{\delta \beta \bar\mu(\rvec_1)}{\delta \rho(\rvec_2)}\\
&= \frac{1}{\rho(\rvec_1)}\delta(\rvec_2 - \rvec_1) - c(|\rvec_2 - \rvec_1|),
\end{split}
\label{eq:inverse_succ_dcf}
\end{equation}
where the last equality was obtained using equation \ref{eq:intrinsic_chem_pot_derivative}.
This result is similar to \ref{eq:density_susceptibility}, providing an alternative interpretation of the total and the direct correlation functions as non-ideal components of the system response to a perturbation.
Completing the parallel between susceptibilities, we express the inverse of susceptibility as a correlation function between intrinsic chemical potentials
\begin{equation}
\chi^{-1}(\rvec_1,\rvec_2) = \left \langle \delta \beta \bar{\mu}(\rvec_1)\delta \beta \bar{\mu}(\rvec_2)\right \rangle,
\end{equation}
where, as usual, $\delta \bar{\mu}(\rvec) = \bar{\mu}_{\delta \rho}(\rvec) - 
\bar{\mu}_0$.
The above expression follows directly from the fluctuation-dissipation theorem \cite{Zwanzig2001ukg}.


These results can be readily generalised to multicomponent mixtures.
The response of the local density of component $i$ to an external field that couples to densities $1\cdots N$ is given by
\begin{equation}
\delta \rho_i(\rvec) = \sum\limits_{j=1}^{N} \int \frac{\delta \rho_j(\rvec)}{\delta \beta \phi_j(\rvec')} \beta \delta \phi_j(\rvec') \diff \rvec',
\end{equation}
or when written in terms of the density-density correlation functions (susceptibilities):
\begin{equation}
\delta \rho_i(\rvec) = -\beta \sum\limits_{j=1}^{N}\int \chi_{ij}(\rvec, \rvec') \delta\phi_j(\rvec') \diff \rvec'.
\end{equation}
As previously, it is easier to work with vectors and matrices when dealing with mixtures.
Using vectors $\delta \vect{\rho} (\rvec) = \left[ \delta \rho_1 (\rvec) \cdots \delta \rho_n (\rvec) \right]$, $\delta \vect{\phi} (\rvec) = \left[ \delta\phi_1 (\rvec) \cdots \delta\phi_n (\rvec) \right]$, and matrix: $\vect{X}(\rvec_1,\rvec_2) = \left[ \chi_{ij}(\rvec_1,\rvec_2)\right]$ we can rewrite the above equation as:
\begin{equation}
\delta \vect{\rho} (\rvec) = -\beta \int \vect{X}(\rvec,\rvec') \delta\vect{\phi} (\rvec') \diff \rvec'.
\end{equation}

Assuming an isotropic system and taking the Fourier transform of all elements of matrices and vectors, we obtain the extensions of the previous results:
\begin{equation}
\begin{split}
\widehat{\delta \boldmath{\rho}} (\kvec) & = -\beta \hat{\vect{X}}(k) \widehat{\delta \boldmath{\phi}}(\kvec),\\
\beta \widehat{\delta \boldmath{\phi}}(\kvec)  & = -\hat{\vect{X}}^{-1}(k)\widehat{\delta  \boldmath{\rho}} (\kvec),\\
\beta \widehat{\delta \boldmath{\bar{\mu}}}(\kvec)  & = \hat{\vect{X}}^{-1}(k)\widehat{\delta  \boldmath{\rho}} (\kvec).\\
\end{split}
\label{eq:inverse_chi_multi}
\end{equation}
Note that unlike equation \ref{eq:inverse_susceptibility}, in which we were dealing with the algebraic inverse of susceptibility, in the above we have the matrix inverse of the susceptibility matrix.
Similarly to the single component case, these results are valid only for small external fields $\delta \vect{\phi}$ and density perturbations $\delta \vect{\rho}$.



\section{Connection to experiment}
\label{sec:pressure_sk}

The integrals of correlation functions are related to a variety of thermodynamic properties of liquids and can be used to construct an equation of state.
First, let us relate direct correlation function to the average fluctuations of particles.

For single and pair local densities we have the following normalization conditions $\int_V \rho(\rvec) \diff \rvec = \langle N \rangle_V$ and $\iint_V \rho^{(2)}(\rvec_1,\rvec_2) \diff \rvec_1 \diff \rvec_2 = \langle N^2 \rangle_V - \langle N \rangle_V$.
Combining these two equations with \ref{eq:h_r_def} we find that:
\begin{equation}
\begin{split}
\iint_V h(\rvec_1, \rvec_2) \rho(\rvec_1) \rho(\rvec_2) \diff \rvec_1 \diff \rvec_2 &= \int \rho^{(2)}(\rvec_1, \rvec_2) - \rho(\rvec_1) \rho(\rvec_2)\diff \rvec_1 \diff \rvec_2\\
&= \langle N^2 \rangle_V - \langle N \rangle^2_V - \langle N \rangle_V\\
&= \left\langle (\delta N)^2 \right\rangle_V - \left\langle N \right \rangle_V,
\end{split}
\label{eq:particle_msf}
\end{equation}
where $\delta N = N - \langle N \rangle$ and $\left\langle (\delta N)^2 \right\rangle_V$ is the mean square fluctuation of a number of particles in volume $V$.  
In a translationally invariant medium the particle density is constant and the direct correlation function depends only on distance $r$, so $\iint \rho(\rvec_1) \rho(\rvec_2) h(|\rvec_2 - \rvec_1|) \diff \rvec_1 \diff \rvec_2 = N \rho \int h(r)\diff r$.
Rearranging the two previous equations we get:
\begin{equation}
1 + \rho \int h(r)\diff r = \frac{\langle N^2\rangle - \langle N\rangle ^2}{\langle N\rangle} = kT\rho \chi_T,
\label{eq:compress_tot_cor_func}
\end{equation}
where $\chi_T$ is the isothermal compressibility.
The last equality can be proven using $\chi_T = (\pdv*{\rho}{\mu})_{V,T}/\rho^2$ and the definition of the grand canonical partition function \cite{Andersen1975tmu}.

We can also express compressibility using the direct correlation functions.
Notice that since the integral over real space is equal to the value of the Fourier-transformed function at $k=0$: $\int f(r) \diff r = \int f(r) e^{-0\cdot r} \diff r = \hat{f} (k=0)$, we can write:
\begin{equation}
kT\rho \chi_T = 1 + \rho \hat{h}(k=0) = \frac{1}{1 - \rho \hat{c}(k=0)},
\label{eq:compres_cor_funcs}
\end{equation}
where the last equality follows from equation \ref{eq:oz_k_space}.

The above equation provides a connection between the integral of the direct correlation function and the pressure of liquid \cite{Smith2013tsd}.
Using the definition of isothermal compressibility we get $\rho \chi_T = \left(\pdv*{\rho}{P}\right)_{T,V}$.
Plugging this back into equation \ref{eq:compres_cor_funcs} we obtain
\begin{equation}
\left(\frac{\partial P}{\partial \rho}\right)_{T,V} = kT \left[ 1 - \rho \hat{c}(k=0; \rho) \right].
\end{equation}
Finally, integrating between initial and final densities $\rho_1$ and $\rho_2$ we obtain the equation of state
\begin{equation}
\Delta P =  kT \Delta \rho - \int\limits_{\rho_1}^{\rho_2} \left[ \rho \hat{c}(k=0; \rho) \right] \diff \rho,
\label{eq:compres_eq_of_state}
\end{equation}
where $\Delta P = P_2 - P_1$ and $\Delta \rho = \rho_2 - \rho_1$.
Thus, we have related experimentally measurable changes in pressure to the changes in the integral of the direct correlation function.
The Kirkwood-Buff theory allows us to extend these results to multicomponent mixtures \cite{Oconnell2016ukd}.
These, more general relationships, have been used for validating both theoretical and experimental measurements of pressure in mixtures of various liquids \cite{Diky2015wxs}.


\begin{figure}
	\centering
	\includegraphics[width=\linewidth]{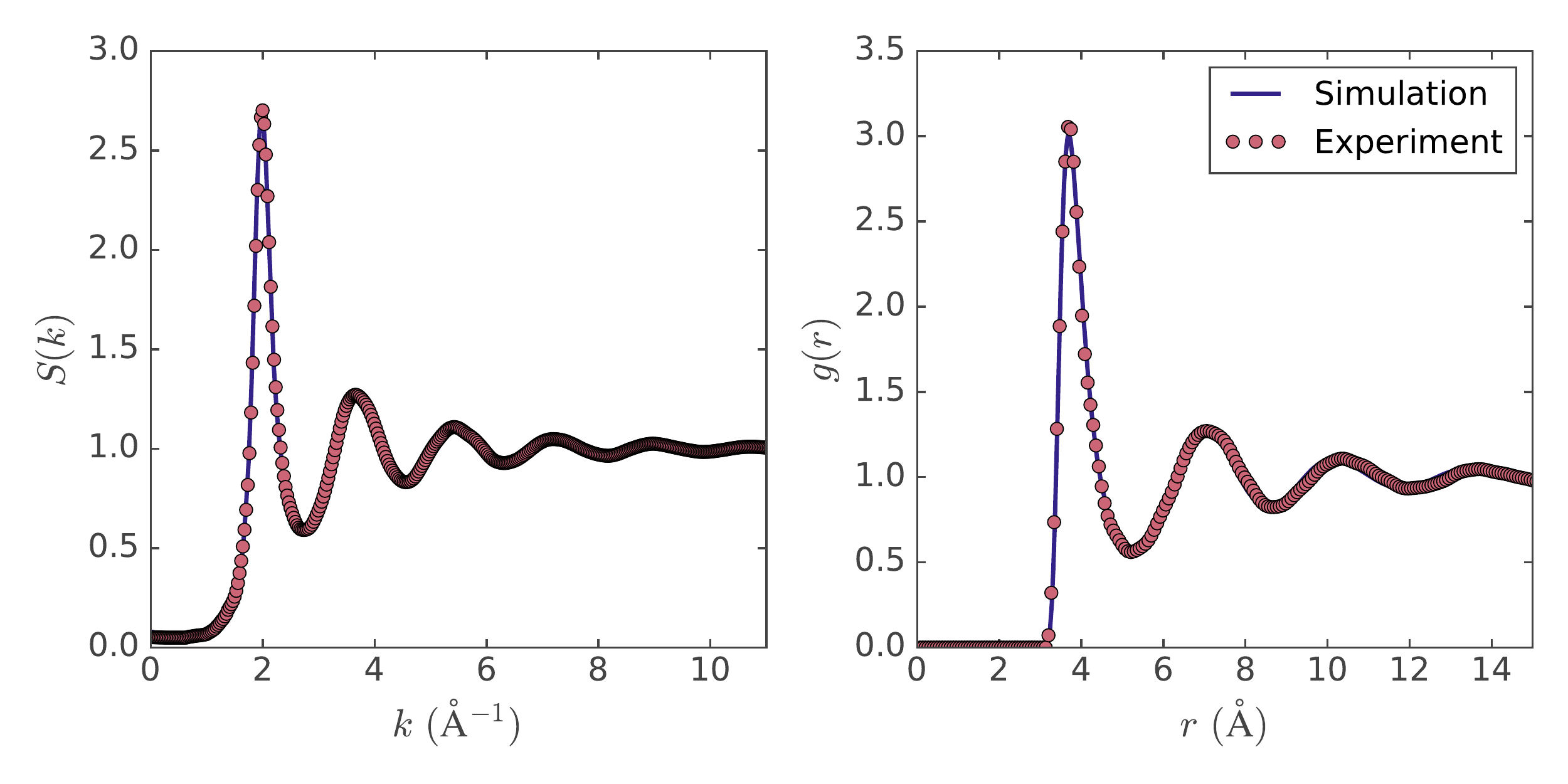}
	\caption{Comparison of experimental and theoretical structure factors (left) and radial distribution functions (right) for argon at 85K (close to its boiling point). Experimental data is taken from Ref. \citenum{Yarnell1973uke}. Theoretical predictions are obtained using molecular dynamics, with simulation parameters taken from Ref. \citenum{Toth2001vdf}. }
	\label{fig:sk_and_gr}
\end{figure}

Finally, we note that it is possible to measure correlation functions directly using spectroscopy \cite{Hansen2000utg}.
The idea is to perturb bulk liquid using small-angle neutron or x-ray scattering (radiation with longer wavelengths is too coarse to probe atomic structure of liquids and describes more macroscopic properties). 
The scattering cross section for the wave vector $k$ is proportional to the liquid structure factor $S(k)$, with the proportionality constant depending on the experimental set-up \cite{Yarnell1973uke}.
In the case of simple liquids, the structure factor $S(k)$ is related to the total correlation function as:
\begin{equation}
S(k) = 1 + \rho \int h(r) e^{-i \kvec \cdot \rvec} \diff \rvec = 1 + \rho h(k).
\label{eq:structure_factor}
\end{equation}
Notice that the structure factor is directly related to the Fourier transform of susceptibility $S(k) = \chi(k)/\rho$.

Figure \ref{fig:sk_and_gr} shows the argon structure factor and radial distribution functions deduced from the former using the above equation.
The experimental results, taken from Ref. \citenum{Yarnell1973uke}, are compared to the molecular dynamics simulation.
Both experimental measurements and theoretical simulations were performed at \SI{85}{K} with argon density $\rho = \SI{0.021}{\angstrom^{-3}}$.
The theory and experiment agree almost precisely, which might be expected, considering that the Lennard-Jones potential constants used in simulations were fitted to reproduce the experimental structure factor.


\chapter{Theory of molecular liquids} 

\label{Chapter4} 

\lhead{Chapter 4. \emph{Molecular liquids}}

This chapter overviews approximations required to compute the free energy of molecular liquids in site-site formalism.
The 3D-RISM equations and closures are derived in detail starting from the results from Refs. \citenum{Chandler1986tqr,Chandler1986uug}.
We finish by introducing the main problem of the thesis.

\section{Hypernetted-chain approximation}
\label{sec:hnc_simple_liq}

All results in the last chapter were exact.
However, they do not provide a tractable way of computing free energies -- to do so one needs to introduce some kind of approximation.
Although a variety of approaches exist \cite{Hansen2000utg}, we are going to consider one of the most basic approximations, built on the expansion of free energy functional around its equilibrium value.
For clarity, in this section we are going to focus on its derivation from the viewpoint of simple liquids; the case of molecular liquids is described for example in Ref. \citenum{Sergiievskyi2014vax}.

In section \ref{sec:free_energy_functionals} we obtained the expression for a grand canonical functional:
\begin{equation}
\Omega[\rho] = \Fcal[\rho] + \int \rho(\rvec) \phi(\rvec) \diff \rvec - \mu \int \rho (\rvec) \diff \rvec,
\label{eq:grand_functional2}
\end{equation}
where the intrinsic free energy functional was given by
\begin{equation}
\begin{split}
\Fcal[\rho] = {} & kT \int \rho(\rvec) \left\{\ln\left[\Lambda^3\rho(\rvec) \right] - 1\right\}\diff\rvec + \Fcal_0^{ex}[\rho] \\
& +\frac{1}{2} \int\limits_{0}^{1} \diff \lambda \iint \rho^{(2)}(\rvec,\rvec'; \lambda) u_{\lambda}(\rvec, \rvec') \diff \rvec \diff \rvec' .
\end{split}
\end{equation}
The coupling constant $\lambda$ linearly interpolates between initial and final interaction strength between individual particles:
\begin{equation}
u_{\lambda} (\rvec, \rvec') = u_0 (\rvec, \rvec') + \lambda u (\rvec, \rvec'),
\end{equation}
similarly to how a single particle was slowly coupled to the rest of the system when computing a chemical potential (see section \ref{sec:chemical_pot}).

While the algebraic expression for the functional is relatively straightforward, it turns out that its application to realistic systems is essentially impossible.
Indeed, to do it one would need to know the dependence of pair density $\rho(\rvec,\rvec'; \lambda)$ on $\lambda$, which is clearly very non-linear \cite{Hansen2000utg}.

A simpler approach is to take the problematic excess part of free energy functional and expand it into Taylor series around an isotropic system with density $\rho_0$.
This approach is often referred to as the (functional) density expansion \cite{Henderson1992vag,Hansen2000utg} and it leads to
\begin{equation}
\begin{split}
\Fcal^{ex}[\rho] = {} & \Fcal^{ex}[\rho_0] + \int \left. \frac{\delta \Fcal^{ex}}{\delta \rho(\rvec)}\right|_{\rho=\rho_0}\Delta \rho(\rvec) \diff \rvec\\
& +\frac{1}{2}\iint \left. \frac{\delta^2 \Fcal^{ex}}{\delta \rho(\rvec)\delta \rho(\rvec')}\right|_{\rho=\rho_0} \Delta \rho(\rvec)\Delta \rho(\rvec')\diff\rvec\diff\rvec' + F^B[\rho],
\end{split}
\end{equation}
where $F^B$ contains all higher order terms.
Note that we defined partial derivatives of excess chemical potential in the section \ref{sec:func_deriv_and_cor}.
Rewriting the expression using the definitions of the direct correlation function we obtain
\begin{equation}
\begin{split}
\Fcal^{ex}[\rho] = {} & \Fcal^{ex}[\rho_0] + \mu_0^{ex}\int \Delta \rho(\rvec) \diff \rvec\\
&-\frac{1}{2}kT\iint \Delta \rho(\rvec)\Delta \rho(\rvec')c_0(|\rvec'-\rvec|)\diff\rvec\diff\rvec' + F^B[\rho],
\end{split}
\end{equation}
where we used the fact that in a homogeneous system $c(\rvec)$ should be constant.
Substituting the above expression back into the intrinsic free energy functional and using $\mu_0 = \mu_0^{ex} + kT \ln \rho_0 \Lambda^3$ we get
\begin{equation}
\begin{split}
\Fcal[\rho] = {} & kT \int \rho(\rvec) \left\{\ln\left[\Lambda^3\rho(\rvec) \right] - 1\right\}\diff\rvec\\
&+\Fcal_0^{ex}[\rho_0] + \mu_0\int \Delta \rho(\rvec)\diff\rvec - kT\int \Delta \rho(\rvec)\ln \rho_0 \Lambda^3 \diff\rvec\\
&-\frac{1}{2}kT\iint \Delta \rho(\rvec)\Delta \rho(\rvec')c_0(|\rvec'-\rvec|)\diff\rvec\diff\rvec' + F^B[\rho].\\
\end{split}
\end{equation}
Finally, we cancel out the terms containing the de Broglie wavelength and plug the above expression into equation \ref{eq:grand_functional2} to obtain
\begin{equation}
\begin{split}
\Omega[\rho] = {} & \Omega_0 + kT \int \left[\rho(\rvec) \ln \frac{\rho(\rvec)}{\rho_0} - \Delta \rho(\rvec)\right]\diff\rvec+\int\rho(\rvec)\phi(\rvec)\diff\rvec \\
&-\frac{1}{2}kT\iint \Delta\rho(\rvec)c_0(|\rvec' - \rvec|)\Delta\rho(\rvec')\diff\rvec\diff\rvec' + F^B[\rho].\\
\end{split}
\label{eq:HNC_grand_cannonical}
\end{equation}
where $\Omega_0 = kT\int \rho_0\ln \rho_0 \Lambda^3\diff\rvec + \Fcal^{ex}[\rho_0] - \mu_0 \int \rho_0 \diff\rvec$.
The five terms in the above expression correspond to reference, ideal, external, second order, and higher order excess contributions to the grand canonical functional.
As one may guess, the $F^B[\rho]$ term, incorporating all higher order functional derivatives is generally unknown \cite{Hirata2003tpg}.


We can find the density distribution $\rho(\rvec)$ which minimizes equation \ref{eq:HNC_grand_cannonical}.
To do it, we take a functional derivative with respect to density
\begin{equation}
\begin{split}
\frac{\delta \Omega[\rho]}{\delta \rho(\rvec)} = {} & kT \ln \frac{\rho(\rvec)}{\rho_0} + \phi(\rvec)-\frac{1}{2}kT \int c_0(|\rvec' - \rvec|)\Delta\rho(\rvec')\diff\rvec'\\
&-\frac{1}{2}kT \int \Delta\rho(\rvec)c_0(|\rvec' - \rvec|)\diff\rvec + \frac{\delta\Fcal^B[\rho]}{\delta \rho(\rvec)} = 0.\\
\end{split}
\end{equation}
It follows that
\begin{equation}
\rho(\rvec) = \rho_0 \exp \left[-\beta \phi(\rvec) + \int \Delta\rho(\rvec')c_0(|\rvec'-\rvec|)\diff\rvec' - \frac{\delta\Fcal^B[\rho]}{\delta \rho(\rvec)} \right].
\label{eq:HNC_density}
\end{equation}
This is an integral equation for $\rho(\rvec)$.
One way of solving it is to make the following approximation:
\begin{equation}
\frac{\delta F^B[\rho]}{\delta \rho(\rvec)} = B(\rvec)= 0,
\end{equation}
where in the context of integral equation theories $B(\rvec)$ is referred to as a bridge function.
The above assumption is called the hypernetted-chain (HNC) approximation (the name comes from the original derivation \cite{Morita1958vuw}, which considered cluster diagrammatic expansion of the configurational partition function).
Extracting $c_0(r)$ from an equilibrium simulation, we can find an equation with a single unknown, which is possible to solve iteratively for a given external potential $\phi(\rvec)$.
When combined with $c_0$, the method is often referred to as the singlet HNC or HNC1 approximation, since both $\rho$ and $u$ depend only on a single coordinate.

\begin{figure}
\centering
\includegraphics[width=0.7\linewidth]{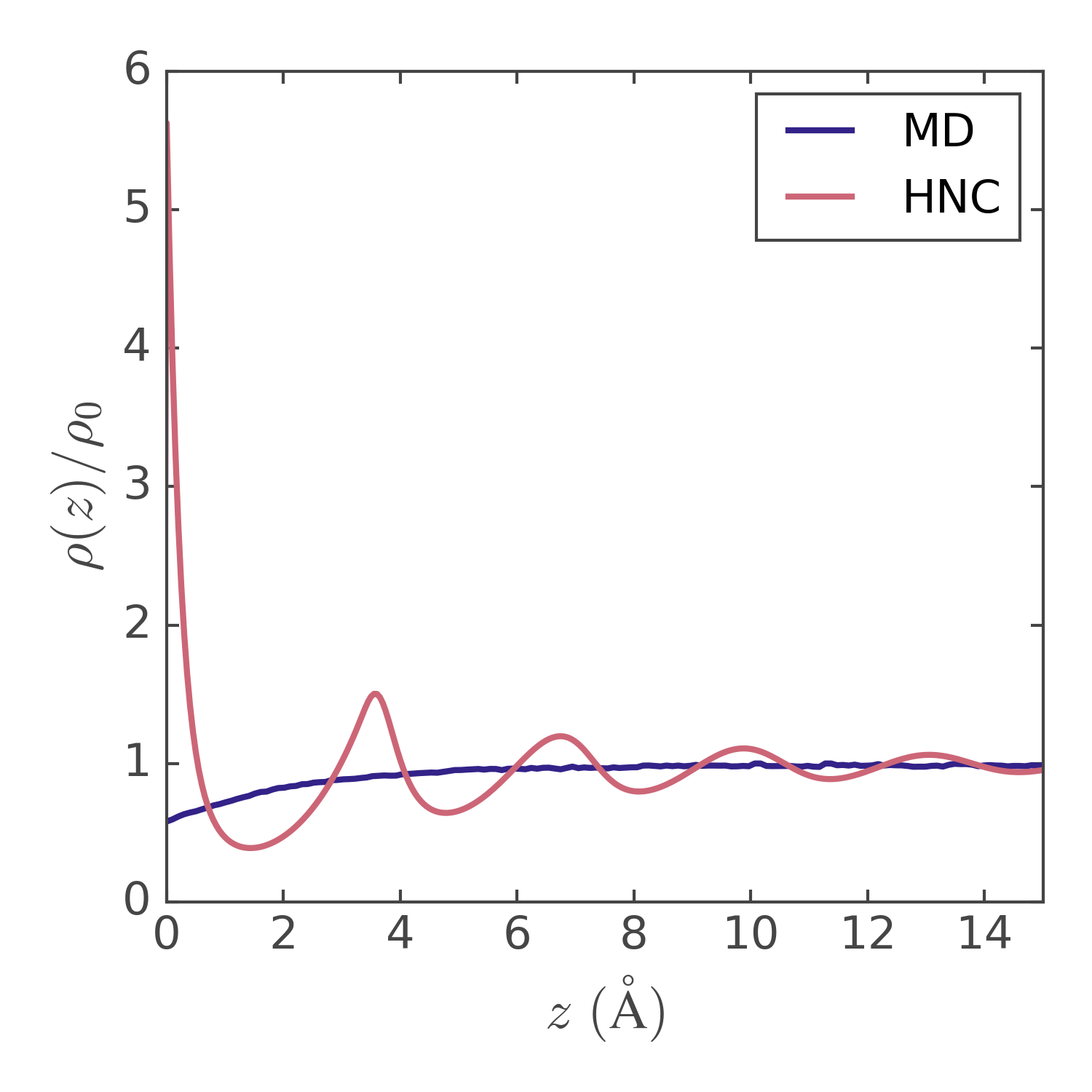}
\caption{Liquid argon local density near hard wall. Both singlet HNC calculations (red line) as well as molecular dynamics simulations (blue line) are performed at standard pressure and \SI{85}{K}. The details of the calculations can be found in appendix \ref{AppendixA}.}
\label{fig:argon_wall}
\end{figure}

An obvious question to ask is how good such an approximation is?
The answer is not too good.
While it can reproduce density oscillations of hard spheres near a wall, when it comes to liquids with both attractive and repulsive forces, HNC fails quite significantly \cite{Henderson1992vag}.
As a demonstration, consider the results obtained for liquid argon near a hard wall (figure \ref{fig:argon_wall}).
While molecular dynamics predicts a slight decrease in density near the wall, caused by the formation of an interface, HNC predicts oscillations similar to what is observed for liquids with purely repulsive interactions.

It is not hard to rationalize the failure of HNC to describe the formation of an interface.
Since HNC only retains second order correlations in liquids, it captures its essential characteristic, namely, the dominant repulsive forces between particles with rather weak attractive components \cite{Widom1967tzb}.
As we shall see, while this does not usually present a problem for describing a bulk liquid structure, the formation of interfaces is mostly due to long ranged attractive potentials which require a more sophisticated description.
Additionally, a hard planar wall creates a huge region of excluded volume.
This can hardly justify the use of a second order Taylor expansion that is valid only for small changes in density.

Before moving on it is useful to relate the singlet HNC approximation to the linear response approach, discussed in section \ref{sec:linear_response}.
From the equation \ref{eq:HNC_grand_cannonical} we find
\begin{equation}
\begin{split}
\Delta \Fcal_{HNC}[\rho] = {} & kT \int \left[\rho(\rvec) \ln \frac{\rho(\rvec)}{\rho_0} - \Delta \rho(\rvec)\right]\diff\rvec\\
&-\frac{1}{2}kT\iint \Delta\rho(\rvec)c_0(|\rvec' - \rvec|)\Delta\rho(\rvec')\diff\rvec\diff\rvec'.\\
\end{split}
\end{equation}
Then the change in chemical potential is given by 
\begin{equation}
\Delta \bar{\mu} (\rvec) = \fdv{\Fcal}{\rho(\rvec)} = kT\ln\frac{\rho(\rvec)}{\rho_0} - kT \int c_0(|\rvec-\rvec')\Delta \rho(\rvec')\diff \rvec'.
\end{equation} 
Contrast this to a linear response change in intrinsic chemical potential obtained via equation \ref{eq:chem_pot_lr}:
\begin{equation}
\begin{split}
\Delta \bar{\mu} (\rvec) & = kT\int \chi^{-1}(|\rvec-\rvec'|) \Delta \rho(\rvec')\diff\rvec'\\
& = kT h(\rvec) - kT\int c_0(|\rvec-\rvec')\Delta \rho(\rvec')\diff \rvec',
\end{split}
\end{equation}
where two terms in the second equality correspond to the local and non-local contributions to the change in intrinsic chemical potential.
We can see that HNC model approximates the excess part of the chemical potential via the linear response approximation, while the ideal (local) contribution is exact.
From this, it is reasonable to suggest that the singlet HNC should give somewhat more accurate results than the standard linear response approach, but they are not going to be significantly different.
At the chapter \ref{chap:pres_cor} we will see that this is exactly what happens.

The failure to describe interfaces does not render HNC useless.
Notably, it can quite accurately describe the bulk structure of simple liquids.
To do so, we use an idea by Percus \cite{Hansen2000utg} and treat equation \ref{eq:HNC_density} as an expression for the pair correlation function $g(\rvec)=\rho(\rvec)/\rho_0$ in the homogeneous liquid
\begin{equation}
g(r) = \exp\left[-\beta u(r) + \rho \int h(r') c(|r-r'|)\diff r' \right],
\end{equation}
where we used $\Delta \rho(r) = \rho h(r)$ and the external potential takes a meaning of a pair potential $u(r)$ between particles.
Using the Ornstein-Zernike equation (\ref{eq:isotropic_OZ}) we obtain
\begin{equation}
\label{eq:hnc_simple_fluids}
g(r) = \exp\left[ - \beta u(r) + h(r) - c(r)\right].
\end{equation}
When the above two equations are solved simultaneously, they are sometimes referred to as the pair HNC approximation (HNC2) to distinguish from its singlet form.
The resulting solution gives direct and total correlation functions for isotropic liquids.
Note that whenever we have equations that are solved for correlation functions (such as equations \ref{eq:hnc_simple_fluids} and \ref{eq:HNC_density}), we view them as an integral equation approach.

\begin{figure}
\centering
\includegraphics[width=1\linewidth]{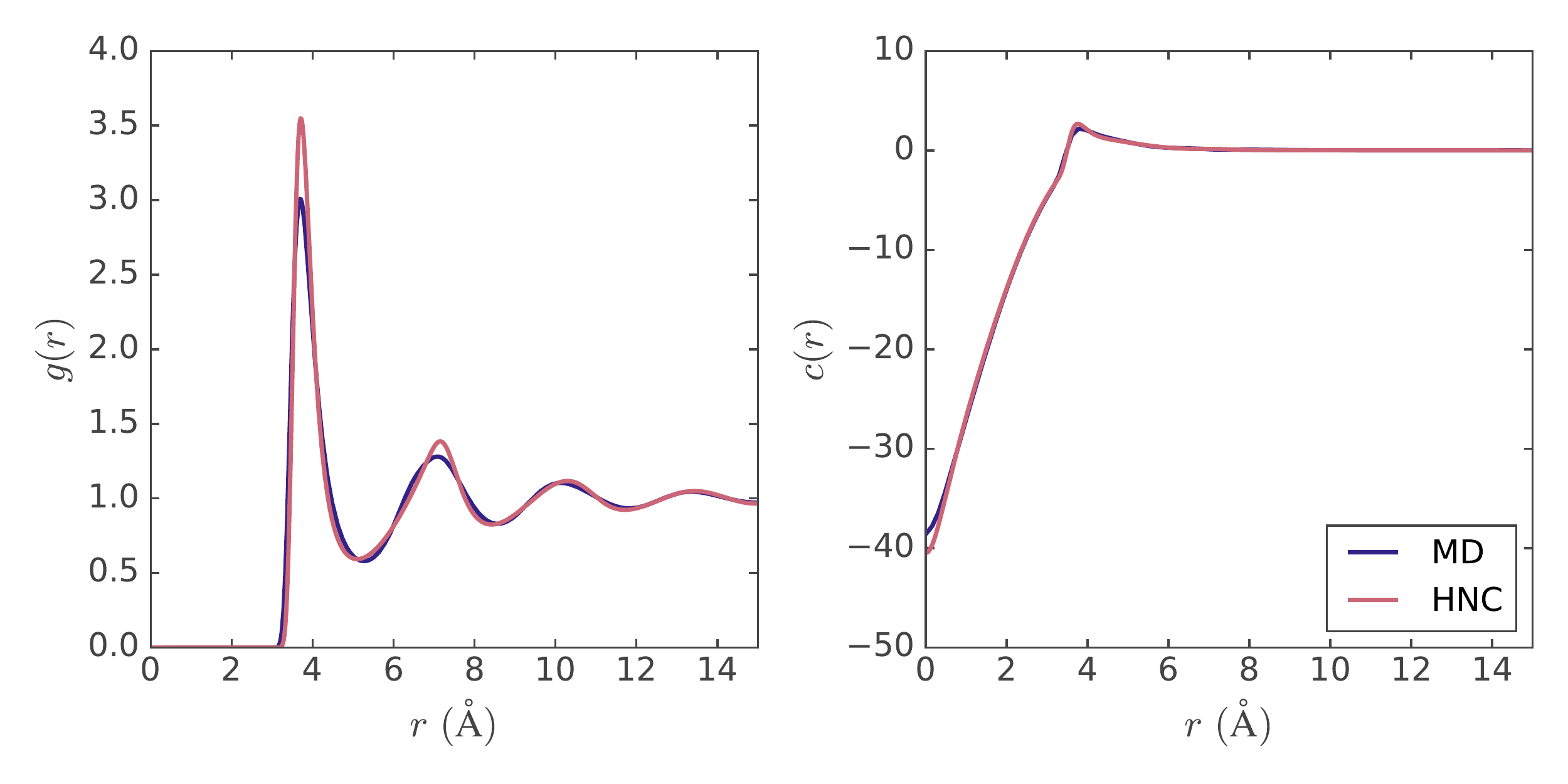}
\caption{Comparison of liquid argon radial distribution functions (left figure) and direct correlation functions (right figure) predicted by molecular dynamics simulation (blue lines) and HNC calculation (red lines).
For both simulation and theory interactions between argon atoms was approximated using Lennard-Jones potential from Ref. \citenum{Toth2001vdf}.}
\label{fig:hnc_theory_vs_sim}
\end{figure}

To test the accuracy of the model we apply the above equation to our model system of liquid argon.
Both molecular dynamics simulation and HNC calculations were performed using the same Lennard-Jones potential and conditions.
The results, shown in figure \ref{fig:hnc_theory_vs_sim}, demonstrate a surprisingly good agreement between the two.
Compared to the hard wall, a single argon atom creates a much less significant perturbation.
Additionally, an attractive $r^{-6}$ part of the pair potential prevents any "drying" of the liquid close to the argon surface, making MD and HNC radial distribution functions quite similar.


\section{Molecular liquids}
\label{sec:CMS}

So far, we have dealt only with simple (monoatomic) liquids.
This made the discussion significantly easier and essentially cut the dimensionality of the problem by two.
Indeed, to fully specify the position of a rigid molecule in space, one not only has to keep track of its spatial location $\rvec$ but also of its orientation, which is typically specified using Euler angles $\vect{\omega}$ \cite{Gubbins1985udt}.
Alternatively, one can keep track of the positions of individual atoms (or more generally, molecular sites) in the molecules that interact with each other.
While this approach can significantly increase the dimensionality in the case of polyatomic liquids, it allows us to write down extensions of the equations used to describe simple monoatomic fluids.
In this thesis, we will be mostly using the second approach.

While the interacting sites approach to predicting liquid structure and properties is relatively old \cite{Chandler1972uqk,Chandler1973wjo,Ladanyi1975utg}, the first papers in which it was formalised as a part of density functional theory were written by Chandler, McCoy, and Singer (CMS) only in 1986 \cite{Chandler1986uug,Chandler1986tqr}.
This theory provides a basis for various interaction site models such as reference interaction site model (RISM) or polymer reference interaction site model (PRISM).

In general, we have a mixture of $M$ different molecules containing $n_1 + n_2 + \cdots + n_M = N$ different sites.
For some molecule $\alpha$, each site has some local density $\rho_{i\alpha}(\rvec)$ that couples to an external field $\phi_{i\alpha}(\rvec)$.
We will use Greek letters $\alpha, \beta, \ldots $ to indicate different types of molecule in the system, and indices $i,j,\ldots$ to specify sites.
For simplicity, we assume that each site in a molecule is unique (thus, water, for example, will have two distinct hydrogens) and has the same (macroscopic) density as the molecule itself $\rho_{1\alpha} = \rho_{2\alpha} = \cdots = \rho_{n_{\alpha}\alpha} = \rho_{\alpha}$.

In CMS theory, we set the chemical potential of molecule $\alpha$ to be equal to the sum of chemical potentials of its individual sites:
\begin{equation}
\mu_{\alpha} = \sum\limits_{i=1}^{n_\alpha} \mu_{i\alpha}.
\end{equation}
This assumption can be regarded as a condition for chemical equilibrium.
We also require all molecules to be rigid.
This condition is not very problematic, as different conformations can be treated as different species.

The Grand canonical functional for such a system can be written in terms of the site densities \cite{Liu2013vpm}
\begin{equation}
\begin{split}
\Omega [\vect{\rho}] &= \Fcal [\vect{\rho}] + \sum_{i=1}^{N} \int \rho_i(\rvec)\phi_i(\rvec)\diff \rvec - \sum_{i=1}^{N}\int \mu_i \rho_i(\rvec) \diff \rvec\\
&=\Fcal [\vect{\rho}] - \sum_{i=1}^{N} \int \rho_i(\rvec)\bar \mu_i(\rvec)\diff \rvec,
\end{split}
\label{eq:molecular_liquids_grand_potential}
\end{equation}
where $\vect{\rho}$ is the vector of all site density distributions and $\Fcal$ is the intrinsic free energy functional for polyatomic liquids.
For clarity, we suppressed molecule subscripts.

We can formally separate $\Fcal$ from equation \ref{eq:molecular_liquids_grand_potential} into an ideal gas part and excess.
Even for a mixture of polyatomic molecules, an ideal gas is still defined to be the uniform mixture of all particles, with absolutely no correlations present \cite{Chandler1986tqr}.
Thus, the following holds
\begin{equation}
\begin{split}
\Fcal [\vect{\rho}]&= \Fcal^{id} [\vect{\rho}] + \Fcal^{ex} [\vect{\rho}]\\
&= \sum_{i=1}^{N} kT \int \rho_i(\rvec) \left\{\ln\left[\Lambda_i^3\rho_i(\rvec) \right] - 1\right\}\diff\rvec + \Fcal^{ex} [\vect{\rho}].
\end{split}
\label{eq:site_intr_functional}
\end{equation}
Here, $F^{ex}$ contains corrections not only for non-ideality, but also, for the fact that the sites belonging to the same molecule are actually bonded.

The intrinsic chemical potentials of each site are obtained by taking the derivative of grand potential with respect to the site densities:
\begin{equation}
\begin{split}
\bar \mu_i(\rvec) &= \frac{\delta \Fcal}{\delta \rho_i(\rvec)}= \mu^{id}_i(\rvec) + \bar \mu^{ex}_i(\rvec) = kT\ln \left[\Lambda_i^3 \rho_i(\rvec) \right] - kT c_i(\rvec),
\end{split}
\label{eq:site_chem_pot}
\end{equation}
where we introduce the direct correlation function in the site formalism, defined as:
\begin{equation}
\label{eq:site_dcf}
c_i (\rvec) = -\beta\frac{\delta \Fcal^{ex}}{\delta \rho_i(\rvec)}.
\end{equation}

The density distributions of individual sites can be obtained by taking the functional derivative of the Grand potential
\begin{equation}
\rho_i(\rvec) = -\frac{\delta \Omega}{\delta \bar \mu_i(\rvec)}.
\end{equation}
Similarly to the simple fluid case, the second derivative of grand potential  gives (site) density-(site) density correlation function or site-site susceptibility \cite{Chandler1986tqr}:
\begin{equation}
\chi_{ij}(\rvec_1,\rvec_2) = -\frac{\delta \beta\Omega}{\delta \beta\bar\mu_i (\rvec_1)\delta\beta\bar\mu_j(\rvec_2)} = \frac{\delta \rho_i(\rvec_1)}{\delta \beta \bar{\mu}_j(\rvec_2)}=\left\langle \delta \rho_i(\rvec_1) \delta \rho_j(\rvec_2) \right\rangle,
\end{equation}
with $\delta \rho(\rvec) = \rho^I(\rvec) - \rho(\rvec)$. 
Repeating manipulations presented in the beginning of section \ref{sec:func_deriv_and_cor}, we can relate $\chi_{ij}(\rvec_1,\rvec_2)$ to the pair distribution function $\rho^{(2)}_{ij}(\rvec_1,\rvec_2)$:
\begin{equation}
\begin{split}
\chi_{ij} & = \rho^{(2)}_{ij}(\rvec_1,\rvec_2) + \delta_{ij}\delta(\rvec_2 - \rvec_1)\rho(\rvec_1) - \rho_i(\rvec_1)\rho_j(\rvec_2)\\
& = \delta_{ij}\delta(\rvec_2 - \rvec_1)\rho_i(\rvec_1) + \rho_i(\rvec_1)\rho_j(\rvec_2) H_{ij}(\rvec_1,\rvec_2),\\
\end{split}
\label{eq:whole_tot_cor_func}
\end{equation}
where $H_{ij} = \rho^{(2)}_{ij}(\rvec_1,\rvec_2)/\rho_i(\rvec_1)/\rho_j(\rvec_2) - 1$, will be referred to as a whole total correlation function.
Equation \ref{eq:whole_tot_cor_func} suggests that $H_{ij}$ should be similar to its simple fluid analogue: total correlation function $h_{ij}$.
However, there is an important difference: while $h_{ij}$ contains only \emph{intermolecular} correlations between different particles, $H_{ij}$ contains both \emph{inter-} and \textit{intramolecular} correlations.

Similarly to atomic liquids, $\chi_{ij}(\rvec_1,\rvec_2)$ determines the linear response of a system to a perturbing external field.
Its functional inverse then characterizes fluctuations of chemical potential (section \ref{sec:linear_response}) $\chi^{-1}_{ij}(\rvec_1,\rvec_2)=\left \langle \delta \beta \bar{\mu}_i \delta \beta \bar{\mu}_j\right \rangle$.
We can split the correlations into ideal/excess parts by defining an analogue of the two-particle direct correlation function, which we will call the \textit{whole} site-site direct correlation function $C_{ij}$ \cite{Chandler1986tqr}:
\begin{equation}
\begin{split}
\chi_{ij}^{-1}(|\rvec_2 - \rvec_1|) &= \frac{\delta \beta \bar \mu_i(\rvec_2)}{\delta \rho_j(\rvec_1)} =  \beta \frac{\delta \mu_i^{id}(\rvec_2)}{\delta \rho_j(\rvec_1) } + \beta \frac{\delta \bar \mu_i^{ex}(\rvec_2)}{\delta \rho_j(\rvec_1) } \\
&= \frac{\delta_{ij} \delta(\rvec_2 - \rvec_1)}{\rho_i} - C_{ij}(\rvec_1, \rvec_2).\\
\end{split}
\label{eq:full_site_site_dcf}
\end{equation}
Here again, $C_{ij}(\rvec_1,\rvec_2)$ is similar to its simple fluid analogue, but in addition to intramolecular also contains intermolecular correlations.

Combining all definitions introduced in this section, we can arrive at an Ornstein-Zernike-like expression for site-site polyatomic fluids.
Let us assume a homogeneous liquid and write the definition of the functional inverse for the site-site susceptibility \cite{Donley1994vre}:
\begin{equation}
\sum\limits_{k=1}^{N} \int \chi_{ik} (\rvec_1 - \rvec') \chi^{-1}_{kj} (\rvec' - \rvec_2)\diff \rvec'  = \delta_{ij} \delta(\rvec_1 - \rvec_2).
\label{eq:site_site_oz}
\end{equation}
In the above equation, the left hand side contains a convolution of two functions.
Taking the Fourier transform of both sides we get $\sum\limits_{k=1}^{N} \hat{\chi}_{ik} (\kvec) \hat{\chi}^{-1}_{kj} (\kvec) = \delta_{ij}$, which can be readily expressed in a matrix form $\hat{\vect{X}}(\kvec) \hat{\vect{X}}^{-1}(\kvec)= \vect{I}$, where $\hat{\vect{X}} = \left[\hat{\chi}_{ij}(\kvec)\right]$ is the $N$ by $N$ matrix of Fourier transformed site-site susceptibilities and $\hat{\vect{X}}^{-1}(\kvec)$ is its matrix inverse.

Taking Fourier transform of expressions \ref{eq:whole_tot_cor_func} and \ref{eq:full_site_site_dcf} we get
\begin{align}
\hat{\chi}_{ij}(\kvec) = \delta_{ij} \rho_i + \rho_i \rho_j \hat{H}_{ij}(\kvec) && \hat{\chi}^{-1}_{ij}(\kvec) = \frac{\delta_{ij}}{\rho_i} - \hat{C}_{ij}(\kvec).
\label{eq:inverse_chi_C}
\end{align}
It follows that the equation \ref{eq:site_site_oz} can be re-written as:
\begin{equation}
\left[\vect{D}\vect{I} + \vect{D}\vect{\hat{H}}(\kvec)\vect{D}\right]\left[\vect{D}^{-1}\vect{I} - \vect{\f C}(\kvec)\right] = \left[\vect{I} + \vect{\f H}(\kvec)\vect{D}\right]\left[\vect{I} - \vect{\f C}(\kvec)\vect{D}\right] = \vect{I}.
\label{eq:site_site_oz_full}
\end{equation}
This is a site-site Ornstein-Zernike equation, written in terms of whole correlation functions \cite{Cummings1982wyk}.

Essentially, all equations in this section were definitions, based on few assumptions.
These equations do not offer any insight into how to compute free energies or correlation.
Similarly to the situation with simple liquids, to make any actual predictions we will have to make some guesses regarding free energy functionals.


\section{Intramolecular correlation function}

There is a special type of correlation that is not present in the case of simple liquids.
These are correlations due to intramolecular bonding.
In this section, we will briefly discuss them and explore their behaviour using the example of water.

Instead of combining all inter- and intramolecular correlations of $\chi_{ij}(\rvec_1,\rvec_2)$ in a single term, we can split them into two separate functions:
\begin{equation}
\chi_{ij}(\rvec_1,\rvec_2) = \rho_i(\rvec_1)\omega_{ij}(\rvec_1,\rvec_2) + \rho_i(\rvec_1)\rho_j(\rvec_2) h_{ij}(\rvec_1,\rvec_2),
\end{equation}
where $\omega_{ij}(\rvec_1,\rvec_2)$ is called the intramolecular correlation function and contains correlations of a site with itself or with its bonded neighbours.
$h_{ij}(\rvec_1,\rvec_2)$ is called the site-site total correlation function and contains site-site correlations between different molecules.

For the next couple of paragraphs we return to the full notation of sites and write $i\alpha$ to indicate site $i$ in the molecule of type $\alpha$.

Formally, $\omega_{i\alpha\,j\beta}(\rvec_1,\rvec_2)$ is defined via the following expression \cite{Hirata2003tpg}:
\begin{equation}
\omega_{i\alpha\,j\beta}(\rvec_1,\rvec_2) = \delta_{i\alpha\,j\beta}\delta(\rvec_2 - \rvec_1) + \delta_{\alpha \beta}(1 - \delta_{ij})\left\langle \sum_{\alpha=1}^{M} \delta(\rvec_1 - \rvec_{i\alpha}) \delta(\rvec_2 - \rvec_{j\alpha})\right \rangle,
\end{equation}
where $M$ stands for total number of distinct molecules in the system and $r_{i\alpha}$ is the position of the site $i$ on the molecule $\alpha$.
For isotropic liquid we can simplify this equation to get
\begin{equation}
\omega_{i\alpha\,j\beta}(r) = \delta_{i\alpha\,j\beta} \delta(r) + \delta_{\alpha\beta}(1 - \delta_{i\alpha\,j\beta}) \frac{\delta(r - L_{i\alpha\,j\alpha})}{4\pi L_{i\alpha\,j\alpha}^2},
\label{eq:isotropic_omega}
\end{equation}
with $r = |\rvec_2-\rvec_1|$, and $L_{i\alpha\,j\alpha}$ is the distance between sites $i$ and $j$ in a rigid molecule of type $\alpha$.
The $4\pi L_{i\alpha\,j\alpha}^2$ term in denominator ensures correct normalisation of the function.

In the absence of any intermolecular interactions $h_{ij}=0$, thus $\chi_{ij}(r) = \rho_j \omega_{ij}(r)$, or in matrix notation $\vect{X}(r) = \vect{\omega}(r)\vect{D}$, where $\vect{\omega}(r) = \left[\omega_{ij}(r)\right]$. 
Then, from the equation \ref{eq:site_site_oz_full} we have $\vect{I} - \vect{\f C}(k)\vect{D} = \vect{\f \omega}^{-1}(k)$, where $\vect{\f \omega}^{-1}(k)$ is the matrix inverse of $\hat{\vect{\omega}}(k) = \left[\hat{\omega}_{ij}(k)\right]$.
Conceptually, $\vect{\f \omega}^{-1}(k)$ is the intramolecular part of $\vect{I} - \vect{\f C}(k)\vect{D}$, similarly to how $\hat{\vect{\omega}}(k)$ contains the intramolecular part of $\vect{I} + \vect{\f H}(\kvec)\vect{D}$ \cite{Cummings1982wyk}.
We can then formally write
\begin{equation}
\vect{I} - \vect{\f C}(k)\vect{D} = \vect{\f \omega}^{-1}(k) - \vect{\f c}(k)\vect{D},
\label{eq:DCF_and_dcf}
\end{equation}
where $\vect{\f c}(k) = \left[\f c_{ij}(k)\right]$ is a matrix of site-site direct correlation function --- an intermolecular part of $C_{ij}(r)$ defined via this equation.

The content of the last couple of paragraphs can be summarised by the following two equations:
\begin{equation}
\begin{split}
\vect{D}^{-1}\vect{X}(r) & = \delta(r)\vect{I} + \vect{H}(r)\vect{D} = \vect{\omega}(r) + \vect{h}(r)\\
\vect{X}^{-1}(r)\vect{D} & = \delta(r)\vect{I} - \vect{C}(r)\vect{D} = \vect{ \omega}^{-1}(r) - \vect{ c}(r)\vect{D}.
\end{split}
\label{eq:intra_dcf}
\end{equation}
This decomposition of correlations allows us to conveniently separate bonding and intermolecular effects on the liquid structure.

%

We illustrate the properties of $\omega(r)$ for the example of water.
The matrix of intermolecular correlations in this case is given by:
\begin{equation}
\renewcommand*{\arraystretch}{1.5}
\vect{\omega} (r)= \begin{bmatrix}
\delta(r)  & \frac{\delta(r-L_{OH})}{4\pi L_{OH}^2}& \frac{\delta(r-L_{OH})}{4\pi L_{OH}^2} \\
\frac{\delta(r-L_{OH})}{4\pi L_{OH}^2} & \delta(r) & \frac{\delta(r-L_{HH})}{4\pi L_{HH}^2} \\
\frac{\delta(r-L_{OH})}{4\pi L_{OH}^2}   & \frac{\delta(r-L_{HH})}{4\pi L_{HH}^2} & \delta(r) \\
\end{bmatrix}.
\end{equation}
The Fourier transform can be taken analytically and produces:
\begin{equation}
\renewcommand*{\arraystretch}{1.5}
\vect{\f \omega} (k)= \begin{bmatrix}
1  & \frac{\sin(kL_{OH})}{kL_{OH}}& \frac{\sin(kL_{OH})}{kL_{OH}} \\
\frac{\sin(kL_{OH})}{kL_{OH}} & 1 & \frac{\sin(kL_{HH})}{kL_{HH}} \\
\frac{\sin(kL_{OH})}{kL_{OH}}   & \frac{\sin(kL_{HH})}{kL_{HH}} & 1 \\
\end{bmatrix}.
\end{equation}
In principle, the matrix inverses of the above functions can be written analytically as well, but they do not have a simple form \cite{Stell1989wav,Cummings1982wyk,Kalyuzhnyi2000umd}.
Additionally, $\vect{\omega}^{-1}(k)$ is undefined at $k\to0$.

\begin{figure}
\centering
\includegraphics[width=1.0\linewidth]{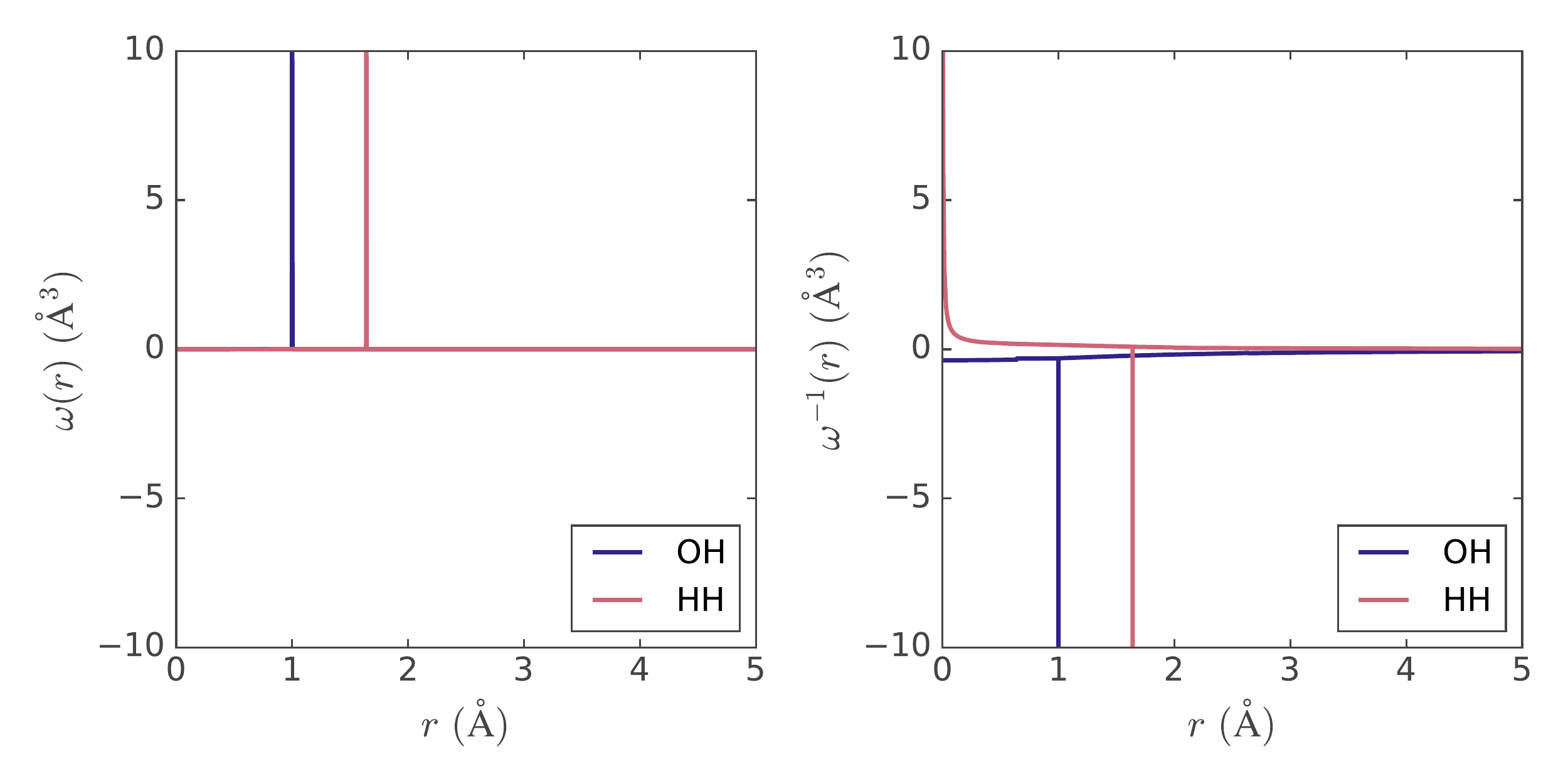}
\caption{Water intramolecular site-site correlation function (left figure) and its functional inverse (right figure).}
\label{fig:omega}
\end{figure}

\begin{figure}
	\centering
	\includegraphics[width=1.0\linewidth]{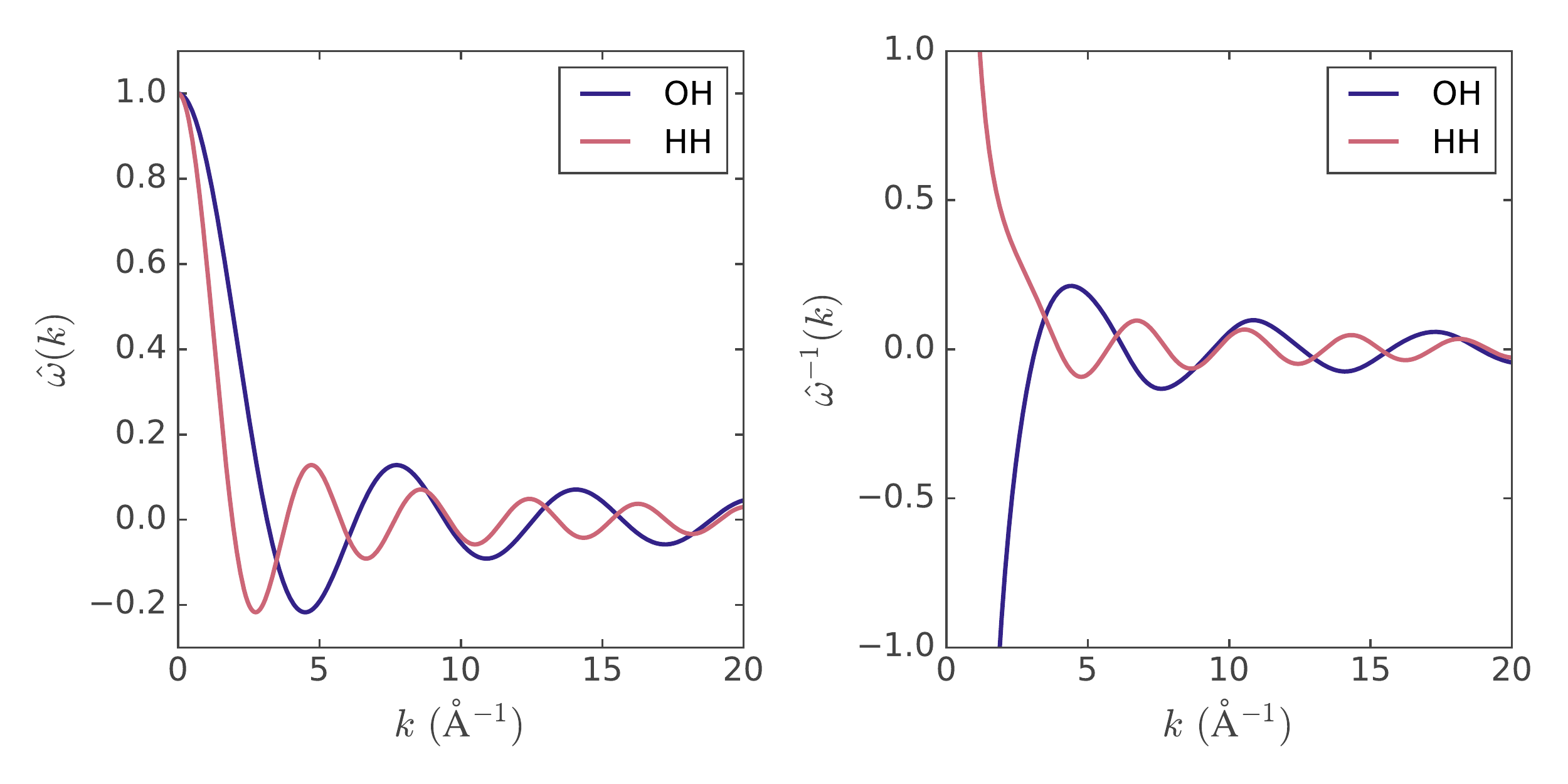}
	\caption{Fourier transforms of water intramolecular site-site correlation function (left figure) and its inverse (right figure).}
	\label{fig:f_omega}
\end{figure}

We can, however, still perform both matrix inversion and inverse Fourier transforms numerically and define $\omega^{-1}(k=0) = \lim\limits_{k\to0} \omega^{-1}(k)$.
Figures \ref{fig:omega} and \ref{fig:f_omega} show the behaviour of all intramolecular correlation functions in both real and momentum spaces.
As you can see,  $\omega^{-1}(r)$ is similar to $-\omega(r)$, although it contains extra components that can be defined by a sum of convolutions of delta functions \cite{Stell1989wav}.
Its sharp spikes at $r=L_{OH}$ and $r=L_{HH}$ can be rationalised as a strongly favourable interaction which enforces a certain distance between bonded atoms:
\begin{equation}
\beta \Delta \mu_i \approx \Delta \rho_j(r) \omega_{ij}^{-1}(r).
\end{equation}
In Fourier space intramolecular correlation functions oscillate with a constant frequency determined by bond distances in the molecule.

\section{1D-RISM}

After long preparations we are now ready to write the site-site Ornstein-Zernike equation for homogeneous liquid in a more common form
\begin{equation}
\left[\vect{\f \omega}(k) + \vect{\f h}(k) \vect{D} \right]\left[\vect{\f \omega}^{-1}(k) - \vect{\f c}(k)\vect{D} \right] = \vect{I}.
\end{equation}
Opening brackets and rearranging terms results in $\vect{\f h}(k) \vect{D}\vect{\f \omega}^{-1}(k) = \vect{\f\omega}(k)\vect{\f c}(k)\vect{D} + \vect{\f h}(k) \vect{D} \vect{\f c}(k)\vect{D}$.
Multiplying both by $\vect{\f \omega}(k)\vect{D}^{-1}$ from the right side gives us the usual form of the site-site Ornstein-Zernike equation \cite{Hansen2000utg}
\begin{equation}
\vect{\hat h}(k) = \vect{\hat \omega} (k) \vect{\hat c} (k) \vect{\hat \omega}(k) +\vect{\hat h}(k) \vect{D} \vect{\hat c} (k) \vect{\hat \omega} (k) ,
\end{equation}
where we utilised $\vect{D} \vect{\hat \omega} (k) \vect{D}^{-1} = \vect{\hat \omega} (k)$.
This equation can be re-written in real space to give
\begin{equation}
\begin{split}
\label{eq:1drism_oz_eq}
h_{ij}(|\rvec_1 - \rvec_2|) =& \sum_{k=1}^{N}\sum_{l=1}^{N} \iint \left[ \omega_{ik}(|\rvec_1 - \rvec'|) c_{kl}(|\rvec' - \rvec''|) \omega_{lj}(|\rvec''-\rvec_2|)\right.\\
&+\left. h_{lj}(|\rvec''-\rvec_2|)\rho_l c_{kl}(|\rvec' - \rvec''|)  \omega_{ik}(|\rvec_1 - \rvec'|)   \right] \diff \rvec' \diff \rvec''.\\
\end{split}
\end{equation}
Although quite cumbersome, this is a direct extension of the Ornstein-Zernike equation for mixtures of simple liquids (\ref{eq:oz_simple_homo_mixtures}), with intramolecular correlation function $\omega(r)$ accounting for additional propagation of interactions through the intramolecular correlations.

To obtain a theory for bulk polyatomic liquids, one has to combine equation \ref{eq:1drism_oz_eq} with another suitable expression.
One possibility is to use an HNC-like closure (eq. \ref{eq:hnc_simple_fluids})
\begin{equation}
\label{eq:site_site_hnc}
h_{ij}(r) + 1 = \exp\left[-\beta u_{ij}(r) + h_{ij}(r) - c_{ij}(r)\right],
\end{equation}
where $u_{ij}(r)$ is a site-site interaction potential energy at separation $r$.
One can also combine site-site Ornstein-Zernike equation with the Percus-Yevick (PY) closure:
\begin{equation}
h_{ij}(r) + 1 = \exp\left[-\beta u_{ij}(r)\right]\left[h_{ij}(r) - c_{ij}(r) + 1\right],
\end{equation}
which is known to be reasonably accurate for hard sphere systems \cite{Hansen2000utg}.

As we saw in the first section of the chapter, for simple liquids, HNC (as well as the PY closure \cite{Hansen2000utg,Attard2002tpb}) can be rationalised from the viewpoint of a density expansion of free energy.
However, the use of these closures for liquids with site-site interactions is harder to justify.
Indeed, the diagrammatic analysis shows that these closures lead to a number of unphysical interactions \cite{Ladanyi1975utg,Chandler1976wcq,Hansen2000utg}.
Nevertheless, experience has shown that both HNC and PY approximations tend to produce relatively reasonable results even for molecular liquids.
Usually, PY describes more accurately hard sphere systems, while HNC tends to be better for liquids interacting via Lennard-Jones and Coulomb potentials \cite{Hansen2000utg}.

For strongly interacting systems, the convergence of HNC closure might become problematic due to the exponent on the right side of \ref{eq:site_site_hnc} becoming increasingly significant.
This problem can be addressed by approximating the greater than one part of the exponential function via a Taylor expansion.
Defining $t^{\ast}_{ij}(r) = -\beta u_{ij}(r) + h_{ij}(r) - c_{ij}(r)$ we can write
\begin{equation}
g_{ij}(r) = \begin{cases}
\mathrm{exp} \left[ t^{\ast}_{ij}(r) \right] & \mbox{if } g_{ij}(r) \leq 1 \\
\sum\limits_{k=0}^{n}\frac{1}{k!} t_{ij}^{\ast}(r)  & \mbox{if } g_{ij}(r) > 1.
\end{cases}
\label{eq:1drism_pse}
\end{equation}
This closure is called a partial series expansion of order n (PSE-n) \cite{Kast2008txr}.
The case $n=1$ is often referred to as Kovalenko-Hirata (KH) closure.

The approach, as we described it, can be successfully applied to liquids interacting only via short-ranged potentials such as oxygen, nitrogen, bromine, \cite{Hsu1976wzh} or \ce{CS2} \cite{Ladanyi1980uyb}.
However, it makes incorrect predictions of an important characteristic of polar molecules: the dielectric constant.
To address this issue, a dielectrically consistent reference interaction site model (DRISM) has been proposed, in which the dielectric constant becomes a fixed input parameter \cite{Perkyns1992wpx}.
The correction somewhat redefines the direct correlation function, but does not change equations \ref{eq:site_site_hnc} or \ref{eq:1drism_pse}.
Thus, to decrease the number of acronyms, we will be referring to both standard as well as dielectrically consistent theory as 1D-RISM, but it will be assumed that polar liquids are treated via the DRISM approach.

\begin{figure}
	\centering
	\includegraphics[draft=false,width=\linewidth]{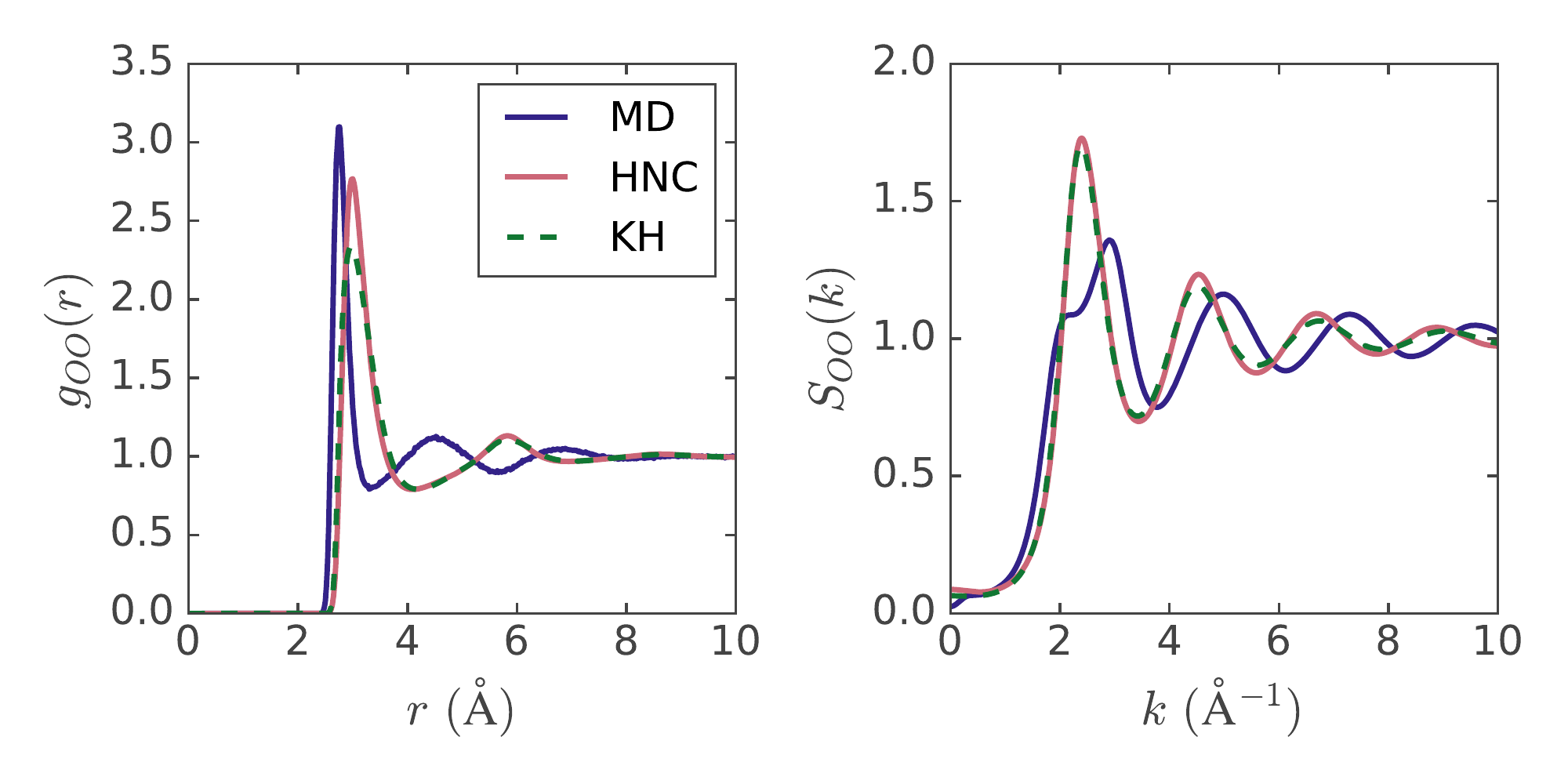}
    \caption{Water oxygen-oxygen radial distribution functions (left figure)  and partial structure factors (right figure) from RISM and molecular dynamics at standard conditions.}
    \label{fig:rism1d_water}
\end{figure}

Since water occupies a central theme in this thesis, it is useful to examine some failures of RISM theory when it is applied to it.
Figure \ref{fig:rism1d_water} demonstrates the site-site radial distribution functions of SPC/E water (cSPC/E in case of RISM \cite{Luchko2010uhh}), obtained at \SI{298}{K} using molecular dynamics (MD) and 1D-RISM with HNC and KH closures.
While the oxygen radial distribution functions look quite similar, the RISM water is radically different from MD one.
A comparison of oxygen partial structure factors $S_{OO}(k) = 1 + \rho \hat h_{OO}(k)$ shows that neither of the closures can predict the characteristic doublet structure of the first peak, which has been linked to the tetrahedral structure of water and voids existing in the actual liquid.
Similarly, comparison of coordination numbers (table \ref{tab:1drism_spce_rdf}) as well oxygen-hydrogen radial distribution functions (figure \ref{fig:rism1d_h}) points towards a lack of hydrogen bonding and a much simpler structure of 1D-RISM water when compared to MD results.

Despite all of the mentioned shortcomings, 1D-RISM is still one of the most useful methods for quickly predicting susceptibilities $\chi$ of molecular liquids.
This might come as a bit surprising, considering all the problems with 1D-RISM we have listed.
However, while 1D-RISM fails to describe short order structure of a liquid ($k>\SI{1}{\per\angstrom}$), it predicts longer wavelength responses relatively accurately.
At the same time, obtaining accurate (and smooth) descriptions of these regions of correlation functions with MD proves to be quite problematic due to the very slow convergence.
Also, most of the information regarding the electrostatic response of water is also connected to the small wavenumber part of the susceptibility.
This makes the DRISM model (with its ability to take experimental dielectric constant as an input parameter) as an arguably better-suited approach for predicting the low-frequency dielectric response of liquid than conventional MD.


\section{3D-RISM as a density functional theory}
\label{sec:3d-rism}

It is not difficult to extend the 1D-RISM model to the situations in which one of the components is present at infinite dilution \cite{Ratkova2015teb}.
This way one can apply the theory to model single molecule solvation. 
However, within this approach, all correlation functions are spherically symmetric, which makes applications to large, non-spherical solutes somewhat problematic.

The three-dimensional reference interaction site model (3D-RISM) provides a clearer picture of the solvation.
The main idea is to treat the solute surrounded by the bulk solvent as an inhomogeneous system, in which external potential is produced by the solute.
The distribution of the solvent sites around the solute is determined by minimizing the total free energy of the system.

The derivation of 3D-RISM proceeds similarly to the derivation of HNC approximation, except that the Grand potential is substituted with its site-site version (equation \ref{eq:molecular_liquids_grand_potential}).
We start by writing down a second order expansion of excess intrinsic free energy functional, defined via equation \ref{eq:site_intr_functional}:
\begin{equation}
\begin{split}
\Fcal^{ex}[\vect{\rho}] = {} & \Fcal^{ex}[\vect{\rho_0}] + \sum_{i=1}^{N} \int \left.\frac{\delta \Fcal}{\delta \rho_i(\rvec)}\right|_{\vect{\rho}=\vect{\rho_0}} \Delta \rho_i (\rvec) \diff \rvec\\
& + \frac{1}{2}\sum_{i=1}^{N}\sum_{j=1}^{N} \iint \left.\frac{\delta^2 \Fcal}{\delta \rho_i(\rvec) \delta \rho_j{\rvec'}}\right|_{\vect{\rho}=\vect{\rho_0}} \Delta \rho_i (\rvec) \Delta \rho_j (\rvec') \diff \rvec \diff \rvec' + F^{B}[\vect{\rho}],\\
\end{split}
\end{equation}
where $\rho_i$ is the number density of site $i$, $\vect{\rho} = [\rho_1, \rho_2, \cdots, \rho_N]$, subscript $0$ indicates reference density, relative to which expansion is being taken, and $F^{B}[\vect{\rho}]$ collects neglected terms in the expansion.

Assuming a homogeneous reference state and using definitions of the whole site-site direct correlation function from equation \ref{eq:full_site_site_dcf}, we get
\begin{equation}
\begin{split}
\Fcal^{ex}[\vect{\rho}] = {} & \Fcal^{ex}[\vect{\rho_0}] + \sum_{i=1}^{N} \bar \mu_i^{ex} \int  \Delta \rho_i (\rvec) \diff \rvec\\
& - \frac{kT}{2} \sum_{i=1}^{N}\sum_{j=1}^{N} \iint  C_{ij}(|\rvec' - \rvec|)\Delta \rho_i (\rvec) \Delta \rho_j (\rvec') \diff \rvec \diff \rvec' + F^{B}[\vect{\rho}].\\
\end{split}
\end{equation}
We write down the full intrinsic free energy functional using equations \ref{eq:site_intr_functional}, \ref{eq:site_chem_pot}, and cancel terms containing the thermal de Broglie wavelength
\begin{equation}
\begin{split}
\Fcal [\vect{\rho}] = {} & \sum_{i=1}^{N} kT \int \rho_i(\rvec) \ln\frac{\rho_i(\rvec)}{\rho_{i,0}}\diff\rvec + \Omega[\vect{\rho}_0] + \sum_{i=1}^{N} \mu_{i,0}\int \rho(\rvec)\diff \rvec \\
& - \frac{kT}{2} \sum_{i=1}^{N}\sum_{j=1}^{N} \iint \Delta \rho_i (\rvec) C_{ij}(|\rvec' - \rvec|) \Delta \rho_j (\rvec') \diff \rvec \diff \rvec' +F^{B}[\vect{\rho}].
\end{split}
\end{equation}

Defining the isotropic grand potential as $\Omega[\vect{\rho}_0] = \sum_{i=1}^{N} \left[ kT\int \rho_{i,0}\ln \rho_{i,0} \Lambda_i^3\diff\rvec - \mu_{i,0} \int \rho_{i,0} \diff\rvec \right]+ \Fcal_0^{ex}[\vect{\rho}_0]$ we can obtain an expression for the change of the grand potential
\begin{equation}
\begin{split}
\Omega[\vect{\rho}] = {} & \Omega[\vect{\rho}_0] + \sum_{i=1}^{N} kT \int \left[\rho_i(\rvec) \ln\frac{\rho_i(\rvec)}{\rho_{i,0}} - \Delta \rho_i(\rvec) \right]\diff\rvec + \sum_{i=1}^{N} \int \rho_i(\rvec) \phi_i(\rvec)\diff \rvec\\
&  - \frac{kT}{2} \sum_{i=1}^{N}\sum_{j=1}^{N} \iint \Delta \rho_i (\rvec) C_{ij}(|\rvec' - \rvec|) \Delta \rho_j (\rvec') \diff \rvec \diff \rvec'+F^{B}[\vect{\rho}].
\end{split}
\label{eq:3d_rism_functional}
\end{equation}
As we can see, this expression differs from the density expansion for simple liquids only via presence of intramolecular correlations, summarised by $C$.

\begin{figure}
\centering
\includegraphics[width=0.7\linewidth,draft=false]{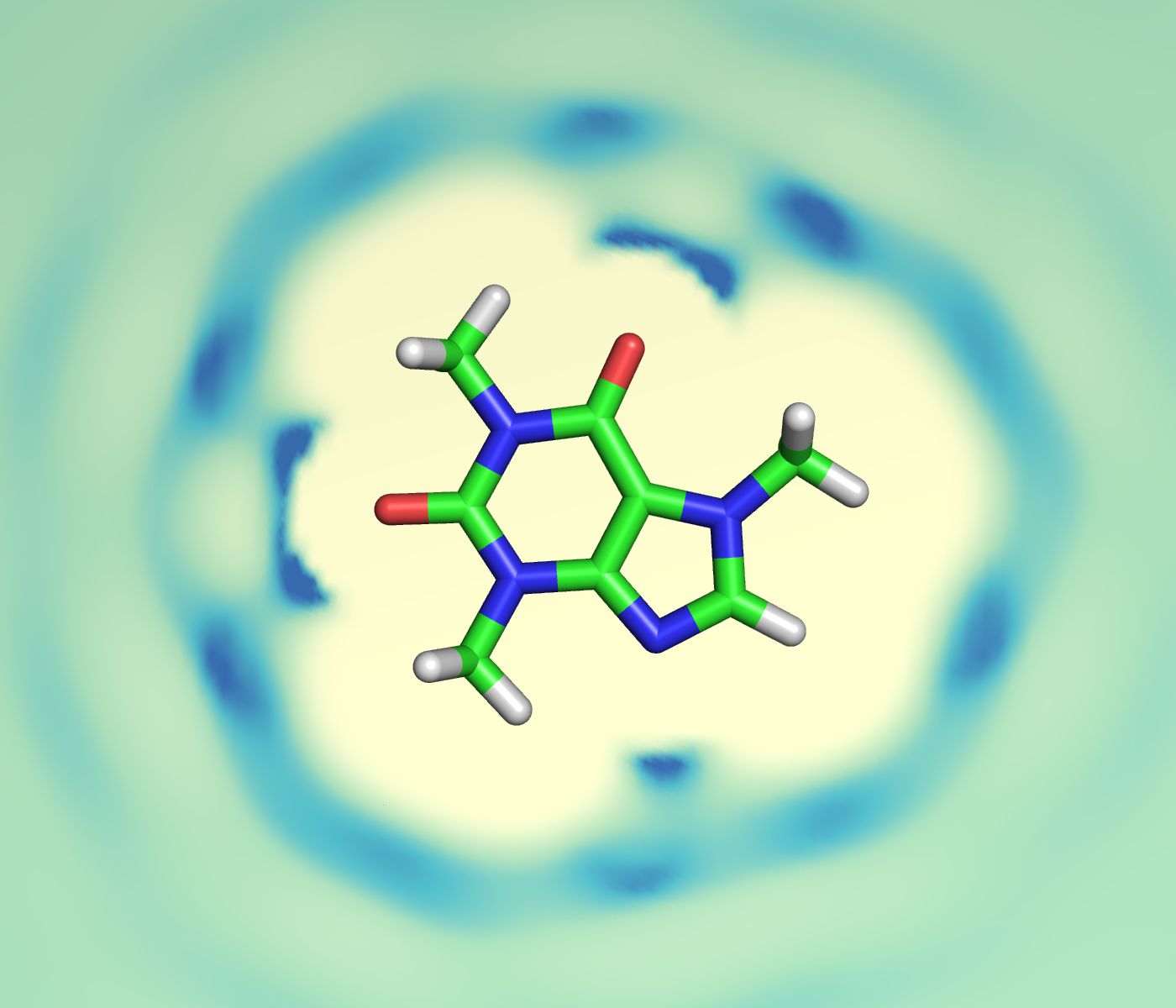}
\caption{Distribution of hydrogens around caffeine, with regions of lower density coloured in yellowish colour and regions of higher in blue.}
\label{fig:H_distrib_caffeine}
\end{figure}

The free energy change from the expression \ref{eq:3d_rism_functional} is obtained by minimizing the $\Omega$ for a given external potential.
Setting $F^B[\vect{\rho}]=\text{const}$ results in a 3D-RISM/HNC approximation, which we will for simplicity call 3D-RISM.

As one would expect from the second order expansion, 3D-RISM is accurate only for small density changes and breaks down for the larger one.
Unfortunately, placing a solute into the liquid makes a region of space inaccessible for solvent, making $\Delta \rho(\rvec) = 0$ inside the solute core.
This does count as a significant perturbation and leads to poor solvation thermodynamics predictions.

Interestingly, despite its shortcomings, the 3D-RISM model is still capable of producing relatively reasonable density distributions around small solutes.
Figure \ref{fig:H_distrib_caffeine} demonstrates a typical result of a 3D-RISM calculation.
You can see a distribution of hydrogen sites around the caffeine molecule.
It is easy to see hydrogen bonds as well as outlines of the first coordination shell.
While it was difficult to compare these results directly to methods such as molecular dynamics, we believe that qualitative pictures are mostly identical.
A number of studies confirm that for solutes described via Lennard-Jones and electrostatic potentials 3D-RISM predicts solvent distributions that are in good agreement with both other computational methods as well as experimental observations, even in the case of macromolecules \cite{Nikolic2012vge,Stumpe2011tam,Imai2009tmb,Imai2007ude,Sindhikara2012uik}.


\begin{figure}
	\centering
	\includegraphics[width=0.7\linewidth,draft=false]{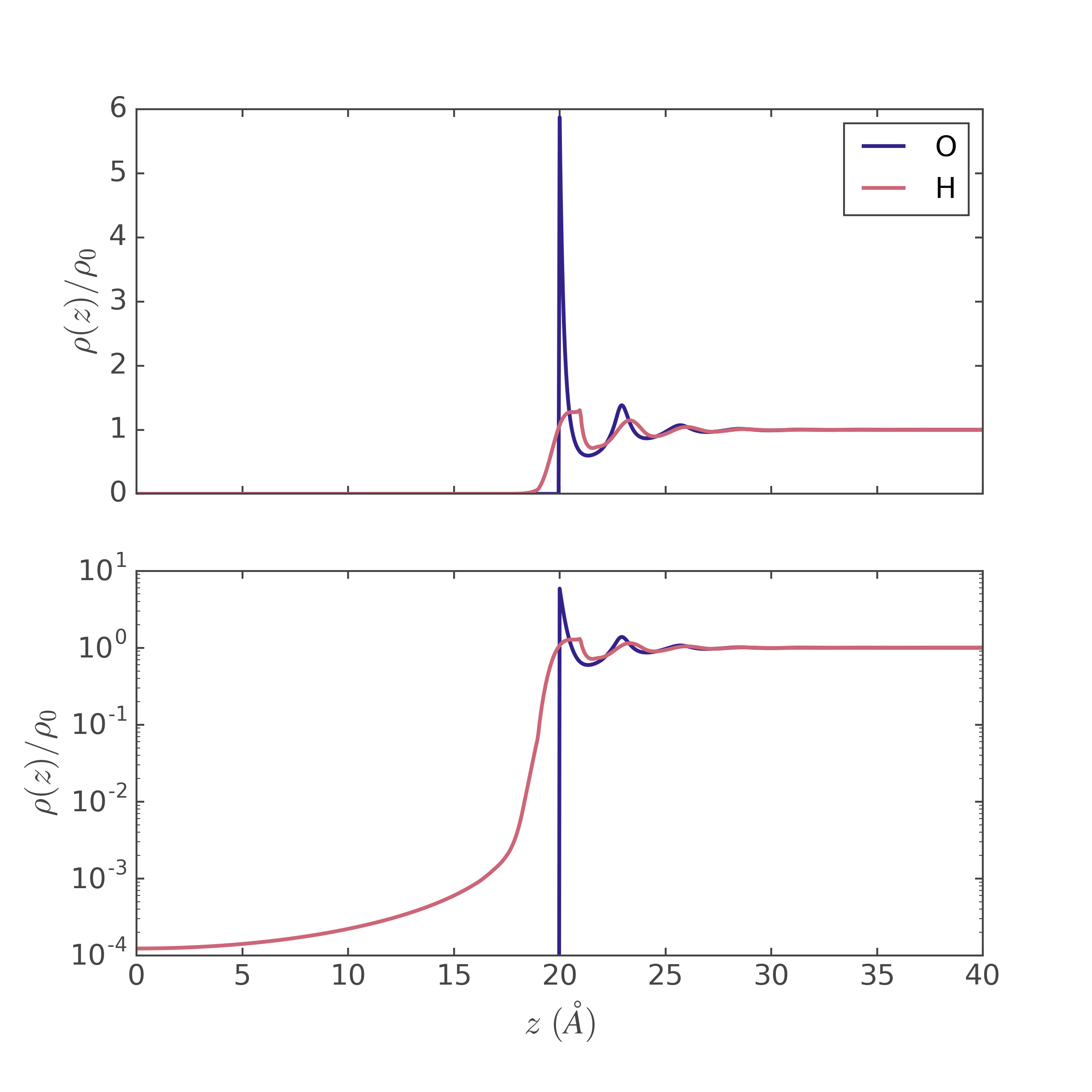}
	\caption{Distribution of water oxygens (blue) and hydrogens (red) in a large box with $u_O = \infty$ for $z<\SI{20}{\angstrom}$. You can see that some hydrogens can be found arbitrarily far away from the oxygens.}
	\label{fig:h_spilling_through_the_wall}
\end{figure}

One of the effects which 3D-RISM fails to account properly for is solvent bonding.
Recall that just as intermolecular correlations, intramolecular correlations are also described only up to second order.
Thus, we expect them to be inaccurate to some extent.
On the figure \ref{fig:h_spilling_through_the_wall} you can see a distribution of water oxygens and hydrogens near a hard wall, defined only for oxygens via $u_O(z)=\infty$ for $z<\SI{20}{\angstrom}$ and $u_O(z) =0$ otherwise.
Hydrogens are not restricted by any potential.
You can see that the density of hydrogens is non-zero at arbitrary distances from oxygen atoms, even though built-in intramolecular correlation functions $\omega$ requires them to be at exactly $\SI{1}{\angstrom}$ from oxygens.

In the previous two sections, we described 1D and 3D-RISM approaches.
While quite similar, it is important to emphasize a major difference between them.
In 1D-RISM one starts with unknown total and direct correlation functions, but the system is assumed to be homogeneous.
It is similar to pair HNC discussed at the beginning of the section.
Within 3D-RISM, the direct correlation functions between solvent molecules are fixed, and the system density is varied.
This approach is closely related to singlet HNC (HNC1).
Note that while HNC1 is capable of describing anisotropic systems, it comes at a cost: the singlet Ornstein-Zernike approaches lose a power of density in accuracy compared to pair approximations \cite{Attard2002tpb}.

\section{3D-RISM as an integral equation theory}

We can also treat 3D-RISM as an integral equation approach \cite{Hirata2003tpg}.
A theoretical background of this approach can be found in Ref. \citenum{Sergiievskyi2013upd} and Ref. \citenum{Cortis1997wll}.
The main idea is to reduce the 6-dimensional Ornstien-Zernike equation \cite{Gubbins1985udt} to get
\begin{equation}
\vect{D} \vect{\hat h} (\kvec) = \vect{\hat X} (k) \vect{\hat c} (\kvec),
\label{eq:3d_oz}
\end{equation}
where $\vect{\hat h} (\kvec)$ is the vector of solute-solvent total correlation functions ($\Delta \rho_i = h_i \rho_i$), and $\vect{\hat c} (\kvec) = \left[\hat c_{1}, \hat c_{2}, \cdots, \hat c_{N} \right]$ is the vector of solute-solvent direct correlation functions.
The above expression is typically called the solute-solvent Ornstein-Zernike equation and assumes that the solute is present at infinitely low concentration.
Similarly to 1D-RISM or HNC theories, this equation needs to be supplemented with a proper closure.

Although one often introduces closures to the above equation by modifying analogous closures from simple liquid theories, we can formally derive a HNC-like closure starting from 3D-RISM free energy functional (equation \ref{eq:3d_rism_functional}).
Recall the equilibrium condition in grand canonical ensemble $\delta \Omega/\delta \rho[\rvec] = 0$.
Applying it to the 3D-RISM/HNC functional, we get:
\begin{equation}
\frac{\Omega[\rho]}{\rho_i(\rvec)} = 0 = \ln g_i(\rvec) + \beta \phi_i(\rvec) - \sum_{j=1}^{N} \int \Delta \rho_j(\rvec') C_{ji}(|\rvec'-\rvec|)\diff \rvec'.
\label{eq:hnc_equilibrium}
\end{equation}
Rearranging this expression we get an integral equation for density distributions of solvent sites:
\begin{equation}
\rho_i(\rvec) = \rho_{i,0} \exp \left[-\beta \phi_i(\rvec) + \sum_{j=1}^{N} \int \Delta \rho_j(\rvec') C_{ji}(|\rvec'-\rvec|)\diff \rvec' \right].
\end{equation}
We use equation \ref{eq:full_site_site_dcf} to simplify the above and get
\begin{equation}
g_i(\rvec) = \exp \left\{ -\beta \phi_i(\rvec) + \sum_{j=1}^{N} \int \Delta \rho_j(\rvec') \left[ \frac{\delta_{ij}\delta(|\rvec' - \rvec|)}{\rho_i} - \chi_{ji}^{-1} (|\rvec'-\rvec|) \right] \diff \rvec' \right\}.
\end{equation}
Integrating the $\delta$ function we obtain
\begin{equation}
g_i(\rvec) =  \exp \left[-\beta \phi_i(\rvec) + h_i(\rvec) - \sum_{j=1}^{N} \int \Delta \rho_j(\rvec') \chi^{-1}_{ji} (|\rvec'-\rvec|) \diff \rvec' \right],
\label{eq:almost_3dhnc}
\end{equation}
which, with the aid of \ref{eq:3d_oz}, is transformed into
\begin{equation}
g_i(\rvec) = \exp \left[-\beta \phi_i(\rvec) + h_i(\rvec) - c_i(\rvec)\right],
\label{eq:3drism_hnc}
\end{equation}
a 3D-RISM equivalent of HNC closure.

For strongly attractive potentials, the 3D-RISM/HNC system of equations can be quite difficult to converge.
Similarly to 1D-RISM theory, we can introduce a partial series expansion of order n (PSE-n) of HNC equation \cite{Kast2008txr}.
Defining $t^{\ast}_i(\rvec) = -\beta u_{i}(\rvec) + h_{i}(\rvec) - c_{i}(\rvec)$ we write
\begin{equation}
g_{i}(\rvec) = \begin{cases}
\mathrm{exp} \left[ t^{\ast}_i(\rvec) \right] & \mbox{if } g_{i}(\rvec) \leq 1 \\
\sum\limits_{k=0}^{n}\frac{1}{k!} t_i^{\ast}(r)  & \mbox{if } g_{i}(\rvec) > 1.
\end{cases}
\label{eq:3drism_pse}
\end{equation}
In practice combining the PSE-3 closure and equation \ref{eq:3d_oz} is the fastest and the most robust way to minimise the 3D-RISM/HNC functional.

3D-RISM equations, written in terms of the solute-solvent correlation functions, are often referred to as an integral equation theory or a molecular theory of liquids.
Using this form, one can also write a somewhat simpler expression for the solvation free energy using thermodynamic integration (equation \ref{eq:kirkwood_mu}).
In the case of 3D-RISM/HNC one arrives at \cite{Hirata2003tpg}
\begin{equation}
\Delta \Omega_{HNC} = kT\sum\limits_{i=1}^{N} \rho_i \int \left[\frac{1}{2} h_i^2(\rvec) - \frac{1}{2} h_i(\rvec)c_i(\rvec) - c_i(\rvec)\right].
\end{equation}
To verify that this expression is consistent with the 3D-RISM/HNC functional one needs to substitute $\ln g_i(\rvec) = - \beta \phi(\rvec) + h_i(\rvec)-c_i(\rvec)$ into equation \ref{eq:3d_rism_functional}.
Applying the same approach to PSE-n closure, one obtains \cite{Kast2008txr}
\begin{equation}
\Delta \Omega_{PSE-n} = \Delta \Omega_{HNC} - k T \sum_{i=1}^{N} \rho_{i} \int \left\{ \Theta \left[ h_{i}(\mathbf{r})\right]  \frac{t^{\ast}_{i} (\mathbf{r})^{n+1}}{(n+1)!} \right\} \diff \mathbf{r},
\end{equation} 
where $\Theta$ is a Heaviside step function:
\begin{equation}
\Theta(x) = \begin{cases}
0 & \mbox{if } x < 0 \\
1 & \mbox{if } x \geq 0.
\end{cases}
\end{equation}

\begin{figure}
\centering
\includegraphics[width=0.7\linewidth,draft=false]{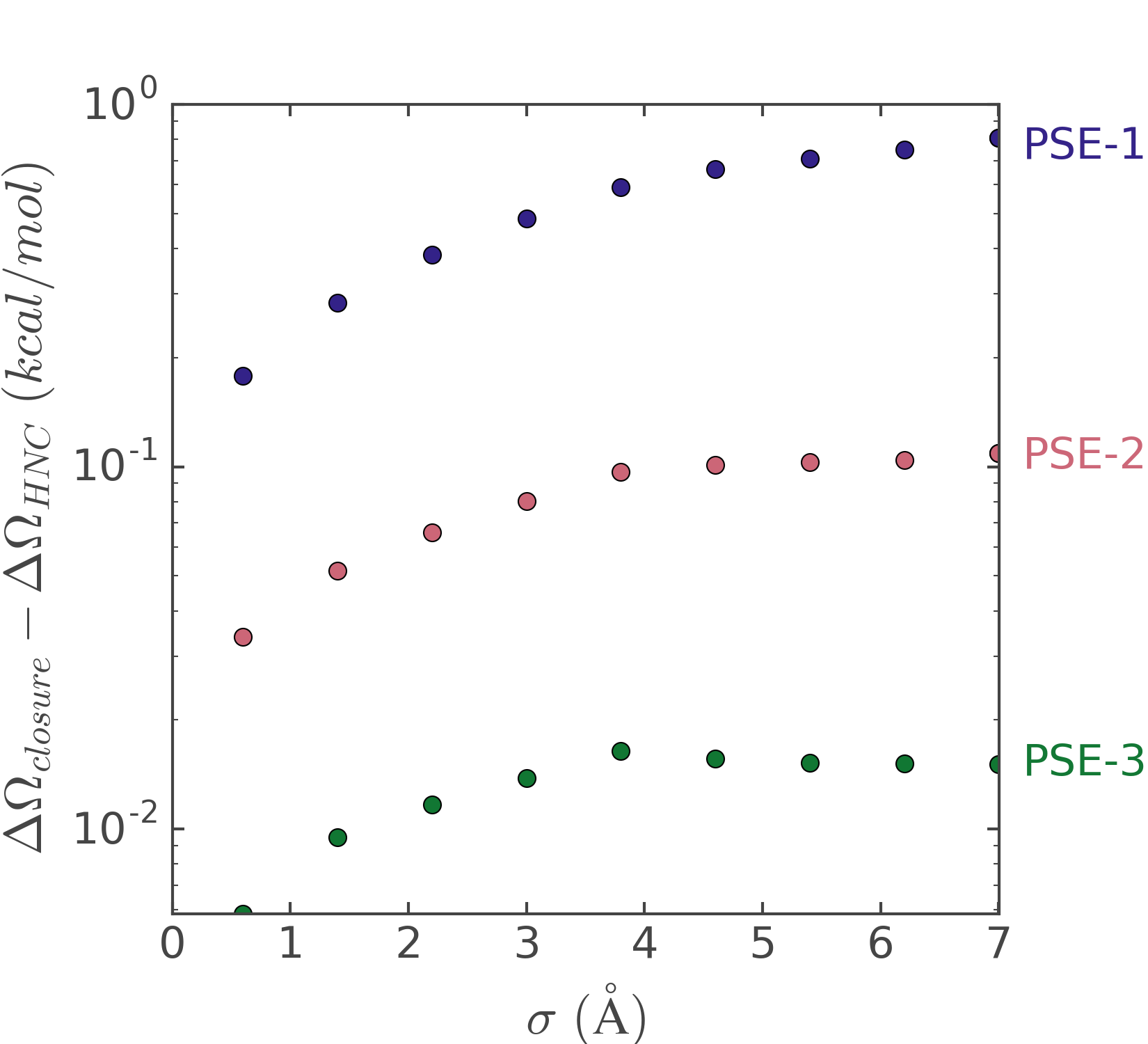}
\caption{Solvation free energies of Lennard-Jones solutes with $\sigma$ given on x axis and $\epsilon=\SI{4}{kcal\per\mole}$. On y axis you can see the difference in free energies by PSE-n and HNC closures.}
\label{fig:closure_acc}
\end{figure}

On figure \ref{fig:closure_acc} you can see how the differences in free energies computed by PSE-n and HNC closures depend on the solute size.
The system is water with Lennard-Jones solutes of different radii and $\epsilon=\SI{4}{kcal/mol}$.
These results should not be generalised, as the precise magnitude of the disparity between closures depends on the type of solute and their size.
However, in general, the differences between free energy changes from PSE-n closures and HNC becomes small quite fast, especially for $n\geq3$.

For this thesis, the majority of calculations with realistic and larger solvents was done with the PSE-3 closure to avoid convergence issues, while calculations on model solutes were mostly done with HNC.
However, it will be assumed that the conclusions which will be drawn for PSE-3, also hold for HNC and vice versa.


\section{Partial molar volume}

Partial molar volume is defined as a change in system's volume upon addition of an $i$-th component at constant pressure \cite{Ratkova2015teb}:
\begin{equation}
\bar{V}_i = \left(\frac{\partial V}{\partial N_i}\right)_{P,T,N_{j\neq i}} = \left(\frac{\partial\mu_i}{\partial P}\right)_{P,T,N_{j\neq i}},
\end{equation}
where the second equality was obtained using Maxwell's relation \cite{Callen1985vqb}.
Usually, the partial molar volume is discussed as a part of solvation thermodynamics, but due to its relation to integral equation theories, we decided to present it here.

Similarly to other thermodynamic quantities, it can be split into ideal and excess contributions.
Using equation \ref{eq:mu-mu_ig-mu_ex} we get:
\begin{equation}
\begin{split}
\bar{V}_i &= \frac{\partial \mu^{ex}_i}{\partial P} + \frac{\partial \mu^{ig}_i}{\partial P}\\
&= \frac{\partial \mu^{ex}_i}{\partial P} + kT\frac{\ln \rho}{\partial P}\\
&= \Delta V_i + kT \chi_T,
\end{split}
\end{equation}
where the last equality was obtained using equation \ref{eq:compress_tot_cor_func}, and $\Delta V_i$ stands for the excess part of partial molar volume.
Thus, $kT\chi_T$ is the molecule volume arising due to its kinetic energy, while $\Delta V_i$ depends purely on intramolecular interactions \cite{Ben-naim2006vpu}.

The excess volume of a solute $\Delta V$ in a single-component solvent can be expressed through a total correlation function \cite{Ben-naim2006vpu,Matubayasi1996wqa}:
\begin{equation}
\Delta V = -\int \left[g(\rvec) - 1\right]\diff \rvec = - \int h(\rvec) \diff \rvec.
\label{eq:pmv}
\end{equation}
This relation can be readily interpreted for a solute that has $h(\rvec) = -1$ for points inside the solute and $h(\rvec)=1$ for other regions of space.
To obtain a more general interpretation we need to introduce the concept of the Gibbs dividing surface.

\begin{figure}
\centering
\includegraphics[width=0.7\linewidth]{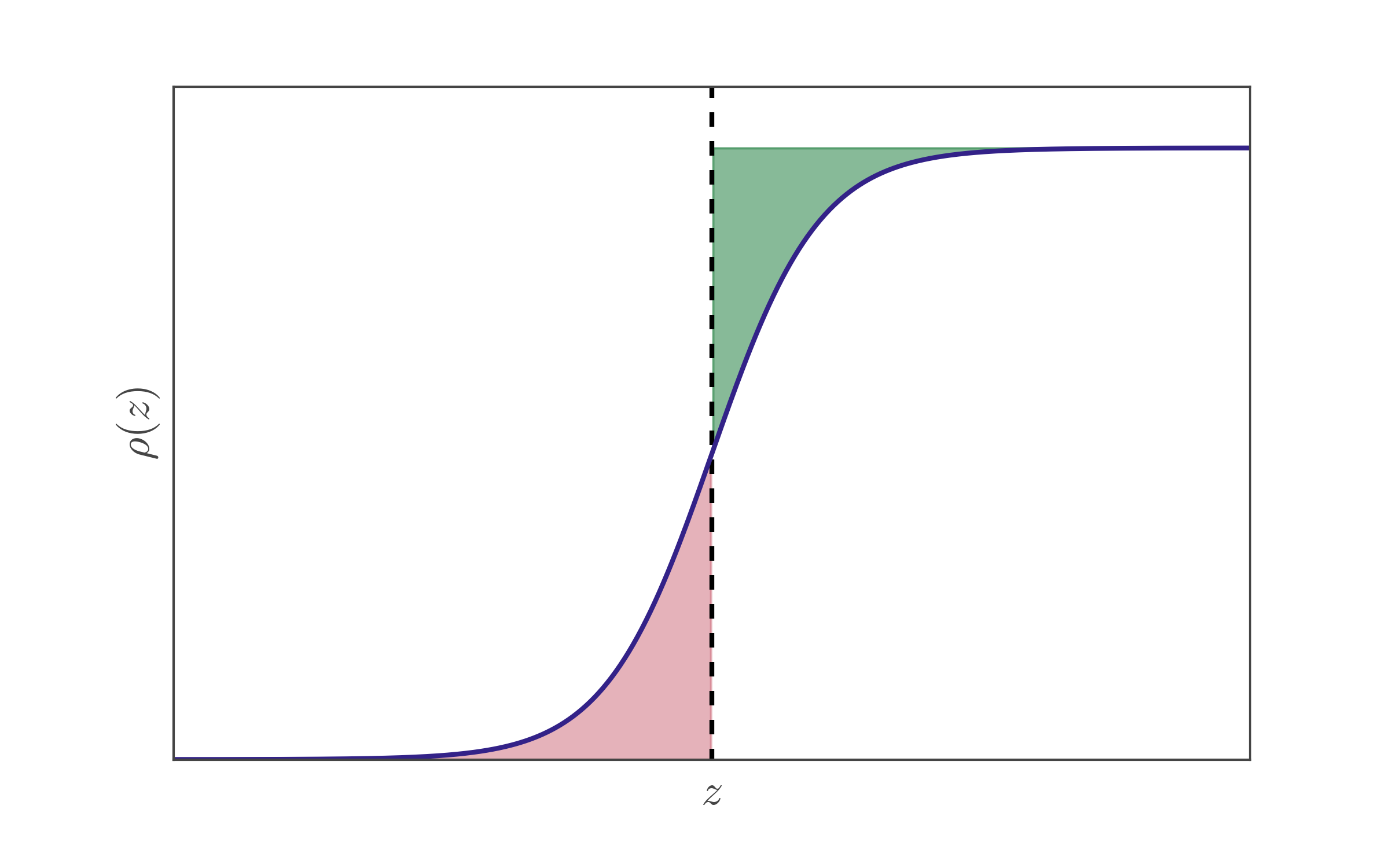}
\caption{A schematic depiction of Gibbs dividing surface for the case of planar interface. The surface, shown as a dashed line divides solvent local density (blue line) in such way that excess of solvent to the left of the surface (region shaded in red) is equal to the depletion of solvent to the right of the surface (region shaded in green).}
\label{fig:gibbs_dividing_surface}
\end{figure}

The Gibbs dividing surface is a two-dimensional boundary, dividing an interface in such a way that the excess of solvent in one phase is equal to its depletion from another (see figure \ref{fig:gibbs_dividing_surface}) \cite{Shimizu2014vtg,Neumann2010tbp}.
For a spherical solute we can express it as
\begin{equation}
4\pi \int\limits_{0}^{R_G} g(r) r^2 \diff r + 4\pi \int\limits_{R_G}^{\infty} \left[ g(r) - 1 \right]r^2\diff r = 0,
\end{equation}
where $R_G$ is a radius of the Gibbs dividing surface.
Adding the above equation to equation \ref{eq:pmv} we find:
\begin{equation}
\begin{split}
\Delta V & = - 4\pi \int\limits_{0}^{\infty} \left[g(r) - 1 \right] r^2 \diff r + 4\pi \int\limits_{0}^{R_G} g(r) r^2 \diff r + 4\pi \int\limits_{R_G}^{\infty} \left[ g(r) - 1 \right]r^2\diff r\\
& = - 4\pi \int\limits_{0}^{R_G} \left[g(r) - 1 \right] r^2 \diff r + 4\pi \int\limits_{0}^{R_G} g(r) r^2 \diff r\\
&= 4\pi \int\limits_{0}^{R_G} r^2 \diff r = \frac{4}{3}\pi R_G^3.
\end{split}
\end{equation}
Thus, for a spheric solute, excess volume is equal to the volume enclosed by the Gibbs dividing surface.
This result also holds for a solute with an arbitrary shape \cite{Shimizu2014vtg}.
It gives us a convenient way to define both volume and surface of a solute that are consistent with each other.

For a multi-component solvent, solute partial molar volume becomes dependent on the solvent component molar volumes.
For instance, in the case of a two-component solvent we have $\bar{V}_s = - \rho_A \bar{V}_A G_{sA} - \rho_B \bar{V}_B G_{sB}$, where $G$ represents the Kirkwood-Buff integral $\bar{G}_{ij} = \int h_{ij}(\rvec)\diff \rvec$.
A simpler expression can be obtained using direct correlation functions
\begin{equation}
\bar{V}_s = kT\chi_T \left[1 - \rho\sum\limits_{i=1}^N \int c_i(\rvec) \diff \rvec \right],
\end{equation}
with excess volume defined by
\begin{equation}
\Delta V_s = -kT\chi_T \rho \sum\limits_{i=1}^N \int c_i(\rvec) \diff \rvec.
\end{equation}
Partial molar volumes in this thesis were computed using the above equations.

\section{Solvation free energy from 3D-RISM}
\label{sec:solv_free_energy_3drism}

Chemical potential (or solvation free energy) of a solute can be expressed as a change in the grand potential of a solvent due to the presence of a single molecule \cite{Liu2013vpm}:
\begin{equation}
\mu^{ex} = \Delta F = \Omega[\vect{\rho}] - \Omega[\vect{\rho}_0],
\end{equation}
where we will reserve the symbol $\Delta F$ to specifically denote solvation free energy in Ben-Naim's definition whenever we do not have to worry regarding the ensemble in which the process takes place.
As was already discussed in section \ref{sec:solv_free_energy_app}, the ability to accurately predict solvation free energies has much practical value.
A major practical advantage of 3D-RISM models, when compared to molecular dynamics, comes from the possibility to obtain free energy changes from a single (or a couple of) end-point calculation, without the need to perform a thermodynamic integration.
It is not surprising that much effort has been put into predicting solvation free energies by interaction site models.
An overview of all proposed models is beyond the scope of the thesis; besides, a number of excellent articles have been published on this subject \cite{Ratkova2015teb,Luchko2016vwg,Johnson2016tta,Tielker2016tdi}.
Here we will only discuss a few corrections related to 3D-RISM.

Molecular dynamics simulations provide a straightforward way for evaluating the accuracy of 3D-RISM free energy functionals.
A comparison of the results with experiment provides a somewhat less clear picture since, in addition to errors arising from 3D-RISM approximations, one has to take into account the accuracy of the potentials, the validity of the classical approximation, experimental errors, etc.
On the other hand, a comparison of 3D-RISM solvation free energies with those from molecular dynamics allows us to directly assess the accuracy of 3D-RISM free energy expression, provided that the same potentials are used.

\begin{figure}
\centering
\includegraphics[width=\linewidth,draft=false]{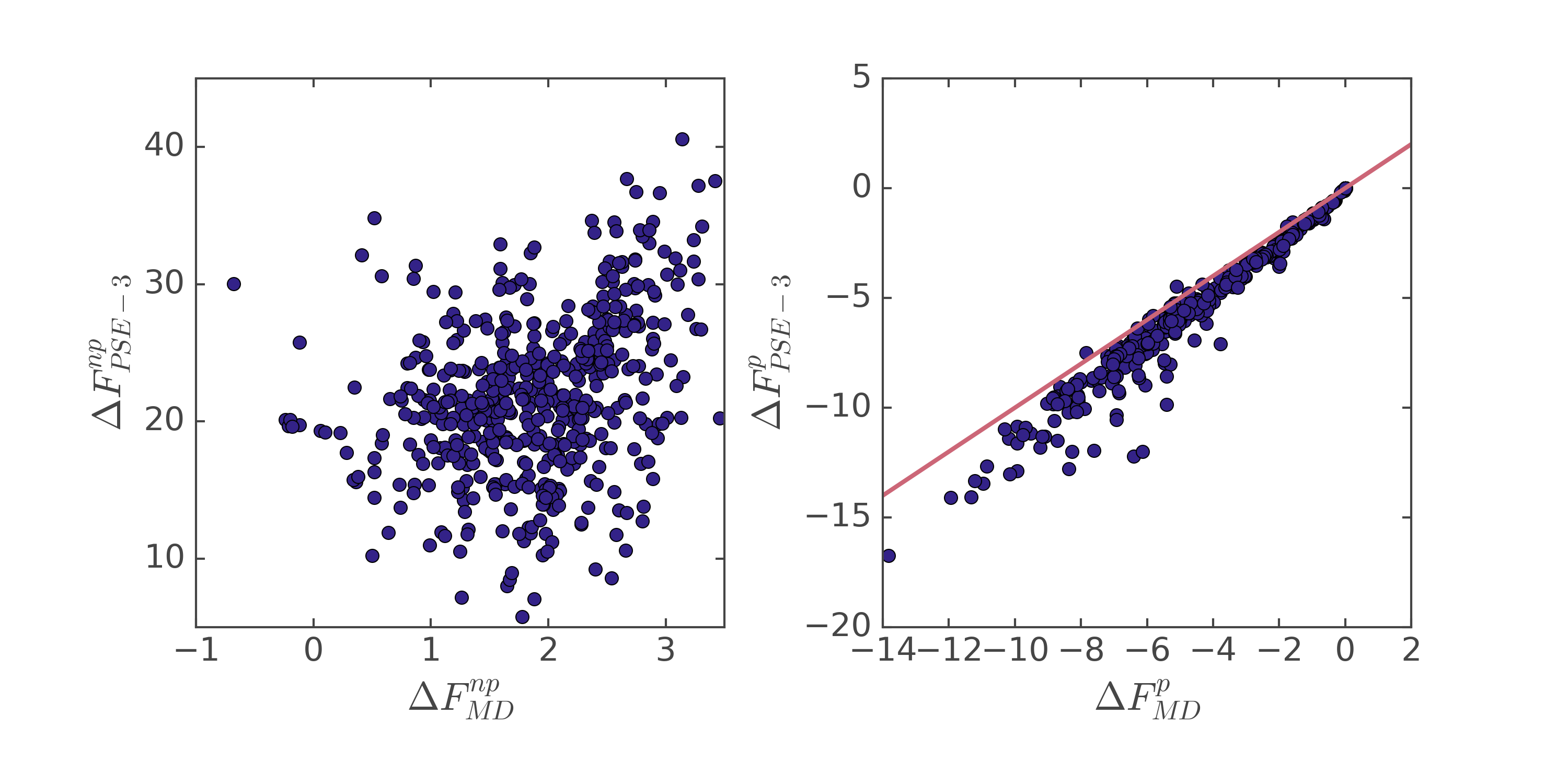}
\caption{Comparison of molecular dynamics and PSE-3 Lennard-Jones components of hydration free energies (left figure) and electrostatic components (right figure) for 504 molecules from Mobley dataset, which is discussed in detail in section \ref{sec:neutral_exp}. Molecular dynamics results are taken from Ref. \citenum{Mobley2009wdk}. All values are in \si{kcal\per\mole}}
\label{fig:pse3_fe}
\end{figure}

We have already discussed (section \ref{sec:numerical_exp}) that it is possible to split solvation free energy into Lennard-Jones and electrostatic contributions formally.
To do so within 3D-RISM theory, we simply compute solvation free energies of a molecule with and without partial charges on atomic sites.
Figure \ref{fig:pse3_fe} demonstrates that while electrostatic components of free energy predicted by 3D-RISM are relatively accurate, the Lennard-Jones components practically do not correlate with molecular dynamics.

Recently, some studies have demonstrated that it is possible to significantly improve the accuracy of 3D-RISM solvation free energies using corrections of the form:
\begin{equation}
\Delta F_{UC} = \Delta F_{3D-RISM} + a \bar{V} + b,
\end{equation}
where $UC$ stands for universal correction, $a$ and $b$ are empirical coefficients that depend on the solvent and closure, and $\Delta F_{3D-RISM}$ is the 3D-RISM free energy, most typically computed with KH closure \cite{Palmer2010upn,Truchon2014vty,Johnson2016tta}.
Another correction, called NgB, was developed specifically for water \cite{Truchon2014vty}
\begin{equation}
\Delta F_{NgB} = \Delta F_{KH} + \frac{kT\rho_O}{2} (1 - \gamma) \int_{V_{in}} c_O^{LJ} (\rvec)\diff \rvec,
\end{equation}
where $\rho_O$ is the density of oxygen sites in water, $\gamma$ is an empircial coefficient, $c^{LJ}_O$ is the direct correlation function for water oxygens, evaluated without the solute charges, and $V_{in}$ is the volume inside the solute, defined via the solute-solvent potential energy.

Both of the corrections introduced above significantly improve the accuracy of 3D-RISM solvation free energies (\ref{fig:uc_ngb_fe}).
However, these corrections were introduced empirically and did not suggest reasons why they might be effective or how empirical coefficients might depend on the solvent or thermodynamic conditions.
In the next chapter, we will introduce correction schemes that do not require prior parametrization and can be applied to a much larger variety of systems.


%
%


\part{Results and Discussion}
\chapter{Pressure corrections} 
\label{chap:pres_cor}

\label{Chapter5}

\lhead{Chapter 5. \emph{Pressure corrections}}

The chapter is dedicated to the main theoretical results of the thesis.
We first show how the HNC functional can be conveniently split into a couple of contributions.
Identifying the overestimated component we eliminate it, introducing simple and advanced pressure corrections.
The remaining sections are dedicated to discussing pressure corrected models in the context of water solvation.
We are using simple model solutes to focus on physical insights instead of the individual peculiarities of realistic molecules.

\section{Dissecting HNC free energies}

For clarity we introduce the following notation \footnote{A similar notation can be found, for example, in Ref. \citenum{Vener2002wkr}.}
\begin{equation}
\begin{split}
\braket{\vect{f}}{\vect{g}} & = \sum_{i=1}^{N} \int f_i(\rvec) g_i(\rvec) \diff \rvec,\\
\mel{\vect{f}}{\vect{K}}{\vect{g}} & = \sum_{i=1}^{N}\sum_{j=1}^{N}\iint f_i(\rvec) K_{ij}(\rvec,\rvec') g_j(\rvec')\diff \rvec \diff \rvec'
\end{split}
\end{equation}
where $\vect{f}$ and $\vect{g}$ are vectors of functions containing $N$ elements, and $\vect{K}$ is an $N$ by $N$ matrix of functions, also referred to as kernel.

Using the above notation we can rewrite the 3D-RISM/HNC functional (equation \ref{eq:3d_rism_functional}) as:
\begin{equation}
\beta \Delta \Omega [\vect{\rho}] = \braket{\vect{\rho}}{\ln \vect{g}} - \braket{\Delta \vect{\rho}}{\vect{1}} + \braket{\vect{\rho}}{\beta\vect{u}} - \frac{1}{2}\mel{\Delta \vect{\rho}}{\vect{C}}{\Delta \vect{\rho}}
\label{eq:hnc_func_change}
\end{equation}
where $\vect{1}$ represents a vector of functions equal to $1$ everywhere and $\ln \vect{g} = \left[\ln g_1(\rvec)\cdots \ln g_N(\rvec) \right]$.
To simplify this expression we also rewrite the condition for equilibrium \ref{eq:hnc_equilibrium}: $\ln \vect{g} + \beta \vect{\phi} = \braket{\vect{C}}{\Delta \vect{\rho}}$.
Plugging this into the equation \ref{eq:hnc_func_change} we obtain the change of grand potential at the equilibrium
\begin{equation}
\begin{split}
\beta \Delta \Omega [\vect{\rho}] & = \braket{\vect{\rho}}{\ln \vect{g} + \beta \vect{u}} - \braket{\Delta \vect{\rho}}{\vect{1}}  - \frac{1}{2}\mel{\Delta \vect{\rho}}{\vect{C}}{\Delta \vect{\rho}}\\
& = \bra{\vect{\rho}}\braket{\vect{C}}{\Delta \vect{\rho}}  - \braket{\Delta \vect{\rho}}{\vect{1}} - \frac{1}{2}\mel{\Delta \vect{\rho}}{\vect{C}}{\Delta \vect{\rho}}\\
& = \frac{1}{2}\mel{\vect{\rho}}{\vect{C}}{\Delta \vect{\rho}} + \frac{1}{2}\mel{\vect{\rho_0}}{\vect{C}}{\Delta \vect{\rho}} - \braket{\Delta \vect{\rho}}{\vect{1}}\\
& = \frac{1}{2}\mel{\vect{\rho}}{\vect{C}}{\vect{\rho}} - \braket{\vect{\rho}}{\vect{1}} - \frac{1}{2}\mel{\vect{\rho_0}}{\vect{C}}{\vect{\rho_0}} + \braket{\vect{\rho_0}}{\vect{1}},\\
\end{split}
\label{eq:hnc_decomp}
\end{equation}
where $\vect{\rho_0}$ is the vector of the site densities in the reference system.

To interpret the above results we recall that $\Omega = -PV$.
To connect the integrals of the correlation functions to the pressure, we can use the compressibility theorem, introduced in the section \ref{sec:pressure_sk}.
Cummings and Stell have derived its extension for the interaction site fluids \cite{Cummings1982wyk}
\begin{equation}
\beta \pdv{P}{\rho} = \sum_{i=1}^{N}\sum_{j=1}^{N} x_i \left[\delta_{ij}- \rho_j \f C_{ij}(0)\right],
\end{equation}
where $\rho_i = x_i \rho$.
After a bit of algebra we can recast this result in the matrix form
\begin{equation}
\beta \left(\pdv{P}{\rho}\right)_{V,T} = \rho \vect{x}^T \left[\vect{D^{-1}} - \vect{\f C}(0)\right] \vect{x} = 1 -  \rho \vect{x}^T \vect{\f C}(0)\vect{x},
\end{equation}
where $\vect{x} = \left[x_1\cdots x_N\right]$ is the vector of the mole fractions of the sites and superscript $T$ denotes a transpose.
To obtain the pressure we need to integrate the above expression.
In general, it is not possible to do it in a straightforward manner, since $\vect{C}$ depends on the density of the system.
However, within the HNC approximation we assume direct correlation functions to be constant (the same as in the reference system), so the integration can be performed analytically from $\rho=0$ to $\rho=\rho_0 = \sum_{i=1}^N \rho_{i0}$ to produce
\begin{equation}
\beta P_0 = \sum_{i=1}^N \rho_{i0}  - \frac{1}{2} \vect{\rho_0}^T \vect{C}(k=0) \vect{\rho_0},
\label{eq:hnc_pressure}
\end{equation}
where $P_0$ indicates the pressure (free energy density) of a homogeneous system.
The grand potential is then
\begin{equation}
\begin{split}
\beta \Omega_0 &= - \beta P_0 V  = -\braket{\vect{\rho_0}}{\vect{1}} + \frac{1}{2}\mel{\vect{\rho_0}}{\vect{C}}{\vect{\rho_0}}\\
&= -V \sum_i \rho_i + \frac{V}{2} \sum_{ij} \rho_i \rho_j \f C_{ij}(0).
\end{split}
\end{equation}
Note that an identical expression for homogeneous pressure in 3D-RISM was obtained by Sergiievskyi et al. \cite{Sergiievskyi2015tab} as well as by a number of others for singlet HNC in general \cite{Evans1983vrg,Attard1991uvb,Attard2002tpb}.

To find the pressure in the case of an inhomogeneous system we can use the exact result obtained by Pozhar et al. \cite{Pozhar1993tfd}
\begin{equation}
P(\rvec; \rho) = \rho(\rvec) \left[\bar{\mu}(\rvec; \rho) - \int_{0}^{1}\diff \lambda \bar{\mu}(\rvec; \lambda \rho) \right],
\end{equation}
where $\bar{\mu}(\rvec; \rho)$ is the intrinsic chemical potential at $\rvec$ and the $\lambda$ parameter controls density $\rho_{\lambda}(\rvec) = \lambda \rho(\rvec)$.
The result can be readily extended to multicomponent systems
\begin{equation}
P(\rvec; \vect{\rho}) = \vect{\rho}(\rvec) \cdot \left[\vect{\bar{\mu}}(\rvec; \vect{\rho}) - \int_{0}^{1}\diff \lambda \vect{\bar{\mu}}(\rvec; \lambda \vect{\rho}) \right],
\end{equation}
in which scalars $\mu$ and $\rho$ are substituted by vector analogues.
Within the HNC approximation
\begin{equation}
\bar{\mu_i} = kT\ln \Lambda^3_i \rho_i(\rvec) + \bar{\mu}^{ex}_{i0}(\rvec) -kT \braket{\vect{C_{i0}}}{\Delta \vect{\rho}}_i,
\end{equation}
where subscript $0$ indicates that both excess quantities were evaluated at some reference system.
Then, using $\braket{\vect{C_{i0}}}{\Delta \vect{\rho}} = \braket{\vect{C_{i0}}}{\vect{\rho}} - \braket{\vect{C_{i0}}}{\vect{\rho_0}}$ and taking the integral, we get
\begin{equation}
\begin{split}
P(\rvec; \vect{\rho}) & = \sum_{i=1}^{N}\rho_i(\rvec)\Big[ kT\ln \Lambda^3_i \rho_i(\rvec) + \bar{\mu}^{ex}_{i0}(\rvec) -kT \braket{\vect{C_{i0}}}{\Delta \vect{\rho}} \\
&\qquad\qquad \left.- kT\ln \Lambda^3_i \rho_i(\rvec) + kT - \bar{\mu}^{ex}_{i0}(\rvec) +\frac{kT}{2} \braket{\vect{C_{i0}}}{\vect{\rho}} - kT\braket{\vect{C_{i0}}}{\vect{\rho_0}}\right]\\
& = kT \sum_{i=1}^{N} \rho_i(\rvec) \left[1 - \frac{1}{2} \braket{\vect{C_{i0}}}{\vect{\rho}} \right].
\end{split}
\end{equation}
It follows that for an inhomogeneous system:
\begin{equation}
\beta \Omega[\vect{\rho}] = -\beta \int P(\rvec; \vect{\rho})\diff \rvec = - \braket{\vect{\rho}}{\vect{1}} + \frac{1}{2}\mel{\vect{\rho}}{\vect{C}}{\vect{\rho}},
\end{equation}
where we dropped the subscript from $\vect{C}$ for consistency with previous results.

The findings of the past couple of paragraphs highlight that the HNC free energy corresponds to nothing else but $\Delta \Omega_{HNC}=-\Delta P V$, which perhaps is not very surprising.
However, these results at least point out that the theory is internally consistent.
Moreover, they readily highlight the problems with the approximation; indeed, using equation \ref{eq:hnc_pressure} one readily finds that liquid water at room temperature and normal density has a pressure of about $\SI{9500}{\bar}$, almost $9500$ times larger than normal \footnote{The results are evaluated using experimentally measured water radial distribution functions, reported by Soper et al.}.
It is apparent that a single set of direct correlation functions evaluated for a bulk system cannot be used to describe regions with low (or high) liquid density (compared to the reference system), which is precisely what HNC does.

Note that the final result in equation \ref{eq:hnc_decomp} can be also expressed as
\begin{equation}
\beta \Delta \Omega[\rho] = \Delta N + \frac{1}{2} \sum_{ij}\iint C_{ij}(|\rvec_1-\rvec_2|)\left[\rho_i(\rvec)\rho_j(\rvec)-\rho_{i0}\rho_{j0}\right]\diff\rvec_1\rvec_2,
\end{equation}
where $\Delta N = \sum_{i=1}^{N}\Delta \rho_i(\rvec)\diff\rvec$.
We can obtain the same result from equation $24$ in reference \citenum{Liu2013vpm}, by setting $F^B=\text{const}$ and using equation \ref{eq:intra_dcf} to relate the whole and site-site direct correlation functions.
We can see that 3D-RISM/HNC is essentially identical to the site density functional theory of Jianzhong Wu and coworkers, provided that one sets the bridge function to zero \cite{Liu2013vpm,Liu2013vac,Zhao2013ujq,Sheng2016ult}.

Instead of expressing solvation free energy purely using the bulk solvent direct correlation functions, we can also split it into somewhat more familiar terms.
Utilizing the HNC equilibrium condition \ref{eq:hnc_equilibrium} and the third equality in equation \ref{eq:hnc_decomp}, we find
\begin{equation}
\beta \Delta \Omega [\vect{\rho}] = \frac{1}{2}\braket{\vect{\rho}}{\ln \vect{g} + \beta \vect{u}} +  \frac{1}{2}\mel{\vect{\rho_0}}{\vect{C}}{\Delta \vect{\rho}} - \braket{\Delta \vect{\rho}}{\vect{1}}.
\end{equation}
To simplify the expression further we note that
\begin{equation}
\iint f(x-x')g(x')\diff x' \diff x = \int g(x') \left[\int f(x-x')\diff x\right] \diff x' = \int g(x')\diff x'\int f(y)\diff y,
\end{equation}
which follows from the Fubini–Tonelli theorem \cite{Dibenedetto2016wtk}.
Using this result we can rewrite the second term as
\begin{equation}
\sum_{ij}\rho_{0i}\iint C_{ij}(|\rvec - \rvec'|)\Delta \rho_j(\rvec')\diff \rvec'\diff\rvec = \sum_{ij} \rho_i\rho_j \f C_{ij}(0) G_j,
\label{eq:fubini_tonelli}
\end{equation}
where we used the definition of Kirkwood-Buff integral $G_j = \int h_j(\rvec)\diff \rvec$ \cite{Ben-naim2006vpu}.
For multicomponent solvent we obtain
\begin{equation}
\Delta \Omega = \frac{kT}{2}\braket{\vect{\rho}}{\ln \vect{g} + \beta \vect{u}} - kT\sum_{i=1}^{N}\rho_i G_i\left(1 - \frac{1}{2} \sum_{j=1}^{N}\rho_j \f C_{ij}(0)\right).
\label{eq:3d_rism_multi_decomp}
\end{equation}
If solvent is a single component liquid with density $\rho_0$ and $N$ sites, $G_j=G_i= -\Delta V$ and the expression can be simplified further
\begin{equation}
\begin{split}
\Delta \Omega &= \frac{kT}{2}\braket{\vect{\rho}}{\ln \vect{g} + \beta \vect{u}} + N \rho_0 kT\Delta V -\frac{kT\rho_0^2}{2}\Delta V\sum_{ij} \f C_{ij}(0)\\
&= \frac{kT}{2}\braket{\vect{\rho}}{\ln \vect{g} + \beta \vect{u}} + P_0 \Delta V.
\end{split}
\label{eq:3d_hnc_decomp}
\end{equation}
In the above equation one can readily identify entropic, enthalpic, and pressure terms, contributing to the total solvation free energy.
However, it is important to note that since we are dealing with the grand potential, $P_0 \Delta V$ does not represent the familiar expansion work for the $NPT$ system.
The origin of this term is effectively entropic in nature.

Two above equations (\ref{eq:3d_rism_multi_decomp} and \ref{eq:3d_hnc_decomp}) are one of the main results of the thesis and can be readily used to both understand the failures of 3D-RISM theory and to formulate reasonable approximations.


\section{Hydrophobic solvation}

The hydrophobic effect is traditionally associated with (a) unusually high solvation free energies of apolar molecules in water, usually several kilocalories compared to organic solvents and (b) the tendency of apolar compounds in water to aggregate to minimize their surface area \cite{Tanford1978wyl,Blokzijl1993wwu}.
Additionally, several other properties became associated with it such as negative solvation entropy, large system heat capacity increases upon solvation, and entropy convergence at higher temperature \cite{Garde1996wjw,Blokzijl1993wwu}.

Here we focus on idealised situation in which our solutes are hard spheres with interaction given by
\begin{equation}
u_{hard-O}(r) = \begin{cases}
\infty & \mbox{if } r < \sigma_{hard-O} \\
0   & \mbox{otherwise.}
\end{cases}
\end{equation}
Then solvation is determined entirely by entropical and pressure effects.
This makes hydrophobic solvation an ideal example using which we can better understand the problems of the 3D-RISM model.

Interestingly, a lot of insight into the failures of 3D-RISM model can be obtained by examining a significantly simpler model of hydrophobic solvation, called information theory (IT) \cite{Hummer1998uva,Chandler2011wsq}.
Let us define $P_V(N)$, a probability that $N$ water molecules can be found in volume $v$.
We assume that water fluctuations can be described by Gaussian distribution
\begin{equation}
P_v(N) \approx \frac{1}{\sqrt{2\pi \sigma_v}}\exp\left[-\frac{(N-\langle N \rangle_v)^2}{2\sigma_v}\right],
\end{equation}
with $\langle N \rangle_v = \rho v$ being an average number of particles, and  $\sigma_v = \left\langle (\delta N)^2 \right\rangle_v$ a mean square fluctuation in volume $v$.

The probability of hard sphere solvation is equivalent to water molecules fluctuating and creating a large enough cavity for the sphere to fit.
Then, using $\Delta F =-kT\ln P_v(0)$ we obtain
\begin{equation}
\Delta F_{IT} =  kT \left[\frac{\rho^2v^2}{2\sigma_v} + \frac{1}{2} \ln(2\pi \sigma_v)\right].
\label{eq:hard_sphere_chem_pot_it}
\end{equation}
Associating the position of water molecules only with the locations of oxygens, we obtain $\sigma_v$ from the equation \ref{eq:particle_msf}:
\begin{equation}
\sigma_v = \iint_v h_{OO}(|\rvec_1 - \rvec_2|) \diff \rvec_1 \diff \rvec_2 + \rho v.
\label{eq:sigma_h}
\end{equation}
From the same equation it follows that at the limit of $v\to \infty$, mean square fluctuations are linked to isothermal compressibility $\lim\limits_{v\to \infty} \sigma_v = kT \rho^2 v \chi_T$ giving
\begin{equation}
\Delta F_{IT} = \frac{v}{2\chi_T} + \frac{kT}{2}\ln(2\pi kT\rho^2 v \chi_T).
\end{equation}

The corresponding expression for hard sphere hydration free energy in 3D-RISM is quite similar:
\begin{equation}
\begin{split}
\Delta F_{HNC} & = \frac{kT}{2}\braket{\vect{\rho}}{\ln \vect{g}} -kT\braket{\Delta \vect{\rho}}{\vect{1}} +\frac{kT}{2}\mel{\vect{\rho_0}}{\vect{C}}{\Delta \vect{\rho}}  \\
& = \frac{kT}{2}\braket{\vect{\rho}}{\ln \vect{g}} - \frac{kT}{2}\braket{\Delta \vect{\rho}}{\vect{1}} - \frac{kT}{2}\mel{\vect{\rho_0}}{\vect{X^{-1}}}{\Delta \vect{\rho}}\\
& = \frac{kT}{2}\braket{\vect{\rho}}{\ln \vect{g}} + \Delta V\left(\frac{kT}{2}\rho_0 + \frac{1}{2 \chi_T}\right),\\
&  = \frac{kT}{2}\braket{\vect{\rho}}{\ln \vect{g}} + P_0 \Delta V,
\label{eq:insertion_prob}
\end{split}
\end{equation}
where to obtain the first equality we used equation \ref{eq:inverse_chi_C}, and the second followed from the result by Imai et al. \cite{Imai2000uxp} $\vect{\rho_0}^T \vect{\f X^{-1}}(0) \vect{\rho_0} = \frac{1}{kT\chi_T}$ and equation \ref{eq:fubini_tonelli}.
You can see that while additional terms are different in two models, essentially both models suggest that hydrophobic solvation free energy scales proportionally to the volume of the solute.
Additionally, in 3D-RISM the proportionality constant is simply the pressure of bulk liquid (equation \ref{eq:3d_hnc_decomp}), which using the above results can be conveniently expressed for a single component liquids as $P_0 = \frac{1}{2}\left(kT\rho_0 + 1/\chi_T\right)$.

\begin{figure}
	\centering
	\includegraphics[width=0.7\linewidth]{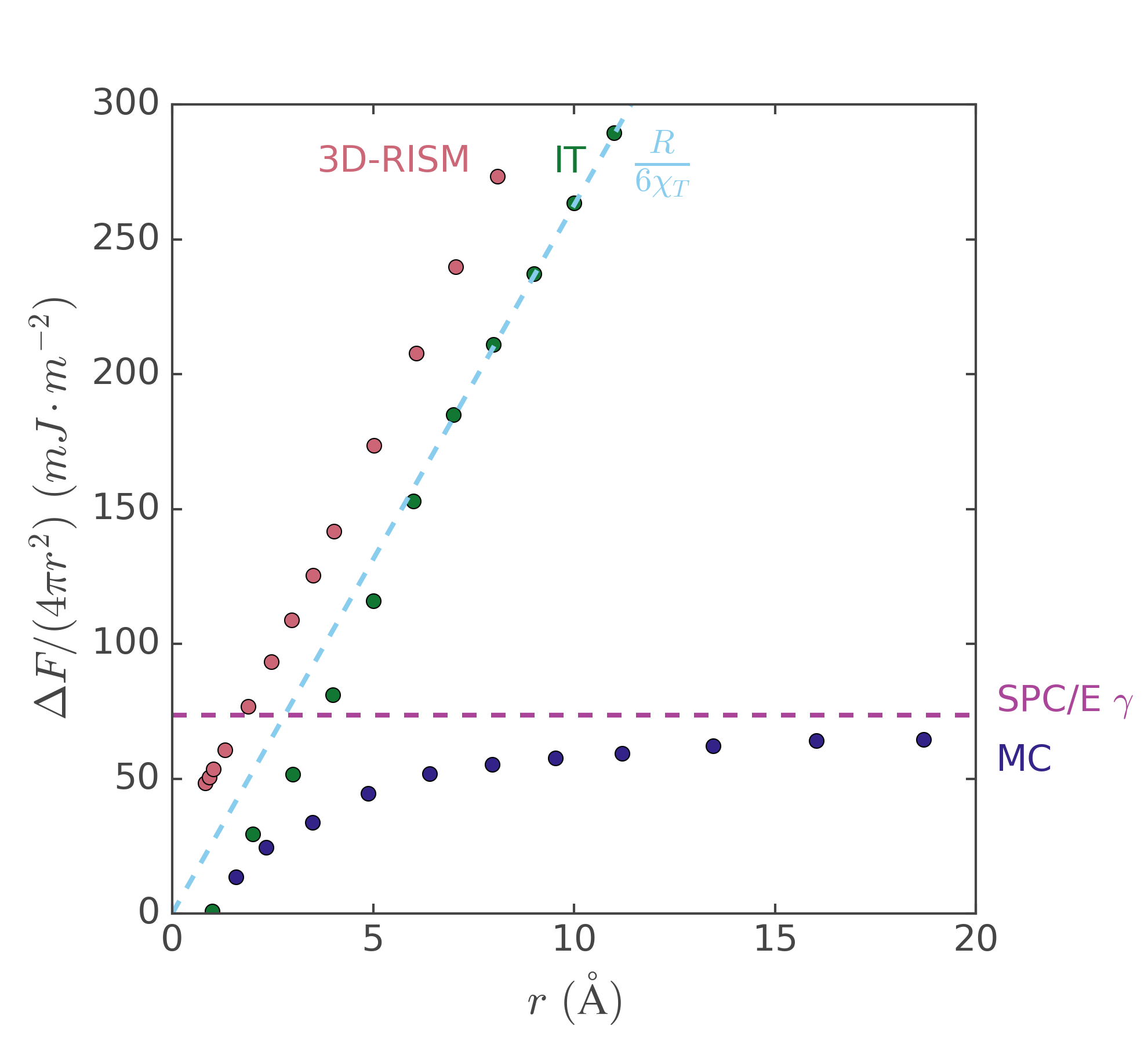}
	\caption{Dependence of surface energy of hard spheres in water depending on their radius. Monte Carlo (MC) results are taken from Ref. \citenum{Chandler2005unq}.}
	\label{fig:hard_tension}
\end{figure}


The figure \ref{fig:hard_tension} shows predictions of the change of solvation free energy per unit area depending on the hard sphere radius, made by three different models.
In principle, as hard solute radius $r$ gets larger and larger $\Delta \Omega/(4\pi r^2) \to \frac{Pr}{3} + \gamma$, where $\gamma$ is the surface tension between hard solute and solvent.
Thus, these type of figure allows us to evaluate both pressure and surface tension within the model.

The results of information theory agree well with Monte-Carlo simulations for small solutes ($r < \SI{4}{\angstrom}$), but become progressively worse for larger solutes.
The 3D-RISM/HNC approach consistently predict surface energy values larger than the two models, but shows trends which are quite similar to information theory.

Monte-Carlo predictions provide a good insight on hydrophobic phenomena.
Before diameter of a hard sphere reaches \SI{1}{nm}, its solvation free energy scales with the volume of the sphere, while afterwards, with its surface area.
The reason for this behaviour has been rationalised by Chandler, Weeks and co-workers in a number of important papers on the hydrophobic effect \cite{Weeks1995tjq,Weeks1998ubj,Lum1999val,Huang2000wlm,Huang2002vzs,Chandler2005unq}.

The hydration free energy of small molecules largely depends on the strength of hydrogen bonding.
Bulk water forms a strong tetrahedral network, which, despite being quite dynamic, rarely breaks down substantially to form solute cavities.
Whenever, cavities do occur, water tries to maintain its bonding network if possible; thus the structure of solvent around small cavities is quite similar to that of bulk water.
This is the reason why approaches such as information theory are able to accurately describe solvation free energy for small molecules using mean square fluctuations obtained without the presence of a solute.

Near larger solutes, which resemble planar interfaces, the water surface layer undergoes substantial reorganization.
The bulk-like hydrogen bonding network is substituted with an interfacial structure similar to water-air interface.
The molecules are oriented with O-H bonds towards the solute and the density of water right next to the solute is lower than that found in the bulk.
The decrease of density occurs due to the force imbalance: the interfacial water molecules do not experience a lot of attraction towards hydrophobic solute, but are strongly drawn in by the bulk water.
These rearrangements help decrease the free energy of hydrophobic solvation, making the creation of larger cavities much more probable than what one would expect from simple Gaussian behaviour.

The dewetting transition is missed by both IT and 3D-RISM.
The hydrophobic effect in these models occurs due to the linear increase of water chemical potential as a consequence of solute excluded volume, and they cannot capture its more subtle details.

\section{Pressure corrections for solvation free energy}
\label{sec:3drism-pressure}

As we saw in the previous section, 3D-RISM largely overestimates bulk solvent pressure (predicting water at normal conditions to have pressure of about \SI{9500}{\bar}).
The high pressure arises because of the truncation of the free energy expansion at the second term.
The dominant forces in water at equilibrium are mostly repulsive, with attractive forces being generally canceled out.
We expect a low order expansion to capture general trends, which are repulsive, and to neglect more subtle attraction interactions, which are described by triplet and higher order correlation functions.
This neglect becomes especially problematic when describing  interfaces, which are dominated by collective, long distance interactions.

\begin{figure}
	\centering
	\includegraphics[width=0.7\linewidth]{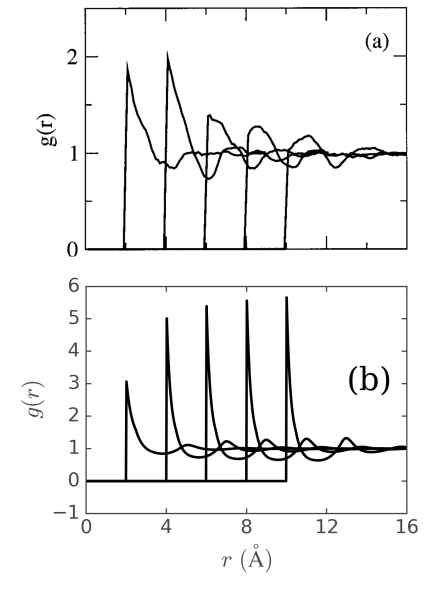}
	\caption{Both figures demonstrates hard sphere--water oxygen radial distribution functions for spheres of various radii. Figure (a) is taken from Ref. \citenum{Huang2001wdw} and was obtained using Monte Carlo simulations. Radial distribution functions in figure (b) were calculated using 3D-RISM/HNC.}
	\label{fig:hnc_phobic}
\end{figure}

Notably, the large compressibility pressure of 3D-RISM (evaluated using \ref{eq:hnc_pressure}) is quite consistent with the behaviour of radial distribution functions.
The contact theorem \cite{Hansen2000utg,Stillinger1972tld} tells us that in the limit of an infinitely large hard sphere, bulk pressure is related to the value of solvent density right next to the hard sphere
\begin{equation}
P_0 = \rho kT g(R),
\end{equation}
where $g(R)$ is the value of oxygen radial distribution function at the surface of the hard sphere.
Thus, at standard conditions, the 3D-RISM contact value of the oxygen radial distribution function with a hard sphere should approach $P_0/(\rho kT) = 7.2$.
From figure \ref{fig:hnc_phobic} we can see the contact value approaches $6$, which is close to what is predicted using compressibility route pressure (although, it is known that singlet HNC only satisfies the contact value theorem at low densities \cite{Attard1991uvb}).
Additionally, the shape of the 1D-RISM partial oxygen structure factor, shown in figure \ref{fig:rism1d_water} resembles experimentally observed structure factors for bulk water at $P=\SI{4000}{\bar}$ \cite{Soper2000vic}.

In view of the above, it is reasonable to try to correct 3D-RISM by subtracting the overestimated pressure work
\begin{equation}
\Delta F_{PC} = \Delta F_{3D-RISM} - P_0\Delta V,
\label{eq:pressure_cor}
\end{equation}
where $P_0$ is the 3D-RISM bulk pressure, $\Delta V$ is the excess volume of solvent, and $PC$ stands for pressure correction.

\begin{figure}
	\centering
	\includegraphics[width=0.7\linewidth]{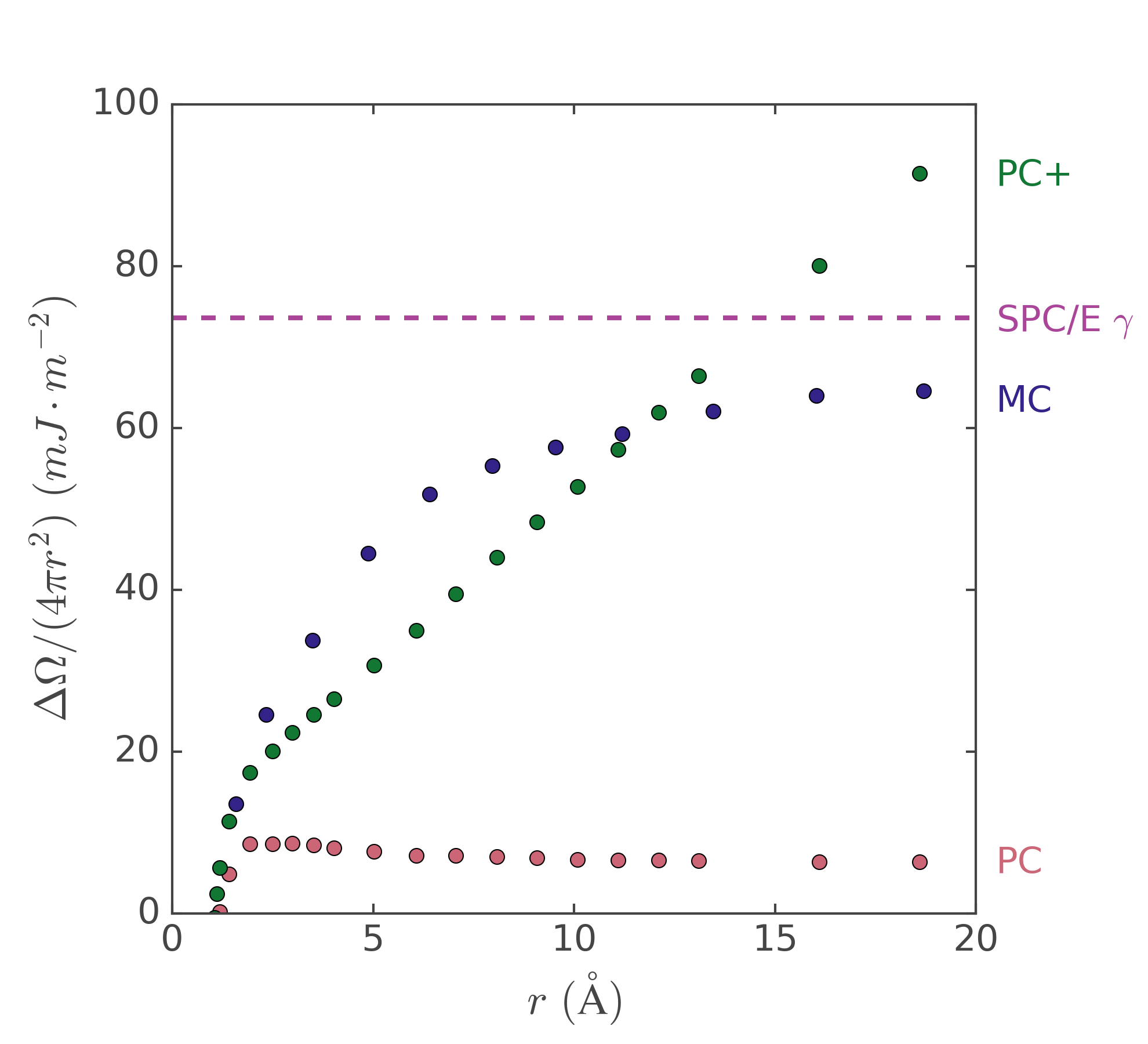}
	\caption{This figure mirrors figure \ref{fig:hard_tension}, except now instead of information theory and HNC we compare results by PC and PC+ models.}
	\label{fig:hard_tension_pc}
\end{figure}

We also introduce another way of correcting 3D-RISM results:
\begin{equation}
\Delta F_{PC+} = \Delta F_{3D-RISM} - P_0\Delta V + P_{id}\Delta V,
\label{eq:pc_plus}
\end{equation}
where $P_{id}= \rho_{id} kT$ is an ideal gas pressure with $\rho_{id} = \sum_{\alpha=1}^{M}\rho_{\alpha}$ being the number density of solvent molecules, not sites.
We call this equation the advanced pressure correction (PC+), and this is one of the main results of the thesis \footnote{Note that the correction was discovered essentially by accident, and was initially referred to as initial state correction (ISc) \cite{Misin2015wdu,Li2015tog}. Only after publication by Sergiievskyi et al. \cite{Sergiievskyi2015tab} it was recognized that the correction was related to 3D-RISM pressure and the name PC+ become popular \cite{Misin2016tqn,Misin2016vee,Johnson2016tta,Luchko2016vwg}.}.
For single component solutions both corrections can be conveniently defined as $\Delta F_{PC} = \frac{kT}{2} \braket{\vect{\rho}}{\ln \vect{g}} + \frac{1}{2} \braket{\vect{\rho}}{\vect{u}}$ and $\Delta F_{PC+} = \frac{kT}{2} \braket{\vect{\rho}}{\ln \vect{g}} + \frac{1}{2} \braket{\vect{\rho}}{\vect{u}} + P_{id}\Delta V$.

Figure \ref{fig:hard_tension_pc} demonstrates the scaling of hydration free energies from PC and PC+ corrections.
As can be seen, the $\braket{\vect{\rho}}{\ln \vect{g}}/2$ term in the PC correction scales with solute surface area and defines its surface tension.
Notice that PC+ effectively sets liquid pressure to its ideal value, which for liquid water at ambient conditions is $P_{id}=\SI{1372}{\bar}$.
It is quite a bit larger than the pressure of water at standard conditions and leads to the overestimation of solvation free energy for larger volumes.
However, the approach, at least for hard spheres, is relatively accurate up to $R\approx\SI{1}{nm}$.
This is twice the size of fullerene \ce{C60} and covers much of the domain of conventional pharmaceutical, analytic and organic chemistries.

Since the beginning of the chapter we have been discussing hard spheres.
While these solutes are quite convenient from a theoretical point of view, almost none of the actual molecules resembles them.
Between pretty much any two materials there would exist dispersion interactions, typically approximated by Lennard-Jones potential.

\begin{figure}
	\centering
	\includegraphics[width=0.7\linewidth]{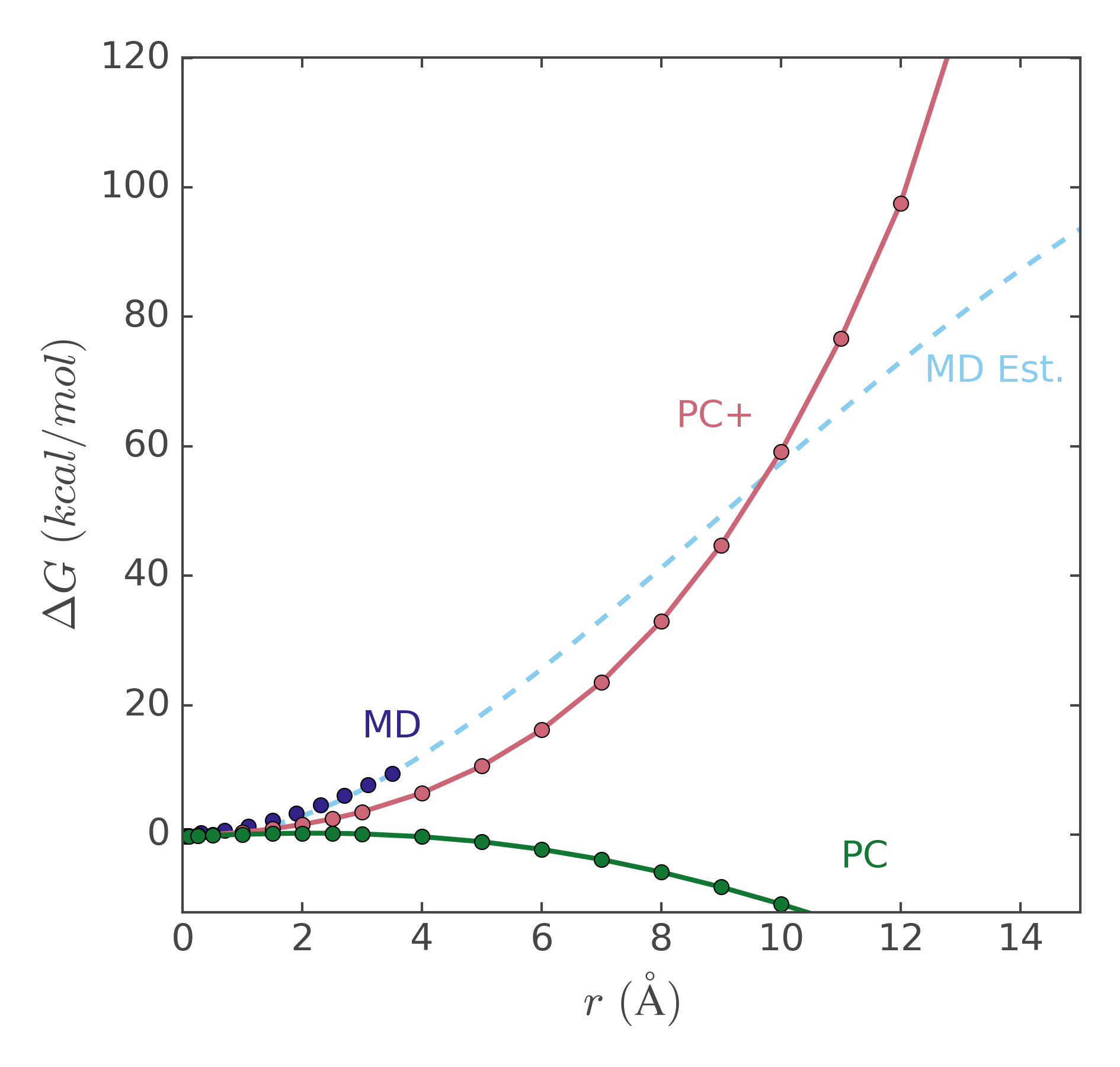}
	\caption{Solvation free energies of Lennard-Jones solutes with different radii and $\epsilon=\SI{0.125}{kcal\per\mole}$ in water. The estimate of MD solvation free energies for larger radii was done using equation \ref{eq:md_estimate}.}
	\label{fig:lj_hfe}
\end{figure}


The attractive forces are usually quite weak, on the order of $0.2 kT$, but are enough to practically remove most dewetting effects between water interface and surface, making pair distribution function predicted by 3D-RISM and molecular dynamics much more similar.
Figure \ref{fig:lj_hfe} shows hydration free energies for a series of Lennard-Jones spheres, predicted by molecular dynamics, PC, and PC+.
To extrapolate results of molecular dynamics to higher radii we estimated the contribution of dispersive interactions assuming that water structure is unperturbed by the sphere outside its exclusion radius (a reasonable approximation for these solutes \cite{Huang2002vzs,Dzubiella2006vjh})
\begin{equation}
U_{disp} = \rho 4 \pi \int_{\sigma_{LJ-O}}^{\infty} 4\epsilon_{LJ-O} \left[\left(\frac{\sigma_{LJ-O}}{r}\right)^{12} - \left(\frac{\sigma_{LJ-O}}{r}\right)^6\right] r^2 \diff r,
\label{eq:md_estimate}
\end{equation}
where both $\sigma_{LJ-O}$ and $\epsilon_{LJ-O}$ were computed using Lorentz-Bertholetz rules.
The total solvation free energy was estimated via $\Delta F_{LR-LJ} = U_{disp} + \gamma A$, where we used $\gamma = \SI{76}{\milli\J \meter^{-2}}\,$ \cite{Huang2001wdw}.

Overall, the results are similar to those that were obtained for hard spheres.
PC+, due to its ideal pressure scales similarly to molecular dynamics up to $r\approx\SI{1}{nm}$.
Because dispersion interactions are relatively weak, their description within both molecular dynamics and 3D-RISM is similar, changing the picture little compared to that of hard spheres.

\section{Free energy of charging}
\label{sec:charging_fe}

Consider a soft sphere with a charge $q$ located at the centre.
Its interaction potential with the surrounding solvent can be expressed as:
\begin{equation}
U_{uv}(q) = U_{LJ} + U_{el}  = U_{LJ} + q \Phi,
\end{equation}
where $\Phi$ is the solvent generated electrostatic potential in the centre.
The charging free energy is the reversible work required to change the solute charge from $0$ to $q$.
It can be found using Kirkwood's charging formula:
\begin{equation}
\begin{split}
\Delta F_{el} = \int\limits_{0}^{q} \left\langle \frac{\partial U_{uv}(q')}{\partial q'}\right\rangle_{q'} \diff q' = \int\limits_{0}^{q} \left \langle \Phi \right \rangle_{q'} \diff q',
\end{split}
\end{equation}
where $\left \langle \Phi \right \rangle_{q'}$ is the electrostatic potential in the centre of the ion with charge $q'$.

It is commonly stated that the charging free energy in water can be well approximated using a linear response relationship.
Technically speaking, it is a bit more complicated; a standard linear response implies $\langle \Phi \rangle_{q'} =  q' \langle \Phi \rangle_{1} + (1-q') \langle \Phi \rangle_{0}$.
However, for water (and other dipolar solvents) one typically encounters a piecewise-linear (piecewise-affine) response \cite{Hunenberger2011vfc,Reif2016vox}; that is:
\begin{equation}
\langle \Phi \rangle_{q'}^{PL} =  \langle \Phi \rangle_{0} - \begin{cases}
q' C_{+} & \mbox{if } q' \geq 0\\
q' C_{-} & \mbox{if } q' < 0.
\end{cases}
\label{eq:partial_linear}
\end{equation}
Both the potential in the uncharged cavity $\langle \Phi \rangle_{0}$ as well as constants $C_{+}$ and $C_{-}$ depend on the size and "stickiness" of the solute.

\begin{figure}
\centering
\includegraphics[width=0.7\linewidth]{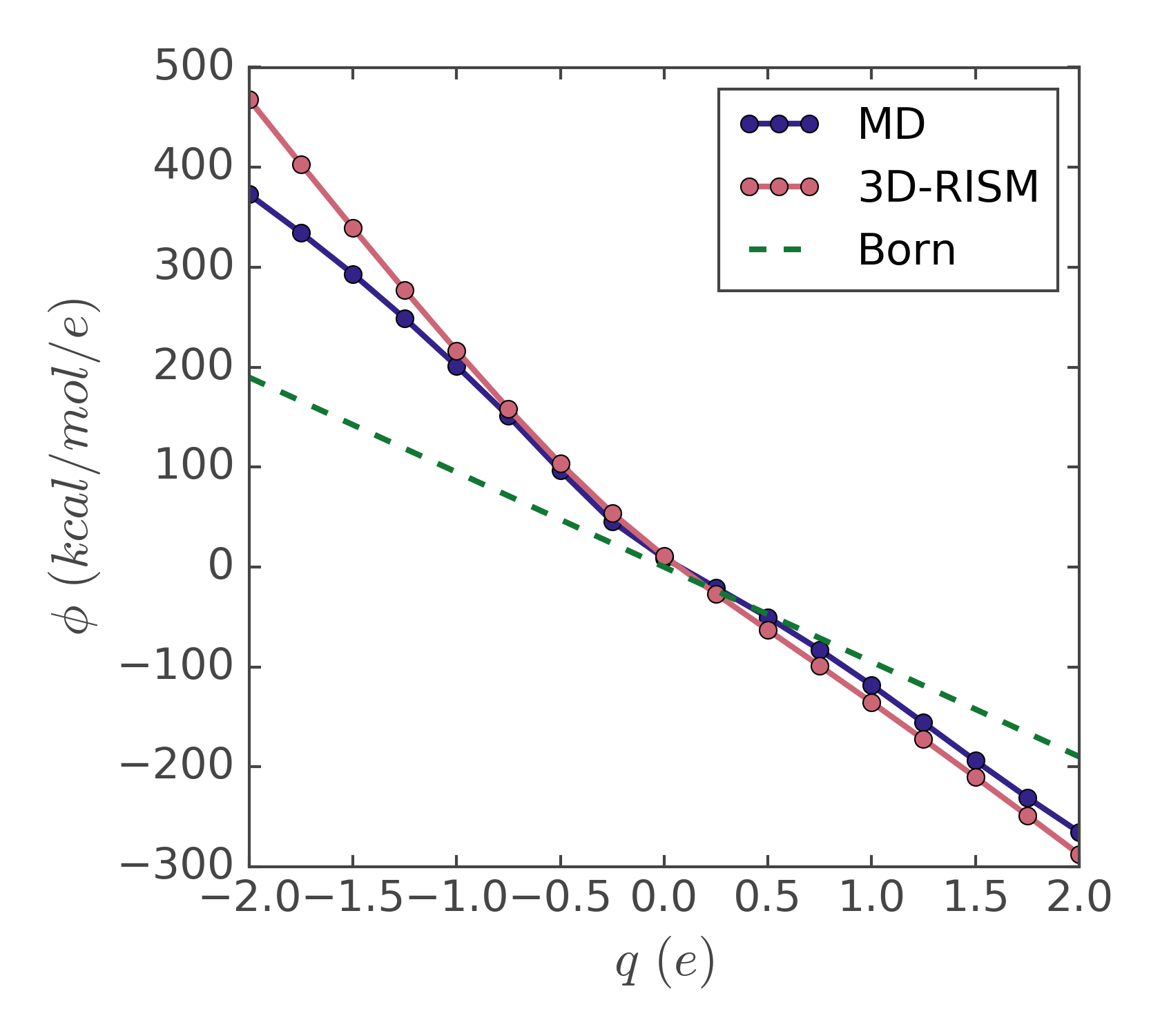}
\caption{Electrostatic potential inside the soft solute with $\sigma=\SI{3.8}{\angstrom}$ and $\epsilon=\SI{0.125}{kcal\per\mole}$ depending on its charge.}
\label{fig:soft_pot}
\end{figure}

Notably, theories which model water as a polarizable continuum cannot predict asymmetry of the solvent response, regardless of whether they take into account local or non-local polarizabilities \cite{Reif2016vox}.
Popular approaches such as the Born model or Poisson-Boltzmann and their modifications incorporate charge asymmetry by scaling the ion radius.
This makes RISM models quite interesting, since they do predict piecewise linear response without any parametrization.

Evidently, the reason for RISM "awareness" of solute charge is due to the use of two separate fields: one for water oxygens and one for hydrogens \cite{Fedorov2007wht}.
Figure \ref{fig:soft_pot} demonstrates the dependence of the water potential on the solute charge, predicted by three different models.
We can see that while 3D-RISM correctly predicts the charge dependence of the water response, its scaling is predicted to be strictly linear, in agreement with the equation \ref{eq:partial_linear}.
At the same time, more precise molecular dynamics simulations do show deviations from it at the higher charges.
The effect is due to dielectric saturation: in discrete solvent after certain point the polarization reaches maximum density and response becomes sublinear \cite{Hunenberger2011vfc}.
Such saturation does not occur in 3D-RISM; similarly, in the case of hard sphere solvation 3D-RISM predicted response was always exactly proportional to the solute's volume.

\begin{figure}
	\centering
	\includegraphics[width=1\linewidth]{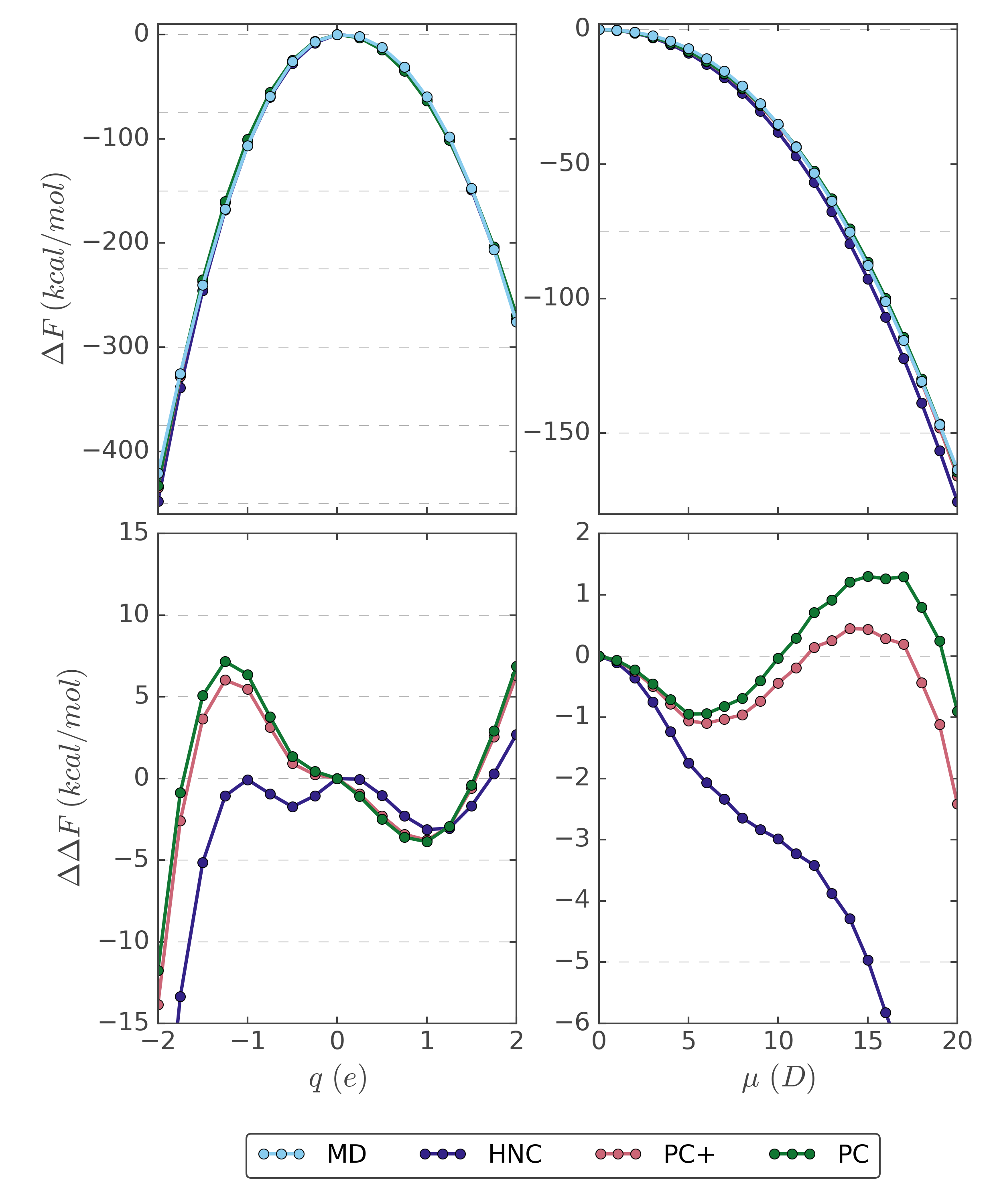}
	\caption{Top figures: charging free energy of Lennard-Jones sphere (left) and Lennard-Jones dipole (right) in water. The dipole consists of two Lennard-Jones spheres, separated by $\SI{2}{\angstrom}$. Both the sphere and dipole sites have $\sigma=\SI{3.8}{\angstrom}$ and $\epsilon=\SI{0.125}{kcal\per\mole}$. Bottom figures: difference in charging free energy predictions between 3D-RISM models and molecular dynamics.}
	\label{fig:chg_vs_fe_charging}
\end{figure}

Even though 3D-RISM does not predict saturation of the dielectric response, the accuracy of its approximation at moderate charges is more than enough to correctly predict charging free energies for the majority of common molecules.
Figure \ref{fig:chg_vs_fe_charging} demonstrates predictions of charging free energies by 3D-RISM for a simple sphere and dipole.
The volumes of van-der-Waals solutes do not depend on the charge, but their excess volumes still change due to the increased attraction.
Thus, it makes sense to use pressure corrections even in the context of charging free energies.
As the figure illustrates, this does not show considerable improvement for single ions, but improves solvation free energy predictions for dipoles.

\section{Effect of corrections on solvation thermodynamics}
\label{sec:pc_pcp_solv_td}

In this section, we will discuss solvation thermodynamics of 3D-RISM/HNC and its pressure corrections.
For simplicity and along with the main goal of the thesis, we restrict the discussion to single-component solvents, although extension to multicomponent mixtures should be relatively straightforward.

Previously (section \ref{sec:solv_thermodynamics}) we have shown that solvation free energy can be split into energetic and entropic contributions.
Specifically, for solvation in the grand canonical ensemble we had $\Delta \Omega_s = E^{uv}_{\mu} - TS^{uv}_{\mu} + \mu \rho \bar{V}$.
Examining the derivation of 3D-RISM one can see that $\mu \rho \bar{V}=\mu \rho \int \Delta \rho(\rvec)\diff\rvec$ gets canceled out because of the way we define excess intrinsic chemical potential (section \ref{sec:3d-rism}).
Thus, one can split 3D-RISM solvation free energy into $\Delta F_{3D-RISM} = E^{uv} - TS^{uv}$, where $E^{uv}$ takes a clear physical meaning, and $TS^{uv}$ is simply defined using the above equation.
We will see that such decomposition provides a sensible way of analyzing 3D-RISM and its pressure corrections. 

We start by defining solute-solvent interactions in the usual way $E^{uv} = \braket{\vect{\rho}}{\vect{u}}$.
Then solute-solvent entropy is given by $S^{uv}_{HNC} = -k \braket{\vect{\rho}}{\ln \vect{g}} + k\braket{\Delta \vect{\rho}}{\vect{1}} + k/2 \mel{\Delta \vect{\rho}}{\vect{C}}{\Delta \vect{\rho}}$.
Using a rough estimate $\mel{\Delta \vect{\rho}}{\vect{C}}{\Delta \vect{\rho}} \approx \Delta \vect{\f \rho} (0) \vect{\f C}(0) \Delta \vect{\f \rho} (0)$ and the fact that for most liquids $\vect{\f C}(0) < 0$, we can see that $S_{HNC}^{uv}$ is negative, consistent with the general result from section  \ref{sec:solv_thermodynamics}.

The $\Delta S^{uv}_{HNC}$ can be split into the ideal gas entropy change, given by $\Delta S^{id}_{HNC} = -k \braket{\vect{\rho}}{\ln \vect{g}} + k\braket{\Delta \vect{\rho}}{\vect{1}}$, as well the excess (or ring entropy, as it has been referred to by some authors \cite{Bush1998vyg,Silverstein2001wbi}) contribution $\Delta S^{ex}_{HNC} =  k/2 \mel{\Delta \vect{\rho}}{\vect{C}}{\Delta \vect{\rho}}$.
The minimization of the grand potential leads to
\begin{equation}
\Delta S_{HNC}^{uv} = -k/2\braket{\vect{\rho}}{\ln \vect{g}}  + \frac{1}{2T}\braket{\vect{\rho}}{\vect{u}} - \frac{1}{T} P_0 \Delta V.
\label{eq:hnc_entrop}
\end{equation}
The above result is interesting since we can readily interpret this entropy as a sum of logarithm of insertion  probability for a hard solute within HNC model (equation \ref{eq:insertion_prob}) and a linear response entropy change occurring due to introduction of attractive interactions.
Recall that $-T\Delta S^{uv}_{LR} = -1/2 E^{uv} + 1/2 E^{uv}_0$, where $1/2 E^{uv}_0$ corresponds to solute-solvent energy without any coupling between solute and solvent; for a hard solute $E^{uv}_0 = 0$ and we recover equation \ref{eq:hnc_entrop}.
Note that the presence of a linear response component in singlet HNC is not surprising, considering that in the section \ref{sec:hnc_simple_liq} we demonstrated that this model essentially treats ideal part of chemical potential exactly and excess part via linear response approximation.

A similar result has been obtained by Sanchez et al \cite{Ozal2006vzx,Sanchez1999wrp} using a more general approach.
They found that
\begin{equation}
\Delta S^{uv} = k\ln P_{ins} - k \ln \left\langle \exp \left[-\beta (E^{uv}_I -E^{uv})\right] \right\rangle_{a} - k \ln P_a,
\end{equation}
where $P_{ins}$ is the probability that a randomly inserted molecule will experience an attractive or zero interaction energy $\Delta E^{uv} \leq 0$; the second term is the familiar solute-solvent fluctuation energy, except the averaging is performed over the states where solute-solvent interactions are attractive.
$P_a$ is the probability that a fully inserted molecule will have an attractive interaction energy, which for the majority of normal molecules $\approx 1$.
We can see that by setting $k \ln \left\langle \exp \left[-\beta (E^{uv}_I -E^{uv})\right] \right\rangle_{a} = - (1/2T) E^{uv}$ (this result is exact in the linear response regime \cite{Ben-amotz2016wap}) one recovers the HNC solute-solvent entropy.

Thus, after analyzing HNC entropy we found it to be consistent with other statistical mechanics theories.
Its main problem is the overestimation of the hard solute insertion free energy, which as we have already discussed, stems from the failure to describe interface formation.
Then we can readily interpret PC and PC+ corrections as adjustments to incorrect $P_{ins}$ from the HNC approximation.
For PC, entropy becomes
\begin{equation}
\Delta S_{PC}^{uv} = -k/2\braket{\vect{\rho}}{\ln \vect{g}} + \frac{1}{2T}\braket{\vect{\rho}}{\vect{u}}
\end{equation}
and for PC+
\begin{equation}
\Delta S_{PC+}^{uv} = -k/2\braket{\vect{\rho}}{\ln \vect{g}} + \frac{1}{2T}\braket{\vect{\rho}}{\vect{u}} + k\braket{\Delta \rho_0}{1},
\end{equation}
where $\Delta \rho_0$ is the change in solvent density.

From the equations above it is not necessarily obvious which approximation should lead to a better estimate of solvation free energy.
A clearer picture can be obtained if we rewrite the expression in terms of ideal/excess contributions.
Both PC and PC+ models have identical excess entropies $\Delta S^{ex}_{PC} = \Delta S^{ex}_{PC+} = k/2 \mel{\vect{\rho_0}}{\vect{C}}{\Delta \vect{\rho}}$.
The ideal entropies are then $\Delta S^{id}_{PC} = -k \braket{\vect{\rho}}{\ln \vect{g}}$ and $\Delta S^{id}_{PC+} = -k \braket{\vect{\rho}}{\ln \vect{g}} + k\braket{\Delta \rho_0}{1}$.
For a single component molecular solvent the solute-solvent entropy can be expanded in terms of n-particle correlation functions, with first terms given by\cite{Paulaitis1994ufm,Ashbaugh1996tui}
\begin{equation}
\begin{split}
\Delta S^{uv} &= -k\rho_0 \int g(\rvec) \ln g(\rvec)\diff \rvec + k\rho_0 \int g(\rvec) - 1 \diff \rvec\\
&\quad  -k\rho \frac{V_i}{\Omega} \int g(\vect{\omega}) \ln g(\vect{\omega}) \diff \vect{\omega} +\cdots\,,
\end{split}
\label{eq:pair_entropy}
\end{equation}
where $\rho$ is the solvent density, $\omega$ is the Euler angle, $\Omega = \int \diff \omega$, and  $V_i$ is the unit volume.
From the expression above we can see that the PC+ model, unlike PC, contains both first terms in the expansion.
Thus, one can expect it to reproduce solute-solvent entropy slightly better.
In all site-site models, the orientational contribution (the second line of equation \ref{eq:pair_entropy}) is partially approximated by summation of $\rho \ln g$ terms over different solvent sites.

\begin{figure}
	\centering
	\includegraphics[width=1\linewidth]{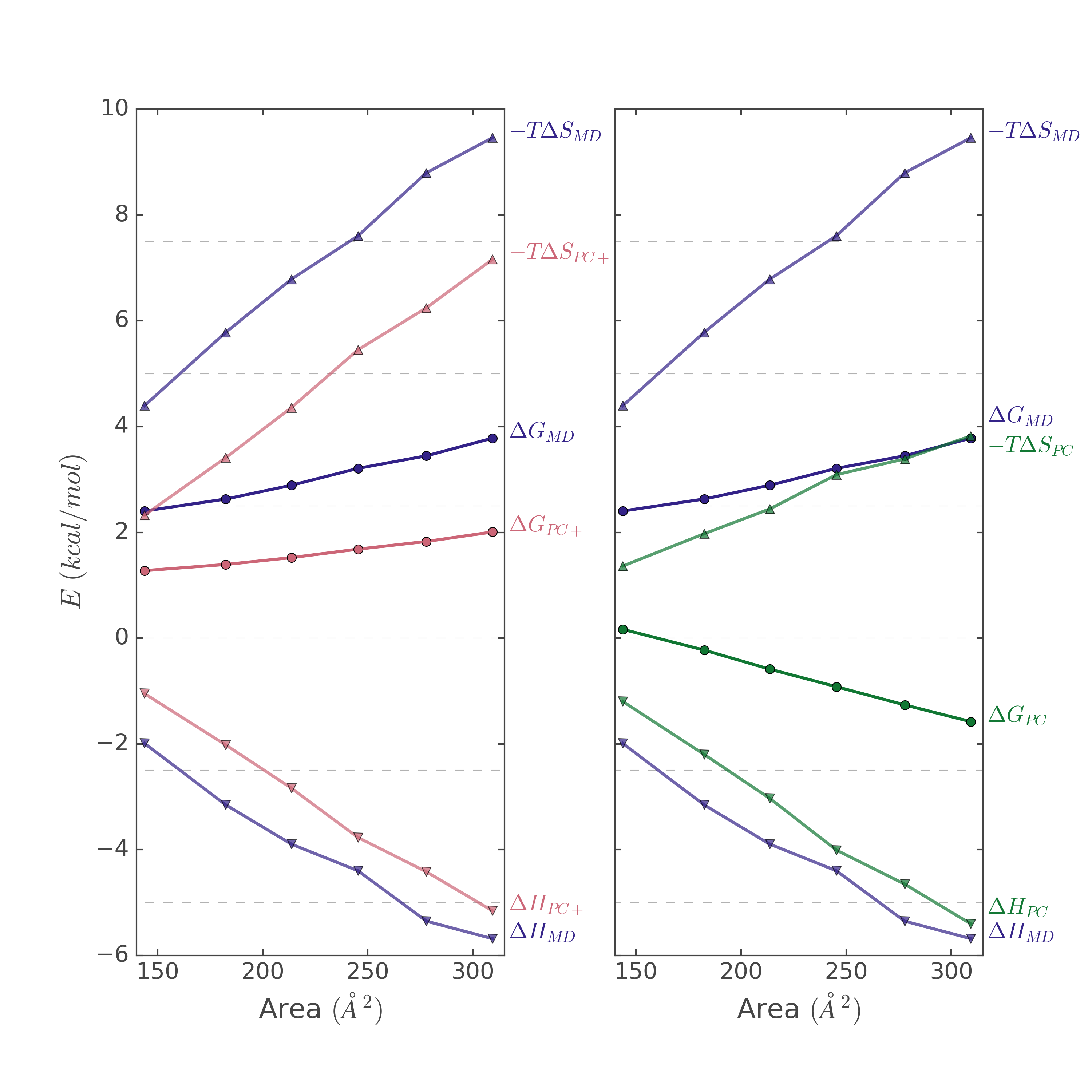}
	\caption{Dependence of free energy, entropy and enthalpy of linear alkanes on their surface area (number of atoms). The molecular dynamics results are taken from Ref. \citenum{Gallicchio2000tfa}.}
	\label{fig:alkanes_therm1}
\end{figure}

A more straightforward way of analyzing 3D-RISM thermodynamics is to simply compare it directly to molecular dynamics.
As usual, we chose water as our solvent and the first six linear alkanes (methane to hexane) as our solutes due to availability of data.
The molecular dynamics simulations were performed by Gallicchio et al. \cite{Gallicchio2000tfa} who used TIP4P water model and OPLS force filed parameters for alkanes.
We use the same force field, combined with cSPC/E water model for 3D-RISM.

Before we proceed it is important to discuss the way we compute solvation entropies and enthalpies within different models.
Since 3D-RISM is formulated in the grand ensemble, that is, under conditions of constant temperature, volume and chemical potential, it can seem that we can only compute properties within this particular ensemble.
However, it is possible to work around this problem by evaluating necessary derivatives numerically over multiple simulations in which only necessary thermodynamic variables are varied and others are kept constant.
Recall that as long as macroscopic thermodynamic parameters are identical, the chemical potential will be independent of the ensemble.
Thus, to evaluate, for example, temperature derivative of chemical potential under constant pressure, we run calculations at two separate temperatures, but identical pressures.

The approach described above will yield the best estimates we can get with RISM for constant pressure enthalpies, entropies as well as their higher derivatives; however, it will also lead to a conceptual problem.
The set of temperatures and densities (input parameters for 3D-RISM calculation) corresponding to a constant pressure in real water actually leads to a variety of different values within the 3D-RISM approximation.
Thus, any 3D-RISM entropies and enthalpies that we obtain via standard formulas will actually contain contributions from the derivatives of 3D-RISM pressure.
This is not a very significant problem since essentially any water model will have a different phase diagram, making these dependences additional error contributions.
For simplicity, we are going to use the symbol $\Delta G$ for both experimental (simulated) and 3D-RISM free energies, even though the later technically corresponds to $\Delta \Omega$.

\begin{figure}
	\centering
	\includegraphics[width=1\linewidth]{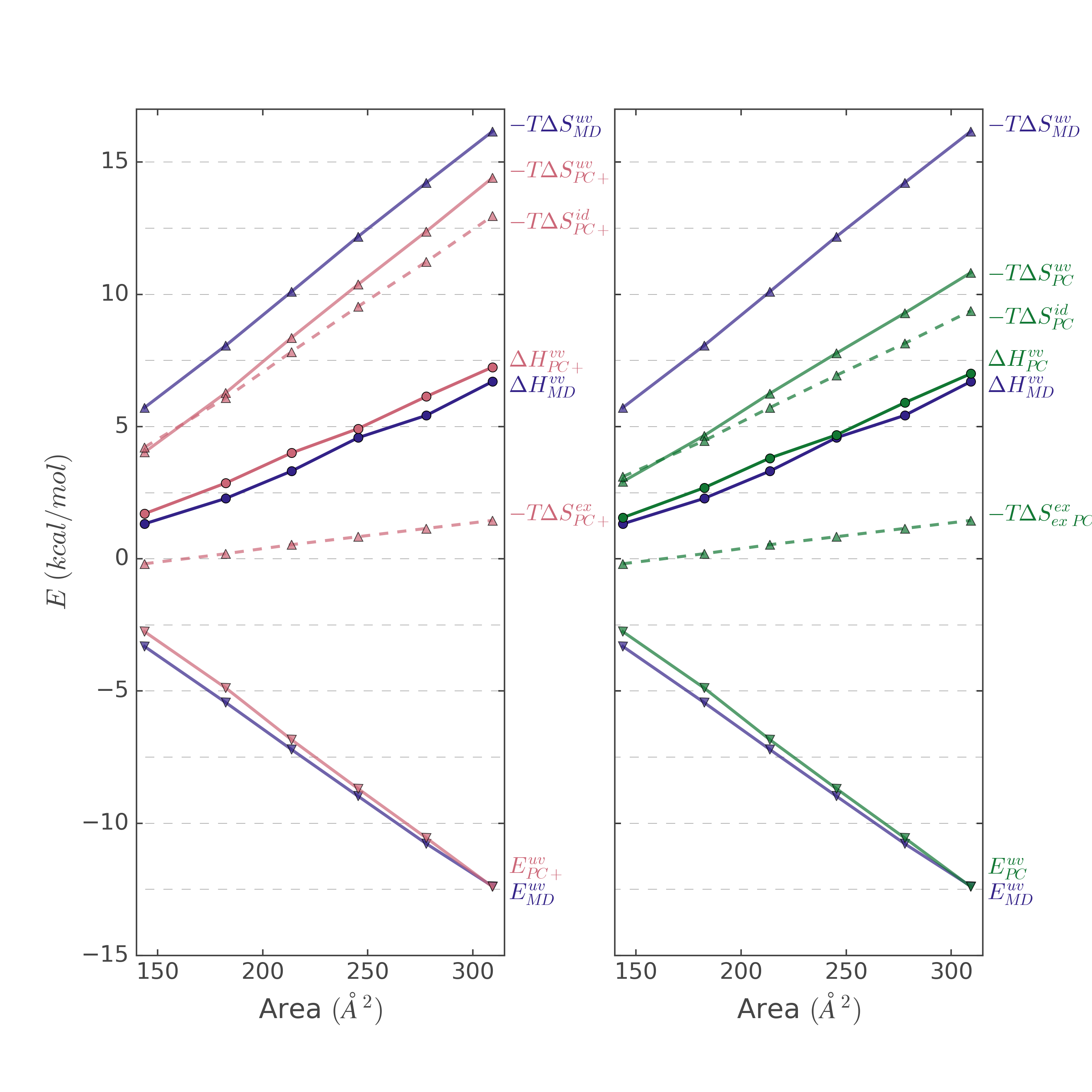}
	\caption{Solute-solvent components of alkanes solvation free energy. The definition of the quantities can be found in the main text. The molecular dynamics results are taken from Ref. \citenum{Gallicchio2000tfa}.}
	\label{fig:alkanes_therm2}
\end{figure}

Figure \ref{fig:alkanes_therm1} demonstrates predictions of solvation free energy, enthalpy, and entropy obtained from molecular dynamics, PC and PC+.
For reference, the same figure with the comparisons between uncorrected 3D-RISM and MD is included in appendix \ref{AppendixB} (figure \ref{fig:alkanes_therm3}).
The solvation entropy from RISM models was calculated numerically via:
\begin{equation}
\Delta S(T) = - \frac{\Delta G(T+\Delta T) - \Delta G(T-\Delta T)}{\Delta T},
\end{equation}
where we used $\Delta T=\SI{2.5}{K}$; to get enthalpy we used
\begin{equation}
\Delta H = \Delta G + T \Delta S.
\end{equation}
The density of bulk water at different temperatures was taken from Ref. \citenum{The_international_association_for_the_properties_of_water_and_steam2011uyk}.
Note that these predictions are for thermodynamic entropies and enthalpies, related to their solute-solvent components as $\Delta S = \Delta S^{uv} + \Delta H^{vv}/T$ and $\Delta H = E^{uv} + \Delta H^{vv}$, where $\Delta H^{vv}$ is the solvent reorganization energy.

The alkanes are essentially chains of fused Lennard-Jones spheres.
The behaviour of solvation free energy that we observed in the case of a single Lennard-Jones sphere (figure \ref{fig:lj_hfe}) is essentially identical for these solutes.
In the case of PC, the dispersion interactions between solute and solvent dominate, making the free energy $\Delta G_{PC}$ become progressively negative with the increase of surface area.
Conversely, the PC+ approximation of insertion probability results in a correct scaling of $\Delta G$ with solute size, although its value is smaller than the one predicted by molecular dynamics.

The predicted solvation enthalpy and entropy are smaller in magnitude for both 3D-RISM models when compared to molecular dynamics.
The differences between PC+ and PC models is mostly due to solvation entropy $\Delta S$.
The enthalpies in the two models differ only slightly due to the temperature dependence of water density.

\begin{figure}
	\centering
	\includegraphics[width=0.6\linewidth]{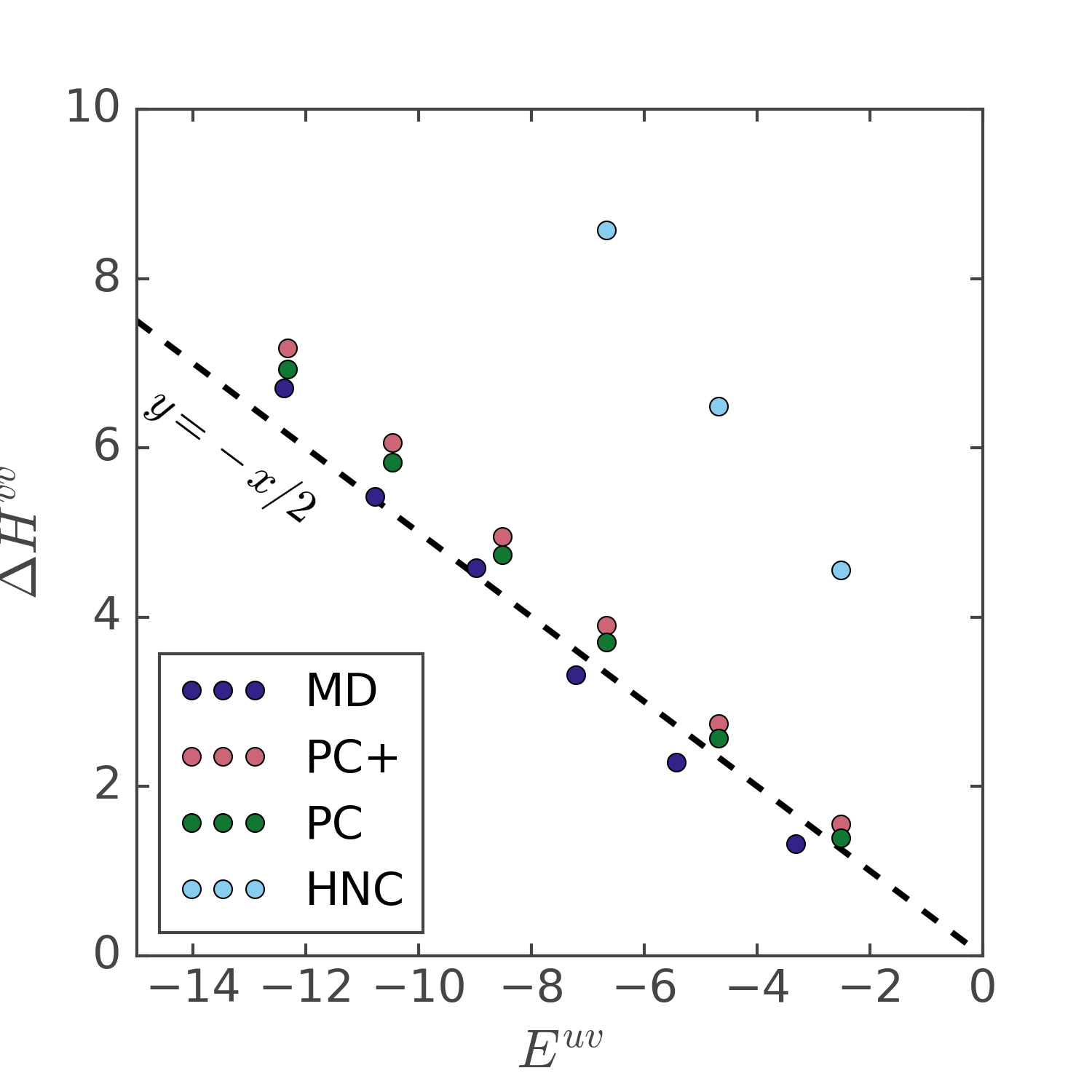}
	\caption{Correlation between solvent reorganization energy and solute-solvent interaction energy for the first 6 linear alkanes. The units are in \si{kcal\per\mole}.}
	\label{fig:reorganization}
\end{figure}

On figure \ref{fig:alkanes_therm2} you can see a further decomposition of previous quantities into solute-solvent and solvent-solvent terms.
The solvent reorganization is obtained from $\Delta H^{vv} = \Delta H - E^{uv}$.
Perhaps surprisingly, the overall agreement of these quantities between 3D-RISM and MD is quite good, despite the fact that all solute-solvent energies and entropies do depend on the ensemble.
Notably, $E^{uv}$ is predicted essentially correctly, considering the fact that water models were not identical in 3D-RISM and MD.
This indicates that for smaller solutes the solvent density distribution within 3D-RISM/HNC is in relatively good agreement with MD, and disagreement primarily comes from the entropic part.
Additionally, as we expected from our analysis in the beginning of the chapter, PC+ model has a better $S^{uv}$ estimate, mostly due to its ideal part.

The accuracy of $\Delta H^{vv}$ predictions is quite interesting, considering that this quantity is significantly overestimated in 3D-RISM/HNC (figure \ref{fig:alkanes_therm3}).
Within the linear response approximation one has $\Delta H^{vv} = - \frac{1}{2} E^{uv}$ \cite{Matyushov1999tgv}.
As figure \ref{fig:reorganization} demonstrates, this result is relatively accurately satisfied by molecular dynamics, as well by PC and PC+ models.
Once we get rid of overestimated insertion free energy contribution to 3D-RISM/HNC, the model starts giving a number of predictions that treat solute-solvent interaction via the linear response approximation.
%
%
%
 

\chapter{Applications} 

\label{Chapter6}

\lhead{Chapter 6. \emph{Applications}}

In contrast to the previous chapter, here we are primarily concerned with realistic solutes and comparison with experimental measurements.
We discuss the solvation of neutral and charged molecules in water at various temperatures.
The accuracy of pressure corrected 3D-RISM models is compared to other approaches.
The chapter is based on two previously published articles: Refs. \citenum{Misin2015wdu} and \citenum{Misin2016tqn}.

\section{Neutral molecules}
\label{sec:neutral_exp}

We start by comparing PC and PC+ predictions to molecular dynamics hydration free energies.
The differences in the results then occur primarily due to the approximations in the model and not due to the inaccuracies of the force field, which would have been a major error source when comparing to experiment.

Unfortunately, the popular 3D-RISM water models are different from water models used in molecular dynamics.
As we saw in section \ref{sec:3d-rism}, the hydrogen atoms' density can behave quite independently from the oxygen one.
Thus, leaving hydrogen atoms without Lennard-Jones parameters, which is commonly done in the water models used for molecular dynamics, will inevitably cause hydrogens to "spill" into the solutes.
To avoid it, all 3D-RISM water models employ small Lennard-Jones parameters on hydrogen atoms.

One of the largest evaluations of hydration free energies using molecular dynamics was done by Mobley et al \cite{Mobley2009wdk}.
They computed and published both Lennard-Jones and electrostatic contribution to hydration free energies of 504 molecules, using quite long simulation runs to ensure low uncertainty.
The set of solute molecules was quite diverse and included all the main functional groups such as alcohols, carboxylates, aromatic compounds, amines, etc.
All solutes were described used general amber force field (GAFF) and AM1-BCC partial charges.
Water was approximated using the standard TIP3P model \cite{Jorgensen1983wtl}.

\begin{sidewaysfigure}
	\centering
	\includegraphics[width=\linewidth,draft=false]{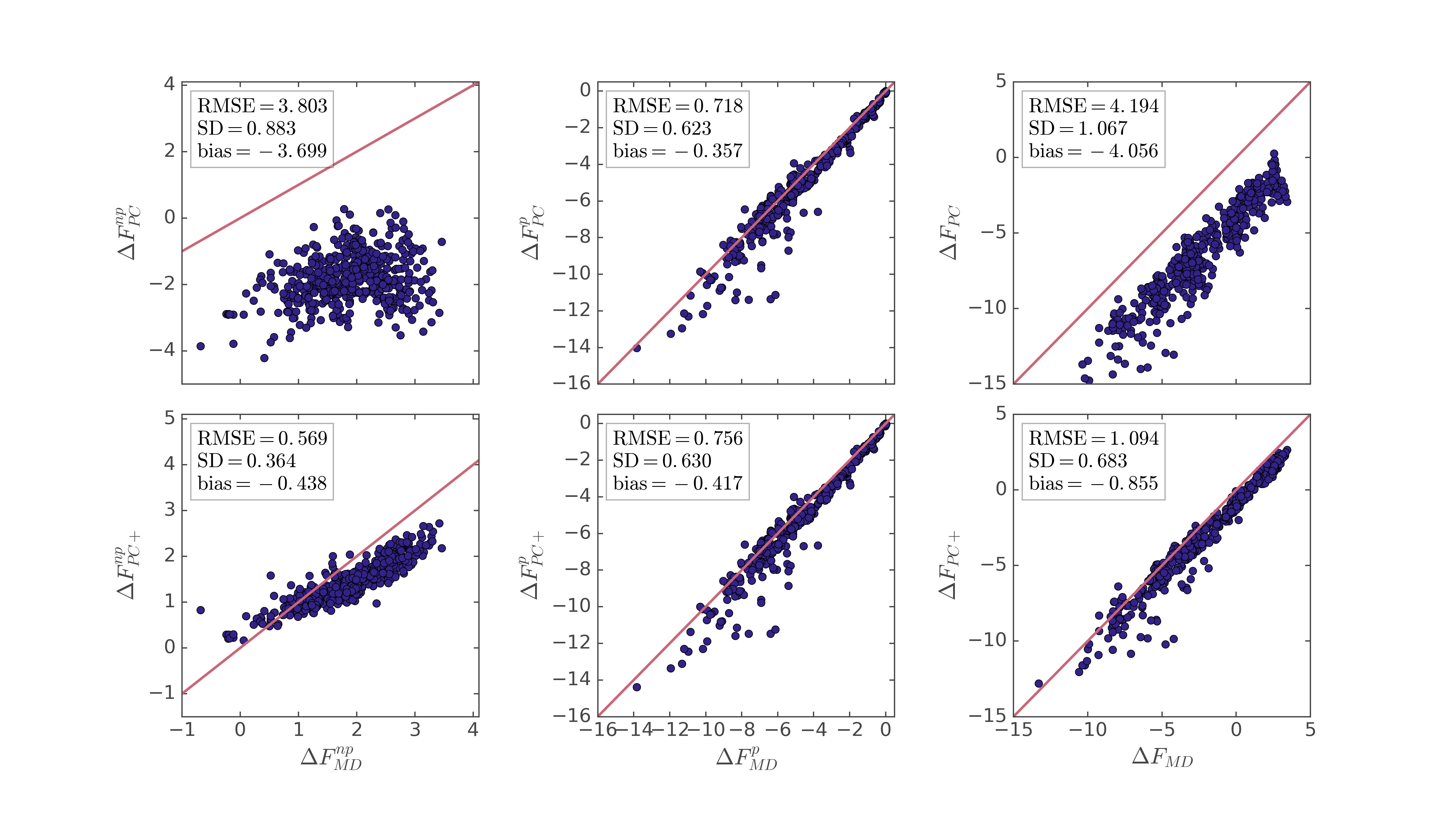}
	\caption{Comparison of PC (top) and PC+ (bottom) free energies with molecular dynamics results, obtained by Mobley et al \cite{Mobley2009wdk}. In the first column we compare nonpolar (Lennard-Jones) components, in the second, electrostatic contributions, and in the third, total hydration free energies. All values are in \si{kcal\per\mole}.}
	\label{fig:pcp_pc_fe}
\end{sidewaysfigure}

For 3D-RISM calculations, we used the same solute potentials.
We did not try to take into account their conformations and simply used a single minimised geometry.
Water was described by the conventional cSPC/E model \cite{Luchko2010uhh}, which differs from ordinary SPC/E model by Lennard-Jones potentials on hydrogen atoms, mentioned earlier.
More technical details are summarised in the appendix \ref{AppendixA}.

The comparison between HNC results and molecular dynamics was already provided in figure \ref{fig:pse3_fe}.
The performance of PC and PC+ models is presented in figure \ref{fig:pcp_pc_fe}.
As we can see, the PC model fails to predict nonpolar free energies $\Delta F^{np}$ (which correlate with cavity creation free energy), while PC+ does capture general trends.
The polar part of hydration free energy $\Delta F^{p}$ is approximated with a reasonable accuracy by both models.
The total hydration free energy is estimated by PC+ quite well.

Both models perform worse for polar molecules that have more negative solvation free energies.
The larger outliers tend to be the molecules containing negatively charged oxygen or hydrogen atoms, such as 2-ethoxyethanol or hydrazine.
It is quite likely that the presence of Lennard-Jones sites on hydrogen atoms becomes increasingly important and thus 3D-RISM describes them somewhat differently.
Another potential source of error is a lack of conformational sampling, but it is unlikely to cause a large effect.

\begin{sidewaysfigure}
	\centering
  	\includegraphics[width=1\linewidth,draft=false]{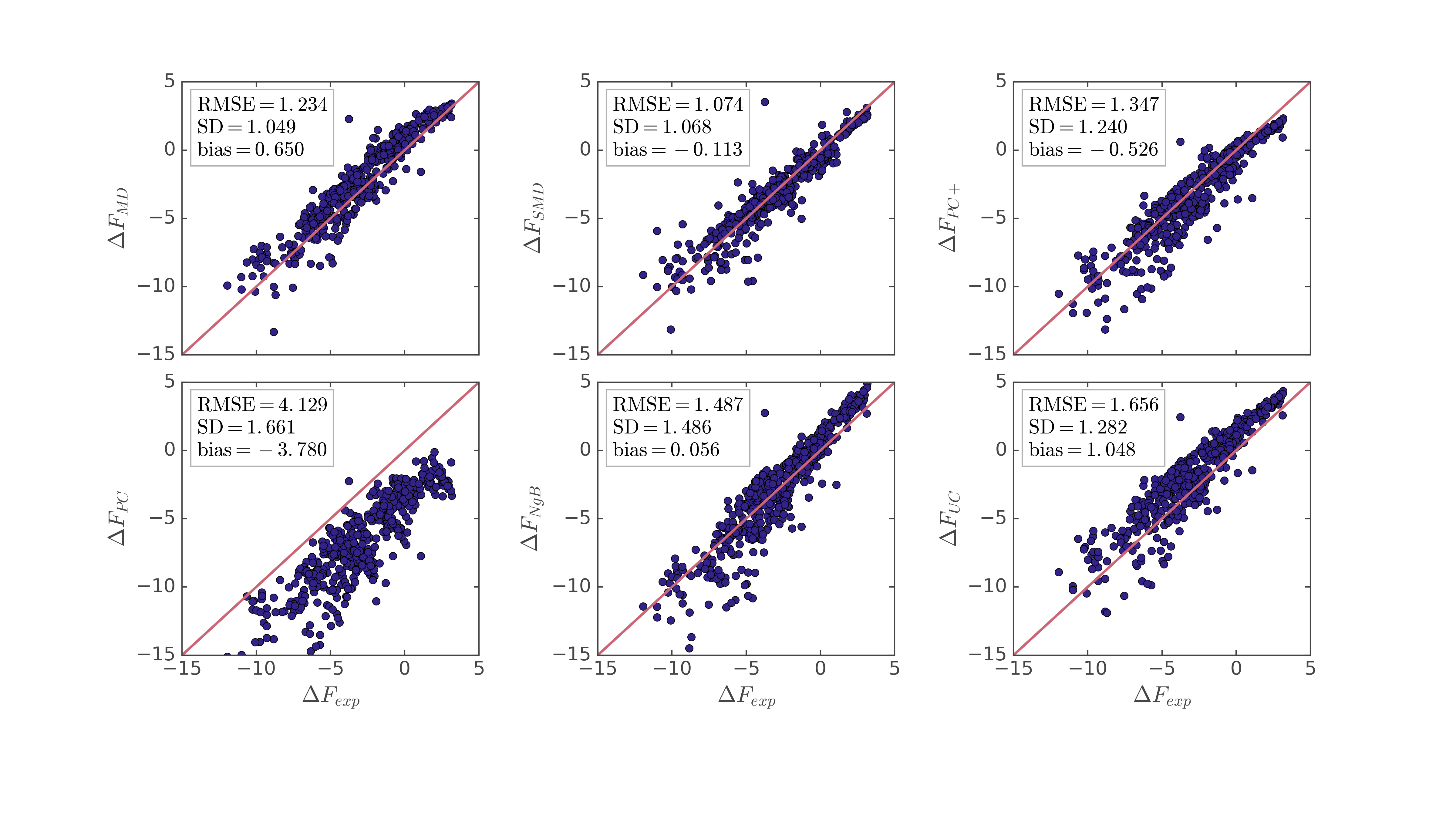}
	\caption{Hydration free energies by different models, compared to experimental data. Both molecular dynamics results and experimental values come from Ref. \citenum{Mobley2009wdk}. All values are in \si{kcal\per\mole}.}
	\label{fig:mobley_exp_all}
\end{sidewaysfigure}

After comparing our corrections to MD, we decided to compare their accuracy against the actual experimental data.
The molecules were taken from the already familiar dataset by Mobley et al.
The uncertainty of experimental values in the set was estimated by authors to be around \SI{0.2}{kcal\per\mole}.
Figure \ref{fig:mobley_exp_all} demonstrates hydration free energy predictions of MD, solvation model density (SMD) \cite{Marenich2009tbw}, PC+, NgB, and UC models to experimental measurements.

SMD is the most accurate model among those compared, with an error of  \SI{1.1}{kcal/mol}.
When using SMD, we followed the recommended protocol, involving running two electronic structure calculations, which we performed in both vacuum and liquid phases \cite{Marenich2009tbw}.
The geometries were optimised in both phases, using the M06-2X functional \cite{Zhao2007unx}, combined with the MG3S basis set \cite{Lynch2003ult}.
The free energy was computed by subtracting the molecules energy in water from the energy in the vacuum.
Thus, SMD is the only model which takes into the account polarization contribution to hydration free energy.
We also expect the continuous charge distribution from quantum calculations to be more accurate than the point charges used in MD and RISM models.

A major shortcoming of SMD is that it estimates the non-polar part of free energy empirically, utilizing an equation based on solvent surface tensions, parametrized to fit experimental hydration free energies \cite{Marenich2009tbw}.
This limits the applicability of the method to compounds without complicated structure and functional groups.
However, for simple and small molecules such as those which were the part of the Mobley's dataset, SMD and related methods are probably the best choices since they can utilize accurate charge distributions from quantum mechanics and their empirical approaches to estimating cavity energies are not too inaccurate.

The PC+ model error is not much larger than the one seen from MD.
The systematic errors due to its approximation of free energy as well as errors of utilised force field (GAFF/AM1-BCC) cancel out favourably, making the overall accuracy quite good.
It performs better than 3D-RISM models with empirical corrections such as NgB and UC, but shows results which are poorer than MD and SMD.
Overall, it seems that while PC+ is moderately accurate, it still performs worse than other common methods for predicting hydration free energies of neutral molecules.

Other 3D-RISM based models such as PC, UC, and NgB perform worse.
We already discussed the issues with PC at the beginning of the section.
The performance of UC on the other hand can likely be improved by a more careful choice of parameters used to fit the model (we used the values provided in section \ref{sec:solv_free_energy_3drism}).

\begin{figure}
	\centering
	\includegraphics[width=1\linewidth]{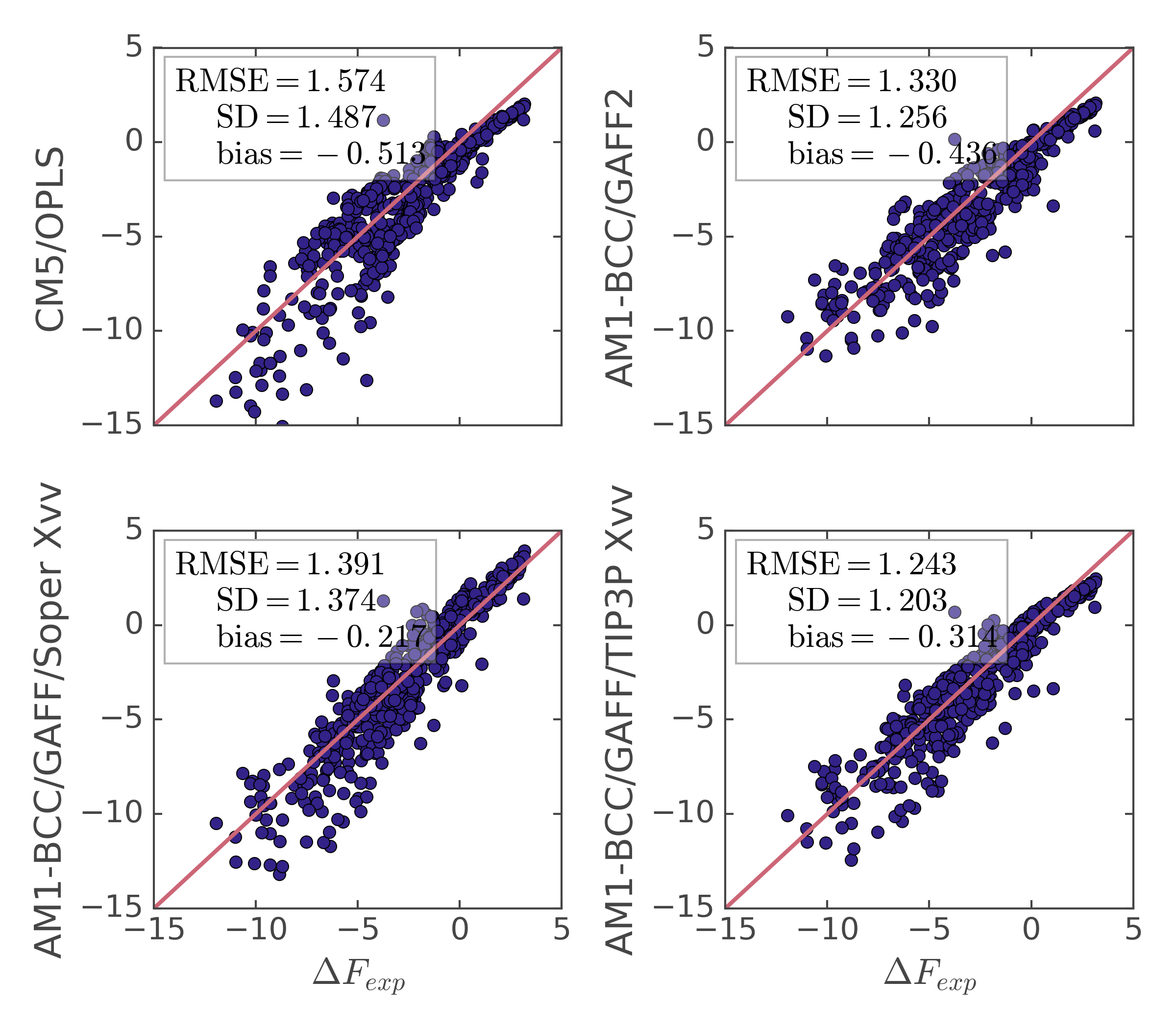}
	\caption{The influence of different force fields on the accuracy of PC+ results when compared to experiment. All values are in \si{kcal\per\mole}.}
	\label{fig:forcefieldeffect}
\end{figure}

Before closing this section, it is interesting to discuss possible effects of the force field on the results.
As molecular dynamics results show, a substantial part of the error can be attributed to the inaccurate solute-solvent interaction energy.
Thus, a more accurate force field might in principle substantially improve the accuracy of the result.

We did not want to explore these possibilities in too much detail as, due to the number of force fields and water models available, such as investigation would warrant a separate study.
However, as a test we performed a few calculations investigating the sensitivity of PC+ results to the choice of interaction potential energy between solute and solvent (figure \ref{fig:forcefieldeffect}.
In the first two tests (demonstrated on the first row) we used the standard cSPC/E water susceptibility functions computed with 1D-RISM, and varied Lennard-Jones and partial charges on solutes.
Neither combination of CM5 and OPLS-2005, nor AM1-BCC and GAFF2 (an updated version of GAFF force field, introduced in AmberTools version 16 \cite{Case2016wcn}) substantially improved the accuracy of free energy predictions, when compared to experiment.
At the same time, standard parameters, combined with other water susceptibility functions, do slightly improve the accuracy of PC+ results (second row in the figure).
We used standard 1D-RISM calculations with the cTIP3P \cite{Luchko2010uhh} water model for one test and carefully smoothed experimental water radial distributions by Soper et al. for another \cite{Soper2013wjx}.
As these results show, more sophisticated or empirical water models might present an easy route towards further improving solvation free energy predictions from PC+ approximation.

\section{Predicting the temperature dependence of $\Delta G$}

It is difficult to predict hydration free energies at non-standard conditions or estimate their derivatives.
First of all, the majority of the empirical/semi-empirical models are parametrized at \SI{298}{K} and are only suited for computing solvation free energy and nothing else.
Even approaches such as molecular dynamics, which can in principle access a wide range of conditions, usually require very extended calculation runs to estimate free energy derivatives with low uncertainty.

The above makes 3D-RISM based models interesting since they can be used at any thermodynamic conditions and can produce results with high numerical accuracy.

To test 3D-RISM at non-standard conditions, we used a set of experimental solvation free energies, measured at many different temperatures.
For each compound, that dataset had at least five hydration free energy measurements, all made at different temperatures between $0$ and $100$ degrees Celsius.
The data was compiled by Chamberlin et al., and presented in Refs. \citenum{Chamberlin2006ueu} and \citenum{Chamberlin2008tqh}; for this reason, we will refer to this data as the Chamberlin dataset.
To extract accurate solvation entropies and heat capacities from this data, we fit the following equation to all measurements, discarding the molecules to which this relationship fits poorly:
\begin{equation}
\Delta G(T) = \Delta G(T^{\ast}) - \Delta S(T^{\ast}) (T-T^{\ast}) + \Delta C_s(T^{\ast}) \left[T - T^{\ast} - T\ln\left(\frac{T^{\ast}}{T}\right)\right],
\label{eq:dg_temp_depend}
\end{equation}
where $T^{\ast}$ is an arbitrary temperature, which we set to \SI{298.15}{K}, $\Delta G(T^{\ast})$ is the solvation free energy at that temperature, $\Delta S(T^{\ast})$ is solvation entropy, and $\Delta C_s(T^{\ast})$ is the solvation heat capacity change, defined as:
\begin{equation}
\Delta C_s(T) = \frac{\partial \Delta H}{\partial T} = \frac{\partial \Delta G + T\Delta S}{\partial T}.
\end{equation}
The equation \ref{eq:dg_temp_depend} is known to accurately fit the temperature dependence of hydration free energy in quite large temperature ranges (since the dependence of solvation heat capacity on temperature is not very significant) and is used in experiments to measure solvation entropies and heat capacities.

For computational evaluation of solvation free energies we used geometries of solutes guessed using Openbabel  software package \cite{Oboyle2011usb,Oboyle2008txc} and further optimised with OPLS\_2005 force field \cite{Banks2005vwm}.
Each experimental solvation free energy at different temperature was matched by a corresponding 3D-RISM calculation.
Water density and dielectric constants at each temperature were evaluated using interpolation functions provided in the Water Society manual \cite{Daucik2011uyk} (the relative uncertainty of the density is around 0.0001\% and for the dielectric constant is $0.01\%$).
After performing all calculation we used equation \ref{eq:dg_temp_depend} to extract solvation thermodynamic parameters from 3D-RISM calculations.
For all solutes, it fit the data with practically perfect accuracy.

\begin{figure}
	\makebox[\textwidth][c]{\includegraphics[width=1.2\textwidth]{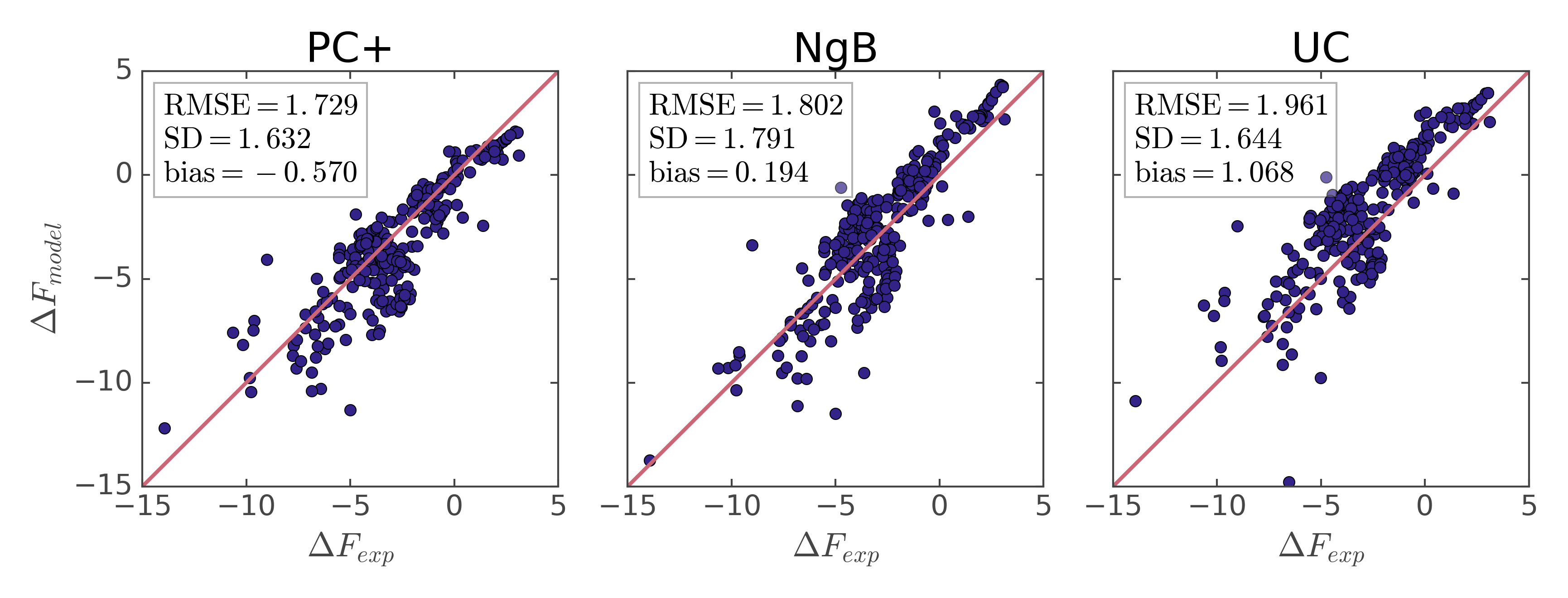}}
	\caption{Comparison experimental and computed hydration free energies at \SI{298}{K}. The data is taken from the Chamberlin dataset. All values are in \si{kcal\per\mole}.}
	\label{fig:chamberlin_fe}
\end{figure}

Before analyzing derivatives, we first checked how accurately PC+, NgB, and UC models could predict standard temperature solvation free energy on this dataset.
The results are shown in figure \ref{fig:chamberlin_fe}.
As you can see, the accuracy of these models on the Chamberlin dataset is lower than on the Mobley dataset.
The reason for that is the larger diversity of the Chamberlin dataset, which covers a broader range of molecules and combinations of functional groups.
Thus, the simple force field used for all 3D-RISM models (GAFF/AM1-BCC) might be somewhat poorly applicable to them.

\begin{figure}
\centering
\includegraphics[width=1.0\linewidth]{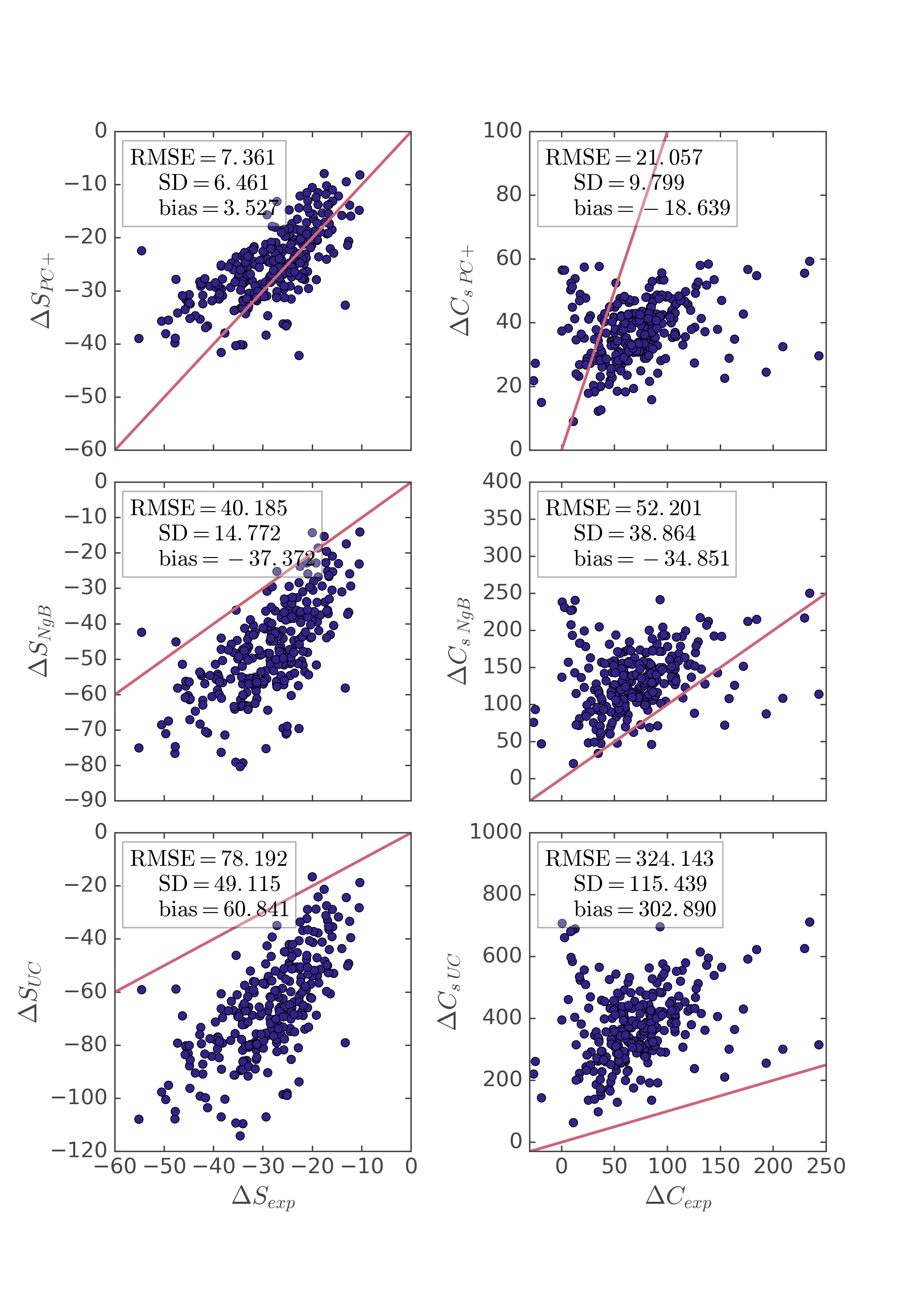}
\caption{Comparison of predicted and experimental solvation entropies and heat capacities, evaluated using equation \ref{eq:dg_temp_depend}. Both entropy and heat capacity are in units of \si{cal\per\mole\per\kelvin}.}
\label{fig:entropy_heat_cap}
\end{figure}

Figure \ref{fig:entropy_heat_cap} demonstrates a comparison between predicted and experimental entropies and heat capacities.
One can immediately see that free energy derivatives are predicted with much poorer accuracy than free energy itself.
The average error in predicted entropies by PC+ accounts for about $20\%$ of its total value.
For NgB and UC it is about $70\%$ and $130\%$ respectively.
For solvation heat capacities, none of the models correlated with experiment.

It is not surprising for a model to give reasonable free energies while failing to predict its derivatives.
Recall that enthalpic and entropic contributions tend to be larger than solvation free energy and have opposing signs (section \ref{sec:solv_thermodynamics}).
Other significant factors are the force fields that are parametrized specifically to reproduce free energies, ignoring both solvation entropies and enthalpies.

\begin{figure}
	\centering
	\includegraphics[width=1.0\linewidth]{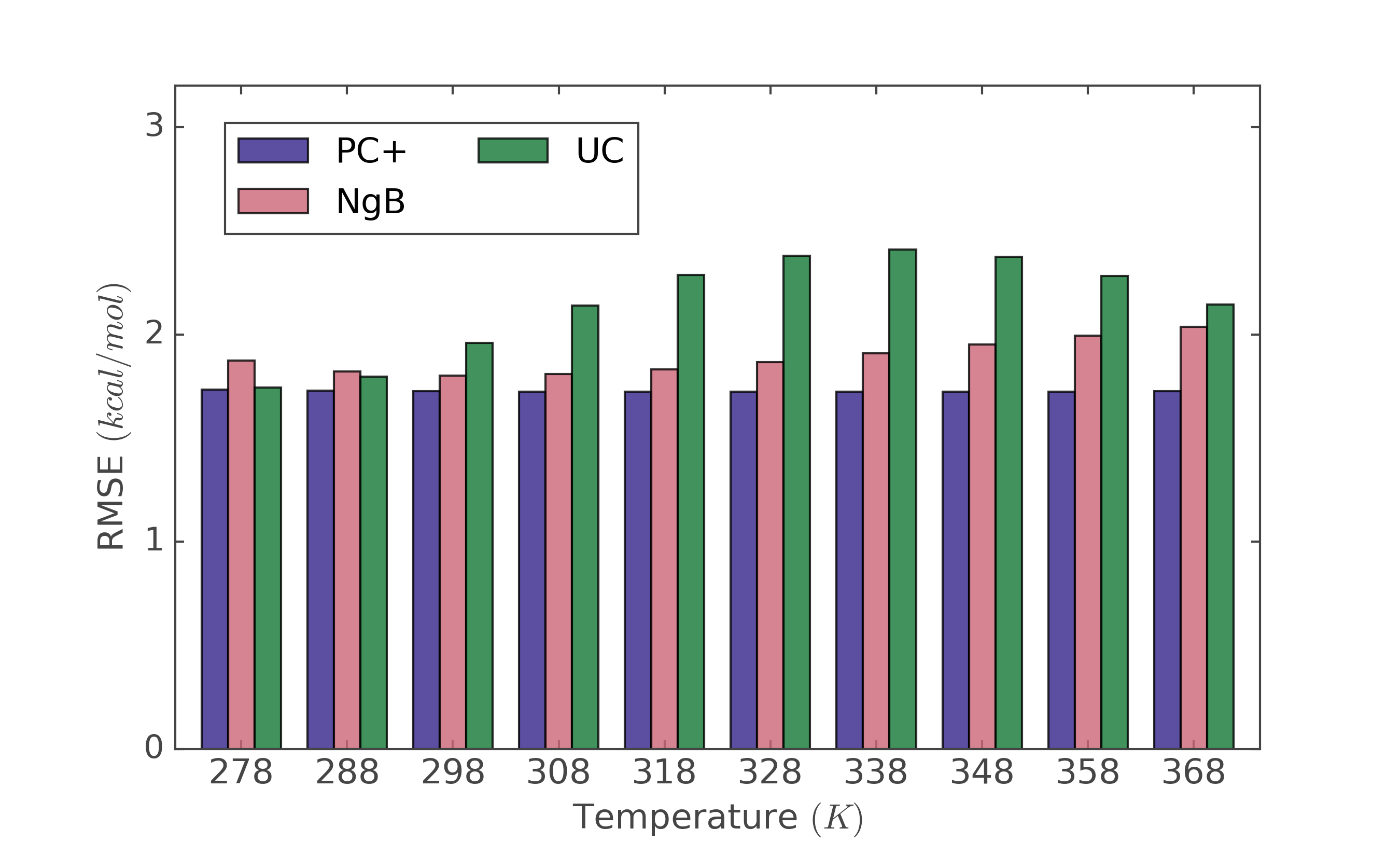}
	\caption{Root mean square error of solvation free energy of 3D-RISM models on Chamberlin dataset depending on temperature.}
	\label{fig:chamberlin_T_fe}
\end{figure}

Despite giving poor estimates for solvation entropies and heat capacities, as figure \ref{fig:chamberlin_T_fe} shows, the accuracy of free energies across the $273$--\SI{373}{K} range remains similar for NgB, and almost constant for PC+.
This is not surprising, considering that the absolute value of solvation free energies at this range, for smaller molecules, changes at most by $1$--\SI{2}{kcal/mol}.
Moreover, since the error in entropy predictions by PC+ is unbiased, half of the solvation free energy improves towards the higher temperatures (for instance, overestimated $\Delta G$ at $\SI{298}{K}$ combined with underestimated entropy will result in more accurate solvation free energy estimates at higher temperature).
The larger problems in solvation heat capacities for all of the models are not significant enough to affect free energies at this temperature range.
It is also worth noting that regardless of the somewhat inflated accuracy of solvation free energies at other temperatures, these predictions are still significantly better than "0 hypothesis" estimates, made under the assumption that solvation free energy does not depend on the temperature at all (figure \ref{fig:uc_ngb_fe}).

\begin{figure}
\makebox[\textwidth][c]{\includegraphics[width=1.2\textwidth]{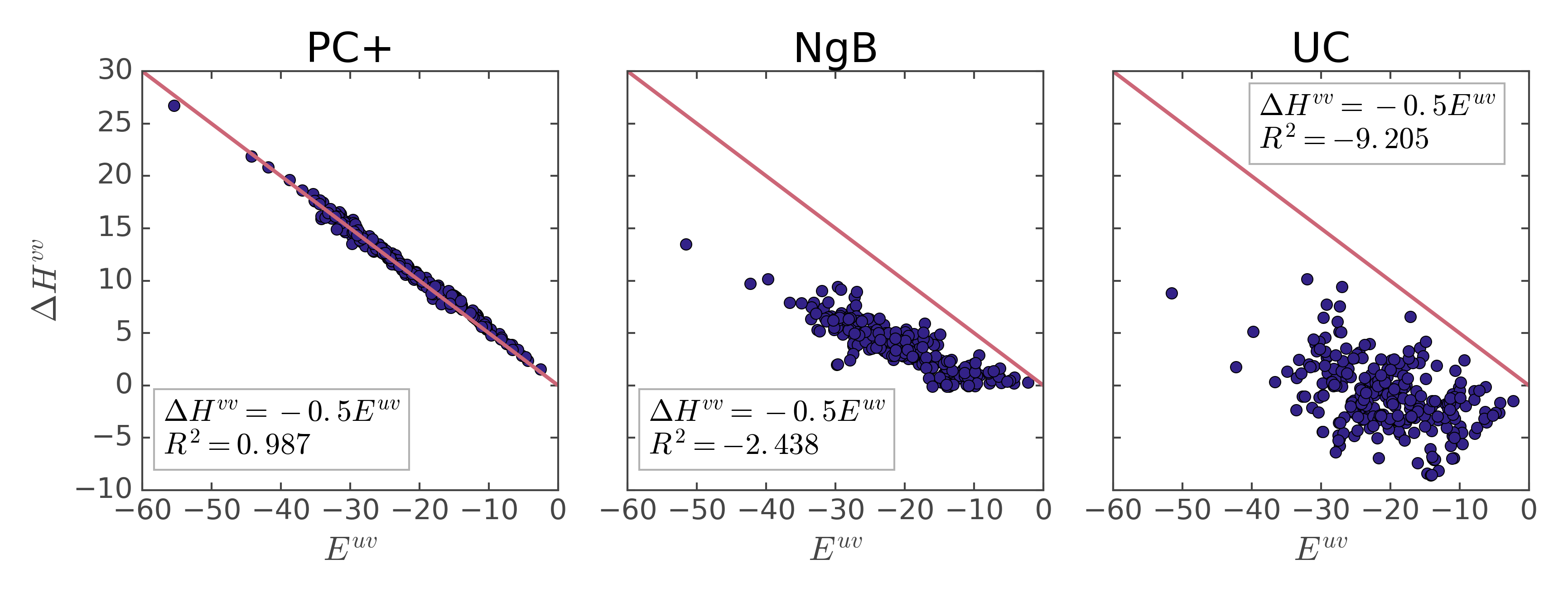}}
\caption{Correlations between solute-solvent and solvent reorganization energies. $R^2$ stands for coefficient of determination. All values are in \si{kcal\per\mole}.}
\label{fig:reorganization_line_response}
\end{figure}

Concluding the discussion of solvation thermodynamics we go back to the linear response relationship $\Delta H^{vv} = -E^{uv}/2$, discussed in the section \ref{sec:pc_pcp_solv_td}.
As one would expect, the relation only holds for the PC+ model, with empirical model failing to reproduce it.
It is possible that re-parametrization of both modes with the aim of recovering this relationship might improve the accuracy of their solvation entropy predictions.

To conclude this section, we analysed the performance of solvation free energy predictions by PC+, NgB, and UC on the extensive dataset of solvation free energies, covering temperatures from $0$ to $100$ degrees Celsius.
Compared to empirically parametrized models, PC+ showed better results across all temperatures.
However, even the model with more theoretical basis still cannot accurately predict the derivatives of free energy due to inherent difficulties associated with this task.
It is also worth mentioning that very recently Johnson et al. extended both UC and NgB by introducing temperature dependence into empirical coefficients \cite{Johnson2016tta}.
This approach might result in improved entropies for empirical models, but it remains to be tested on a larger dataset.

\section{Ionic solvation}

As we demonstrated in the first section of the chapter, using continuum models such as SMD, it is possible to predict solvation free energies of small neutral molecules with around $\SI{1}{kcal/mol}$ accuracy.
The situation is quite different in the case of charged compounds, for which even the most accurate implicit models show relatively poor results (typical accuracies of about $\SI{5}{kcal/mol}$ for water solvation).
The reason for these difficulties is associated with much larger interaction energies between solute and solvent, as well as some effects (such as charge asymmetry) that cannot be described by continuum models.

Considering the above information, applications of models such as PC+ to ionic solvation seems quite promising.
In section \ref{sec:charging_fe} we demonstrated that 3D-RISM is capable of predicting reasonable charging free energy for Lennard-Jones solutes.
To test its accuracy in a more realistic setting, we turned to polyatomic ions.
Both experimental hydration free energies and solute geometries were taken from the 2012 version of the Minnesota solvation database \cite{Marenich2012vuj,Kelly2006tom}.
These values are based on the hydration free energy of the proton = \SI{265.9}{kcal/\mol} \cite{Tissandier1998tvt}.
When selecting compounds from the database, we avoided water clusters as well as ions that were structurally similar to other chosen molecules.
We ended up selecting $70$ compounds in total: $36$ anions and $34$ cations.

\begin{figure}
\centering
\includegraphics[width=1.0\linewidth]{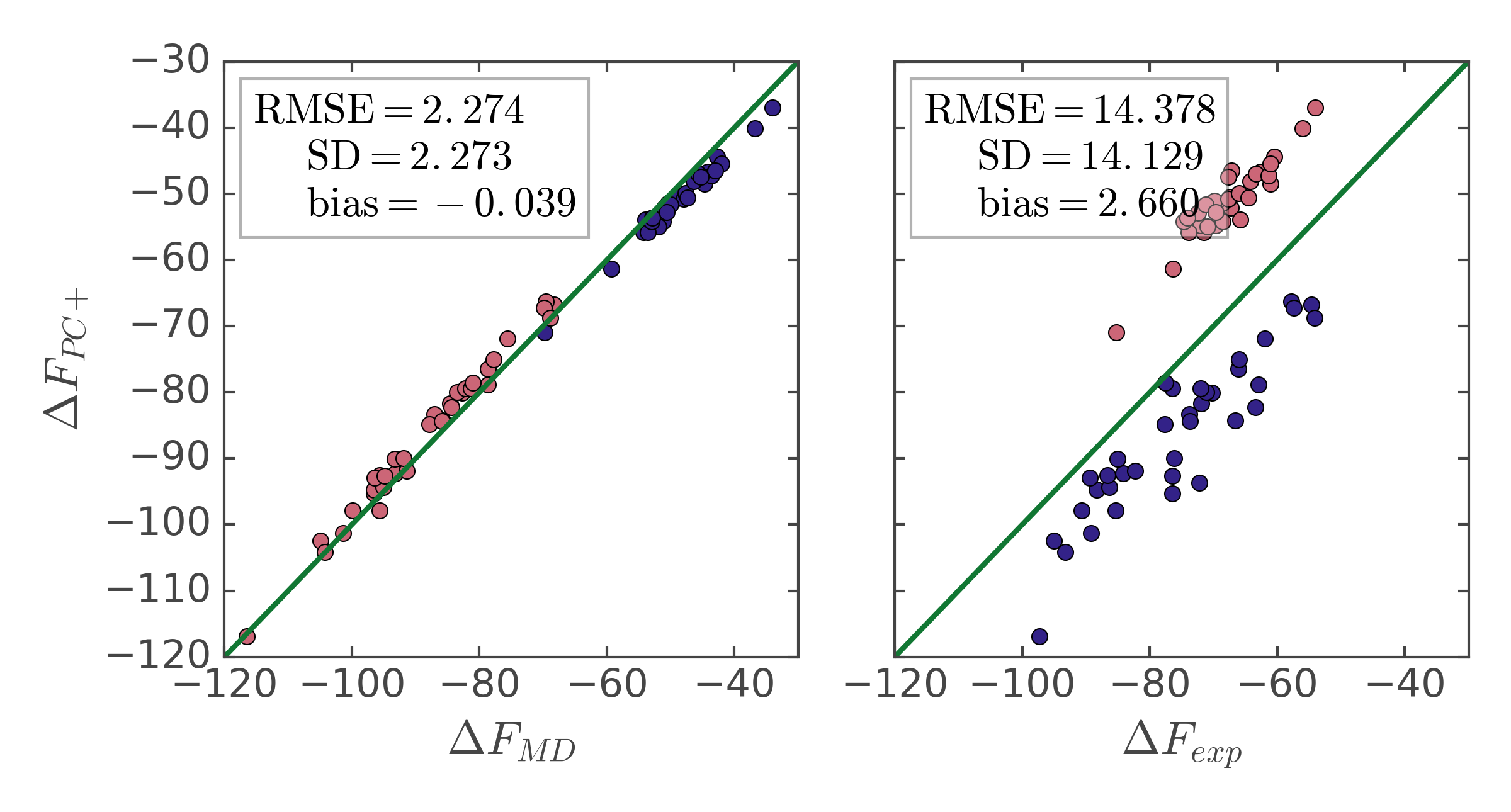}
\caption{Comparison of 3D-RISM/PC+ ionic hydration free energies with molecular dynamics results (left) and experimental values (right). Galvani potential is not taken into account. All values are in \si{kcal/\mole}.}
\label{fig:ionic_uncorrected}
\end{figure}

The non-bonding parameters for ions were derived from GAFF and were combined with AM1-BCC charges.
We started by performing all free energy calculations with PC+ and then repeated calculations using molecular dynamics.
The results are shown in figure \ref{fig:ionic_uncorrected}.

\begin{figure}
	\centering
	\includegraphics[width=0.7\linewidth]{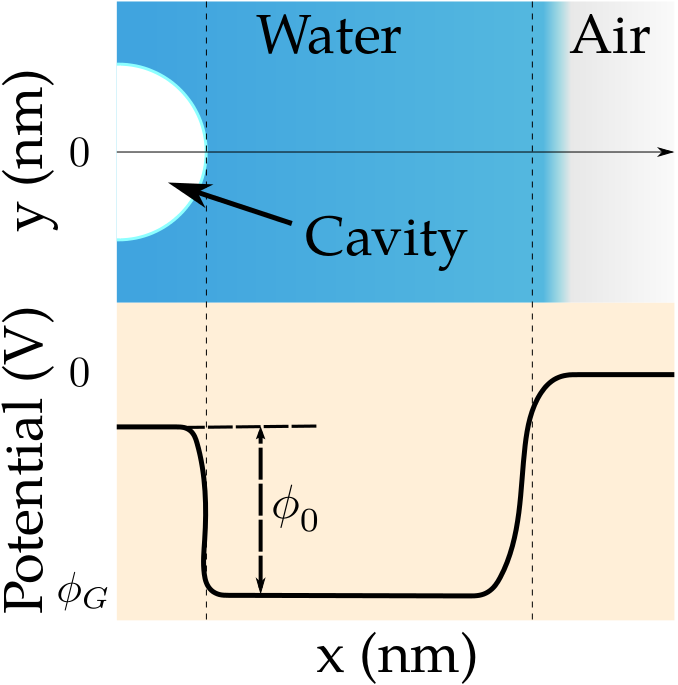}
	\caption{Schematic demonstration of interfacial potential jumps in solution.}
	\label{fig:galvani_bethe}
\end{figure}

As you can see, the agreement between MD and PC+ is quite good.
However, neither of the methods agreed well with experimentally measured values.
The reason for that is the lack of explicit water-air boundary in both MD and 3D-RISM.
The whole system is schematically demonstrated at the figure \ref{fig:galvani_bethe}.
While in simulations the potential in the empty cavity is $\phi_0$, in experimental settings it has an extra contribution from the Galvani potential $\phi_G$, making it $\phi_0 + \phi_G$.

To correct the solvation free energies from both models, we need to add an extra $q \phi_G $ term, accounting for the transfer from one phase to another.
The question is, which $\phi_G$ to use?
Simulations using various models have shown that it is extremely sensitive to the water representation.
It has a value of \SI{~3000}{mV} when measured using ab initio approaches, while atomistic simulations give estimates close to \SI{600}{mV}.
The experimental data are also quite conflicting \cite{Hunenberger2011vfc}.

To understand the disparities in $\phi_G$ estimates it is important to consider different contributions towards the interfacial potential.
It has been shown that electrostatic potential across any interface is given by \cite{Remsing2014vhq,Lin2014tgi}
\begin{equation}
\begin{split}
\phi_{vl} & =  -\frac{1}{\epsilon_0}\int\limits_{z_v}^{z_l} \,dz' z' \langle \rho_q (z') \rangle \\
&\approx -\frac{1}{\epsilon_0}\int\limits_{z_v}^{z_l}\left\langle P_z(z') \right \rangle \diff z' + \frac{1}{\epsilon_0} \left[ \left\langle Q_{zz}(z_l)\right \rangle - \left \langle Q_{zz}(z_v)\right \rangle \right]\\
& = \phi_D + \phi_Q,\\
\end{split}
\label{eq:interfacial_pot}
\end{equation}
where $\phi_{vl}=\phi_l - \phi_v$ is the electrostatic potential difference between a liquid and vacuum, $z$ is the direction perpendicular to the interface towards vacuum, $z_l$ and $z_v$ are positions sufficiently deep into the liquid and vacuum (we assume that liquid vapour contribution is negligible), $\rho_q$ is the charge density, and $\epsilon_0$ is the vacuum permittivity.
In the above, the first line corresponds to the solution of the one-dimensional Poisson equation, while the second line is obtained by the Taylor expansion of the charge density in terms of molecular multipoles.
The average polarization at $z'$ is given by
\begin{equation}
\left \langle P_z(z')\right\rangle = \left\langle \sum_m \delta (z' -z_m) \left(\sum_i q_{im}z_{im}\right) \right \rangle,
\end{equation}
with indices $m$ and $i$ indicating molecules and sites respectively, $z_m$ being the $z$ coordinate of molecular centre, $q_{im}$ is the charge of the site $i$ of molecule $m$, and $z_{im}$ is the $z$ component of distance $\rvec_{im}$ from the molecular centre to site $i$. 
Similarly, the quadrupole contribution is
\begin{equation}
\left\langle Q_{zz}(z_l)\right\rangle = \left\langle \sum_m \delta (z' -z_m) \left(\frac{1}{2}\sum_i q_{im}z^2_{im}\right)\right \rangle.
\end{equation}
For clarity, we combined dipolar and quadrapolar contributions to the potential into $\phi_D$ and $\phi_Q$ respectively.
Note that from equation \ref{eq:interfacial_pot} it follows that $\phi_Q$ does not depend on the structure of the interface and instead is only determined by the values of quadrupole contributions inside the liquid (since it is zero in the vacuum).

As was demonstrated by Remsing et al. \cite{Remsing2014vhq}, the $\phi_D$ is reasonably similar across models, and also is quite small for the water-vacuum interface, while $\phi_Q$ is a larger quantity and is the one responsible for disparities across the models.
However, it gets canceled out from the sum of the cavity and Galvani potentials since it does not depend on the structure of the interface
\begin{equation}
\phi_0 + \phi_G = -\phi_{D}^0 - \phi_{Q}^0 + \phi_{D}^{vl} + \phi_Q^{vl} = -\phi_{D}^0 + \phi_{D}^{vl},
\end{equation}
where superscripts $0$ and $vl$ denote cavity-liquid or liquid vacuum interfaces respectively.
Thus, to compare solvent potentials as well as ionic solvation free energies across different models, one has to compare solvation free energies with Galvani potential contribution taken into account.
Moreover, the Galvani potential has to be evaluated within the model so that quadrupole contribution to the solute-solvent potential is cancelled by the corresponding contribution to $\phi_G$ \cite{Hunenberger2011vfc}.

While evaluation of the Galvani potential in MD does not present a lot of difficulties (one can simply use the Poisson equation defined in the first line of equation \ref{eq:interfacial_pot}), it is hard to obtain it from 3D-RISM.
In principle, one can construct a planar air-liquid interface in 3D-RISM; however, as we have seen in section \ref{sec:3drism-pressure}, such an interface will have a density distribution quite different from the one observed in the experiment and will have quite a large dipole across it.
An alternative approach was suggested by Reif and Hunenberger for molecular simulations \cite{Hunenberger2011vfc}.
A potential inside a small cavity will be mostly determined by $\phi_Q$, since $\phi_D \approx 0$.
Then the quadrupole contribution can be evaluated directly via
\begin{equation}
\phi_Q \approx \phi_0  = \frac{1}{4\pi\epsilon_0} \sum\limits_{i=1}^{N} \int \frac{\rho_i (\rvec) q_{i}}{r} \,d \mathbf{r},
\end{equation}
where $r = |\rvec-\rvec_0|$, with $\rvec_0$ being the center of cavity.

Since the MD calculations were done with SPC/E water, we used the results of Beck who found its Galvani potential to be  \SI{-14.9}{kcal\per\mol/e} (\SI{-650}{\milli\volt}) \cite{Beck2013vnn}.
Within 3D-RISM, setting the size of hard sphere cavity to $r=\SI{0.5}{\angstrom}$ (same as grid size), we found the cSPC/E water Galvani potential to be  \SI{-13.43}{kcal\per\mol/e} (\SI{-606}{\milli\volt}).

\begin{figure}
	\centering
	\includegraphics[width=1.0\linewidth]{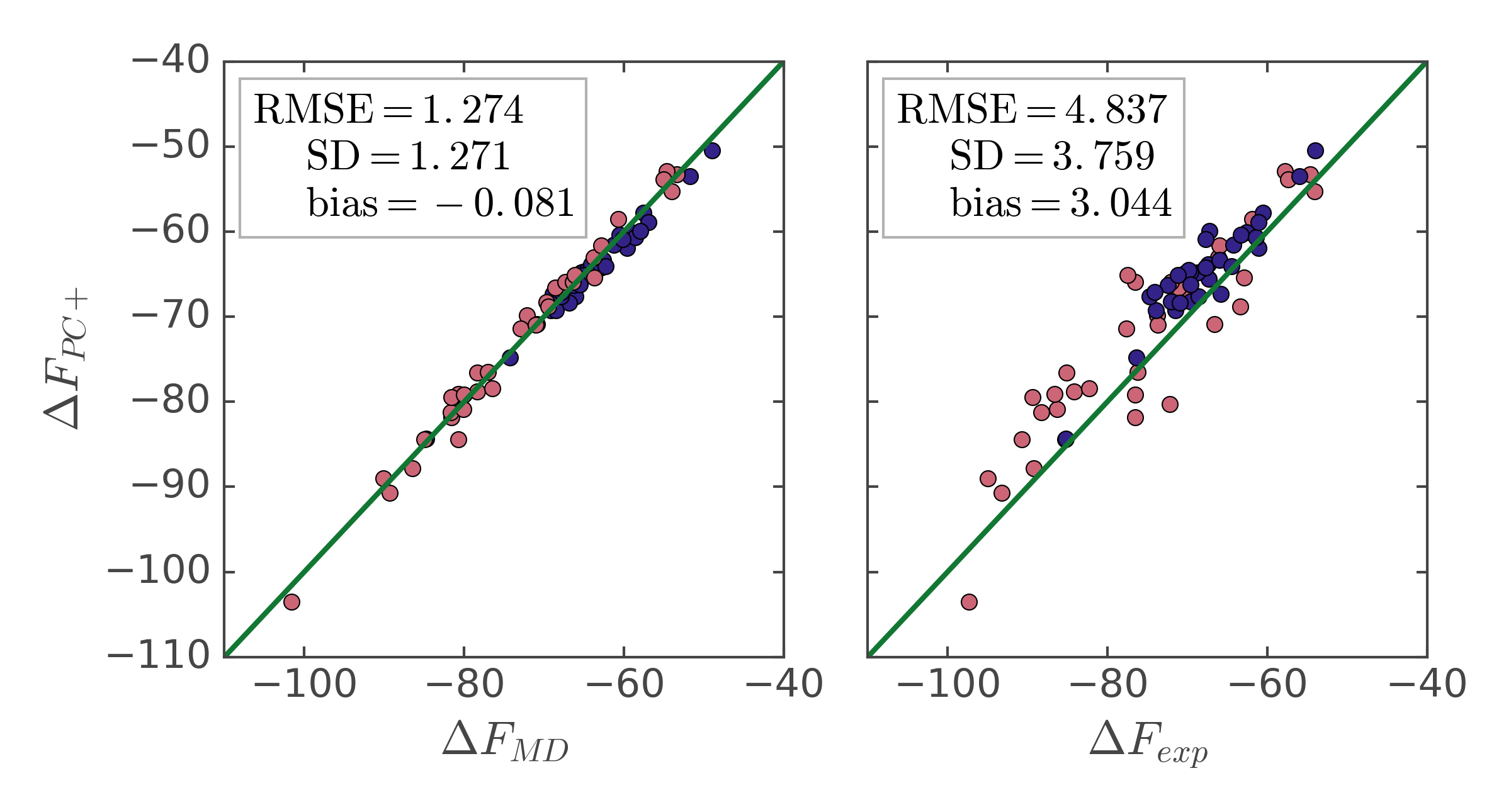}
	\caption{The same comparison as in figure \ref{fig:ionic_uncorrected}, but with model Galvani potentials taken into account. All values are in \si{kcal/\mole}.}
	\label{fig:ionic_corrected}
\end{figure}

The figure \ref{fig:ionic_corrected} demonstrates the comparison of real free energies (with Galvani contributions included) from PC+, MD, and experiment.
Not only accounting for Galvani potential improves the agreement between RISM and MD, bringing it to the level of neutral particles, but it also dramatically improves the agreement of PC+ predictions with experiment.

\begin{figure}
	\centering
	\includegraphics[width=1.0\linewidth]{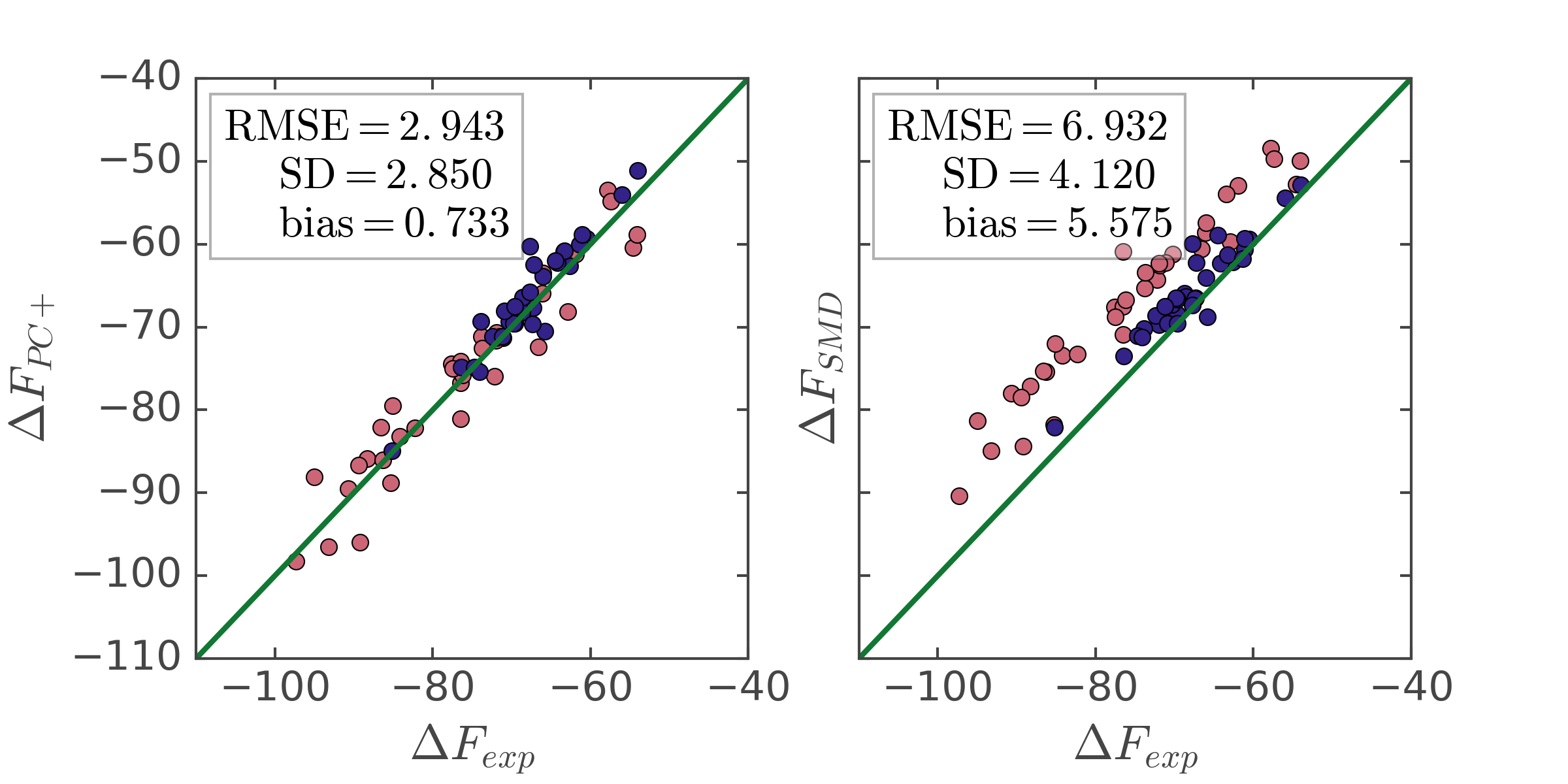}
	\caption{Left figure: 3D-RISM/PC+ ionic hydration free energies compared to experiment. The force field is OPLS/CM5. Right figure: SMD ionic hydration free energies; electronic structures are computed with MG3S/M06-2X level of theory. The values are in \si{kcal\per\mole}.}
	\label{fig:ionic_opls}
\end{figure}

While the agreement between the PC+ and experimental ionic solvation free energies was relatively good, the error was still larger than in the case of neutral compounds.
At least part of the error originates from the approximate solute charges that we obtained using the simple AM1-BCC scheme.
While it does not introduce much of a problem in the case of neutral compounds (section \ref{sec:neutral_exp}), for charged molecules an accurate charge distribution is crucial since their solvation free energy is dominated by the charging free energy term.
To accurately estimate the partial charges on solute atoms we used CM5 charges, obtained via electronic structure calculations.
These partial charges in combination with OPLS non-bonded parameters, decreased the error of the PC+ model quite significantly (figure \ref{fig:ionic_opls}).

In the figure \ref{fig:ionic_opls} we have also shown the prediction of the same ionic hydration free energies obtained by SMD.
The SMD calculations were run with the same basis set and theory level as  calculations we performed to evaluate CM5 charges.
Thus, the charge distribution of solutes in both PC+ and SMD calculations were practically identical.
The observed difference between the two models is then likely related to the asymmetry of water electrostatic response, which is missed by continuum methods such as SMD but is captured by 3D-RISM.


\textit{}

\chapter{Beyond pure water} 
\label{chap:beyond_water}

\lhead{Chapter 7. \emph{Beyond pure water}}

In this chapter, we show a few applications of 3D-RISM/PC+ to systems other than pure water.
The major difficulty here is not theoretical but rather finding an appropriate model of the solvent.
In the case of salt solution, the problem is solved by essentially a brute-force approach, while for non-aqueous solvents we develop a coarse-grained approximation.
The chapter extends Refs. \citenum{Misin2016wlh} and \citenum{Misin2016vee}.

\section{Setschenow constant}

Dissolving salts in water significantly affects its structure and polarity.
Consequently, it changes solvation free energies, solubilities, activities, and other thermodynamic parameters of solutes \cite{Tielrooij2010vlt,Marcus2009uap,Robinson1965uea,Mcdevit1952thg}.
Understanding and modeling these effects is quite important since both the natural water reservoirs as well as water in biological tissues will have a considerable amount of ions.
Thus, to accurately determine the nvironmental fate of compounds as well as their distribution in cellular environments one has to take into account effects of dissolved ions \cite{Cacace1997wvj,Readman1984ukk,Mabey1978upr,Park2008tsu}.
Additionally, techniques such as purification, polymorph control, and yield improvement all utilize salt related effects \cite{Holmberg2003uhx,Sasaki2001wdv,Melander1977tbc}.

In the context of solvation, the effects of salt on partition coefficient can be quantified using Setschenow's equation \cite{Endo2012vqa}:
\begin{equation}
\log_{10} \left(\frac{K_{1/\mathrm{water}}}{K_{1/\mathrm{salt \, water}}}\right) = k_S C,
\label{eq:setch}
\end{equation}
where $C$ is the molar concentration of salt in solution, $k_S$ is the Setschenow's (or salting out) constant, and $K$ is a partition coefficient of a compound between two phases, given by
\begin{equation}
K_{1/water} = \left[\mathrm{solute}\right]_{\mathrm{water}}/\left[\mathrm{solute}\right]_{1},
\end{equation}
in which square brackets denote equilibrium concentrations.
Setting phase 1 to a dilute gas, we can express the above equation in terms of corresponding solvation free energies \cite{Ben-naim2006vpu} to get
\begin{equation}
\Delta G_C = \Delta G_0 + k_S R T C \ln(10),
\label{eq:gibbs-setsch}
\end{equation}
where $\Delta G$ stands for solvation free energy, subscripts $C$ and $0$ denote salt concentrations in water, $R$ is the universal gas constant, and $T$ is temperature.

\begin{figure}
	\centering
	\includegraphics[width=0.7\linewidth]{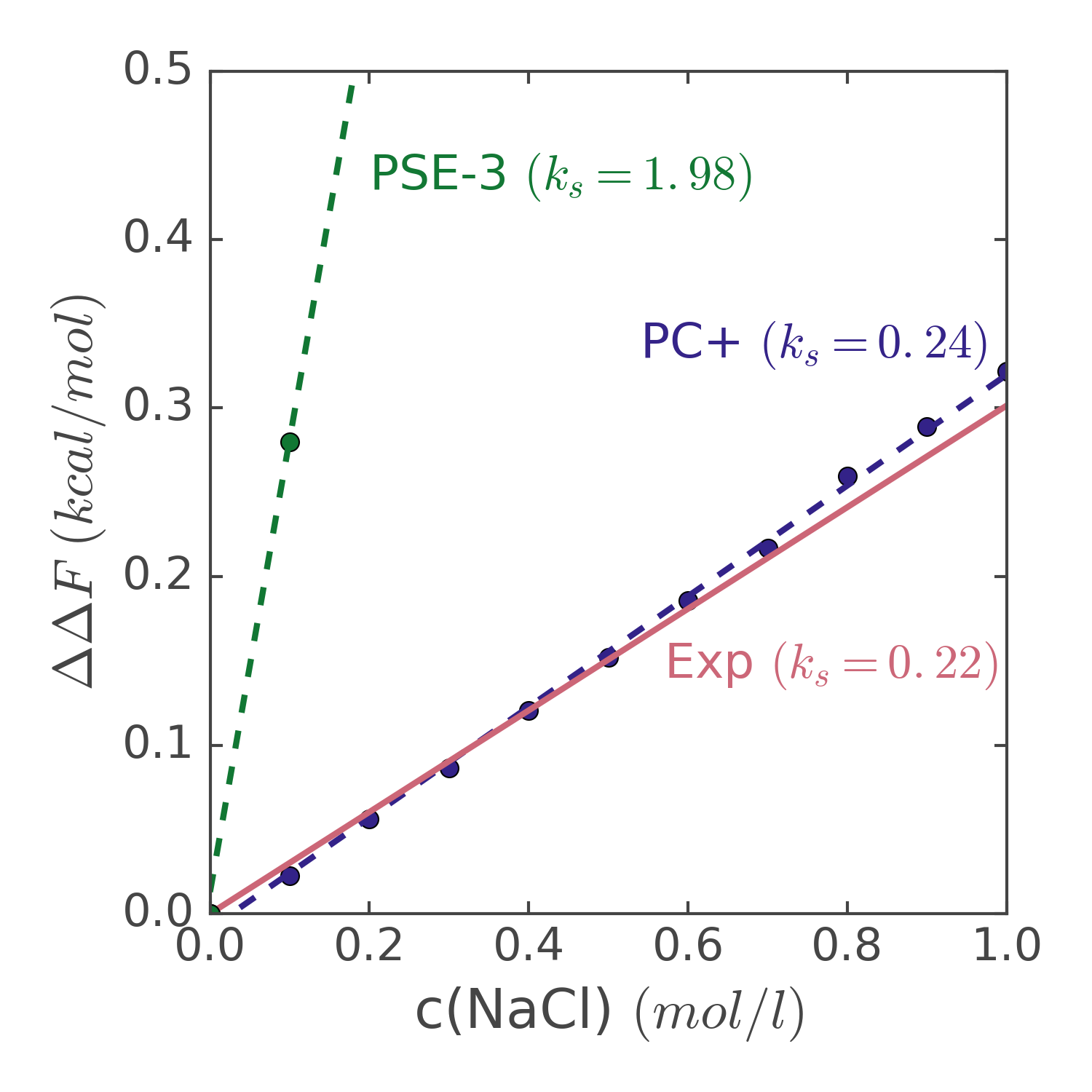}
	\caption{The change in hexanol hydration free energy depending on \ce{NaCl} concentration. The experimental trend is estimated from the experimental Setschenow's constant.}
	\label{fig:setsch_demo}
\end{figure}

$k_S$ is largely determined by molecular size.
In sodium chloride solutions, the surface tension of water increases proportionally to the salt concentration \cite{Marcus2013uzj}, and thus, provides a positive contribution to the solvation free energy of a molecule.
However, this is not the only factor contributing to the Setschenow's constant \cite{Schlautman2004vgh}.
Polar regions of molecules interact strongly with salts, and this can provide negative contributions to $\Delta G$ \cite{Li2011wjk}.
Thus, to accurately predict $k_S$, one has to take into account the change of solvent surface tension, favourable interactions between solute and salts, and correlations between anions and cations.

The figure \ref{fig:setsch_demo} demonstrates how much the solvation free energy of a compound is affected by the dissolved \ce{NaCl}.
It also demonstrates predictions by PSE-3 and PC+, which we will discuss in greater detail in the following section.

\section{Predicting solvation free energy in salt solutions}

Despite the importance of salt effects, the majority of computational approaches struggle to incorporate them.
The difficulties stem from the fact that effects of salts on solvation free energy are largely non-electrostatic and arise mainly due to the changes in $\Delta F^{LJ}$.

\begin{figure}
	\centering
	\includegraphics[width=1\linewidth]{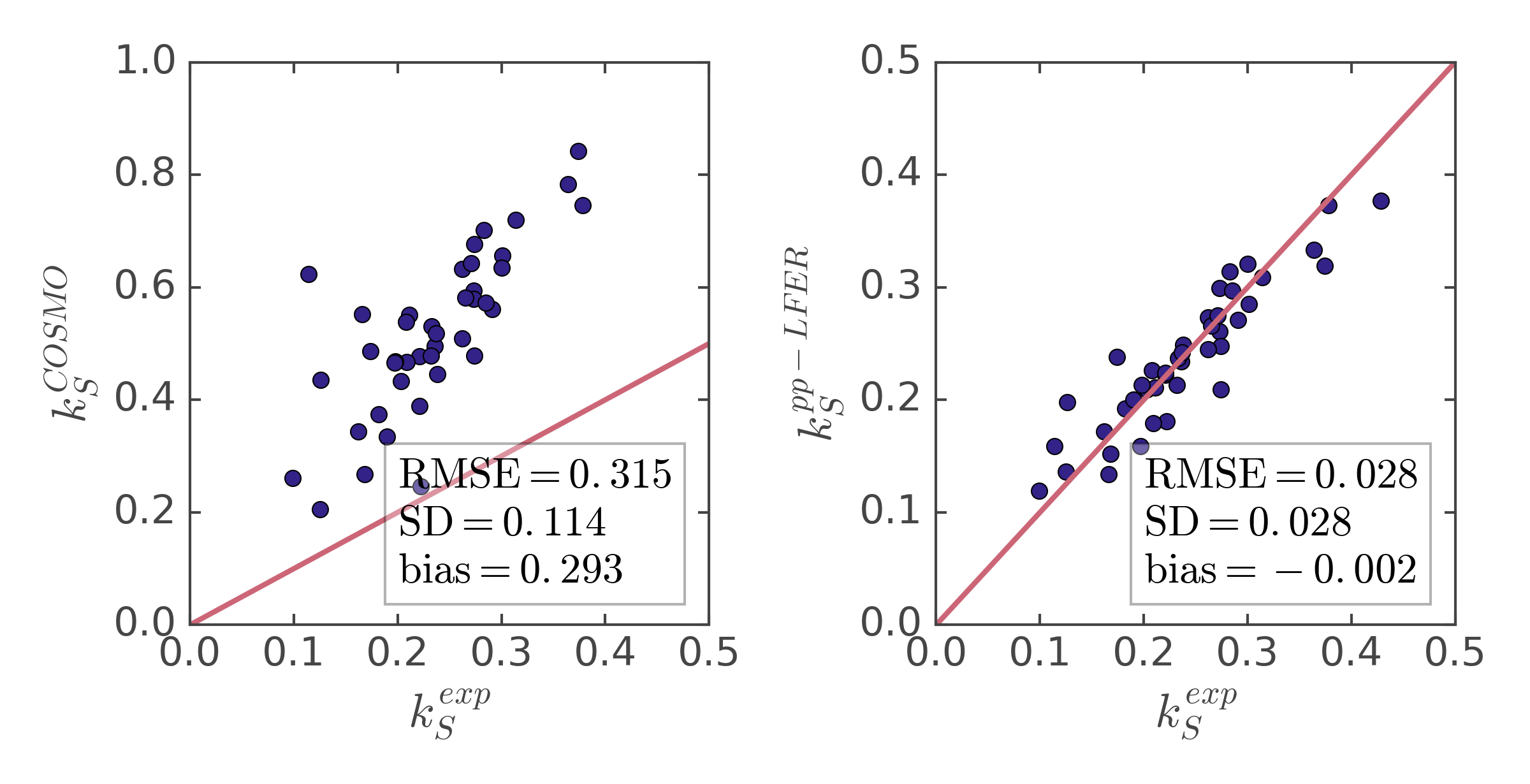}
	\caption{Accuracies of Setschenow's constants predictions by COSMO-RS (left) and pp-LFER model (right). Both experimental and computed values are taken from Ref. \citenum{Endo2012vqa}. All values are in \si{\l\per\mole}.}
	\label{fig:cosmo_chemoinf_setsch}
\end{figure}

Figure \ref{fig:cosmo_chemoinf_setsch} compares experimental values of Setschenow's constants to predictions from two computational models.
The experimental measurements were performed by Endo et al., for a set of 42 environmentally relevant compounds.
Estimates of $k_S$ made with a conductor like screening model for real solvents \cite{Klamt2005whr} (COSMO-RS) are relatively poor.
The model's inaccuracy is likely caused by its assumption of piecewise surface interactions between surface elements.
Better predictions are made by the polyparameter linear free energy relationship (pp-LFER) approach, developed by Abraham and co-workers \cite{Abraham2004txn,Stenzel2013wpe}.
The model typically uses an empirical equation of the type:
\begin{equation}
\log_{10} K = c + eE + sS + aA + bB + vV,
\end{equation}
where $K$ is the partition coefficient between two phases, $E$ is solute excess molar refraction, $S$ is polarizability, $A$ is solute H-bond acidity, $B$ is solute H-bond basicity, $V$ is the solute molar volume, and lowercase letters are adjustable parameters that depend on the phases between which solutes are distributed.
If one applies the above equation to the distribution of molecules between pure water and \SI{1}{M} solution of sodium chloride, $\log_{10} K = k_S$.
While the accuracy of the model is quite remarkable, the coefficients $c$, $e$, $s$, $a$, $b$, $v$ were determined using this dataset.
Thus, to at least some extent this agreement reflects the success of the linear regression.

\begin{table}
	\centering
	\caption{Lennard-Jones parameters of \ce{NaCl} models used in this study. The values of $\sigma$ are in Angstroms and $\epsilon$ in \si{kcal/mol}.}
	\begin{tabular}{crrrrc}
		\toprule
		Abbreviation &  $\sigma_{Na}$ &  $\epsilon_{Na}$ &  $\sigma_{Cl}$ &  $\epsilon_{Cl}$ & Ref.\\
		\midrule
		da &          2.584 &            0.100 &          4.401 &            0.100 &  \citenum{Dang1995uka} \\
		jc &          2.160 &            0.353 &          4.830 &            0.013 & \citenum{Joung2008tha}\\
		de &          1.890 &            0.199 &          4.410 &            0.199 &\citenum{Deublein2012won} \\
		$\mathrm{ho_a}$ &          2.130 &            1.540 &          4.400 &            0.100 &\citenum{Horinek2009wwm} \\
		$\mathrm{ho_b}$ &          2.230 &            0.650 &          4.400 &            0.100 &\citenum{Horinek2009wwm} \\
		\bottomrule
	\end{tabular}
	\label{tab:parameters}
\end{table}

To predict the Setschenow's constants using 3D-RISM we first need to obtain susceptibility functions $\chi$ for the bulk salt solutions.
We decided to calculate them using the 1D-RISM approach with the HNC closure.
As an input, the 1D-RISM calculations require site interaction potential energies.
Pretty much all \ce{NaCl} force fields use the same charges for ions: plus and minus one.
However, there are quite a few options for Lennard-Jones parameters \cite{Moucka2013vgf,Hess2006vrj,Patra2004wxq,Asmadi2013vot}.

For simplicity, we limited our attention to only five sodium chloride models that were compatible with SPC/E water \cite{Berendsen1987twg} and developed with Lorentz-Berthelot combination rules \cite{Allen1987uqm} in mind (table \ref{tab:parameters}).
The Lennard-Jones parameters in different models were fit to different experimental observables.
Dang's \ce{NaCl} force-field parameters (da) were developed by fitting interaction energy, the first peak of the radial distribution function, and coordination number \cite{Dang1995uka}.
Joung and Cheatham's model (jc) is based on fitting the experimental hydration free energies of ions, as well as lattice constants and energies \cite{Joung2008tha}.
Deublein and co-workers (de) adjusted \ce{NaCl} Lennard-Jones parameters to reproduce experimental density at a range of concentrations \cite{Deublein2012won}.
Finally, Horinek et al. developed multiple force fields, by taking Dang's \ce{Cl-} ion parameters and adjusting \ce{Na+} parameters to match the solvation free energy of the ion pair \cite{Horinek2009wwm}.
Since this approach does not lead to a unique pair of $\epsilon$ and $\sigma$, the authors proposed models based on small $\epsilon$ (we could not converge this model in 1D-RISM), large $\epsilon$ ($\mathrm{ho_a}$), and medium $\epsilon$ values ($\mathrm{ho_b}$).

\begin{table}
	\caption{Accuracies of different models for predicting Setschenow's constant. The units are \si{l/mol}.}
	\centering
	\begin{tabular}{lrrrr}
		\toprule
		Model &   RMSE &    SDE &   bias &     $r^2$ \\
		\midrule
		\multicolumn{5}{c}{OPLS/CM5}\\
		da &  0.028 &  0.028 &  0.005 &  0.840\\
		$\mathrm{ho_b}$ & 0.036 &  0.029 & -0.021 &  0.823 \\
		jc &  0.051 &  0.034 &  0.038 &  0.789  \\
		$\mathrm{ho_a}$ & 0.058 &  0.029 & -0.050 &  0.818 \\
		de & 0.085 &  0.037 & -0.076 &  0.713\\
		\midrule
		\multicolumn{5}{c}{GAFF/AM1-BCC}\\
     da &  0.032 &  0.032 & -0.004 &  0.800 \\
     jc &  0.043 &  0.034 &  0.025 &  0.772  \\
     $\mathrm{ho_b}$ &  0.050 &  0.035 & -0.035 &  0.750 \\
     $\mathrm{ho_a}$ &  0.078 &  0.042 & -0.066 &  0.650 \\
     de &  0.119 &  0.062 & -0.102 &  0.300 \\     
		\midrule
		\multicolumn{5}{c}{Other models}\\
		pp-LFER &  0.028 &  0.028 & -0.002 &  0.844 \\
		SEA \textsuperscript{\emph{a}} &  0.051 &  0.035 &  0.037 &  0.706 \\
		MD/TIP3P \textsuperscript{\emph{a}}&  0.120 &  0.029 &  0.116 &  0.848 \\
		COSMO-RS &  0.315 &  0.114 &  0.293 &  0.670 \\
		\bottomrule
	\end{tabular}
	
	\textsuperscript{\emph{a}} The accuracy of the model was evaluated on a different dataset.
	\label{tab:results} 
\end{table}

Table \ref{tab:results} compares accuracies of Setschenow's constant predictions by different salt models.
The results by the polyparameter linear free energy relationship (pp-LFER), semi-explicit assembly (SEA), molecular dynamics simulations with TIP3P water and Joung-Cheetham ions (MD/TIP3P), and COSMO-RS \cite{Klamt2005whr} are taken from previous studies \cite{Li2014tbv,Endo2012vqa}.
In the literature, one can also find a few more chemoinformatics methods for Setschenow's constant prediction based on other descriptors or various machine learning methods \cite{Ni2003uit,Li2004wkp,Xu2011vkx,Yu2013taq}, but the accuracy of these models did not significantly exceed the accuracy of the pp-LFER approach.
We found that predictions made with Dang's salt force field (da), combined with OPLS/CM5 force field for solutes, had the best agreement with experimental data among the studied 3D-RISM models, both in terms of its accuracy and the correlation.
RISM calculations based on other salt models had similarly low random error, but larger biases.

In 3D-RISM/PC+ calculations we used two different sets of force fields to describe solutes: OPLS/CM5 and GAFF/AM1-BCC.
Only the salt model by Joung-Cheetham (jc) showed better results when paired with GAFF/AM1-BCC solutes.
In all other cases the use of OPLS/CM5 parameters improved predictions by various extents.

The results from PC+ with Dang's salt model are similar to the pp-LFER model that was fit on Endo's dataset using six adjustable descriptors.
It outperforms both SEA and COSMO-RS models, which are both partially based on the idea that summing surface elements of a solute is a useful strategy for predicting solvation free energies.
3D-RISM, on the other hand, takes into account correlations between densities of solvent at the surface of solute, which most likely contributes to its better accuracy.
Notice that accuracies of both SEA and MD/TIP3P models are evaluated on a different dataset for which 3D-RISM with Dang's \ce{NaCl} force field has RMSE = \SI{0.038}{l/mol}, SDE = \SI{0.029}{l/mol}, bias = \SI{-0.025}{l/mol}, and $r^2=0.798$.
The slight decrease of accuracy is likely explained by the use of less reliable experimental data.

It is useful to note that we defined PC and PC+ models using equations \ref{eq:pressure_cor} and \ref{eq:pc_plus}.
Then, for solvents consisting of multiple different species $\Delta G_{PC} \neq kT/2 \braket{\vect{\rho}}{\ln\vect{g}} + 1/2 \braket{\vect{\rho}}{\vect{u}}$.
Instead using equation \ref{eq:3d_rism_multi_decomp} we have
\begin{equation}
\begin{split}
\Delta \Omega_{PC} & = \frac{kT}{2}\braket{\vect{\rho}}{\ln \vect{g} + \beta \vect{u}} - kT\sum_{i=1}^{N}\rho_i G_i\left(1 - \frac{1}{2} \sum_{j=1}^{N}\rho_j \f C_{ij}(0)\right) \\
& - kT \Delta V \sum_{i=1}^{N}\rho_i \left(1 - \frac{1}{2} \sum_{j=1}^{N}\rho_j \f C_{ij}(0)\right),
\end{split}
\end{equation}
which cannot be simplified a lot further due to $G_i \neq  - \Delta V$.
We can define the PC+ prime model as:
\begin{equation}
\Delta \Omega_{PC+'} = \frac{kT}{2}\braket{\vect{\rho}}{\ln \vect{g} + \beta \vect{u}} - kT \braket{\Delta \vect{\rho}}{\vect{1}}.
\label{eq:pc_plus_prime}
\end{equation}
However, this definition of PC+ leads to worse results (see table \ref{tab:results2}).

Molecular dynamics based predictions of Setschenow's constant, despite achieving impressive correlation with experimental data ($r^2=0.848$), have a large positive bias.
We believe that the origin of this bias is likely related to the chosen salt model (jc) and force fields: GAFF with TIP3P water.
It is likely that a combination of Dang salt model and SPC/E water would reduce the bias in the prediction and make molecular dynamics simulations one of the most accurate ways of predicting Setschenow's constant, although, quite time consuming.

\begin{figure}
	\centering
	\includegraphics[width=1\linewidth]{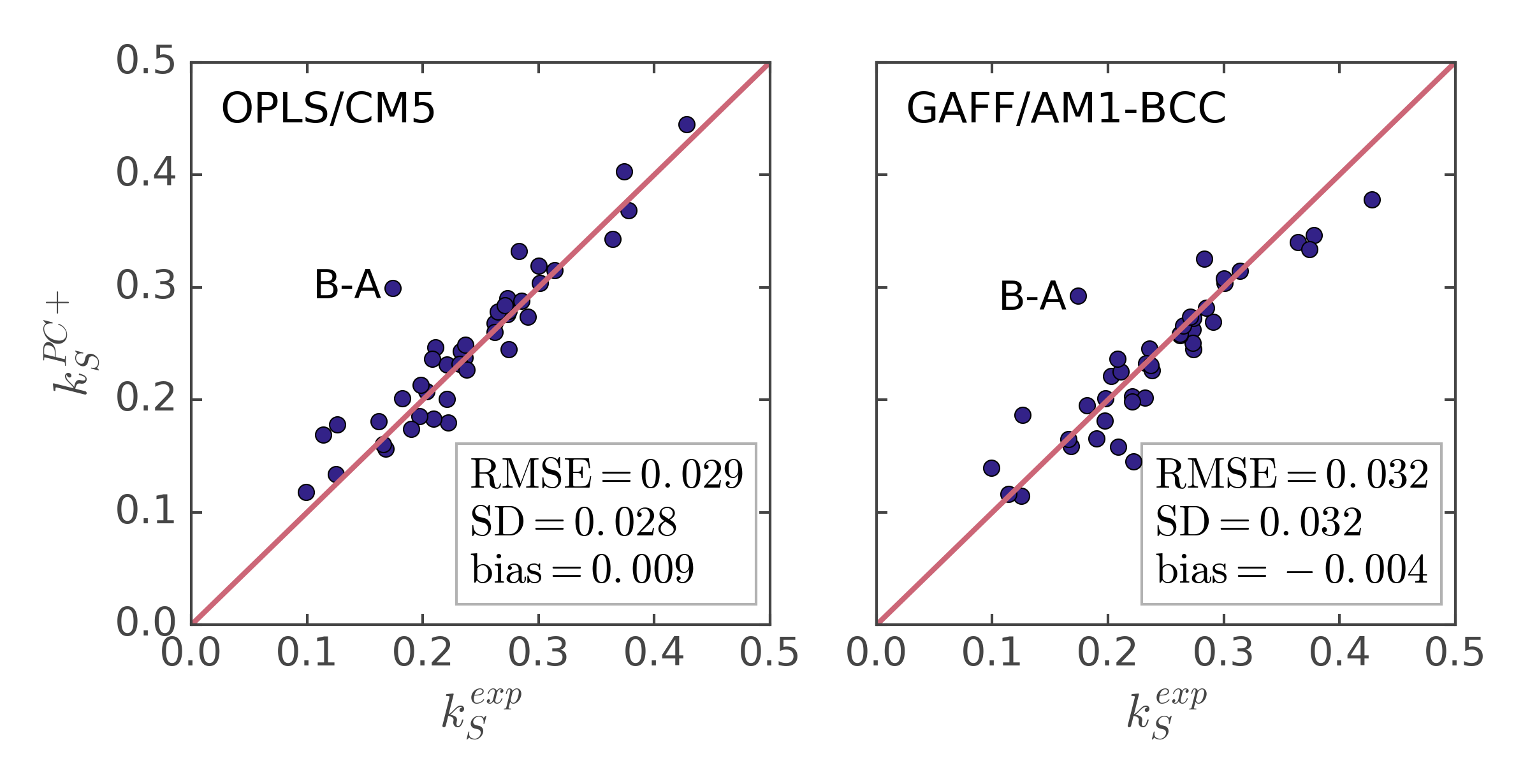}
	\caption{Setschenow's constants by 3D-RISM/PC+ with Dang salt model compared to experimental measurements by Endo et al. Results on the left and right figures are obtained with OPLS/CM5 and GAFF/AM1-BCC force fields respectively. B-A stands for bisphenol A. The values are in \si{l/mol}.}
	\label{fig:dang_model}
\end{figure}

\begin{figure}
\centering
\includegraphics[width=0.6\linewidth]{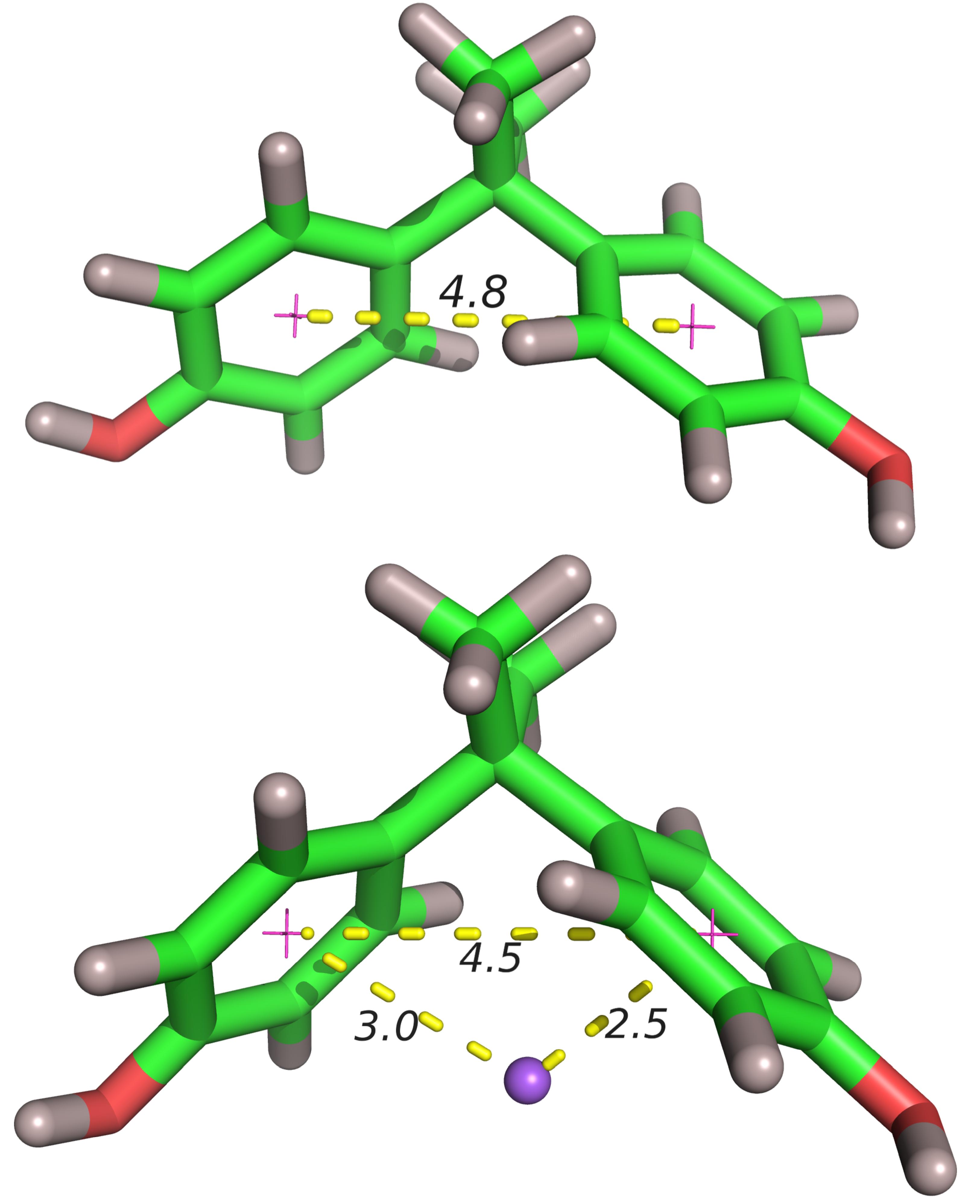}
\caption{The optimised geometries of bisphenol A with and without \ce{Na+} ion. The distances, shown in Angstroms, are measured between the centres of benzene rings and the ion. SMD model was used to take into account solvent effects.}
\label{fig:bisphenols}
\end{figure}

Figure \ref{fig:dang_model} has a comparison between Setschenow's constants predicted by 3D-RISM with Dang's \ce{NaCl} model and those from experimental measurements.
For both calculations made with OPLS/CM5 and GAFF/AM1-BCC parameters, one major outlier is bisphenol A (B-A).
In both cases 3D-RISM calculations overestimate its $k_S$ by \SI{0.13}{l/mol}: more than four times greater than the average prediction error for Dang's model.
This molecule was also an outlier in 3D-RISM calculations with other salt models.
While potentially this might be the result of measurement error, we believe that the reason for this lies in the fact that bisphenol A binds relatively strongly to sodium ions via $\pi$-cation interactions.
We performed electronic density functional theory calculations to test this hypothesis.
The optimization was done using the same level of theory and software as for the initial molecule preparation.
Optimised geometries for bisphenol A with and without sodium atoms, shown in figure \ref{fig:bisphenols}, indicate significant structural rearrangement as well as considerable bonding between \ce{Na+} and both phenol rings.
These type of interactions are difficult to characterize using conventional force fields \cite{Archambault2009uxq}, and would require a quantum mechanics approach to dispersion interactions.
Additionally, conventional 3D-RISM operates with rigid solutes and does not capture salt-induced changes in solute conformation.

\section{Non-aqueous solvents and the corresponding state principle}

3D-RISM is generally poorly suited for a description of solvents with a large number of sites.
As solvent molecules gets larger, it gets harder to generate susceptibility files, as well as the speed and convergence of 3D-RISM calculations become slower.
Additionally, since intramolecular correlations are also approximated in 3D-RISM, as the number of sites increases, the description of bonding within solute becomes poorer, which also negatively impacts 3D-RISM performance.

A solution to this problem is to coarse-grain organic solvents.
In principle, there a number of conventional schemes available, but they all tend to be quite slow.
However, when reviewing the literature we discovered a scheme based on the corresponding state principle that lets one predict coarse-grained interaction parameters in a straightforward manner.

According to the corresponding states principle, reduced critical temperature
\begin{equation}
T^{\ast}_{c} = \frac{kT_c}{\epsilon}
\label{eq:eps}
\end{equation}
and reduced critical density
\begin{equation}
\rho^{\ast}_{c} = \rho_c \sigma^3
\label{eq:sigma}
\end{equation}
are constants for all classical fluids with orientation-independent interaction potentials \cite{Leland1968uns,Guggenheim1945www}.
Here $k$ is the Boltzmann constant, $T_c$ and $\rho_c$ are critical temperature and density, $\sigma$ is effective particle diameter and $\epsilon$ is a constant that determines the strength of intramolecular interactions.
This principle can be further extended to non-spherical molecules by assuming $T^{\ast}_c$ and $\rho^{\ast}_c$ are functions of molecular shape and electrostatic properties \cite{Leland1968uns}.

It follows that knowing $T^{\ast}_c$ and $\rho^{\ast}_c$ for a single reference fluid, one can easily obtain intermolecular interaction parameters $\epsilon$ and $\sigma$ for many others from their critical properties.
This idea has been used by a number of authors to construct coarse-grained models of real fluids and estimate their properties at a wide range of conditions \cite{Hirschfelder1954wzx,Flynn1962vix,Chung1984wuq,Van_loef1984vbh,Cuadros1996trq,Ben-amotz2002wih,Zhu2002ttt,Galliero2005wyu,Galliero2007uqb,Galliero2009vfd,Mejia2014vab}.
Most of them came to the conclusion that with the exception of a few simple fluids such as argon, nitrogen, or methane, the majority of the real fluids cannot be adequately described by just two simple parameters and require either additional fittings or more complicated interaction potentials.

\begin{figure}
	\centering
	\includegraphics[width=0.6\linewidth]{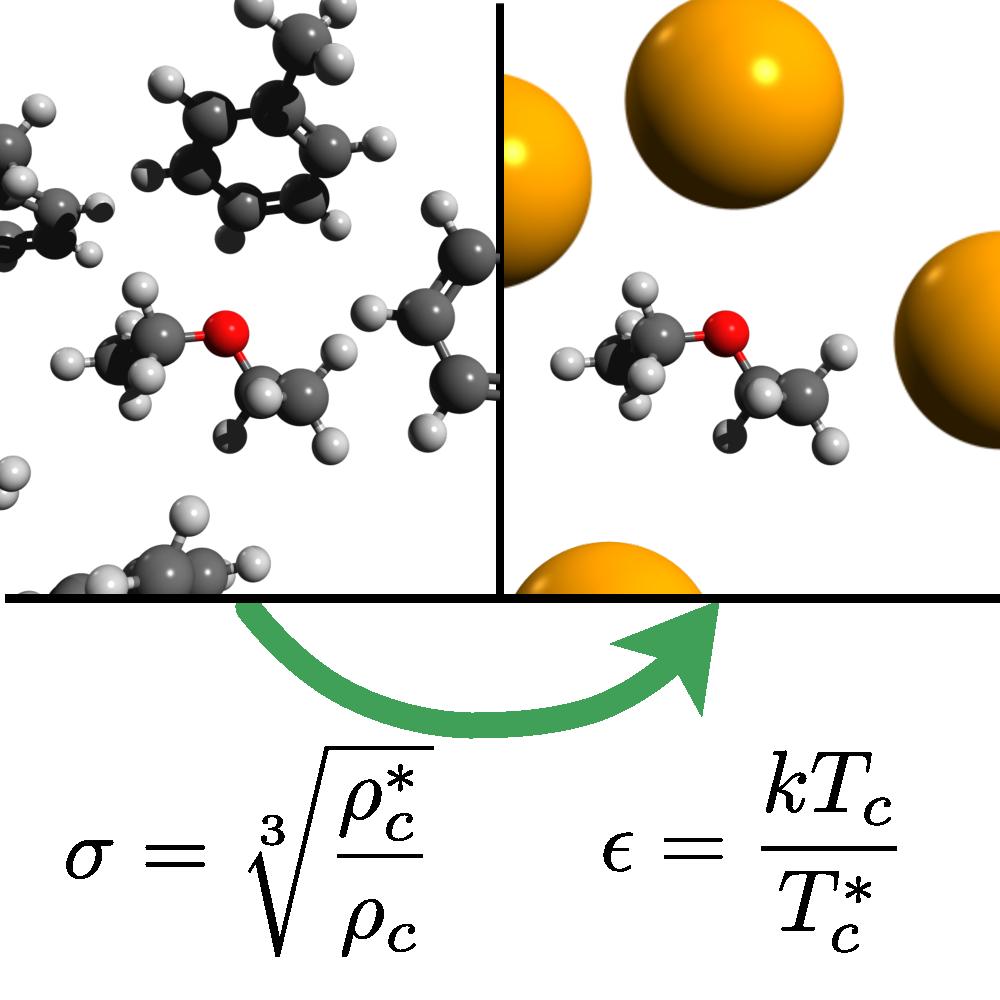}
	\caption{The basic idea behind coarse graining used for 3D-RISM calculations. The real liquids are approximated as spheres interacting via Lennard-Jones potentials with parameters deduced using equations shown on the figure above.}
	\label{fig:coarse_graining}
\end{figure}

However, a precise description of solvent behaviour and phase diagram is frequently not necessary for an accurate estimation of solvation free energy, as can be seen by a number of successful implicit solvation models \cite{Marenich2009tbw,Klamt2005whr}.
As we demonstrated in section \ref{sec:solv_thermodynamics}, solvation free energy is independent of solvent reorganization energy, so it most likely can be estimated using a rather simple coarse-grained model.
Thus, to construct an approximation for a number of organic solutes we simply used the reduced critical parameters of Lennard-Jones fluid: $T^{\ast}_{c,LJ} = 1.313$, $\rho^{\ast}_{c,LJ} = 0.304$ that were obtained by Okumura et al. using molecular dynamics \cite{Okumura2000uqx}.

Since we were mainly interested in predicting solvation free energies, we decided to focus on popular solvents for which a large amount of data are available.
The Minnesota solvation database contains a large collection of measurements made at standard conditions. 
From it we picked 17 non-associating solvents, listed in the table \ref{table:nb}.
The table lists critical properties liquids, taken from Ref. \citenum{Yaws2014wbx}, as well as parameters of Lennnard-Jones spheres approximating these liquids (obtained via equations \ref{eq:sigma} and \ref{eq:eps}).
Note that the solvent xylene is a mixture of isomeric ortho-, meta- and para-forms of xylene.

\begin{table}
	\centering
	\caption{Critical properties and Lennard-Jones parameters.}
	\begin{tabular}{llrrr}
		\toprule
		Name &  $T_{c}$ [\si{K}] &  $\rho_{c}$ [\si{nm^{-3}}] &  $\epsilon$ [\si{kcal/mol}] &  $\sigma$ [\si{nm}] \\
		\midrule
		1,2-dichloroethane &       561.60 &                     2.74 &                      0.85 &              0.48 \\
		acetonitrile &       545.00 &                     4.11 &                      0.82 &              0.42 \\
		benzene &       562.05 &                     2.35 &                      0.85 &              0.51 \\
		bromobenzene &       670.15 &                     1.86 &                      1.01 &              0.55 \\
		carbon disulfide &       552.00 &                     3.76 &                      0.84 &              0.43 \\
		carbon tetrachloride &       556.35 &                     2.18 &                      0.84 &              0.52 \\
		chloroform &       536.40 &                     2.52 &                      0.81 &              0.49 \\
		cyclohexane &       553.80 &                     1.96 &                      0.84 &              0.54 \\
		diethyl ether &       466.70 &                     2.15 &                      0.71 &              0.52 \\
		dimethyl sulfoxide &       729.00 &                     2.65 &                      1.10 &              0.49 \\
		ethyl acetate &       523.30 &                     2.11 &                      0.79 &              0.52 \\
		isooctane &       543.80 &                     1.29 &                      0.82 &              0.62 \\
		isooctane (2-mer) &           &                       &                      0.61 &              0.49 \\
		n-decane &       617.70 &                     1.07 &                      0.93 &              0.66 \\
		n-decane (4-mer) &           &                       &                      0.54 &              0.39 \\
		n-heptane &       540.20 &                     1.41 &                      0.82 &              0.60 \\
		n-heptane (3-mer) &           &                       &                      0.52 &              0.40 \\
		octanol &       652.50 &                     1.21 &                      0.99 &              0.63 \\
		toluene &       591.75 &                     1.91 &                      0.90 &              0.54 \\
		xylenes &       624.57 &                     1.59 &                      0.95 &              0.58 \\
		\bottomrule
	\end{tabular}
	\label{table:nb}
\end{table}

Of course, a spherical Lennard-Jones fluid is a poor reference system for most of these solvents.
For this reason, isooctane, heptane, and decane were also modelled as chains of Lennard-Jones spheres composed of $m$ segments, each separated by a bond of length $\sigma$.
The choice of $m$ was motivated by an equation employed in Statistical Associating Fluid Theory (SAFT) and in some molecular dynamics studies \cite{Lafitte2006tnk,Galliero2009vfd,Galliero2014wms}
\begin{equation}
m = 1 + \frac{n(C) - 1}{3},
\label{eq:beads}
\end{equation}
where $n(C)$ is the number of carbons in the linear alkane.
$m=3$ for heptane and $m=4$ for decane follow directly from the equation.
We also assumed that $m=2$ would be a reasonable choice for isooctane.
The $\sigma$ and $\epsilon$ parameters for chain beads were similarly obtained using equations \ref{eq:eps} and \ref{eq:sigma},
but using critical points for the 2-mer ($T^{\ast}_{c,LJC2} = 1.78$, $\rho^{\ast}_{c,LJC2} = 0.149$, Ref. \citenum{Dubey1993uxj}), 3-mer ($T^{\ast}_{c,LJC3} = 2.063$, $\rho^{\ast}_{c,LJC3} = 0.088$, Ref. \citenum{Blas2001vhv}) and 4-mer ($T^{\ast}_{c,LJC4} = 2.26$, $\rho^{\ast}_{c,LJC4} = 0.0625$,  Ref. \citenum{Escobedo1996wyh}) Lennard-Jones chain fluids.

\section{Solvation of model solutes in Lennard-Jones fluids}
\label{sec:solvation_lj_in_lj}

Now that we have defined the solvents, we can start predicting solvation free energies of various solutes in them.
However, before doing it, we first test the applicability of pressure corrections to these liquids by comparing results of HNC and PC+ approximations to molecular dynamics.

\begin{figure}
\centering
\includegraphics[width=0.7\linewidth]{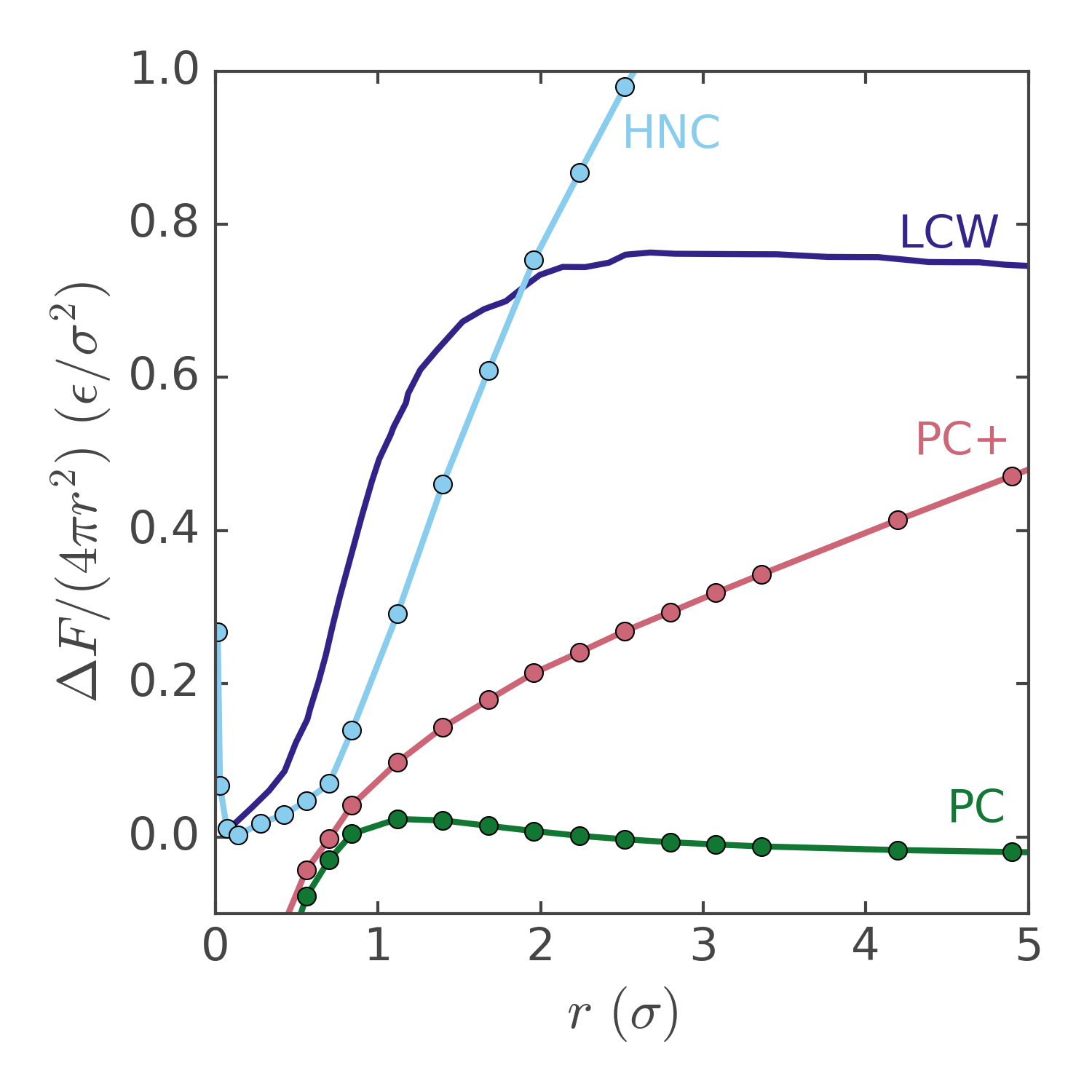}
\caption{Solvation free energy per unit area for solutes of radii $r$ in the Lennard-Jones liquid with $T^{\ast}=0.701$ and $\rho^{\ast}=0.843$. LCW results are taken from Ref. \citenum{Huang2001wdw}.}
\label{fig:hard_sphere_in_lj_fe}
\end{figure}

The figure \ref{fig:hard_sphere_in_lj_fe} compares predictions of hard sphere solvation free energies in a Lennard-Jones fluid at reduced temperature $T^{\ast} = kT/\epsilon = 0.701$ and reduced density $\rho^{\ast}=\rho \sigma^3 = 0.005$.
The 3D-RISM results are evaluated against estimations from the Lum-Chandler-Weeks (LCW) model that is known to be quite accurate and agrees well with MD predictions.
The trends in errors of 3D-RISM models are similar to those observed for water (figures  \ref{fig:hard_tension} and \ref{fig:hard_tension_pc}), however, now PC+ model underestimates the insertion free energy of the solute, while HNC gives relatively good predictions up to the point where the interface forms.
It seems that the compressibility-based estimate of insertion free energy that is employed in HNC works well for Lennard-Jones fluids.

\begin{figure}
\centering
\includegraphics[width=1\linewidth]{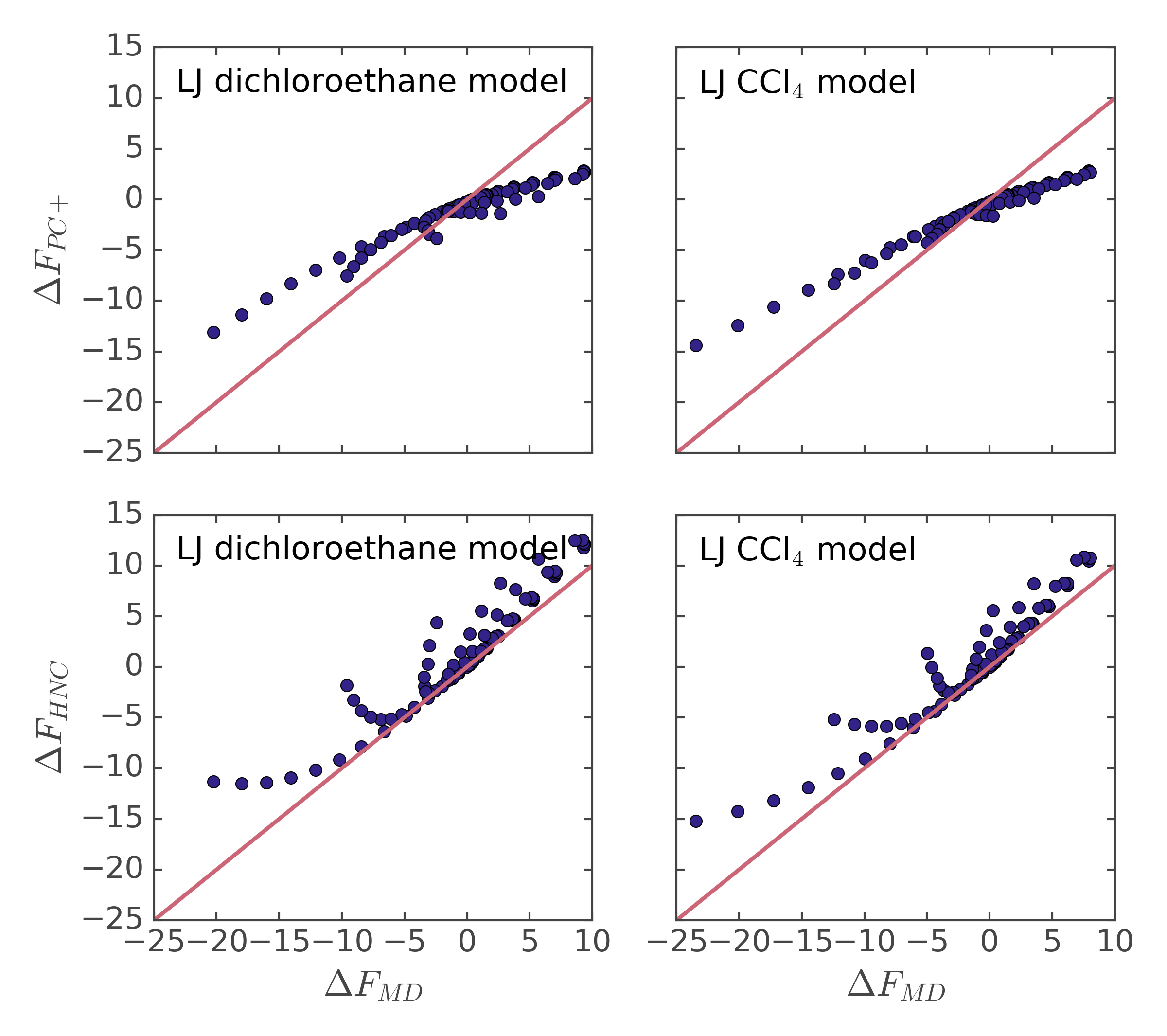}
\caption{Calculated Lennard-Jones solutes solvation free energies against MD data. Dichloroethane and \ce{CCl4} models are defined in table \ref{table:nb}. All values are in \si{kcal\per\mole}.}
\label{fig:lj_in_lj_fe}
\end{figure}

We also wanted to check how well can 3D-RISM predict solvation free energies of Lennard-Jones solutes that offer a somewhat more accurate representation of typical molecules.
We performed a number of molecular dynamics free energy simulations in model dichloroethane and tetrachloromethane.
In total we used 90 different solutes that had all possible combinations of $\sigma = 0.6$, $1.4$, $2.2$, $\cdots$, \SI{7.0}{\angstrom}, and $\epsilon = 2^{-7}$, $2^{-6}$, $2^{-5}$, $\cdots$, $2^2$ \si{kcal\per\mol}.
The simulations were performed at NVT conditions to make sure that the density of Lennard-Jones spheres matched the density of the real liquids at \SI{298}{K}.
Note that all molecules in the system are uncharged; we are primarily comparing the accuracy of insertion free energy estimates.

Comparison of MD solvation free energies with both PC+ and HNC predictions is shown in figure \ref{fig:lj_in_lj_fe}.
When compared to water, the agreement of PC+ with MD was poorer, while agreement of HNC with molecular dynamics was significantly better.
Interestingly, while absolute values were predicted slightly more accurately by HNC, the trends were captured more faithfully by PC+.

These findings demonstrate that 3D-RISM, when applied to Lennard-Jones fluids, cannot predict interface formation.
Similarly to what happens in the case of water, the results between two models agree only up to a certain solute size.
Moreover, the PC+ correction seems to work significantly worse and does not approximate the insertion free energy too well.

\section{Comparison with experimental values}
\label{sec:nonaq_exp}

As we discussed previously, we also evaluated the accuracy of our solvent approximation using experimental data from Minnesota solvation database \cite{Marenich2012vuj,Marenich2009tbw}.
We selected data for 17 popular solvents, presented in table \ref{tab:parameters}.
Chosen solvents have 1247 associated experimental measurements for 482 unique solute molecules.

The solvent susceptibility functions were generated using 1D-RISM, with a bulk density of model solvents set to experimental number densities at \SI{298}{K} \cite{Yaws2014wbx}.
The solute geometries were obtained from the Minnesota database; non-bonding parameters were taken from the OPLS-2005 force field.
All partial charges were set to zero.

\begin{figure}
\centering
\includegraphics[width=1\linewidth]{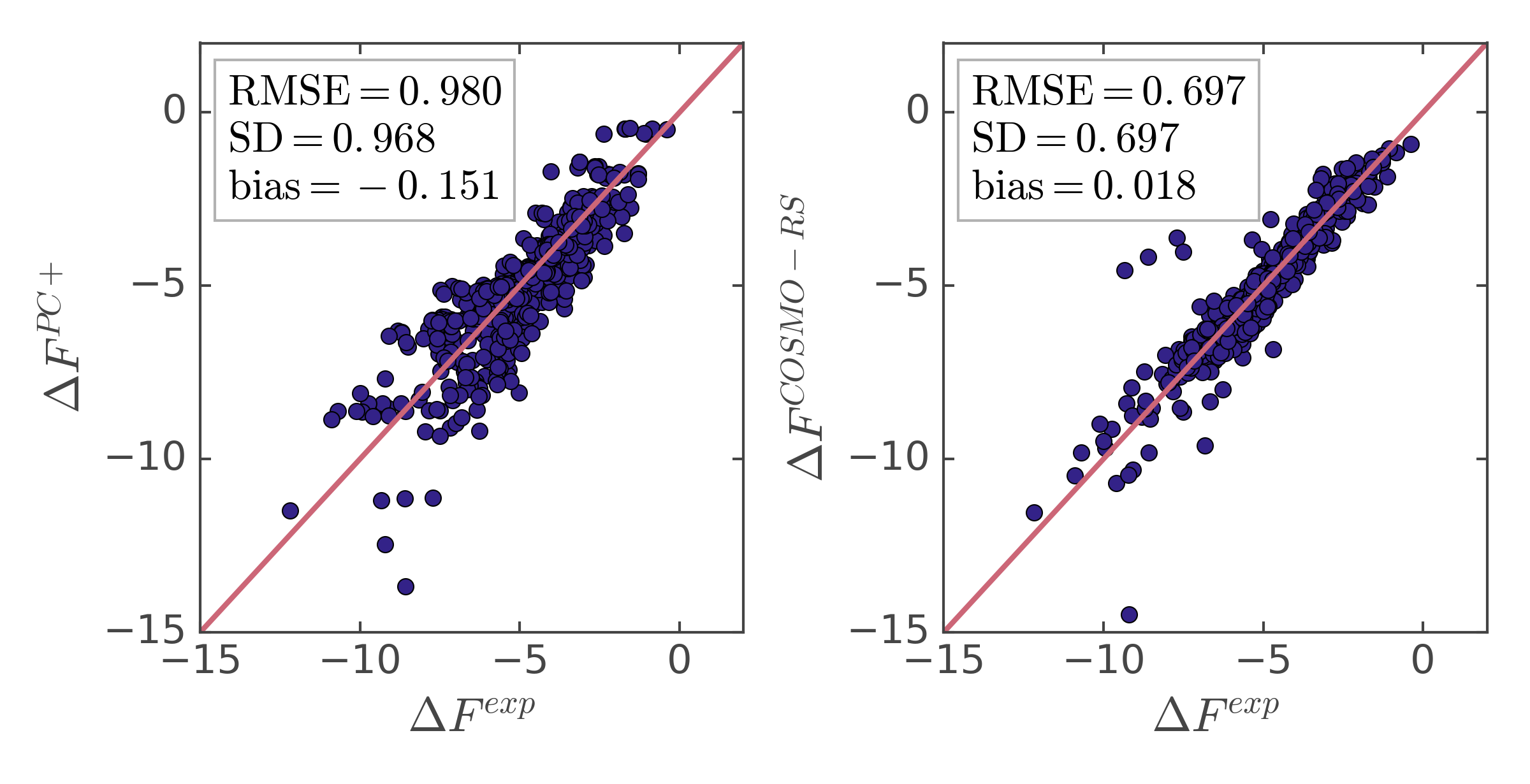}
\caption{Computed solvation free energies of a number of compounds in apolar solvents (as defined in table \ref{table:results}) against experimental data. COSMO-RS results are taken from Ref. \citenum{Klamt2015uqj}. All values are in \si{kcal\per\mole}.}
\label{fig:nonpolar_pc_vs_cosmo}
\end{figure}

The figure \ref{fig:nonpolar_pc_vs_cosmo} demonstrates the accuracy of PC+ predictions for apolar solvents, with table \ref{tab:results} showing a more detailed breakdown.
For comparison, we also plotted COSMO-RS predictions made for the same set of solvents.
You can see that while PC+ results are poorer than those made by more advanced model, they are still within acceptable \SI{1}{kcal/mol} range.

\begin{figure}
\centering
\includegraphics[width=1\linewidth]{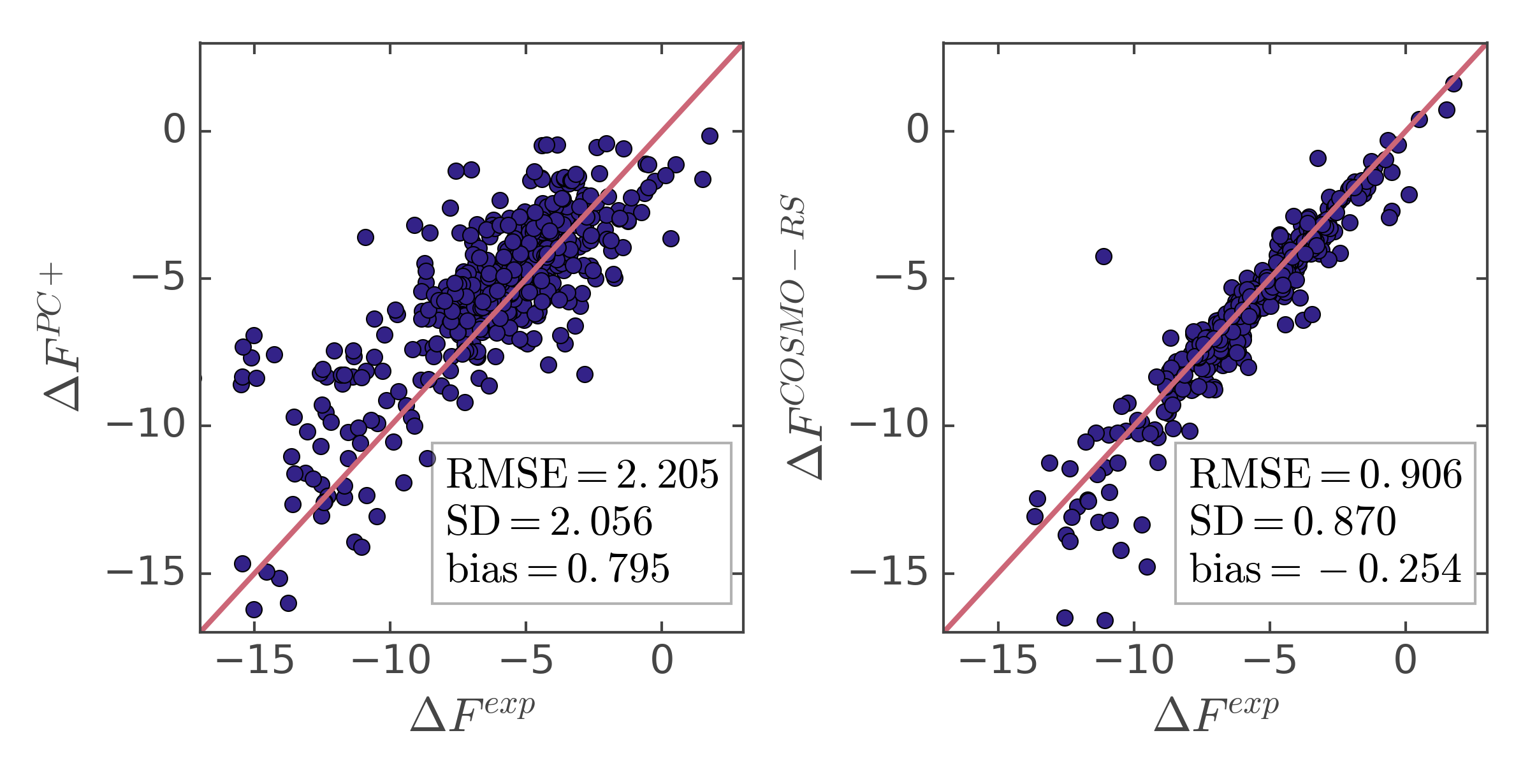}
\caption{Computed solvation free energies of a number of compounds in polar solvents (as defined in table \ref{table:results}) against experimental data. COSMO-RS results are taken from Ref. \citenum{Klamt2015uqj}. All values are in \si{kcal\per\mole}.}
\label{fig:polar_pc_vs_cosmo}
\end{figure}

\begin{table}
	\centering
	\begin{tabular}{lrrrr}
		\toprule
		Solvent &    N &  RMSE &   SDE &  bias \\
		\midrule
		\multicolumn{5}{c}{Apolar} \\
		1,2-dichloroethane &   39 &  1.16 &  1.07 &  0.47 \\
		benzene &   71 &  1.28 &  1.28 &  0.04 \\
		bromobenzene &   27 &  1.17 &  1.15 & -0.23 \\
		carbon disulfide &   15 &  0.94 &  0.89 & -0.30 \\
		carbon tetrachloride &   79 &  0.85 &  0.84 & -0.11 \\
		cyclohexane &  103 &  1.01 &  0.75 & -0.67 \\
		isooctane &   32 &  0.98 &  0.68 & -0.70 \\
		isooctane (2-mer) &   32 &  0.63 &  0.60 & -0.21 \\
		n-decane &   39 &  1.70 &  1.23 & -1.17 \\
		n-decane (4-mer) &   39 &  0.68 &  0.56 & -0.38 \\
		n-heptane &   67 &  0.95 &  0.86 & -0.42 \\
		n-heptane (3-mer) &   67 &  0.74 &  0.74 &  0.05 \\
		olive oil &  218 &  1.30 &  1.06 & -0.75 \\
		toluene &   51 &  1.00 &  0.99 &  0.08 \\
		xylenes &   48 &  1.00 &  0.99 & -0.10 \\
		\multicolumn{5}{c}{Polar} \\
		acetonitrile &    7 &  2.23 &  2.13 &  0.67 \\
		chloroform &  107 &  1.86 &  1.37 &  1.25 \\
		diethyl ether &   70 &  2.21 &  1.65 &  1.47 \\
		dimethyl sulfoxide &    7 &  2.65 &  2.65 &  0.01 \\
		ethyl acetate &   22 &  3.02 &  2.18 &  2.09 \\
		octanol &  245 &  2.24 &  2.22 &  0.31 \\
		\bottomrule
	\end{tabular}
	\caption{Accuracies of solvation free energy predictions by 3D-RISM/PC+ for various solvents. RMSE stands for root mean square error, SDE is standard deviation of error. Energies are in kcal/mol.}
	\label{table:results}
\end{table}

The same approach is significantly less successful for polar solvents, which often interact with the solutes via strong specific interactions.
As figure \ref{fig:polar_pc_vs_cosmo} demonstrates, the accuracy of PC+ is almost two times worse when compared to the predictions in polar solutes.
This decrease of accuracy is not surprising since our coarse-grained models of solvents lack electrostatic charges.
Still, the existence of any correlations between PC+ and experimental values suggests that this approximation still allows us to roughly estimate solvation free energies in these solvents.

In view of the results in section \ref{sec:solvation_lj_in_lj}, the overall accuracy and correlation of PC+ predictions with experimental data seems surprising.
The HNC model, which showed better agreement with MD values of solvation free energy for Lennard-Jones fluids, actually correlated worse with experiment (figure \ref{fig:polar_nonpolar_hnc}).
The PC+ solvation free energy estimates made with coarse-grained solvents agree better with experimental values than predictions made with the same solvent and molecular dynamics.
In other words, making an approximation in solvation free energy improves the result!
It is not entirely clear why exactly this occurs; it is likely that a simplistic, mean-field estimate of solvent parameters made with the corresponding state principle, works best when combined with a simple, linear-response like free energies, given by pressure-corrected models.

%
%
%
%
%
%
%
%
%

\addtocontents{toc}{\protect\addvspace{2.25em plus 1pt}}
\bookmarksetup{startatroot}

\chapter{Conclusion} 

\label{Chapter8}

\lhead{Chapter 8. \emph{Conclusion}}

The main goal of this work was to develop a 3D-RISM-based advanced pressure correction model, PC+, and to investigate its scope of application and accuracy.
The key findings of the thesis can be summarised as follows:
\begin{itemize}
	\item From the theoretical point of view, the PC+ model is based on a linear response approximation to solvation free energy, combined with an estimate of cavity creation work. It was shown that the PC+ model could provide accurate predictions of the solvation free energy as long as both of these approximations hold. Chapter \ref{Chapter5}.
	\item For a pure aqueous solvent,  the approach turns out to be quite useful; it predicts hydration free energies with an accuracy of about \SI{1.3}{kcal\per\mole} and \SI{3.0}{kcal\per\mole} for small neutral and charged molecules correspondingly. Chapter \ref{Chapter6}.
	\item The model can also be applied to aqueous \ce{NaCl} solutions.
	It was shown that the model could significantly improve estimates of the Setschenow's constant for molecular compounds compared to commonly used models; with a properly chosen salt representation, pressure-corrected 3D-RISM can achieve a good accuracy in quantitative predictions of the Setschenow's constant. Chapter \ref{chap:beyond_water}.
	\item Finally, we demonstrated that PC+ could also be applied to non-aqueous solvents. A major problem of such systems, a large amount of sites and flexibility, was solved by introducing a consistent coarse-grained approximation. This approach in combination with PC+ led to an accuracy of about \SI{1}{kcal\per\mole} for a range of non-polar solvents. Chapter \ref{chap:beyond_water}.
	\item Notable failures of the model include its relatively poor prediction of the solvation entropies and a failure to accurately estimate the changes in the heat capacity occurring due to the solute insertion. Additionally, the model provides poor performance for polar non-aqueous solvents such as DMSO or methanol. Chapters \ref{Chapter6} and \ref{chap:beyond_water}.
\end{itemize}

The work warrants further investigations.
From the theoretical point of view, it is still not entirely clear why the pressure-correction approach works well for multicomponent mixtures.
We also left for future investigations a  possibility of defining a self-consistent pressure corrected functional and a detailed analysis of corresponding density distributions obtained via it.
Additionally, it is also worthy to test a combination of pressure corrections with advanced free energy functionals such as anisotropic HNC or hydrostatic approximations.
Even without focusing on the theory itself, we believe that extra improvements of the model can be achieved by using more sophisticated solute-solvent potentials, as well as testing more optimal models of solvents, designed specifically with 3D-RISM/PC+ in mind.


\addtocontents{toc}{\vspace{2em}} 

\appendix 



\chapter{Methodology}
\label{AppendixA} 

\lhead{Appendix A. \emph{Methodology}} 

Throughout the thesis, the majority of the calculations were performed using the same software and settings.
For this reason, we decided to summarize the general methodology in this appendix, mentioning specifics of each calculation in appropriate parts of the main text.

\section{1D-RISM}
\label{sec:methods_1drism}

In the thesis, 1D-RISM calculations were primarily used to generate the susceptibility functions for 3D-RISM calculations.
The calculation inputs are molecule geometry, its Lennard-Jones parameters, and partial charges, as well as bulk solvent density and dielectric constant, which we obtained from the experimental data.

The majority of the actual calculations were performed with the \texttt{rism1d} program \cite{Kovalenko2000vzf,Hirata2003tpg,Luchko2010uhh} included in the AmberTools 15 package \cite{Case2012tff}. 
Calculations performed with AmberTools 14 or 16 versions only differed in the additional output, while the susceptibility functions remained the same.
The 1D-RISM equations were solved with a tolerance set to $1 \times 10^{-12} $ and grid spacing to $0.025$ \AA.
Note that whenever solvent had partial charges, and thus an associated dielectric constant, we used the dielectrically consistent formulation of 1D-RISM (DRISM).

Most commonly we used either HNC or PSE-3 closures, which gave practically identical results.
For solvents other than pure water it was often impossible to converge susceptibility functions starting with these closures.
Typically, we obtained an initial solution using the KH closure, then tried to converge using the PSE-2 closure, using the KH susceptibility function as the initial guess.
Only after obtaining PSE-2 solution did we move to perform PSE-3 or HNC calculations.
Quite often, we also had to adjust parameters such as force field, temperature or density to obtain a good initial guess.
Weaker interacting systems provided reasonable starting guesses for further calculations.

\section{3D-RISM}

The input for 3D-RISM calculations includes external field specification (typically in the form of force field potential of a solute molecule) as well as bulk solvent susceptibility functions.

Most of the 3D-RISM calculation in the thesis were performed using \texttt{rism3d.snglpnt} program from AmberTools 15 package.
Similarly to 1D-RISM, part of the calculations were done using different versions; however, it did not affect the results.
For the majority of the calculations we used a grid spacing set to \SI{0.5}{\angstrom}, buffer to \SI{25}{\angstrom}, and tolerance to $1 \times 10^{-5}$.
Calculations performed in the early stages of the thesis used finer grids with a \SI{0.3}{\angstrom} spacing, \SI{30}{\angstrom} buffer and \num{1e-10} tolerance.
However, we found that solvation free energy was largely unaffected by the change in grid settings and the additional precision obtained with the finer grid was offset by a significantly larger computational time (often about $10$ longer CPU times).

\begin{figure}
	\centering
	\includegraphics[width=1\linewidth]{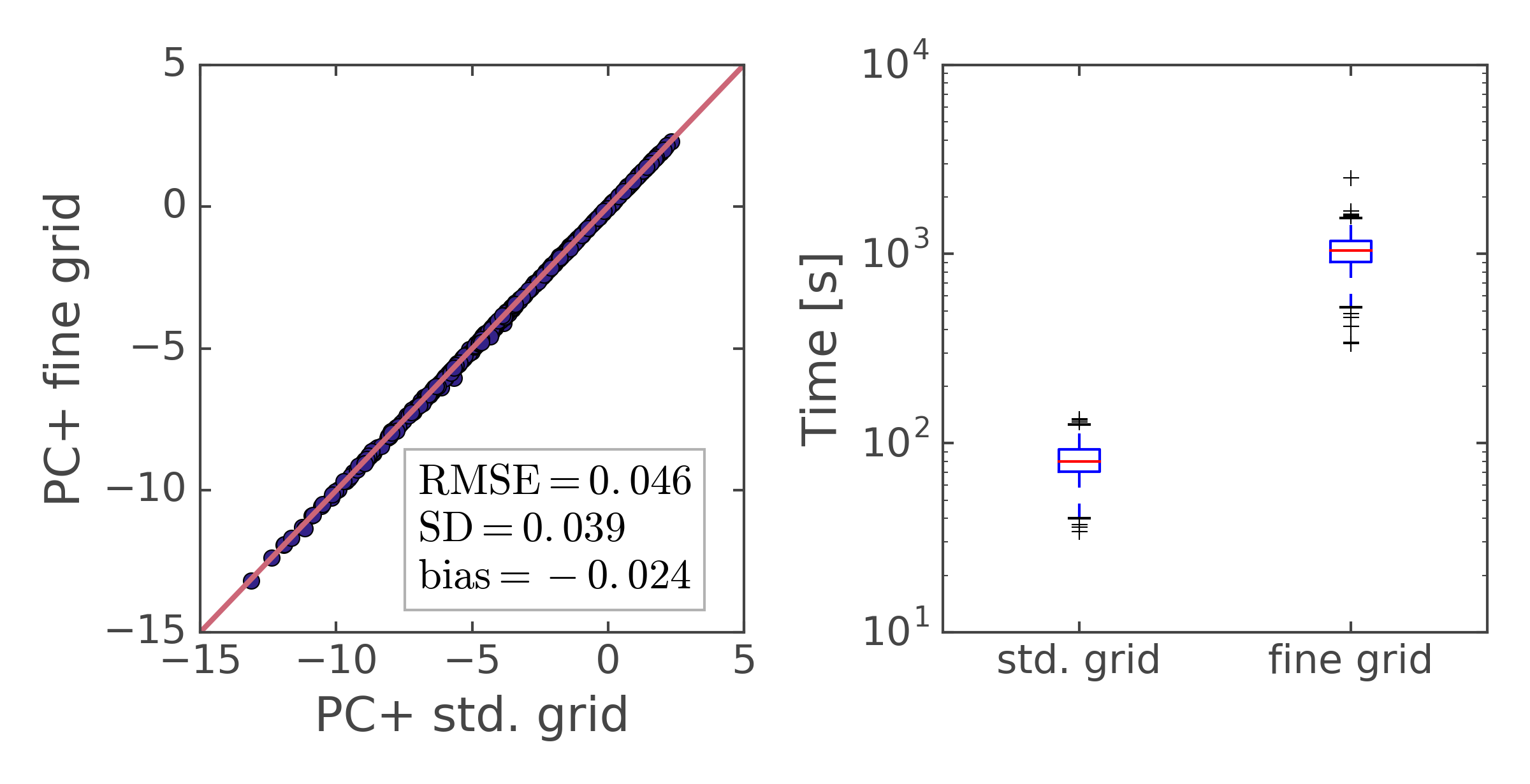}
	\caption{We performed two sets of calculations on the Mobley dataset (described at the beginning of section \ref{sec:neutral_exp}) using standard and fine grid settings. The comparison of free energies (\si{kcal\per\mole}) is shown on the left, and in run time on the right figure.}
	\label{fig:finegridstdgrid}
\end{figure}

To simplify the calculation of pressure corrections as well as solvation free energy workflow we created a small script, hosted on \url{https://github.com/MTS-Strathclyde/PC_plus}.

The majority of the 3D-RISM calculations were performed with the PSE-3 closure.
The solvation free energies were almost unaffected by the choice of closure, with differences between HNC and PSE-3 often being smaller than the uncertainty due to the grid spacing.
At the same time, PSE-3 calculations converged quicker and more reliably, which led us to prefer this particular closure.

\section{Molecular dynamics}

Molecular dynamics simulations in this thesis were performed using Gromacs 5.04 software \cite{Pronk2013vhm}.
We used a cubic box with periodic boundary conditions.
All bonds with hydrogens were kept rigid using LINCS algorithm of 12-th order.
Dynamics was simulated using the Langevin integrator, with a reference temperature of $298.15$ K and a friction constant of \SI{1.0}{\per\pico\second}.

For short-range interactions, a pair list was generated using a Verlet cut-off scheme.
Lennard-Jones interactions were smoothly switched off between 9 and \SI{12}{\angstrom}. 
The cut-off artifacts were accounted for using long-range pressure and dispersion corrections as implemented in Gromacs.
Electrostatics interactions were treated using particle-mesh Ewald (PME) method \cite{Darden1993wfb} with \SI{12}{\angstrom} real space cutoff, \SI{1.2}{\angstrom} Fourier spacing, 6-th order spline interpolation, and tolerance set to $10^{-6}$.
For uncharged solvents, we used simple cut-off electrostatics.

The solvation free energy was typically computed using $20$ separate calculations at each $\lambda$, decoupling first electrostatics and then Lennard-Jones interactions between solute and solvent.
Intramolecular interactions within a solute were kept the same at all lambda values.
Calculations with modified electrostatics interactions were performed at $\lambda = 0$, $0.25$, $0.5$, $0.75$, $1$.
The decoupling of Lennard-Jones interactions was done using calculations at $\lambda = 0$, $0.05$, $0.1$,  $0.2$,  $0.3$, $0.4$, $0.5$, $0.6$, $0.65$, $0.7$, $0.75$, $0.8$, $0.85$, $0.9$, $0.95$, $1.0$.
This setup has been shown to give good convergence \cite{Klimovich2015wbq}.

Before running MD simulations at each $\lambda$, we performed 5000 steps of steepest descent optimization. After that, we performed \SI{200}{ps} equilibration and \SI{1300}{ps} production runs.
Typically, we used either \SI{1}{fs} or \SI{2}{fs} time steps.
For NPT runs, the pressure was kept constant at 1 bar using Berendsen barostat \cite{Berendsen1984vio}, with time constant set to \SI{1}{ps} and compressibility to \SI{4.5e-5}{\per\bar}.

After completing the simulations, the intrinsic hydration free energy was evaluated using Multistate Bennett Acceptance Ratio (MBAR) \cite{Shirts2008upm}.
The actual calculation was performed using a python script \texttt{alchemical-analysis.py} \cite{Klimovich2015wbq}.

\section{Force fields and geometry}

To generate the initial geometry of molecules we typically used the Openbabel software package \cite{Oboyle2011usb,Oboyle2008txc}.
For some molecules, further refinement of geometry was performed using the quantum chemical package Gaussian 09, Revision D.01 \cite{Frisch2009tfw}.
The calculations were performed with the M06-2X functional \cite{Zhao2007unx}, and MG3S basis set \cite{Lynch2003ult}.
Molecules that were not further optimized using the Gaussian package were simply optimized using assigned force field parameters.

For the majority of the solutes, we assigned either GAFF/AM1-BCC non-bonded parameters or the OPLS\_2005 force field, combined with CM5 charges.
Due to the amount of data we used only software which did the assignment automatically.
The GAFF/AM1-BCC workflow mainly relied on the \texttt{antechamber} program found in the AmberTools package.
The program was used to both assign parameters as well as to evaluate partial charges.
For OPLS force field assignment we used the Maestro package \cite{2014tph}.

Similarly to quantum chemical geometry optimization, CM5 charges were evaluated with  Gaussian 09 software at the MG3S/M06-2X level of theory.
The solvent was represented using the SMD model \cite{Marenich2009tbw} and charges were extracted from output files using the CM5PAC program \cite{Marenich2013tuj}.

\chapter{Additional Results} 

\label{AppendixB} 

\lhead{Appendix C. \emph{Additional Results}} 

\section{Partial molar volume and grand canonical ensemble}
\label{sec:pmv_gc}

We want to represent the following partial derivative $\left(\pdv*{N}{N'}\right)_{T,V,\mu}$ in terms of partial molar volume $\bar{V} = \left(\pdv*{V}{N'}\right)_{T,P,N}$.
For this we are going to rely on the following mathematical relationship
\begin{equation}
\left(\frac{\partial X}{\partial Y}\right)_Z = \left. - \left(\frac{\partial Z}{\partial Y}\right)_X \left(\frac{\partial X}{\partial Z}\right)_Y \right.\,.
\end{equation}
Applying it to the initial derivative we obtain
\begin{equation}
\left(\frac{\partial N}{\partial N'}\right)_{T,V,\mu} = - \left(\pdv{\mu}{N'}\right)_{T,V,N} \left(\pdv{N}{\mu}\right)_{T,V,N'}.
\label{eq:partial_n_partial_v}
\end{equation}
To proceed further we use the Gibbs-Duhem relationship $\mu = - s dT + v dP$, where $s$ is molar entropy and $v=1/\rho$ is molar volume.
Applying it to the first derivative in the above equation we get
\begin{equation}
\begin{split}
\left(\pdv{\mu}{N'}\right)_{T,V,N} & = -s \cancelto{0}{\left(\pdv{T}{N'}\right)_{T,V,N}} + v \left(\pdv{P}{N'}\right)_{T,V,N}\\
& = -v \left(\pdv{V}{N'}\right)_{T,P,N}\left(\pdv{P}{V}\right)_{T,N,N'}.
\end{split}
\end{equation}
Applying the same procedure for the second derivative in equation \ref{eq:partial_n_partial_v} we get
\begin{equation}
\begin{split}
 \left(\pdv{N}{\mu}\right)_{T,V,N'} & = \left(\pdv{\mu}{N}\right)^{-1}_{T,V,N'} = \left[-s \cancelto{0}{\left(\pdv{T}{N}\right)_{T,V,N'}} + v \left(\pdv{P}{N}\right)_{T,V,N'}\right]^{-1}\\
 & = \rho \left(\pdv{N}{P}\right)_{T,V,N'} = -\rho \left(\pdv{V}{P}\right)_{T,N,N'} \left(\pdv{N}{V}\right)_{T,P,N'} = -\rho^2 \left(\pdv{V}{P}\right)_{T,N,N'},
\end{split}
\end{equation}
since $\left(\pdv*{N}{V}\right)_{T,P,N'}=\rho$.
Finally, we plug in the above two results into original equation to obtain
\begin{equation}
\left(\frac{\partial N}{\partial N'}\right)_{T,V,\mu} = -v \left(\pdv{V}{N'}\right)_{T,P,N}\left(\pdv{P}{V}\right)_{T,N,N'} \rho^2 \left(\pdv{V}{P}\right)_{T,N,N'} = -\rho \bar{V},
\end{equation}
which is our final relationship.
%
%
%

\section{Extra figures and tables}

In this section, we included a few extra figures and tables that did not make it into the main thesis.
All of them are referenced and discussed in the main text.

\begin{figure}
	\centering
	\includegraphics[width=1\linewidth]{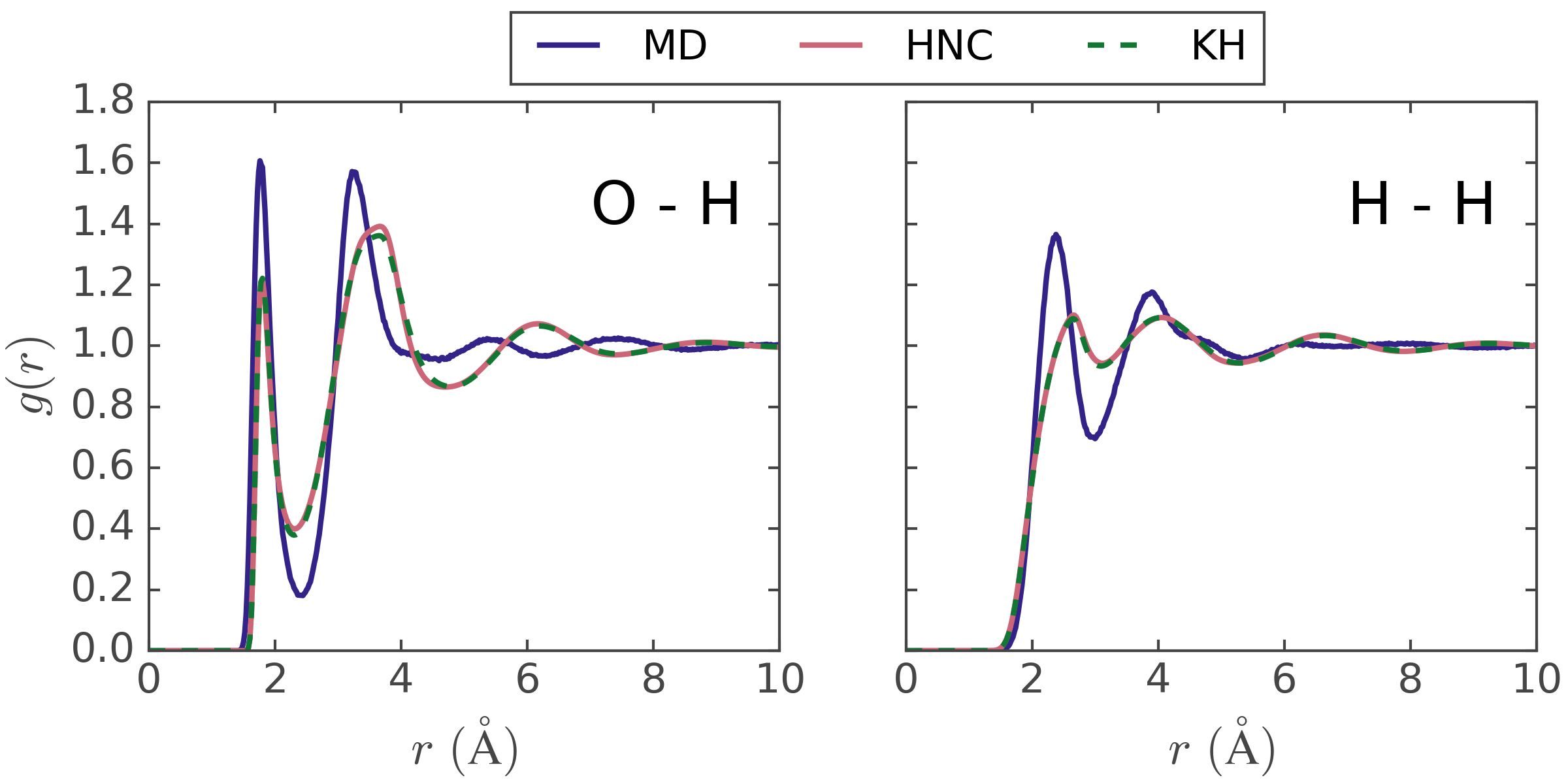}
	\caption{Site-site radial distribution functions of SPC/E water (cSPC/E in case of DRISM calculations).}
	\label{fig:rism1d_h}
\end{figure}

\begin{table}
	\centering
	\begin{tabular}{@{}cccccccccc@{}}
		\toprule
		\multirow{2}{*}{Model} & \multicolumn{3}{c}{O - O} & \multicolumn{3}{c}{O - H} & \multicolumn{3}{c}{H -H} \\ \cmidrule(l){2-4} \cmidrule(l){5-7} \cmidrule(l){8-10}
		& $r$    & $g(r)$    & CN   & $r$    & $g(r)$    & CN   & $r$    & $g(r)$   & CN   \\
		\midrule
		HNC &  3.000 &  2.767 &  9.371 &  1.800 &  1.217 &  0.734 &  2.650 &  1.101 &  6.400 \\
		PSE-3 &  3.000 &  2.738 &  9.374 &  1.800 &  1.218 &  0.734 &  2.650 &  1.101 &  6.398 \\
		KH &  2.975 &  2.338 &  9.988 &  1.800 &  1.221 &  0.728 &  2.650 &  1.089 &  6.155 \\
		MD &  2.760 &  3.098 &  4.327 &  1.760 &  1.607 &  0.951 &  2.380 &  1.365 &  5.518 \\
		\bottomrule
	\end{tabular}
	\caption{Location of the first peak in \si{\angstrom} $r$, the value of radial distribution function in the first peak $g(r)$, and coordination number CN for site-site radial distribution functions in SPC/E (cSPC/E in case of DRISM) water model.}
	\label{tab:1drism_spce_rdf}
\end{table}

\begin{sidewaysfigure}
	\centering
	\includegraphics[width=\linewidth,draft=false]{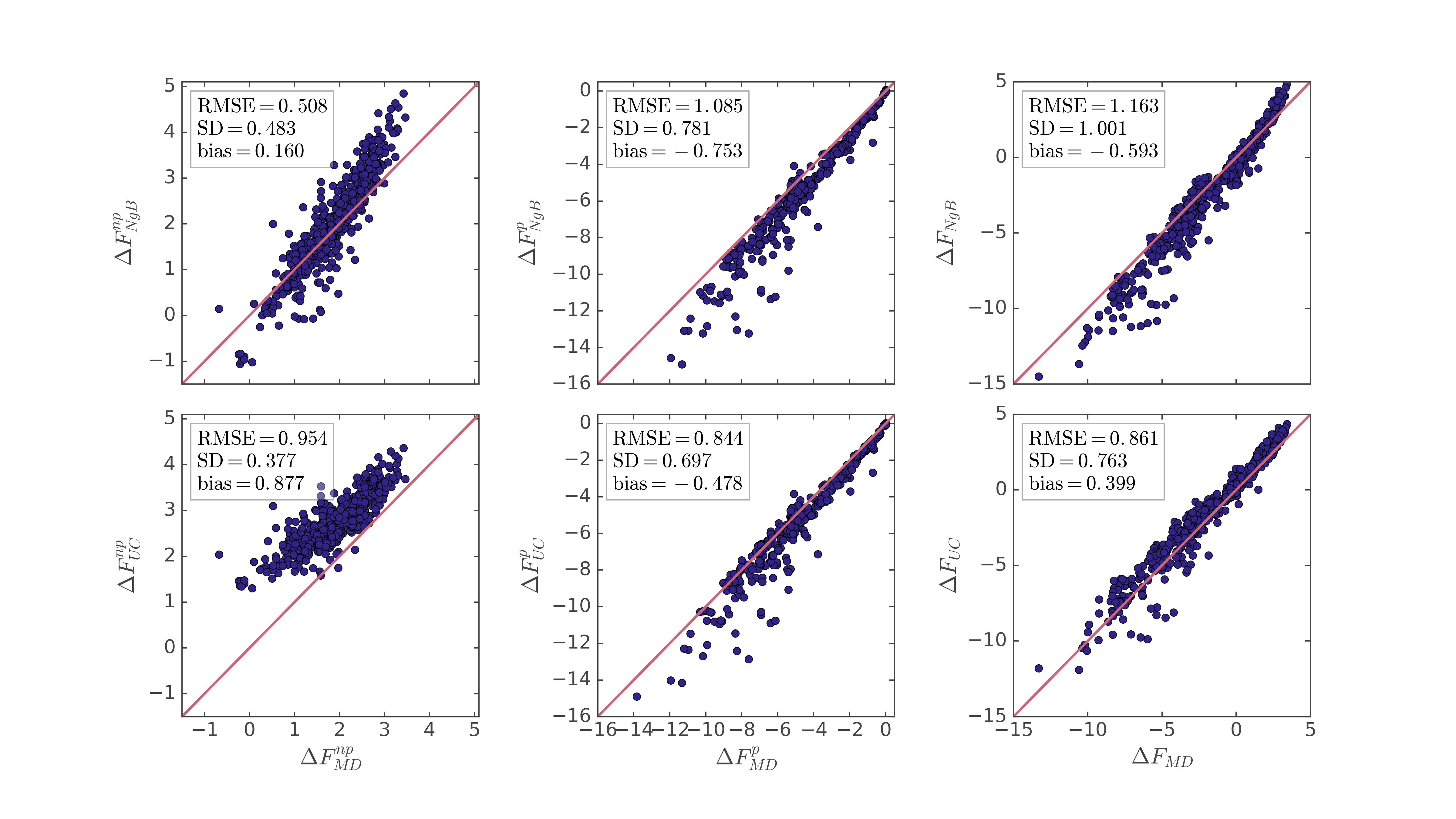}
	\caption{The figure mirrors figure \ref{fig:pcp_pc_fe}, but compares UC and NgB models, instead of PC and PC+.}
	\label{fig:uc_ngb_fe}
\end{sidewaysfigure}

\begin{figure}
	\centering
	\includegraphics[width=1\linewidth]{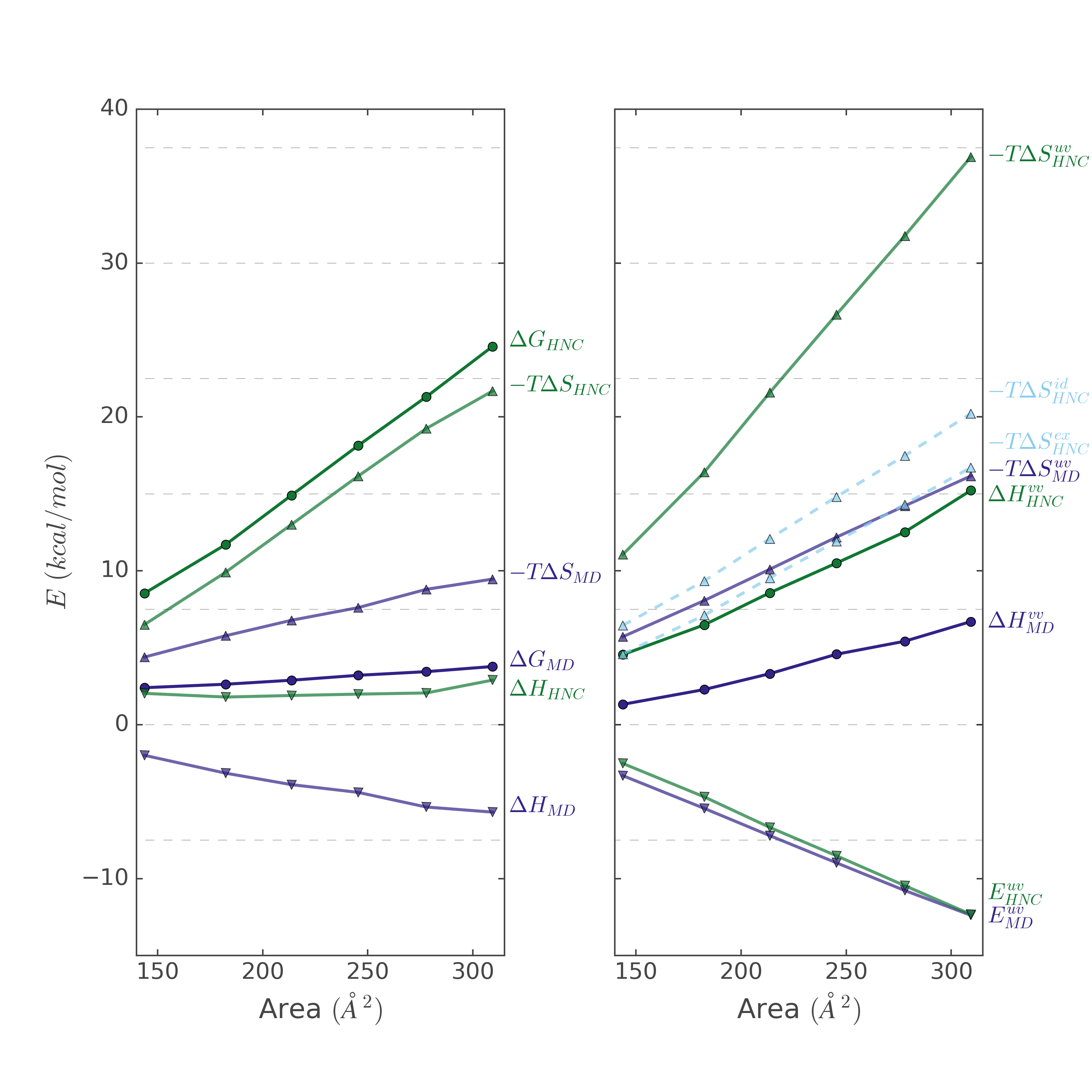}
	\caption{The figure mirrors figures \ref{fig:alkanes_therm1} and \ref{fig:alkanes_therm2}, but instead of pressure corrected models uses HNC results. The molecular dynamics results are taken from Ref. \citenum{Gallicchio2000tfa}.}
	\label{fig:alkanes_therm3}
\end{figure}

\begin{figure}
	\centering
	\includegraphics[width=1\linewidth]{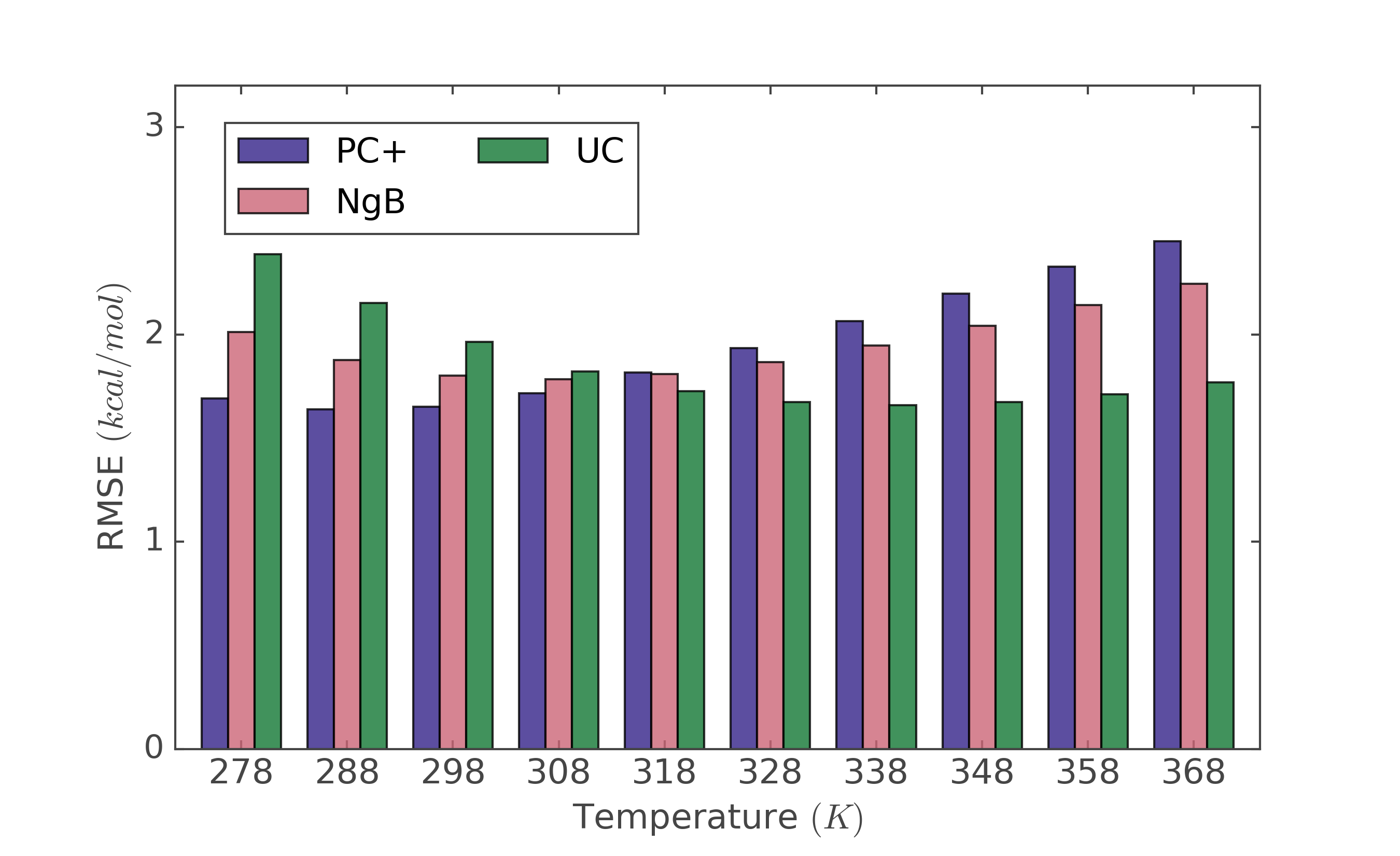}
	\caption{Root mean square error of solvation free energy of 3D-RISM models on the Chamberlin dataset. Unlike figure \ref{fig:chamberlin_T_fe}, here we do not vary temperature in 3D-RISM calculations, using the value of $\Delta G$ at \SI{298}{\kelvin} for all comparisons.}
	\label{fig:chamberlinconstt}
\end{figure}

\begin{table}
	\caption{Accuracies of different models for predicting Setschenow's constant using PC+' correction (defined using equation \ref{eq:pc_plus_prime}). The units are \si{l/mol}.}
	\centering
	\begin{tabular}{lrrrr}
		\toprule
		Model &   RMSE &    SDE &   bias &     $r^2$ \\
		\midrule
		\multicolumn{5}{c}{OPLS/CM5}\\
		da & 0.225 &  0.061 &  0.217 &  0.722 \\
		$\mathrm{ho_b}$ & 0.195 &  0.055 &  0.187 &  0.752 \\
		jc &   0.366 &  0.091 &  0.354 &  0.617 \\
		$\mathrm{ho_a}$ &0.136 &  0.050 &  0.126 &  0.796\\
		de &0.079 &  0.049 &  0.062 &  0.744 \\
		\bottomrule
	\end{tabular}
	\label{tab:results2} 
\end{table}

\begin{figure}
	\centering
	\includegraphics[width=1\linewidth]{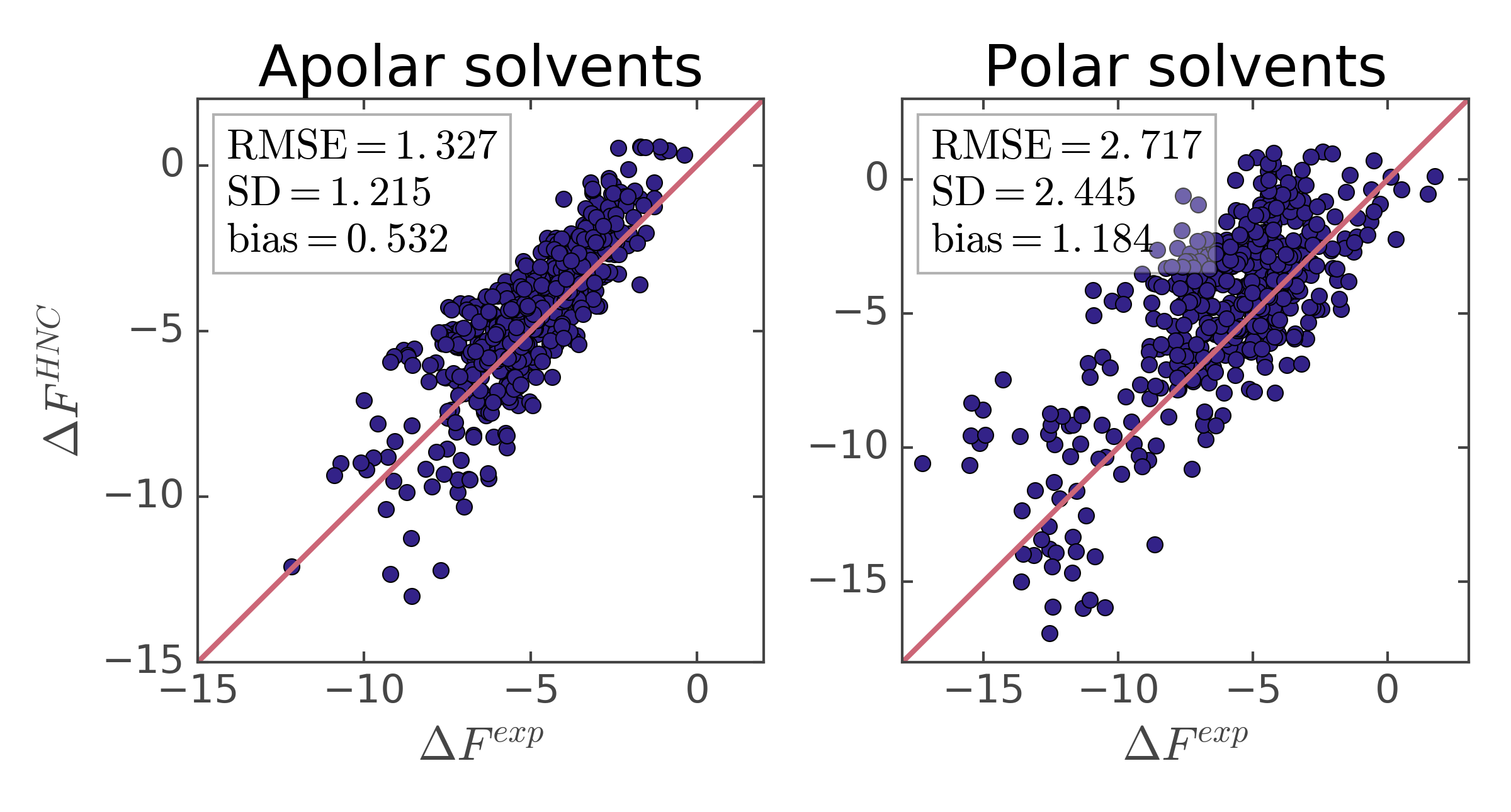}
	\caption{Comparison of computed (3D-RISM/HNC) and experimental solvation free energies in polar and apolar solvents (as defined in section \ref{sec:nonaq_exp}). All values are in \si{kcal\per\mole}.}
	\label{fig:polar_nonpolar_hnc}
\end{figure}

\clearpage
\addtotoc{List of original publications} 
\lhead{\emph{List of original publications}}
\btypeout{List of original publications}

\begin{spacing}{1}{
		\setlength{\parskip}{1pt}
		\chapter*{List of original publications}
}\end{spacing}

\makeatletter
\newlength{\bibhangMy}
\setlength{\bibhangMy}{1em} 
\newlength{\bibsepMy}
{\@listi \global\bibsepMy\itemsep \global\advance\bibsepMy by\parsep}
\newenvironment{bibsectionMy}%
{\begin{enumerate}{}{%
			\setlength{\leftmargin}{\bibhangMy}%
			\setlength{\itemindent}{-\leftmargin}%
			\setlength{\itemsep}{\bibsepMy}%
			\setlength{\parsep}{\z@}%
			\setlength{\partopsep}{0pt}%
			\setlength{\topsep}{0pt}}}
	{\end{enumerate}\vspace{-.6\baselineskip}}
\makeatother

\begin{bibsectionMy}
	
	\item {\bf Misin, M.}, Fedorov, M.V., Palmer, D.S. Accurate Hydration Free Energies at a Wide Range of Temperatures from 3D-RISM. \emph{J. Chem. Phys.}, 142(9):091105, March 2015. 
	
	\item Palmer, D.S., {\bf Misin, M.}, Fedorov, M.V., Llinas, A. Fast and General Method To Predict the Physicochemical Properties of Druglike Molecules Using the Integral Equation Theory of Molecular Liquids. \emph{Mol. Pharm.}, 12(9):3420--3432, August 2015. 
	
	\item {\bf Misin, M.}, Fedorov, M.V., Palmer, D.S. Hydration Free Energies of Ionic Species by Molecular Theory and Simulation. \emph{J. Phys. Chem. B}, 120(5):975--983, February 2016. 
	
	\item {\bf Misin, M.}, Palmer, D.S., Fedorov, M.V. Predicting Solvation Free Energies Using Parameter-Free Solvent Models. \emph{J. Phys. Chem. B}, 120(25):5724--57313, June 2016. 
	
	\item {\bf Misin, M.}, Petteri, V., Fedorov M.V., Palmer, D.S. Salting-out effects by pressure-corrected 3D-RISM. \emph{J. Chem. Phys.}, 145(19):194501, October 2016.
	
\end{bibsectionMy}

\clearpage 

\addtocontents{toc}{\vspace{2em}} 

\backmatter


\label{Bibliography}

\lhead{\emph{Bibliography}} 

\newpage




\end{document}